\documentclass[final,onecolumn]{elsart3p}

\usepackage{graphicx,rotating,color}
\usepackage{amssymb}
\usepackage{amsmath}
\usepackage{url}
\usepackage{wick}

\newcommand{\figw}{0.7}

\newcommand{\la}{\langle}
\newcommand{\ra}{\rangle}
\newcommand{\lla}{\left\langle}
\newcommand{\rra}{\right\rangle}

\newcommand{\rmi}{\mathrm{i}}
\newcommand{\ii}{\rmi}
\newcommand{\TT}{{\hat T}}
\newcommand{\OO}{{O}}
\newcommand{\dd}{\mathrm{d}}
\newcommand{\Ione}{\mathbf{1}}

\newcommand{\tre}{{\rm tr}_{\rm e}}
\newcommand{\trc}{{\rm tr}_{\rm c}}

\newcommand{\eps}{\varepsilon}

\def\>{\rangle}
\def\<{\langle}
\def\ket#1{|#1\>}
\def\bra#1{\<#1|}
\def\braket#1#2{\< #1 | #2 \>}
\def\bracket#1#2#3{\< #1 | #2 | #3 \>}
\def\ave#1{\< #1\>}
\def\aave#1{\<\!\< #1 \>\!\>}
\def\ve#1{{\vec{#1}}}
\def\mat#1{{\rm #1}}
\def\tr{\,{\rm tr}\,}
\def\one{\mathbbm{1}}
\newcommand{\op}[1]{\hat{#1}}
\def\beq{\begin{equation}}
\def\eeq{\end{equation}}
\def\beqa{\begin{eqnarray}}
\def\eeqa{\end{eqnarray}}

\usepackage{bbm,bm}

\def\veb#1{{\bm#1}}
\def\oC{\bar{C}}
\def\oV{\bar{V}}
\def\Z{{\mathbbm{Z}}}
\def\Vres{V_{\rm res}}

\def\Md{M_\lambda(t)}
\def\Fn{F(t)}
\def\Fr{F_{\rm R}(t)}
\def\Fp{F_{\rm P}(t)}
\def\tre#1{\tr_{\rm e}{\lbrack #1 \rbrack}}
\def\trc#1{\tr_{\rm c}{\lbrack #1 \rbrack}}

\begin{document}

\begin{frontmatter}



\title{Dynamics of Loschmidt echoes and fidelity decay}


\author[FR]{Thomas Gorin},
\ead{gorin@mpipks-dresden.mpg.de}
\author[LJ]{Toma\v z Prosen\corauthref{cor}},
\ead{tomaz.prosen@fmf.uni-lj.si}
\author[MX,MX2]{Thomas H. Seligman}, and
\ead{seligman@fis.unam.mx}
\author[LJ]{Marko \v Znidari\v c}
\ead{marko.znidaric@fmf.uni-lj.si}

\corauth[cor]{Corresponding author}

\address[FR]{Max-Planck-Institut f\" ur Physik komplexer Systeme, 
   N\" othnitzer Str. 38, D-01187 Dresden, Germany}
\address[LJ]{Physics Department, Faculty of Mathematics and Physics,
   University of Ljubljana, Jadranska 19, SI-1000 Ljubljana, Slovenia}
\address[MX]{Centro Internacional de Ciencias, Apartado postal 6-101,
   C.P.62132 Cuernavaca, Morelos, Mexico}
\address[MX2]{Centro de Ciencias F\'\i sicas, University of Mexico (UNAM),
   C.P.62132 Cuernavaca, Morelos, Mexico}

\begin{abstract}
Fidelity serves as a benchmark for the relieability in quantum information
processes, and has recently atracted much interest as a measure of the 
susceptibility of dynamics to perturbations. A rich variety of regimes for 
fidelity decay have emerged. The purpose of the present review is to describe 
these regimes, to give the theory that supports them, and to show some 
important applications and experiments. While we mention several approaches we
use time correlation functions as a backbone for the discussion.
Vanicek's uniform approach to semiclassics and random matrix theory provides
an important alternative or complementary aspects. Other methods will be 
mentioned as we go along. Recent experiments in micro-wave cavities and in 
elastodynamic systems as well as suggestions for experiments in quantum optics 
shall be discussed.
\end{abstract}


\end{frontmatter}

\tableofcontents

% main text

%\newpage
%\tableofcontents
%\newpage

\section{Introduction}
\label{I}

Irreversibility of macroscopic behavior has attracted attention ever since 
Boltzmann introduced his $H$-theorem in a seminal paper in 
1872~\cite{Boltzmann:72} (for English translation of some of Boltzmann's papers 
see book~\cite{Brush:66}), together with the equation bearing his name. 
%The $H$-theorem is a precursor of what is today known as the second law of 
%thermodynamics, stating that entropy can not decrease in the course of time. 
The problem of irreversibility is how to reconcile the apparent ``arrow of time'' with the underlying 
{\em reversible} microscopic laws. How come that macroscopic systems always 
seem to develop in one direction whether the underlying dynamics is symmetric 
in time? This so called reversibility paradox is usually attributed to Josef 
Loschmidt. He mentioned it briefly at the end of a paper published in 
1876~\cite{Loschmidt:76} discussing the thermal equilibrium of a gas subjected 
to a gravitational field, in an attempt to refute Maxwell's distribution of 
velocities for a gas at constant temperature. He also questioned Boltzmann's 
monotonic approach towards the equilibrium. Discussing that, if one would 
reverse all the velocities, one would go from equilibrium towards the initial 
non equilibrium state, he concludes with : {\em ``\ldots Das ber\" uhmte 
Problem, Geschehenes ungeschehen zu machen, hat damit zwar keine L\" osung 
\ldots''}. Bolzmann was quick to answer Loschmidt's objections in a paper from 
1877~\cite{Boltzmann:77}, pointing out the crucial importance of the initial 
conditions and of the probabilistic interpretation of the second law. The story 
goes that Boltzmann's reply to Loschmidt's question regarding velocity reversal 
was: "Then try to do it!". It is obvious that such velocity reversal is for all 
practical purposes impossible due to large number of particles involved and 
also due to high sensitivity to small errors made in the reversal.    

The first written account of the reversibility paradox is actually not due to 
Loschmidt but due to William Thompson (later Lord Kelvin), although it is 
possible that Loschmidt mentioned the paradox privately to Boltzmann before. 
Boltzmann and Loschmidt became good friends while working at the Institute of 
Physics in Vienna around 1867, for a detailed biography of Boltzmann see 
book~\cite{Cercignani:98}. In a vivid paper~\cite{Thompson:74} from 1874 
Thompson gave a very modern account of irreversibility. When discussing the 
heat conduction and the equalization of temperature he says: {\em ``\ldots If 
we allowed this equalization to proceed for a certain time, and then reversed 
the motions of all the molecules, we would observe a disequalization. However, 
if the number of molecules is very large, as it is in a gas, any slight 
deviation from absolute precision in the reversal will greatly shorten the time 
during which disequalization occurs... Furthermore, if we take account of the 
fact that no physical system can be completely isolated from its surroundings 
but is in principle interacting with all other molecules in the universe, and 
if we believe that the number of these latter molecules is infinite, then we 
may conclude that it is impossible for temperature-differences to arise 
spontaneously \ldots''}. The interesting question is then, how short is this 
{\em short disequalization time}? 

An analysis of the disequalization time is very near to the modern concept of 
echo dynamics. From discussions of Boltzmann, Loschmidt and Thompson it is 
clear that this disequalization time will depend on the imperfections made in 
the reversal, {\it i.e.} the change in the initial condition for the backward 
evolution, and on the {\em perturbations} to this backward evolution, {\it i.e.} on 
how much the backward evolution after the reversal differs from the forward 
evolution. The quantity measuring sensitivity to perturbations of the backward 
evolution is nowadays known as the Loschmidt echo after 
Ref.~\cite{Jalabert:01}, or the fidelity (we shall interchangingly use both 
terms, Loschmidt echo and fidelity). It is the overlap of the initial state 
with the state obtained after the forward unperturbed evolution followed by the 
backward perturbed evolution (Loschmidt echo) or equivalently as the overlap of 
state obtained after the forward unperturbed evolution and the state after the 
forward perturbed evolution (fidelity). It can be considered in classical 
mechanics as well as in quantum mechanics. If unperturbed and perturbed 
evolutions are the same the Loschmidt echo is obviously equal to one while for 
unequal evolutions it generally decreases with time. The decay time of the 
Loschmidt echo will then be the time of disequalization in question. 
Surprisingly, despite its importance for statistical mechanics and 
irreversibility it was not considered until some years ago, even though various 
``problems'' regarding irreversibility appear again and again. Echo dynamics 
and with it connected the Loschmidt echo is the subject of the present report. 

Echo dynamics has its applications beyond statistical mechanics. Performing 
echo evolution is a standard experimental technique in NMR where also the 
earliest measurements of the Loschmidt echo were 
performed~\cite{Hahn50,Rhi70,Zha92,LevUsa98}. It can also be related to other 
situations such as the seismic response in volcanos~\cite{GreSni05}. The 
quantum information community has adopted the autocorrelation function of echo 
dynamics, known as fidelity, as a standard benchmark for the quality of any 
implementation of a quantum information device~\cite{Nielsen:01}. 
Quantum information theory enables one to do things not possible by classical 
means, e.g., perform quantum computation. The main obstacle in producing quantum 
devices that manipulate individual quanta are errors in the evolution, either 
due to unwanted coupling with the environment or due to internal imperfections. 
Therefore, the goal is to build a device that is resistant to such 
perturbations. For this one ought to understand the behavior of fidelity in 
different situations to know how to maximize it. The theoretical framework for 
decreasing the unwanted evolution in quantum information devices, called
dynamical decoupling~\cite{Viola:99,Viola:05}, is intimately connected to 
fidelity theory, especially to the interesting phenomenon of quantum 
freeze~\cite{Prosen:03,Prosen:05,GKPSSZ05}. But the ultimate motivation for 
theoretical study of fidelity neither came from quantum information theory nor 
from statistical mechanics but from the field of quantum 
chaos~\cite{Haake:91,Stockmann:99}. The exponential instability of classical 
systems is a well known and much studied subject. As the underlying laws of 
nature are quantum mechanical the obvious question arises how this 
``chaoticity'' manifests itself in quantum systems whose classical limit is 
chaotic. The field of quantum chaos mainly dealt with stationary properties of 
classically chaotic systems, like spectral and eigenvector statistics. Despite 
classical chaos being defined in a dynamical way it was easier to pinpoint the 
``signatures'' of classical chaos in stationary properties. There were not that 
many studies of the dynamical aspects of quantum evolution in chaotic systems, 
some examples being studies of the reversibility of quantum
evolution~\cite{Shepelyansky:83,Casati:86}, the dynamical
localization~\cite{Fishman:82,Grempel:84}, energy 
spreading~\cite{BSW93,Cohen:00} or wave-packet evolution~\cite{Heller:91,Kottos:2K}. 
Classical instability is usually defined as an exponential separation of two 
nearby trajectories in time. In quantum mechanics one could be tempted to look 
at the sensitivity of quantum mechanics to variations of the initial wave 
function. But quantum evolution is unitary and therefore preserves the dot 
product ({\it i.e.} the distance) between two states and so there is no 
exponential sensitivity with respect to the variation of the initial state. The 
sensitivity to perturbations as measured by the Loschmidt echo on the other 
hand naturally lends to the comparison between quantum and classical situation. 
For classical systems Loschmidt echo gives the same exponential sensitivity to 
perturbations of the evolution as to perturbations of initial conditions. The 
quantum fidelity though can behave in a very different way, displaying a rich 
variety of regimes. Some other theoretical questions like quantization 
ambiguity, {\it i.e.} the difference between different quantizations having the 
same classical limit, can also be connected to fidelity 
decay~\cite{Kaplan:02}. Considering that fidelity lies at the crossroad of 
three very basic areas of physics: statistical mechanics, quantum information 
theory and quantum chaos, and that its understanding is crucial for building 
successful quantum devices it is not surprising that it received a lot of 
attention in recent years. Let us give a brief historical overview, listing the 
most important discoveries. The list of references is by no means complete, for 
detailed references see the later sections of the paper.

\subsection{Historical overview}

\label{sec:history}
Fidelity has been first used as a measure of stability by Peres in 
1984~\cite{Peres:84}, see also his book~\cite{Peres:95}. Peres reached the 
conclusion that the decay of fidelity is faster for chaotic than for regular 
classical dynamics. As we shall see the general situation can be exactly the 
opposite. Non decay of fidelity for regular dynamics in Peres's work was 
due to a very special choice of the initial condition, namely that a coherent 
wave packet was placed in the center of a stable island. Such a choice is 
special in two ways, first the center of an island is a stationary point and 
second the number of constituent eigenstates of the initial state is very 
small. After Peres's work the subject lay untouched for about a decade. In
1996 Ballentine and Zibin~\cite{Ballentine:96} numerically studied a quantity 
similar to fidelity. Instead of perturbing the backward evolution, they took 
the same backward evolution but instead perturbed the state after forward 
evolution by some instantaneous perturbation, like shifting the whole state by 
some $\delta x$. They also looked at the corresponding classical quantity. The 
conclusion they reached was that for chaotic dynamics quantum stability was 
much higher than the classical one, while for regular dynamics the two agreed. 
All these results were left mainly unexplained. Gardiner et 
al.~\cite{GCZ97,Gardiner:98erratum} proposed an experimental scheme for 
measuring the fidelity in an ion trap. In a related work, Schack and 
Caves~\cite{Schack:92,Schack:93,Schack:96} studied how much information about 
the environment is needed to prevent the entropy of the system to increase. 
Fidelity studies received fresh impetus from a series of NMR experiments
carried out by the group of Levstein and Pastawski.

In NMR echo experiments are a standard tool. The so called spin echo
experiment of Hahn~\cite{Hahn50} refocuses free induction decay in
liquids due to dephasing of the individual spins caused by slightly
different Larmor frequencies experienced due to magnetic field
inhomogeneities. By an appropriate electromagnetic pulse the Zeeman
term is reversed and thus the dynamics of {\em non-interacting} spins
is reversed. The first {\em interacting} many-body echo
experiment was done in solids by Rhim et al.~\cite{Rhi70}. Time reversal,
{ \it i.e.} changing the sign of the interaction, is achieved for a dipolar
interaction whose angular dependence can change sign for a certain
``magic'' angle, that causes the method to be called magic
echo. Still, the magic echo showed strong irreversibility. Much
later a sequence of pulses has been devised enabling a {\em local} 
detection of polarization~\cite{Zha92} ({\it i.e.} magnetic moment). They used a
molecular crystal, ferrocene ${\rm Fe}({\rm C}_5{\rm H}_5)_2$, in
which the naturally abundant isotope $^{13}{\rm C}$ is used as an
``injection'' point and a probe, while a ring of protons $^1{\rm H}$
constitutes a many-body spin system interacting by dipole forces. The
experiment proceeds in several steps: first the $^{13}{\rm C}$ is
magnetized, then this magnetization is transferred to the neighboring
$^1{\rm H}$. We thus have a single polarized spin, while others are in
``equilibrium''. The system of spins then evolves freely,  {\it i.e.} spin
diffusion takes place, until at time $t$ the dipolar interaction is
reversed and by this also spin diffusion. After time $2t$ the echo is
formed and we transfer the magnetization back to our probe $^{13}{\rm
C}$ enabling the detection of the {\em polarization echo}. Note that
in the polarization echo experiments the total polarization is
conserved as the dipole interaction has only ``flip-flop'' terms like
$S^j_+S^{j+1}_-$, which conserve the total spin. To detect the spin
diffusion one therefore needs a local probe. With the increase of the
reversal time $t$ the polarization echo -- the fidelity -- decreases in 
approximately exponential way. The nature of this decay has been further 
investigated by Pastawski et al.~\cite{Pasta95}. The group of Pastawski 
performed a series of NMR experiments where they studied in more detail the 
dependence of the polarization echo on various 
parameters~\cite{LevUsa98,Usaj:98,Pastawski:00}. They were able to control the 
strength of the residual part of the Hamiltonian, which was not reversed in the 
experiment and is assumed to be responsible for the polarization echo decay. By 
diminishing the residual interactions they obtained a Gaussian decay with a 
decay rate, saturated and independent of the perturbation strength, {\it i.e.} the 
strength of the residual interaction. That would imply the existence of a 
perturbation independent regime. While there is still no complete consensus on 
the interpretation of these experimental results they triggered a number of 
theoretical and even more numerical investigations. We shall briefly list just 
the most important ones, first regarding quantum fidelity.

Using the semiclassical expansion of the quantum propagator Jalabert and 
Pastawski~\cite{Jalabert:01} derived a {\em perturbation independent} quantum 
fidelity decay for localized initial states and chaotic dynamics, also called 
``Lyapunov decay'' due to its dependence on the Lyapunov exponent.
This regime has been numerically observed for the first time in Ref.~\cite{Cucchietti:02a}. 
Studying fidelity turned out to be particularly fruitful in terms of the 
{\em correlation function}~\cite{Prosen:02}, see also Ref.~\cite{Prosen:01}, 
which is also the approach we use in the present review. Perturbation dependent 
exponential decay of quantum fidelity observed in 
Ref.~\cite{Prosen:02,Jacquod:01,Cerruti:02,Prosen:02corr}, 
also called the Fermi golden rule decay, has been derived in 
Refs.~\cite{Jacquod:01} using random matrix theory and in Ref.\cite{Cerruti:02} using semiclassical methods, and independently in Refs.~\cite{Prosen:02,Prosen:02corr} using Born expansion in
terms of correlation functions. Perhaps the most interesting result of this approach is that one can, by 
{\em increasing} chaoticity of the corresponding classical system, 
{\em increase} quantum fidelity, {\it i.e.} improve the {\em stability} of the 
quantum dynamics. Note that classical correlators already appeared in
Ref.~\cite{Jalabert:01}, and also later in Ref.~\cite{Cerruti:02}.
For sufficiently small perturbations fidelity decays as a 
Gaussian, which is the so-called 
perturbative decay~\cite{Jacquod:01,Cerruti:02,Prosen:02corr}. 
Refs.~\cite{Jacquod:01,Cerruti:02} were the first which presented a unified theoretical treatment of
all main regimes of fidelity decay, and identified the scales of perturbation strength which
controlled the transitions among them.

Decoherence, as characterized by the decay of off-diagonal matrix elements of 
the reduced density matrix, has been connected to the fidelity in 
Ref.~\cite{Karkuszewski:02,GPSS04}. Echo purity, which is a generalization of 
purity to echo dynamics, has been introduced in Refs.~\cite{Prosen:02spin} and 
the so-called reduced fidelity, measuring the stability on a subsystem, in 
Ref.~\cite{Znidaric:03}. A rigorous inequality between fidelity, reduced 
fidelity and echo purity has been proved in 
Refs.~\cite{Znidaric:03,Prosen:03corr}. It is well known that the quantization
of classical system is not unique. Kaplan~\cite{Kaplan:02} compared the 
quantization ambiguity in chaotic and regular systems, reaching the conclusion 
that in chaotic systems the quantization ambiguity is {\em suppressed} as 
compared with regular ones. A detailed random matrix (RMT) formulation of fidelity has 
been presented in Ref.~\cite{GPS04}, although random matrix theory for describing fidelity
decay has been used before\cite{Jacquod:01,Cerruti:03,Cerruti:03a}.
A supersymmetric method has been used to 
derive an analytic expression for fidelity decay in RMT 
models~\cite{StoSch04,StoSch04b}. Quantum fidelity decay in regular systems 
has been discussed in Ref.~\cite{Prosen:02corr}. An interesting phenomenon of 
prolonged stability for certain type of perturbations, called quantum freeze, 
has been discovered in Refs.~\cite{Prosen:03,Prosen:05,GKPSSZ05}. Using a
semiclassical propagator fidelity has been expressed as a Wigner function 
average of phases due to action differences~\cite{Vanicek:04,Vanicek:05}, the 
so-called dephasing representation, which is particularly handy for numerical 
evaluation of quantum fidelity in terms of classical orbits.   

Classical fidelity as compared to the quantum one received much less attention. 
It has been first defined and linear response derived in 
Ref.~\cite{Prosen:02corr}. Numerical results on the classical fidelity and its 
correspondence with the quantum fidelity in chaotic systems and in systems 
exhibiting diffusion have been presented in Ref.~\cite{Benenti:02}. Classical 
fidelity in regular and chaotic systems has also been theoretically 
discussed~\cite{Eckhardt:03}. A detailed explanation of the asymptotic decay in 
chaotic systems has been given in Ref.~\cite{Benenti:03b}. Short time Lyapunov 
decay of classical fidelity has been  obtained in Ref.~\cite{Veble:04}, showing 
that the Lyapunov decay of quantum fidelity is a consequence of 
quantum-classical correspondence. Classically regular systems on the other hand 
have been worked out in Ref.~\cite{Benenti:03}.

\subsection{Outline of the paper}   
  
\label{sec:outline}
A wide variety of methods have been used to treat echo dynamics, and shall be 
discussed in this report, yet a linear response treatment, which relates 
fidelity decay to the decay of the correlation function of the perturbation in 
the interaction picture, is sufficient to understand most of the results. The 
method is based on writing the so-called echo propagator in the interaction 
picture, thereby focusing attention only on the perturbation that actually 
causes fidelity to decay. By employing the Baker-Campbell-Hausdorff formula or
a linear response expansion, one is then able to obtain approximations in a 
perfectly controllable way. In this review we shall therefore use this 
method as skeleton and discuss other methods as they are needed or have been 
used in the literature. All the theoretical work has been accompanied by 
extensive numerical experiments. Finally we shall see, that a random matrix 
model represents real and numerical experiments under chaotic conditions quite 
well.
    
Throughout this review we shall pay special attention to the role of the 
initial state (localized packets versus random states) and the type of dynamics 
(integrable versus chaotic dynamics). Furthermore, we shall discuss random 
matrix models.
   
We shall now describe the organization of this review. In the process we shall 
point out a hierarchical structure, that will allow the reader interested in 
some particular aspect, to read only the next section and a few subsections to 
obtain the information he needs.

Thus Section~\ref{D} contains the basic definitions of echo dynamics and 
fidelity, which will be indispensable throughout the paper. Also we shall see, 
that the interaction picture provides the ideal representation to understand 
echo dynamics, because it isolates the effect of the perturbation on the wave 
function from that of the unperturbed dynamics. To get a first understanding we 
shall in Section~\ref{DF} see how dynamical correlations of the perturbation 
control the basic behavior of fidelity decay, in a linear response analysis. 
This subsection is essential for the entire paper, because we shall use it as a 
skeleton for most considerations, and all new results will be based on its 
contents. The section will give definitions of other measures corresponding to 
different experimental and theoretical situations, such as scattering, 
polarization echo, or the evolution of entanglement as measured by purity under 
echo dynamics. The analysis of the precision of different quantization 
schemes is also connected to fidelity decay.
    
Section~\ref{Q2} deals with dynamical chaotic systems, and we shall explicitely 
see the effects of correlations of the perturbation on the fidelity decay. 
Special attention will be given to perturbations with zero diagonal, which 
result in the so-called quantum freeze. We shall discuss echo measures for 
composite systems.

In Section~\ref{R} random matrix theory of echo dynamics will be developed, and 
we shall discuss when we expect the results found by RMT to represent a chaotic 
system well. The quantum freeze will be analyzed in the context of RMT. New
results on quantum freeze and decoherence will be presented.

Section~\ref{Q1} will deal with integrable systems and has similar internal 
structure as the one on chaotic dynamics. Though, due to the lack of 
universality the behavior is richer. In this context the difference between 
coherent states and radom states will play a significant role. Some previously 
unpublished results are presented in this section. These includes 
Section~\ref{sec:ch4avgF} describing decay of fidelity averaged over position 
of initial packets and Section~\ref{sec:freezeHO} describing freeze in a 
harmonic oscillator.

The following Section~\ref{C} deals with the decay of classical fidelity. First 
linear response is derived already showing marked difference between classical 
and quantum fidelity. Then long time Lyapunov decay is discussed as well as the 
decay in regular systems. The section concludes with the discussion of 
classical fidelity decay in many-body systems.

Section~\ref{sec:timescales} on time scales is an overview of various 
decays derived for chaotic and regular systems. It can serve as a brief summary 
of results obtained in Sections~\ref{Q2} and~\ref{Q1}. The scaling of decay 
time scales on different parameters is given and the decay is compared between 
chaotic and regular dynamics. In certain situations quantum regular systems 
can be more sensitive to perturbations than chaotic ones and additionally, 
increasing the chaoticity can improve the stability. The decay of fidelity is 
also illustrated in terms of Wigner functions. Unpublished material consists of 
some figures illustrating and comparing decay time scales in chaotic and regular 
systems. 

Application of fidelity to quantum information is described in 
Section~\ref{sec:AppQI}. First we give an incomplete list of various studies of 
the influence of errors on quantum computation and point out how different 
results obtained in previous sections can be employed to understand results 
obtained in the literature. We give an illustrative example of how more 
randomness can improve stability of quantum computation. A connection between 
dynamical decoupling and fidelity freeze is also pointed out.
    
Finally in Section~\ref{E} we shall describe experiments measuring fidelity. 
We shall discuss NMR experiments, microwave experiments and elasticity 
experiments, for which fidelity measurements have already been performed as 
well as possible atom optics experiment.

Although our exposition is intended as a unified and self-contained review of 
the results which have mainly been published before in the quoted literature, 
there is however a substantial body of new results which have never been 
published. Let us here pin down the main original results presented in this 
report: (i) The results about quantum freeze of fidelity in the random matrix 
framework (Section~\ref{Q2QF}), (ii) The extension of the results on scattering 
fidelity to the weak coupling regime (Appendix C), (iii) Random matrix 
treatment of purity decay in composite systems, both in the framework of echo 
dynamics and of forward dynamics of a central system weakly coupled to an 
environment, (iv) Treatment of the quantum freeze of fidelity for quadratic 
Hamiltonians (Section~\ref{sec:freezeHO}).

\section{General theoretical framework}
\label{D}

\subsection{Fidelity}

\label{DF}

We consider a general quantum Hamilton operator $H_\eps(t)$ which may 
depend explicitly on time $t$, as well as on some external parameter 
$\eps$, like magnetic field, interaction strength, shape of potential well, etc.
Without essential loss of generality we shall assume that the parameter 
dependence is linear, namely
\begin{equation}
H_\eps(t) = H_0(t) + \eps V(t).
\end{equation} 
Let $U_\eps(t)$ be the propagator
\begin{equation}
U_\eps(t) = \TT\exp\left(-\frac{\ii}{\hbar}\int_0^t \dd t' H_\eps(t')\right)
\end{equation}
where $\TT$ designates the time ordering.
If $\ket{\Psi}$ is some arbitrary initial state, then the time
evolution, depending on the time variable $t$ as well on the perturbation 
parameter $\eps$ of the Hamiltonian, is given by
\begin{equation}
\ket{\Psi_\eps(t)} = U_\eps(t)\ket{\Psi}.
\end{equation}
Having the tools defined above, {\em the fidelity amplitude}, 
with respect to the
unperturbed evolution $U_0(\eps)$, is defined as the overlap of the
perturbed and unperturbed time-evolving states as
\begin{equation}
f_\eps(t) = \braket{\Psi_0(t)}{\Psi_\eps(t)} = \bra{\Psi}U_0(-t)U_\eps(t)\ket{\Psi}.
\label{eq:fiddef}
\end{equation}
The square of its modulus is {\em fidelity} 
\begin{equation}
F_\eps(t) = |f_\eps(t)|^2.
\label{eq:fiddef2}
\end{equation}

At this point we wish to stress the dual interpretation of fidelity, 
which is the central
characteristic of echo dynamics often called 'Loschmidt echo':
on one hand $F_\eps(t)$ is the probability that the states of 
unperturbed and perturbed
time evolution are the same, but on the other hand, due to unitarity of 
time evolutions,
$F_\eps(t)$ is also the probability that after an echo - composition of
forward perturbed and backward unperturbed dynamics - we arrive back to 
the initial state.

In the following we shall write the expectation value in a given initial 
state simply as $\ave{A} = \bra{\Psi}A\ket{\Psi}$. 
The fidelity amplitude may now be compactly written in terms of the
{\em expectation value} of the so-called {\em echo-operator}
\begin{equation}
M_\eps(t) = U_0(-t)U_\eps(t),
\label{eq:eodef}
\end{equation}
as
\begin{equation}
f_\eps(t) = \ave{M_\eps(t)}.
\label{eq:fidech}
\end{equation}
One may also 
use a non-pure initial state, namely a statistical mixture 
described by a density matrix $\rho$, 
with $\ave{A} = \tr \rho A = \sum_k p_k \bra{\phi_k}A\ket{\phi_k}$,
%{\bf THS haow about using M(t) directly instead of A}
where $p_k$ and $\ket{\phi_k}$ are, respectively, 
the eigenvalues and eigenvectors of the density matrix $\rho$.
One might be lead to an intuitive belief that incoherent superpositions of fidelity
amplitude, such as the above for non-pure initial states, 
would not have a clear physical significance.
Nevertheless, it turns out that this is not the case and that such a generalized object may
be experimentally observable \cite{Fazio:05} and exhibit certain interesting 
theoretical
features \cite{Sokolov:05}.

Writing the perturbation operator in the interaction picture
\begin{equation}
\tilde{V}(t) = U_0(-t) V(t) U_0(t)
\end{equation}
one can straightforwardly check by differentiating Eq.~(\ref{eq:eodef})
that the echo-operator solves the evolution equation
\begin{equation}
\frac{\dd}{\dd t} M_\eps(t) = -\frac{\ii}{\hbar}
\eps \tilde{V}(t) M_\eps(t)
\label{eq:VN}
\end{equation}
with the (effective) Hamiltonian $\eps \tilde{V}$.
The echo-operator is nothing but the propagator in the interaction
picture, which can be written in terms of a formal solution of
Eq.~(\ref{eq:VN})
\begin{equation}
M_\eps(t) = \TT\exp\left(-\frac{\ii}{\hbar}\eps\int_0^t
\dd t' \tilde{V}(t')\right).
\label{eq:Mformal}
\end{equation}
Again, a time-ordered product has to be used since $\tilde{V}(t)$ 
for different times does not in general
commute. The computation of fidelity becomes rather straightforward 
if one plugs the 
expression for the echo-operator (\ref{eq:Mformal}) into the definition
(\ref{eq:fidech}). 

In particular, this approach is ideally suited for
a perturbative treatment when the perturbation strength $\eps$ can
be considered as a small parameter. One can write explicitely the 
Born series for the echo operator, which has an infinite radius of convergence
provided that the perturbation $V(t)$ is an operator, uniformly bounded for
all $t$:
\begin{equation}
M_\eps(t) = \Ione + 
\sum_{m=1}^\infty \frac{(-\ii\eps)^m}{\hbar^m m!}
\int_0^t \dd t_1 \dd t_2 \cdots \dd t_m \TT \tilde{V}(t_1)\tilde{V}(t_2)\cdots \tilde{V}(t_m).
\label{eq:born}
\end{equation}
If we truncate the above series at the second order, $m=2$, and insert the 
expression into Eq.~(\ref{eq:fidech}), we obtain
\begin{equation}
f_\eps(t) = 1 - \frac{\ii\eps}{\hbar}
\int_0^t \dd t' \ave{\tilde{V}(t')}
- \frac{\eps^2}{\hbar^2} \int_0^t \dd t'\int_{t'}^t \dd t''
\ave{\tilde{V}(t')\tilde{V}(t'')} + \OO(\eps^3).
\end{equation}

Taking the square modulus we obtain the fidelity
\begin{equation}
F_\eps(t) = 1 - \frac{\eps^2}{\hbar^2}\int_0^t\d t'\int_0^t\d t'' C(t',t'') 
+ \OO(\eps^4)
\label{eq:fidlr}
\end{equation}
where
\begin{equation}
C(t',t'') = \ave{\tilde{V}(t')\tilde{V}(t'')} - \ave{\tilde{V}(t')}\ave{\tilde{V}(t'')},
\label{eq:CFdef}
\end{equation}
is the 2-point time-correlation function of the perturbation.
This high-fidelity approximation (\ref{eq:fidlr}) shall be called 
{\em linear response} expression for fidelity.
We stress that the validity of the linear response formula (\ref{eq:fidlr}) 
is by no means restricted to short times. The only condition is to have 
high-fidelity, namely $\eta:=1-F_\eps(t)$ has to be small. The error of the
linear approximation is typically of order $\OO(\eps^4)$.
However, using the same technique one can go beyond the linear response approximation for certain particular cases of dynamics, for example we shall derive Fermi golden rule decay for strongly mixing
dynamics (Subsect.~\ref{Q2D}) and various non-universal decays for integrable dynamics 
(e.g. Subsect.~\ref{sec:ch4Vneq0}).

Note that the linear-response formula (\ref{eq:fidlr}) already establishes a 
very important physical property of quantum echo-dynamics. One observes an 
intimate connection between fidelity decay and temporal-correlation
decay, namely faster decay of the correlation function $C(t',t'')$, as 
$|t'-t''|$ grows, implies slower decay of fidelity and vice versa. For example, 
for quantum systems in the semiclassical regime, whose classical limit is 
chaotic and fast mixing, such that $C(t',t'')$ decays faster than 
${\rm const}\, /|t'-t''|$, fidelity is expected to decay as a linear function 
of time $F_\eps(t) = 1 - {\rm const}\, \eps^2 t$, whereas for regular systems, 
with integrable classical limit, $C(t',t'')$ is expected to have oscillating 
behavior with generally {\em nonvanishing} time average, hence fidelity is 
expected to decay as a quadratic function of time 
$F_\eps(t) = 1 - {\rm const}\, \eps^2 t^2$. This implies the seemingly 
paradoxical conclusion that the fidelity of regular dynamics may decay faster than 
the fidelity of chaotic ones. Detailed discussion and comparision between chaotic and 
regular dynamics is given in Sect.~\ref{sec:timescales}.

The linear response formula (\ref{eq:fidlr}) can be cast into another 
intuitively useful form. Let us define an {\em integrated perturbation 
operator} $\Sigma(t)$
\begin{equation}
\Sigma(t) = \int_0^t \dd t' \tilde{V}(t').
\label{eq:defsig}
\end{equation}
Then, the RHS of Eq.~(\ref{eq:fidlr}) rewrites in terms of an {\em uncertainty} of operator 
$\Sigma(t)$:
\begin{equation}
F_\eps(t) = 1 - \frac{\eps^2}{\hbar^2}\left\{\ave{\Sigma^2(t)}-
\ave{\Sigma(t)}^2\right\}
+ \OO(\eps^4)
\label{eq:lrsig}
\end{equation}
Now, another interpretation of the quantum fidelity decay in the quantum 
chaotic - or stochastic - versus regular regime can be given. 
For quantum chaotic dynamics, one expects to have {\em diffusive}%
\footnote{Of course, for a finite quantum system, diffusive behavior can only 
  be observed on finite time scales, typically below the Heisenberg time. The 
  issue of time scales is discussed extensively in later sections, in 
  particular in Section~\ref{sec:timescales}.} 
behavior of typical observables, like $V$, hence 
$\ave{\Sigma^2}-\ave{\Sigma}^2 \propto t$, whereas for regular dynamics one 
expects {\em ballistic} behavior, $\ave{\Sigma^2}-\ave{\Sigma}^2\propto t^2$.

Another point has to be stressed: The fidelity amplitude is in general a 
c-number whose phase can be shifted by a perturbation which is a a multiple of 
an identity operator. So, adding a constant to the perturbation, e.g. forcing 
it to have a vanishing expectation value $V' = V - \ave{V}\Ione$, rotates the 
fidelity amplitude by a unimodular factor 
$f'_\eps(t) = \exp(\ii \eps t \ave{V}/\hbar)f_\eps(t)$ while it leaves the
fidelity $F_\eps(t)$ unchanged. This follows trivially from the representation 
of fidelity amplitude (\ref{eq:fidech}) in terms of the echo operator 
(\ref{eq:Mformal}).

\subsubsection{Quantum Zeno regime}

\label{sec:zeno}

Note, however, that for very short times, below a certain time scale 
$t_{\rm Z}$, namely before the correlation function starts to decay, 
$|t'|,|t''|< t_{\rm Z}$, $C(t',t'') \approx C(0,0) = \ave{V^2}$, the fidelity 
always exhibits (universal) quadratic decay 
\begin{equation}
F(t) = 1 - \frac{\eps^2}{\hbar^2} \ave{V^2} t^2, \quad {\rm for} \; |t| < 
t_{\rm Z} = \left(\frac{C(0,0)}{\dd^2 C(0,t)/\dd t^2}\right)^{1/2}
= \hbar \left(
\frac{\ave{V^2}}{\ave{[H_0,V]^2}}\right)^{1/2}
\end{equation}
This short-time regime (also discussed in \cite{Peres:84}) 
may be identified with the quantum Zeno effect 
\cite{zeno:77,zeno:80}, 
and the time scale $t_{\rm Z}$ referred to as the Zeno time.

\subsubsection{Temporally stochastic perturbations}

\label{sec:noisypert}

It is instructive to explain what happens in the case when the perturbation 
$V(t)$ is explicitly time dependent and stochastic, {\it i.e.} being a Gaussian 
delta correlated white noise, with variance
\begin{equation}
\overline{V(t')V(t'')} = v^2 \delta(t'-t'') \Ione.
\end{equation} 
$\overline{\bullet}$ here denotes an average over an ensemble of stochastic
perturbation operators.
The dynamical correlation $C(t',t'')$ is obviously insensitive to the nature of
unperturbed dynamics; hence the correlation function is again a delta function
$C(t',t'') = v^2 \delta(t'-t'')$. Furthermore, due to the Gaussian nature of 
the noise, higher-order correlation functions, say of order $2n$, in the 
expression for the
echo operator (\ref{eq:born}) factorize with the multiplicity $(2n-1)!!$ due to
the Wick theorem, yielding a simple exponential decay
\begin{equation}
F(t) = \exp\left(-\frac{\eps^2}{\hbar^2} v^2 t\right).
\end{equation}
For stochastic uncorrelated perturbations fidelity thus decays
exponentially with the rate which depends on the magnitude of perturbation
only and not on dynamics of the unperturbed system.

\subsubsection{Effect of conservation laws}

\label{sec:conslaws}

One should note that the correlation integral 
\begin{equation}
C_{\rm i}(t) =
\int_0^t \dd t' \int_0^t \dd t'' C(t',t'')=\ave{\Sigma^2(t)}-\ave{\Sigma(t)}^2
\label{eq:CI}
\end{equation}
always increases quadratically, even in the case of chaotic quantum dynamics,
if the perturbation has a nonvanishing component in the direction of some 
{\em (trivial) conserved quantities} such as the energy, the Hamilton operator $H_0$ (if
time-independent) or the angular momentum, {\it etc}.
For example, let $\{Q_n,n=1,2\ldots M\}$ be an orthonormalized set of
conserved quantities with respect to the initial state $\ket{\Psi}$, 
such that $\ave{Q_n Q_m} = \delta_{n m}$. 
\footnote{Any set of physical conservation laws $Q'_m$ can be orthogonalized using a
standard Gram-Schmidt procedure. In order to prevent degeneracy of the scalar product we only
have to assume {\em independence of conservation laws} in the sense 
that all the $M$ states $Q'_m\ket{\Psi}$ should be linearly independent.}
Then any time-independent perturbation can be decomposed uniquely as
\begin{equation}
V = \sum_{m=1}^M c_m Q_m + V'
\end{equation}
with coefficients $c_m = \ave{V Q_m}$ and $V'$ being the remaining non-trivial part 
of the perturbation, by construction orthogonal to {\em all} trivial
conservation laws,
\begin{equation}
\ave{Q_m V'} = 0,\quad {\rm for\; all\; }m.
\end{equation}
In such a case the correlation integral will always grow asymptotically 
as a quadratic function
\begin{equation} 
C_{\rm i}(t) \to \left(\sum_{m=1}^M c_m^2\right) t^2;
\end{equation}
hence fidelity will decay quadratically. Therefore it is desirable to subtract
this trivial effect by always choosing perturbations which are orthogonal
to all known trivial conservation laws.

For example, if $H_0$ is time-independent, then it is
a trivial invariantbe . The fact that a general perturbation $V$ may not be
orthogonal to $H_0$, $\ave{H_0 V}\neq 0$ is equivalent to the fact that $V$ will
change the density of states (in the first order in $\eps$) in the region
of eigenstates of $H_0$ populated by the initial state $\Psi$.
This means that the Heisenberg time will be different for forward (perturbed) 
and backward (unperturbed) dynamics, so fidelity will decay quadratically due
to this trivial effect. This effect is simply removed by replacing the perturbation
by $V' = V - \ave{H_0 V} H_0/\ave{H_0^2}$, or equivalently by measuring forward 
and backward time evolution in units of their respective Heisenberg times. We 
note that this effect has actually been discussed and taken care of in beautiful 
experimental studies of fidelity by Lobkis and Weaver~\cite{LobWea03} and 
St\" ockmann et al.~\cite{SGSS05,SSGS05} (see Section~\ref{ES}). 
Nevertheless it is important  to add that 
such "trivial" perturbations are non-trivial in the framework of quantum information 
applications, and indeed this shows, that the control of such perturbations is 
particularly important. 
%Another mechanism, where the maximal echo may appear for 
%unequal times of forward and backward evolutions has been considered by 
%Hiller et al. \cite{Hiller:04}. 

There is another consequence of conservation laws, namely they divide the 
Hilbert space of unperturbed evolution $U_0(t)$ into blocks specified by 
eigenvalues of the conserved operators, or quantum numbers. As a result the 
effective Heisenberg time - which will be one of the key time scales in later 
discussions - is reduced due to a reduced average density of states with 
{\em fixed} quantum numbers.

\subsubsection{Baker-Campbell-Hausdorff expansion of the echo operator}

\label{sec:BCH}

For the purposes of semi-classical analysis in subsequent sections it will be 
useful to perform another formal manipulation on the echo operator 
(\ref{eq:Mformal}), namely we may want to rewrite it approximately in terms of 
a single exponential. For this aim, we apply the well known 
Baker-Campbell-Hausdorff expansion 
$e^A e^B = \exp(A + B + (1/2)[A,B] + \ldots)$, to the infinite product 
(\ref{eq:Mformal}) yielding
\begin{eqnarray}
M_\eps(t) &=& \exp\left\{-\ii\frac{\eps}{\hbar}\int_0^t \dd t' \tilde{V}(t') + 
\frac{\eps^2}{2\hbar^2}\int_0^t\dd t'\int_{t'}^t \dd t'' [\tilde{V}(t'),\tilde{V}(t'')] 
+ \ldots \right\} \notag\\
&=& \exp\left\{-\frac{\ii}{\hbar}\left(
\Sigma(t)\eps + \frac{1}{2}\Gamma(t)\eps^2 + 
\ldots\right)\right\}
\label{eq:BCH}
\end{eqnarray}
where
\begin{equation}
\Gamma(t) = \frac{\ii}{\hbar}
\int_0^t \dd t' \int_{t'}^t \dd t'' [\tilde{V}(t'),\tilde{V}(t'')].
\label{eq:defGamma}
\end{equation}
Similar algebraic manipulations are well known in quantum field theory and are
sometimes known as the Magnus expansions, see for example 
Ref.~\cite{Magnus:54}. Two general remarks concerning the above asymptotic BCH 
expansion (\ref{eq:BCH}) are in order. First, the double integral of the 
commutator, which defines the operator $\Gamma(t)$, does in general grow only 
linearly in time, namely one can show that for arbitrary pair of states 
$\ket{\psi}$ and $\ket{\phi}$,
\begin{equation}
\left|\bra{\psi}\Gamma(t)\ket{\phi}\right| < C_{\psi\phi} t.
\label{eq:lineargrowth}
\end{equation}
where $C_{\psi\phi}$ is some constant, which may depend on the choice of the 
states $\ket{\psi}$ and $\ket{\phi}$.

Second, the matrix elements of the {\em third} and {\em fourth} order terms of 
the BCH expansion (\ref{eq:BCH}) can be estimated by ${\rm const}\, \eps^3 t$
and ${\rm const}\, \eps^4 t^2$, respectively. Therefore, Eq.~(\ref{eq:BCH}) 
provides in general a good approximation of the echo operator up to times 
$t < {\rm const}\, \eps^{-1}$, which can be made arbitrary long for 
sufficiently weak perturbations.

\subsection{Perturbation with vanishing time average}

\label{sec:freeze}

Let us define the {\em time average} of the perturbation operator
\begin{equation}
\bar{V} = \lim_{t\to\infty}\frac{\Sigma(t)}{t} = 
\lim_{t\to\infty} \frac{1}{t}\int_0^t \dd t' \tilde{V}(t').
\label{eq:timeaverage}
\end{equation}
Generally, an arbitrary perturbation $V$ can be decomposed into its time 
average $\oV$ and the rest, denoted by $\Vres$ and called the 
{\em residual part},
\begin{equation}
V=\oV+\Vres.
\label{eq:Vres_def}
\end{equation}
Let us, for the rest of this discussion, assume  for simplicity  that the unperturbed Hamiltonian $H_0$ and
the perturbation $V$ are time-independent operators. 
Let $E_k$ and $\ket{E_k}$ denote, respectively, eigenenergies and eigenvectors of the
unperturbed Hamiltonian $H_0$ and assume that the spectrum $\{E_k\}$ is
non-degenerate, which is typical for non-integrable and often true even for integrable
systems. In this case the time average is the {\em diagonal part} of 
the perturbation $\oV={\rm diag} V$, 
and the residual part $\Vres$ is the 
{\em off-diagonal part}.
The operator $\Sigma(t)$ can be asymptotically written as $\Sigma(t)=\oV t + \mathcal{O}(t^0)$, 
where the second term has off-diagonal matrix elements only, if expressed in the eigenbasis of
$H_0$. A further important condition is that $\Vres$ does not grow 
asymptotically with time. 
Therefore, the second order correlation integral
$C_{\rm i}(t)$ when expressed as (\ref{eq:CI}) does not grow with time either and remains bounded for 
all times if $\oV = 0$ or, more precisely, if and only if $\ave{\oV^2}-\ave{\oV}^2=0$.

Even though one may argue that the case of vanishing time-averaged perturbation 
$\oV=0$ may be rather special and non-generic, it deserves to be studied 
separately as it exhibits very unusual and perhaps surprisingly stable behavior 
of echo-dynamics. Furthermore, it can be realized in several important physical 
situations:
\begin{itemize}
\item When the perturbation $V$ can be written as time-derivative of another 
  observable or equivalently as a commutator with the Hamiltonian then 
  obviously, by construction the diagonal part $\oV$ vanishes.
\item When the unperturbed system is invariant under a certain unitary symmetry 
  operation $P$, say parity, $P H_0 = H_0 P$, whereas the symmetry changes sign 
  of the perturbation $P V = -V P$. This means that the perturbation breaks the 
  unitary symmetry in a maximal way and matrix elements of $V$ are nonvanishing 
  only between the states of opposite parity.
\item When the unperturbed system is invariant under a certain anti-unitary 
  symmetry operation $T$, say time-reversal, $T H_0 = H_0 T$, whereas the 
  symmetry changes sign of the perturbation $T V = -V T$. This means that the 
  perturbation breaks the antiunitary symmetry in an optimal way and the matrix 
  of $V$ in the eigenbasis of $H_0$ can be generally written as
  $V_{mn}={\ii }A_{mn}$ where $A$ is a {\em real antisymmetric} matrix.
\item Sometimes diagonal elements of the perturbation can be taken out 
  {\em by hand} and treated as a trivial first order perturbation of the 
  unperturbed part. In few-body and many-body physics this approach is 
  commonly known as the mean-field approximation. One may be generally 
  interested in the effect of residual perturbations and analyze it through 
  echo-dynamics.
\end{itemize}

In the present subsection we discuss some general properties of echodynamics 
for perturbations with $\oV=0$, also called {\em residual}
perturbations because $V=\Vres$. As we shall see, the short time fidelity 
will still be given by the operator $\Sigma(t)$. Because its norm does 
not grow with time, the fidelity will {\em freeze} at a constant value for the time
of validity of the linear response approximation. 
After a sufficiently long time it will again start to decay due to the 
operator $\Gamma(t)$.
We claim that this behavior is rather insensitive 
to the nature of unperturbed dynamics, for example whether it is regular or chaotic. 

Provided that the spectrum of $H_0$ is non-degenerate, 
any residual perturbation can be defined in terms of another operator $W$ 
by the following prescription~\footnote{Note that in the classical 
limit one can replace the commutator with a Poisson bracket, 
$(1/\ii \hbar)[A,B]\to\{A,B\}$.}
\begin{equation}
V=(d/dt)W(t)=\frac{\ii}{\hbar}[H_0,W],\qquad W(t)=U_0(-t) W U_0(t).
\label{eq:Wdef}
\end{equation}
Indeed, given a residual perturbation one easily determines the matrix 
elements of $W$ as
\begin{equation}
W_{jk}:=-\ii \hbar \frac{V_{jk}}{E_i-E_j},
\label{eq:Wmatrix}
\end{equation}
where $E_i$ are eigenenergies of $H_0$. For the corresponding definition in a 
discrete-time case, e.g. for kicked systems, see~\cite{Prosen:03,Prosen:05}.
From the definition of $W$ in Eq.~(\ref{eq:Wmatrix}) we can see that the matrix 
elements $W_{jk}$ are large for near degeneracies. For $W$ to be well 
behaved the matrix elements $V_{jk}$ of the perturbation must 
decrease smoothly approaching the diagonal. Setting diagonal elements of 
$V$ to zero will typically [e.g. in classically chaotic systems 
(Section~\ref{sec:chfreeze}) or in models of random matrices 
(Section~\ref{Q2QF})] produce singular behavior and reduce the effect of 
fidelity freeze. However, for regular unperturbed systems the effect of 
quantum freeze is even more robust due to abundance of selection rules, namely 
the matrix $V_{jk}$ is typically sparse so singularities due to near 
degeneracies may not occur even if we take out the diagonal part by hand.
With this newly defined operator $W$, the expression for the integrated 
perturbation $\Sigma(t)$ is extremely simple, 
\begin{equation}
\Sigma(t)=W(t)-W(0).
\label{eq:SW}
\end{equation}
Similarly, the expression for $\Gamma(t)$ (\ref{eq:defGamma}) is considerably 
simplified to
\begin{equation}
\Gamma(t)=\Sigma_R(t)-\frac{\ii}{\hbar}[W(0),W(t)],\qquad R=\frac{\ii}{\hbar}[W,(d/dt)W(t)],
\label{eq:GR}
\end{equation}
where
\begin{equation}
\Sigma_R(t)=\int_{0}^{t}{\!R(t')dt'},\qquad R(t)=U_0(-t) R U_0(t).
\label{eq:SR_def}
\end{equation}
Apart from the ``boundary term'' $\frac{\ii}{\hbar}[W,W(t)]$, the
operator $\Gamma(t)$ has, similarly to $\Sigma(t)$, the structure of a 
time integral of another operator $R$. From this representation it follows 
that any matrix element or the norm of operator $\Gamma(t)$ grows at most 
linearly with time. The aforementioned boundary term can be used to factorize 
the expression for the echo operator which will be the form most useful for 
applications in the following sections. Namely, to order $\OO(t \eps^3)$ it 
can be written as
\begin{equation}
M_\eps(t) = \exp\left(-\frac{\ii}{\hbar}W(t)\eps\right)
\exp\left(-\frac{\ii}{\hbar}\Sigma_R(t)\frac{\eps^2}{2}\right)
\exp\left(\frac{\ii}{\hbar}W(0)\eps\right).
\label{eq:BCHfreeze}
\end{equation}
From the expression (\ref{eq:BCHfreeze}) we see that,
since $||W(t)||=||W(0)||$, a time scale $t_2 = \OO(1/\eps)$ should exist
such that  for times $t<t_2$, the term in the second exponential 
$\Sigma_R(t)\eps^2$ can be neglected as compared to the one in 
the first and the
third, $W(t)\eps,W(0)\eps$.
Therefore, for $t<t_2$ we can write
\begin{equation}
F(t)=\left|\ave{\exp\left(-\ii \frac{\eps}{\hbar}W(t)\right)
\exp\left(\ii\frac{\eps}{\hbar}W(0)\right)}\right|^2.
\label{eq:Fnplateau}
\end{equation}
As we shall see later, the RHS of the above equation will typically not depend on time, beyond some
timescale $t_1$, much shorter than $t_2$, namely for $t_1 < t < t_2$, but we can claim quite generally that RHS of 
Eq.~(\ref{eq:Fnplateau}) has a strict lower bound.
Expanding the formula (\ref{eq:Fnplateau}) to the second order in $\eps$ one can 
show~\cite{Prosen:05} that the fidelity has a lower bound irrespective 
on the nature of dynamics, given by
\begin{equation}
F(t) \ge 1-4\frac{\eps^2}{\hbar^2}r^2,\qquad t < t_2,
\end{equation}
where
\begin{equation}
r^2 = \sup_t \left[\ave{W(t)^2} - \ave{W(t)}^2\right].
\end{equation}
For a residual perturbation the
fidelity therefore stays high up to a classically long time $t_2 \sim 1/\eps$ and only 
then starts to decay again. We call this phenomena fidelity freeze. 
The freeze happens for {\em arbitrary} quantum dynamics, irrespective of 
the existence and the nature of the classical limit.  After $t_2$ the term 
$\Gamma(t) \sim t \bar{R}$ in the second exponential of (\ref{eq:BCHfreeze}) 
will become important and eventually
this will cause the fidelity to decay. In the regime $t > t_2$ the expression of
the echo operator is formally similar to the original form (\ref{eq:Mformal}) where the
perturbation $\eps V$ has to be replaced by an effective perturbation operator $\frac{1}{2}\eps^2 R$.
This point shall be discussed in more detail later. We have to stress that the freeze of fidelity is of 
purely quantum origin. The classical fidelity discussed in Section~\ref{C}
does not exhibit a freeze.

\subsection{Average fidelity}

\subsubsection{Time averaged fidelity}
\label{sec:timeaveraged}

In a finite Hilbert space the fidelity will not decay to zero but fluctuate 
around some small plateau value, which is equal to the time averaged fidelity. 
For ergodic systems this time averaged value is in turn equal to the phase 
space averaged one. For a finite Hilbert space of size $N$, fidelity will 
start to fluctuate for long times due to the discreteness of the spectrum of 
the evolution operator. The size of this fluctuations can be calculated by 
evaluating time averaged fidelity $\bar{F}$
\begin{equation}
\bar{F}=\lim_{t\to \infty} \frac{1}{t} \int_0^t \dd t' F(t').
\label{eq:Fnavgdef}
\end{equation}
We expand the initial state in the eigenstates $\ket{E_n}$ of the unperturbed 
Hamiltonian $H_0$, write the eigenenergies and eigenstates of the perturbed 
Hamiltonian as $E^\eps_n$ and $\ket{E^\eps_n}$, and denote the matrix elements 
between unperturbed and perturbed eigenstates by 
\begin{equation}
O_{kl} = \braket{E_k}{E^\eps_l}.
\end{equation}
The transition matrix
$O$ is unitary, and if both eigenvectors can be chosen real it is orthogonal. This happens if $H_0$ and $H_\eps$ commute
with an antiunitary operator $T$ whose square is identity.
%{\bf THS -- I believe both footenotes are superflouous, but if you want to keep them 
%explain me why in 5 an additional rotation symmetry will du the trick --n I guess it could be 
%a reflection??} 
%\footnote{Antiunitary operator must satisfy
%$\braket{T\psi}{T\phi}=\braket{\psi}{\phi}^*$.} operator
%
%~\footnote{Square of an arbitrary
%antiunitary operator is $T^2=\pm \mathbbm{1}$. Time reversal operator for
%a system with spin is
%$T=\exp{(-\ii \pi S_{\rm y})}K$, with a complex conjugation
%operator $K$. If the system has an integer spin (or half-integer spin
%and an additional rotational invariance symmetry) we have $T^2=\mathbbm{1}$.}. 
The fidelity amplitude can now be written
\begin{equation}
f(t)=\sum_{lm}{(O^\dagger \rho)_{lm} O_{ml} \exp{(-\ii (E_l^\eps-E_m)t/\hbar)}},
\label{eq:fnP}
\end{equation}
with $\rho_{lm}=\bracket{E_l}{\rho(0)}{E_m}$ being the matrix elements of the 
initial density matrix in the 
unperturbed eigenbasis. To calculate the average fidelity $\bar{F}$ we have to take the 
absolute value square of $f(t)$. 
Averaging over time $t$ we shall assume that the phases are non-degenerate and 
find
\begin{equation}
\overline{\exp{(\ii(E_l^\eps-E_{l'}^\eps+E_m-E_{m'})t/\hbar)}}=
\delta_{m\,m'}\delta_{l\,l'}.
\label{eq:timeavg}
\end{equation}
This results in the average fidelity
\begin{equation}
\bar{F}=\sum_{ml}{|(\rho O)_{ml}|^2 |O_{ml}|^2}.
\label{eq:Fnavg}
\end{equation}
The time averaged fidelity therefore understandably depends on the initial state $\rho$ as well as on the ``overlap'' matrix $O$. 
\par
For {\em small perturbation} strengths, say $\eps$ smaller than some critical $\eps_{\rm rm}$, the unitary matrix $O$ 
will be close to identity. Using $O \to \mathbbm{1}$ for $\eps \ll \eps_{\rm rm}$ (\ref{eq:Fnavg}) gives us
\begin{equation}
\bar{F}_{\rm weak}=\sum_l{(\rho_{ll})^2} 
\label{eq:Fnavgweak}
\end{equation}
One should keep in mind that for the above result $\bar{F}_{\rm weak}$ we 
needed the eigenenergies to be non-degenerate,
$E_l^\eps \neq E_m$ (\ref{eq:timeavg}), and at the same time 
$O\to \mathbbm{1}$. This approximation is justified in lowest order in 
$\eps$, if the offdiagonal matrix elements are $|O_{ml}|^2 \propto \eps^2$. For pure initial states $\sum_l \rho_{ll}^2=\sum_l |\braket{E_l}{\Psi}|^4$ is 
just the inverse participation ratio of the initial state expressed in terms of the unperturbed eigenstates. Note that these expressions should be symmetric 
with respect to the interchange of the unperturbed and perturbed basis. On the other hand, for sufficiently {\em large} perturbation strength $\eps$ and 
complex perturbations $V$, such that the two bases -- the perturbed one and the unperturbed one --
become practically unrelated, one might assume 
$O$ to be close to a random matrix, unitary or orthogonal. In the limit 
$N\to\infty$, the matrix elements $O_{ml}$ can be treated as independent
complex or real Gaussian random numbers.
Then we can average the expression 
(\ref{eq:Fnavg}) over a Gaussian distribution $\propto \exp{(-\beta { N} |O_{ml}|^2/2)}$ of matrix elements $O_{ml}$, where 
we have $\beta=1$ for orthogonal $O$ and $\beta=2$ for unitary $O$. This averaging gives $\< |O_{ml}|^4 \>=(4-\beta)/{ N}^2$ 
and $\langle |O_{ml}|^2 \rangle=1/{ N}$ for the variance of $O_{ml}$ (brackets $\< \bullet \>$ denote here averaging over the 
distribution of matrix elements and not over the initial state). The average fidelity for strong perturbation can therefore be 
expressed as
\begin{equation}
\bar{F}_{\rm strong}=\frac{4-\beta}{N}\sum_l{\rho_{ll}^2}
+\frac{1}{N}\sum_{l,n}^{l\neq m}{|\rho_{lm}|^2}.
\label{eq:Fnavgstrong}
\end{equation}
More details on various cases of initial states as well as on crossover of the
average fidelity from the case of weak to the case of strong perturbations can be
found in Appendix~\ref{sect:appTA}.

\subsubsection{State averaged fidelity}  
\label{sec:stateavg}

Sometimes the average fidelity is of interest, {\it i.e.} the fidelity averaged over 
some ensemble of initial states. Such an average fidelity is also more amenable 
to theoretical treatment. Easier to calculate is the average fidelity 
amplitude $f(t)$ which is of second order in the initial state $\ket{\Psi}$ 
while fidelity $F(t)$ is of fourth order in $\ket{\Psi}$. We shall show that the 
difference between the absolute value squared of the average fidelity amplitude 
and the average fidelity is small for large Hilbert space dimensions $N$.

Let us look only at the simplest case of averaging over random initial
states, denoted by $\aave{\bullet}$. In the asymptotic limit of large
Hilbert space size ${N}\to \infty$ the averaging is simplified by the fact 
that the expansion coefficients $c_m$ of a random initial state in an arbitrary 
basis become independent Gaussian variables, with variance $1/{N}$. Quantities 
bilinear in the initial state, like the fidelity amplitude or the correlation 
function, result in the following expression
\begin{equation}
\aave{\Psi |A|\Psi}=:\aave{A}=\aave{\sum_{ml} c_m^* \, A_{ml}\, c_l}=\frac{1}{N}\tr{A},
\label{eq:Avgtrace}
\end{equation}
where $A$ is an arbitrary operator. The averaging is done simply by means of a trace over the whole Hilbert space. 
For the fidelity $F(t)$ which is of fourth order in $\ket{\Psi}$ we get after
some analysis

\begin{equation}
\aave{F(t)}=\sum_{mlpr}
\aave{ c_m^* [M_\eps(t)]_{ml} \, c_l\, c_p\, [M_\eps(t)]_{pr}^* \,c_r^* }= 
\frac{\left| \aave{f(t)} \right|^2 + 1/N}
{1 + 1/N}.
\label{eq:Avg4order}
\end{equation}
The difference between the average fidelity and the average fidelity amplitude is therefore of order $1/N$~\cite{Prosen:03ptps}. 
Note that the random state average (\ref{eq:Avg4order}) is in fact exact for
any $N$.
\par
There are two reasons why averaging over random initial states is of
interest. First, in the field of quantum information processing these
are the most interesting states as they have the least structure, {\it i.e.} 
can accommodate the largest amount of information. Second, for ergodic
dynamics and sufficiently long times, one can replace expectation
values in a specific generic state $\ket{\Psi}$ by an ergodic
average.
\par
Assuming ergodicity
for sufficiently large Hilbert spaces there should be no difference between averaging the fidelity 
amplitude or the fidelity or taking a typical single random initial state. For mixing dynamics the long time fidelity 
decay is independent of the initial state even if it is non-random, whereas in the regular regime it is state 
dependent. For instance, the long time Gaussian decay (Section~\ref{Q1}) depends on the position of the 
initial coherent state. The fidelity averaged over this position of the initial coherent state might be of interest 
and will not be equal to the fidelity averaged over random initial states. In the special case of very strong perturbation, $\eps > \hbar$, the average fidelity depends on the way we average even for mixing dynamics. This happens due to large fluctuations of fidelity for different initial coherent states, see Ref.~\cite{Silvestrov:03} and also~\cite{Wang:02}. 
Systematic study of fluctuation of fidelity with respect to an ensemble of initial coherent states has
been performed in Ref.~\cite{Petitjean:05} and showed that variance of fidelity is a non-monotonous function of time with a well defined maximum, where the standard
deviation of fidelity can dominate its average value.

%%%%%%%%%%%%%%%%
\subsection{Estimating fidelity}

For relatively short times or sufficiently weak perturbations, there exists an inequality giving a lower 
bound on fidelity in terms of the quantum uncertainty of the time-evolving perturbation operator. This is a time dependent version of Mandelstam-Tamm 
inequality~\cite{Mandelstam:45} (sometimes also called Fleming's 
bound~\cite{Fleming:73}) which is usually used to derive bounds on the 
time necessary for a given state to evolve into an orthogonal 
state~\cite{Uffink:93}. In the context of fidelity it was used by 
Peres~\cite{Peres:84}, see also Ref.~\cite{Karkuszewski:02}. In order to derive the inequality one starts with an observation 
that the time derivative of fidelity can be written as
\begin{equation}
\frac{\dd}{\dd t}F(t) = 
-\frac{\ii\eps}{\hbar}\bra{\Psi_0(t)}[P_\eps,V]\ket{\Psi_0(t)}
\end{equation}
where $P_\eps = \ket{\Psi_\eps(t)}\bra{\Psi_\eps(t)}$ is the projector 
onto the time-evolving state of the 
perturbed evolution. Using the Heisenberg uncertainty relation for the 
operators $P_\eps$ and $V$,
\begin{equation}
\delta V(t) \delta P_\eps(t) \le \frac{1}{2}\left|\bra{\Psi_0(t)}
[P_\eps,V]\ket{\Psi_0(t)}\right|
\end{equation}
with the time-evolving quantum uncertainty of operator $A$ defined as 
\begin{equation}
\delta A(t) = 
\left(\bra{\Psi_0(t)}A^2\ket{\Psi_0(t)}-\bra{\Psi_0(t)}A\ket
{\Psi_0(t)}^2\right)^{1/2},
\end{equation}
we can estimate the time-derivative of fidelity as
\begin{equation}
-\frac{\dd}{\dd t}F(t) \le \left| \frac{\dd}{\dd t}F(t)\right| 
\le \frac{2\eps}{\hbar}\delta V(t) \delta P_\eps(t) = 
\frac{2\eps}{\hbar}\delta V(t) F(t)(1-F(t)).
\end{equation}
Separating the variables and integrating we arrive at the final
inequality
\begin{equation}
F(t) \ge \cos^2(\phi(t)), \quad \phi(t) = \frac{\eps}{\hbar}\int_0^t 
\dd t' \delta V(t')
\label{eq:zurekineq}
\end{equation}
We should note, however, that the inequality (\ref{eq:zurekineq}) can 
be used only in the first 
 of $\cos^2$, namely 
for $|\phi|\le \pi/2$.

\subsection{Errors in approximate quantization schemes}

Let us sidetrack our discussion for a moment by mentioning a closely
related subject.

An early and very fundamental application of fidelity decay was given by 
L. Kaplan, who analyzed extensively the discrepancy between semi-classical 
quantization and the exact one \cite{Kap:98,Kaplan:02,Kap:04,Kaplan:05}. While most 
research concentrates directly on the behavior of the spectra, here the 
temporal deviation between the exact and the approximate solution are 
discussed. Considering the deviation at Heisenberg time will then yield 
information about the correctness of the level sequence and also shed light on 
the question whether localization can be explained in the framework of 
semi-classics.

Iterating Bogomolny's semi-classical solution for a quantum Poincar\' e 
map~\cite{Bog:92} provides a crude approximation. The fidelity taken 
between this approximation and the real quantum time evolution will show a 
significant decay of fidelity long before the Heisenberg time. On the other 
hand it is by no means practical to pursue semi-classical evolution up to 
Heisenberg time, because the exponential growth in the number of classical 
orbits, which we have to consider, precludes such a task. What Kaplan shows, 
is that semi-classics can be performed for times long compared to the return 
time $T$  of a simple Poincar\' e map, but short compared to Heisenberg time 
$T_H$  {\it i.e.} for a time $T_K$  such that $T\ll T_K \ll T_H$. Iterating 
this map we can then reach Heisenberg time with a good semi-classical 
approximation. With other words this iterated dynamics is an
arbitrarily good approximation to the true semiclassics, the difference falling
off exponentially as $\exp(- {\rm const}\, T_K / T)$. Therefore the fidelity 
decay between true quantum evolution and exact semiclassics is essentially the 
same as between quantum evolution and iterated semiclassics as long as 
$T_K \gg T$.

The upshot of these investigations is in agreement with our general findings. 
Fidelity decay of even the best semi-classical approximation will go as 
$1-{\rm const} t^2$ for integrable systems and as $1-{\rm const} t$ 
for chaotic ones. Thus 
semi-classical approximations are found to be  more stable for chaotic 
systems than for integrable ones. In particular the level sequence becomes 
marginally correct in two dimensions for integrable systems, while for chaotic 
ones three dimensions seem to be just about the limit. Note that this does not 
imply that semi-classics cannot be usefully applied for some purposes in 
higher  dimensional cases. Yet if we use a correct level sequence as the 
criterion, then this limitation exists for sure.

\subsection{Echo measures beyond the fidelity}

\label{embf}

The quantum state overlap - the fidelity amplitude - is only one of the 
measures of quality of the echo, or of the deviation between two quantum time 
evolutions. For example in real experiments, there is often different 
information at our disposal to make the comparison between two quantum states. 
The information about the system's state is based on the measurements of 
certain observables whose outcomes, in turn, only partially determine the state 
in question. It is thus relevant to define other measures of echo-dynamics 
based on the partial informations between two quantum states.

\subsubsection{\label{DOS} Scattering fidelity}
% 3.2.1
% \subsection{\label{DOS} Scattering fidelity}

Fidelity, as it is usually defined, also applies to scattering systems.
A wave packet can be evolved with two slightly different scattering 
Hamiltonians. This would be the standard fidelity of a scattering system.
In contrast, ``scattering fidelity'' stands for a quantity which can be 
obtained from simple scattering data, though under certain conditions it agrees
with the standard fidelity~\cite{SGSS05,SSGS05,GSW05}.

Typically scattering theory is developed around the scattering or 
S-matrix, and in this context it is only logical to inquire on the stability 
of S-matrix elements under perturbations. If we take into account that the 
S-matrix is due to some Hamiltonian that now describes an open system, we can 
again consider an unperturbed Hamiltonian and its perturbation, which define 
the S-matrices $S$ and $S'$. Usually the S-matrix
is given in the energy domain,  {\it i.e.} an S-matrix element is written as 
$S_{ab}(E)$, where $a$ and $b$ denote the scattering channels. By taking the 
Fourier transform 
\begin{equation}
 \hat S_{ab}(t)= \int\d E\; \e^{-2\pi\rmi\, E t}\; S_{ab}(E) \; ,
\label{DOS:hatS}\end{equation}
of any S-matrix element, we obtain the S-matrix $\hat S$ in the time domain. 

It now seems natural to consider the correlation function in the time domain
\begin{equation} 
\hat C[S_{ab}{}^* ,S'_{ab}] (t) \propto \hat S_{ab}(t)^*\; \hat S'_{ab}(t)
\end{equation}
as an appropriate measure of scattering fidelity. Yet this is not the case, 
because this quantity is dominated by the behavior of the autocorrelation 
functions. This can be amended by normalizing with the autocorrelation 
functions to obtain the scattering fidelity amplitude
\begin{equation}
f^s_{ab}(t)= \hat C[S_{ab}{}^* ,S'_{ab}](t) \; \big [\, 
 \hat C[S_{ab}{}^*, S_{ab}](t)\; \hat C[S'_{ab}{}^*, S'_{ab}](t) 
 \, \big ]^{-1/2} \; .
\end{equation}

While we formally started out in the energy domain, to conform with the usual 
language of scattering theory, it must be noted, that Lobkis and 
Weaver~\cite{LobWea03} reach essentially the same definition when analyzing 
scattering data taken in the time domain. The connection with fidelity 
was established in \cite{GSW05}.

We may ask how the scattering fidelity amplitude is related to the usual 
concept of fidelity. In principle, we could do this by considering the 
scattering process in terms of wave packet dynamics, where fidelity can be 
defined in the usual sense, {\it i.e.} as the overlap between two forward 
evolutions. Yet, the relation between the scattering matrix and the scattering 
of wave packets is somewhat involved and not very practical, even if it is 
standard textbook knowledge (see \cite{Taylorbook72}). 

We shall therefore ask a question which looks more practical: Under what 
circumstances scattering fidelity will be equivalent to fidelity or maybe 
state averaged fidelity in the bound system used to describe the interaction 
region? For such a concept to be well defined we need a fairly weak coupling 
to the asymptotic channels. We shall analyze this question below, but it is 
important to note, that the interest of scattering fidelity is by no means 
limited to such a situation, because a perturbation of ideal couplings to the 
continuum is very relevant, and then no simple analogy to standard fidelity 
exists.

Scattering theory usually distinguishes between short time signals usually 
referred to as direct reactions and long time signals which are termed very 
fittingly {\it coda} in elasticity. It is the latter, that tends to be 
universal and associated to some form of equilibration in the system. It is 
thus for this part, that we expect a close analogy with standard fidelity 
decay. This is fortunate, because the theory of fidelity decay that is 
available is also restricted to this part. This becomes clear, in the 
formulation of correlations of the perturbation in the interaction picture. 
The particular shape of this decay is usually not considered, and will give 
very specific properties that may be strongly system dependent. It will also 
depend on the initial state even if we consider chaotic or mixing systems. On 
the other hand we expect the signal of scattering fidelity to become 
independent of the channels for the coda.

To investigate the analogy of scattering fidelity with standard fidelity we 
shall use the simplest and most common model for scattering with a resonant 
part, which we present in the following subsection exclusively for the 
discussion of this analogy~\cite{SGSS05,SSGS05}.

\subsubsection*{The effective Hamiltonian approach}

In the effective Hamiltonian approach to scattering~\cite{MahWei69,Guhr:98}, 
the S-matrix reads
\begin{equation}
S_{ab}(E) = \delta_{ab} - {V^{(a)}}^\dagger\; \frac{1}{E- H_{\rm eff}}\;
   V^{(b)} \qquad H_{\rm eff} = H_0 - \rmi\Gamma/2 \qquad
   \Gamma= V\, V^\dagger + \gamma_W \; ,
\label{DOS:defS}\end{equation}
where $V^{(a)}$ is the column vector of $V$ corresponding to the scattering channel $a$. For later use, 
we introduce the absorption width $\gamma_W$, which is simply a scalar in the present context. This is 
equivalent to model absorption with infinitely many very weakly coupled channels, whose partial widths add 
up to $\gamma_W$~\cite{Sch03}. The Fourier transform of $S_{ab}(E)$ reads (for $t\ne 0$)
\begin{equation}
\hat S_{ab}(t)= \int\d E\; \e^{-2\pi\rmi\, E t}\; S_{ab}(E)
 = -2\pi\rmi\; \theta(t)\; {V^{(a)}}^\dagger\; \e^{-2\pi\rmi\, H_{\rm eff} t}
   \; V^{(b)} \; .
\end{equation}

Assume, $H_0$ and $H'_0= H_0 + \epsilon\, W$ are two slightly different Hamiltonians. 
According to Eq.~(\ref{DOS:defS}) these define two effective Hamiltonians $H_{\rm eff}$ and $H'_{\rm eff}$, 
as well as two scattering matrices $S_{ab}(E)$ and $S'_{ab}(E)$. We consider the cross-correlation function
\begin{equation}
\hat S_{ab}(t)^*\, \hat S'_{ab}(t) = 4\pi^2\; \theta(t)\; 
   {V^{(b)}}^\dagger\; \e^{2\pi\rmi\, H_{\rm eff}^\dagger t} \; V^{(a)} \; 
   {V^{(a)}}^\dagger\; \e^{-2\pi\rmi\, H'_{\rm eff} t} \; V^{(b)} \; .
\end{equation}
The singular value decomposition of $V$ provides an orthogonal (unitary) transformation of the $S(E)$ and 
$S'(E)$ where the channel vectors are orthogonal to each other (Engelbrecht-Weidenm\" uller 
transformation~\cite{EngWei73}). Thus, we may restrict ourselves to the case of orthogonal channel 
vectors; this implies that all direct or 
fast processes will happen in the elastic channels. We then have 
\begin{equation}
\hat S_{ab}(t)^*\, \hat S'_{ab}(t) = 4\pi^2\; w_a\, w_b\; \theta(t)\;
   \la v_b|\, \e^{2\pi\rmi\, H_{\rm eff}^\dagger t} \, |v_a\ra\;
   \la v_a|\, \e^{-2\pi\rmi\, H'_{\rm eff} t} \, |v_b\ra \; .
\label{DOS:CCfun}\end{equation}
$V^{(a)}= \sqrt{w_a}\; |v_a\ra$ with normalized vectors $|v_a\ra$. The general idea of what follows is 
to assume the state vectors $|v_a\ra$ and $|v_b\ra$ to be random and independent. This equation looks 
very suggestive to obtain the desired relation with the standard fidelity for a random state  $ |v_b \rangle$
\begin{equation}
f(t)= \la v_b|\; \e^{2\pi\rmi\, H_0\, t}\; 
   \e^{-2\pi\rmi (H_0+\epsilon W)\, t}\; |v_b\ra \; .
\end{equation}
In the eigenbasis of $H_0$ we have
\begin{equation}
f(t)= \sum_j \la |v_{jb}|^2\ra\; \e^{2\pi\rmi\, E_j\, t}\; U_{jj}(t)
 = \frac{1}{N}\sum_j \e^{2\pi\rmi\, E_j\, t}\; U_{jj}(t) 
 = \frac{1}{N}\sum_{jk} O_{jk}^2\; \e^{2\pi\rmi\, (E_j-E'_k)\, t}\; ,
\end{equation}
where 
\begin{equation}
U_{jj}(t) = \sum_k \la j|o_k\ra \e^{-2\pi\rmi\, E'_k\, t}\; \la o_k|j\ra
 = \sum_k O_{jk}^2\; \e^{-2\pi\rmi\, E'_k\, t}\; .
\end{equation}

Comparing with Eq.~(\ref{DOS:CCfun}) we see that the only remaining problem resides in the fact, 
that we have the effective Hamiltonian in the exponent, which includes the coupling term. This can 
be handled for weak and intermediate coupling. While the former will be presented in what follows,  
more involved cases of intermediate coupling are presented in appendix C, still leading to the same result. 

We assume that all coupling parameters $w_a$ go to zero.
Then the anti-Hermitian part of $H_{\rm eff}$ becomes a scalar and we are
essentially left with transition amplitudes of Hermitian Hamiltonians.
\begin{align}
\hat S_{ab}(t)^*\, \hat S'_{ab}(t) &\sim 4\pi^2\; w_a\, w_b\; \theta(t)\;
   \e^{-\gamma_W\, t}\;\la v_b|\, \e^{2\pi\rmi\, H_0\, t} \, |v_a\ra\;
   \la v_a|\, \e^{-2\pi\rmi\, (H_0+\epsilon W)\, t} \, |v_b\ra \notag\\
&= 4\pi^2\; w_a\, w_b\; \theta(t)\; \e^{-\gamma_W\, t}\; 
   \sum_{j,klm} v_{jb}^*\; 
   \e^{2\pi\rmi\, E_j\, t}\; v_{ja}\; v_{ka}^*\; \la k|o_l\ra\;
   \e^{-2\pi\rmi\, E'_l\, t}\; \la o_l|m\ra\; v_{mb} \; .
\end{align}
For $a\ne b$, we find by averaging over the initial states:
\begin{align}
\la\hat S_{ab}(t)^*\, \hat S'_{ab}(t)\ra &\sim 4\pi^2\; w_a\, w_b\; \theta(t)\;
   \e^{-\gamma_W\, t}\; \sum_{j,l} \la |v_{jb}|^2\ra\; \la |v_{ja}|^2\ra\; 
   \e^{2\pi\rmi\, E_j\, t}\;
   \la j|o_l\ra\; \e^{-2\pi\rmi\, E'_l\, t}\; \la o_l|j\ra \notag\\
&= 4\pi^2\; \frac{w_a\, w_b}{N}\; \theta(t)\; \e^{-\gamma_W\, t}\;
   \frac{1}{N}\sum_{jl}
   O_{jl}^2\; \e^{2\pi\rmi (E_j-E'_l) t} 
 = 4\pi^2\; \frac{w_a\, w_b}{N}\; \theta(t)\; \e^{-\gamma_W\, t}\;
   f(t)\; .
\label{DOS:sfres}\end{align}
The prefactor, which decays exponentially in time, is exactly the normalization 
by the two autocorrelation functions, which we postulated at the beginning, as 
the two agree in this approximation up to the factor $w_a$ or $w_b$, 
respectively.

An average over initial states will often be impractical. However, for chaotic
(ergodic) systems it may not be necessary. We may use a spectral average 
and/or average over different samples, instead. Indeed the results obtained for 
stronger coupling in the appendix assume chaotic behavior of the long time 
dynamics in the interaction region more explicitly, and probably the relation 
to standard fidelity is of practical importance only in such a situation. 
Yet it must be emphasized, that scattering fidelity is a relevant concept in 
its own right, if we take into account the fundamental role played by the 
S-matrix in many approaches to quantum systems.

\subsubsection{Polarization echo}

In NMR experiments the quantity that can naturally be measured  
is the echo in the polarization of the local
nuclear spin. One prepares the initial state $\ket{\Psi_0}$ of a many-spin system in an eigenstate of a 
projection of the spin (1/2) of a specific nucleus, say $s^{\rm z}_0$
\begin{equation}
s^{\rm z}_0 \ket{\Psi_0} = m_0 \ket{\Psi_0}, 
\end{equation}
where $m_0 = \pm 1/2$. 
Next one performs the echo experiment, {\it i.e.} we apply the echo operator 
and then measures local polarization of the same nucleus. 
It has the expectation value 
\begin{equation}
m_\eps(t) = \ave{M^\dagger_\eps(t) s^{\rm z}_0 M_\eps(t)}.
\end{equation}
The {\em polarization echo} $P_\eps(t)$ is defined as the probability that 
the local polarization of 
the spin is restored after the echo dynamics
\begin{equation}
P_\eps(t) = \frac{1}{2} + 2 m_0 m_\eps(t) = \frac{1}{2} + 
2 \ave{s^{\rm z}_0 M^\dagger_\eps(t) s^{\rm z}_0 M_\eps(t)}
\label{eq:defPE}
\end{equation}
This quantity has been extensively measured in a series of NMR experiments 
performed by the group 
of Levstein and Pastawski~\cite{Pasta95,Pastawski:00}

One can try to generalize this quantity to an echo with respect to some 
arbitrary quantum observable $A$.
Again, in order that the quantity makes sense as the echo of a 
observable $A$, the system has to be prepared
in the initial state which is an eigenstate $\ket{\Psi}$, or eventually 
a statistical mixture $\rho$ of eigenstates, of observable $A$. 
Due to this properties, the averages have the
property $\ave{AB}=\ave{BA}$ for any other observable $B$.
Then we define the echo of an observable $A$ as its correlation 
function with respect to echo-dynamics
\begin{equation}
P^A_\eps(t) = \frac{\ave{A M_\eps^\dagger(t) A M_\eps(t)}}
{\ave{A^2}}.
\end{equation}
One can write a general linear response expression for this quantity. 
Inserting into the previous expression 
the echo operator to second order
\begin{equation}
M_\eps(t) = \Ione - \ii \frac{\eps}{\hbar}\Sigma(t) - \frac{\eps^2}{2\hbar^2}\hat{T}\Sigma^2(t)
+ \OO(\eps^3)
\label{eq:echoop2}
\end{equation}
and retaining terms to second order, we obtain a simple expression to order$\OO(\eps^4)$
\begin{equation}
P^A_\eps(t) = 1 - \frac{\eps^2}{\hbar^2}\frac{\ave{A^2\Sigma^2(t)} 
- \ave{A\Sigma(t)A\Sigma(t)}}{\ave{A^2}}.
\end{equation} 
This can for example be used for the polarization echo.

\subsubsection{Composite systems, entanglement and purity}

In various studies of decoherence and effects of external degrees of freedom to the system's dynamics one studies
composite systems. Coupling with the environment is usually unavoidable so
 that the evolution of our system is no longer Hamiltonian. To preserve the Hamiltonian formulation we have to include the 
environment in our description. We therefore have a ``central system'', denoted by a subscript ``${\rm c}$'', and an environment, 
denoted by subscript ``${\rm e}$''. The names {\em central system} and {\em environment} will be used just to denote two pieces 
of a composite system, without any connotation on their properties, dimensionality etc. The central system will be that part 
which is of actual interest. The Hilbert space is a tensor product 
${\mathcal H}={\mathcal H}_{\rm c} \otimes {\mathcal H}_{\rm e}$ and the evolution of a whole system is determined by a Hamiltonian or a unitary
propagator on the whole Hilbert space ${\mathcal H}$ of dimension ${\mathcal N}={\mathcal N}_{\rm c} {\mathcal N}_{\rm e}$. The unperturbed 
state $\ket{\psi(t)}$ and the perturbed one $\ket{\psi_\eps(t)}$ are obtained with propagators $U(t)$ and $U_\eps(t)$. 
Fidelity would in this case be the overlap of two wave functions on the {\em whole space} ${\mathcal H}$. But if we are 
not interested in the environment, this is clearly not the relevant quantity. Namely, the fidelity will be low even if the 
two wave functions are the same on the subspace of the central system and differ only on the environment.      

We can define a quantity analogous to the fidelity, but which will measure the overlap 
just on the subspace of interest {\it i.e.} 
on the subspace of the central system. Let us define the reduced density matrix of the central subsystem
\begin{equation}
\rho_{\rm c}(t):=\tre{\rho(t)},\qquad \rho_{\rm c}^{\rm M}(t):=\tre{\rho^{\rm M}(t)},
\label{eq:reduced_rho}
\end{equation}
where $\tre{\bullet}$ denotes a trace over the environment and $\rho^{\rm M}(t)=\Md \rho(0)\Md^\dagger$ is the so-called echo 
density matrix. Throughout this chapter we shall assume that the initial state is a pure 
product state {\it i.e.} a direct product,
\begin{equation}
\ket{\psi(0)}=\ket{\psi_{\rm c}(0)}\otimes\ket{\psi_{\rm e}(0)}=:\ket{\psi_{\rm c}(0);\psi_{\rm e}(0)},
\label{eq:psiprod}
\end{equation}
where we also introduced a short notation $\ket{\psi_{\rm c};\psi_{\rm e}}$ for pure product states. The resulting 
initial density matrix $\rho(0)=\ket{\psi(0)}\bra{\psi(0)}$ is of course also pure. Fidelity can be written as 
$F(t)=\tr{\lbrack \rho(0)\rho^{\rm M}(t)\rbrack}$ and in a similar fashion we shall define a 
{\em reduced fidelity}~\cite{Znidaric:03} denoted by $\Fr$,
\begin{equation}
\Fr:=\trc{\rho_{\rm c}(0) \rho_{\rm c}^{\rm M}(t)}.
\label{eq:Frdef}
\end{equation}
The reduced fidelity measures the distance between the initial reduced
density matrix and the reduced density matrix after the echo. Note
that our definition of the reduced fidelity agrees with the
information-theoretic fidelity~\cite{Uhlmann:76,Jozsa:94,Nielsen:01} on a central
subspace ${\mathcal H}_{\rm c}$ only if the initial state is a pure product state, so that $\rho_{\rm c}(0)$ is also a pure state.  
\par
One of the most distinctive features of quantum mechanics is entanglement. 
Due to the
coupling between the central system and the environment the initial
product state will evolve after an echo into the pure entangled state $\Md \ket{\psi(0)}$ and therefore the 
reduced density matrix $\rho^{\rm M}_{\rm c}(t)$ will be a mixed one. For a pure state $\ket{\psi(t)}$ there is a simple 
criterion for entanglement. It is quantified by {\em purity} $I(t)$, defined as
\begin{equation}
I(t):=\trc{\rho_{\rm c}^2(t)},\qquad \rho_{\rm c}(t):=\tre{\ket{\psi(t)}\bra{\psi(t)}}.
\label{eq:I}
\end{equation}
Purity, or equivalently von Neumann 
entropy $\tr{(\rho_{\rm c}\ln{\rho_{\rm c}})}$, is a standard quantity used in 
decoherence studies~\cite{Zurek:91}. If the purity is less than one, $I<1$, 
then the state $\ket{\psi}$ is entangled (between the environment and the 
central system), otherwise it is a product state. Similarly, one can define a 
purity after an echo, called {\em purity fidelity} in~\cite{Prosen:02spin} by
\begin{equation}
\Fp:=\trc{\{ \rho^{\rm M}_{\rm c}(t) \}^2}.
\label{eq:Fpdef}
\end{equation}
We shall rename this quantity more appropriately {\em echo purity}. All three 
quantities, the fidelity $\Fn$, the reduced fidelity $\Fr$ and the echo purity  
$\Fp$ measure stability with respect to perturbations of the dynamics. If the 
perturbed evolution is the same as the unperturbed one, they are all equal to 
one, otherwise they are less than one. Fidelity $\Fn$ measures the stability of 
a whole state, the reduced fidelity gives the stability on the subspace 
${\mathcal H}_{\rm c}$ and the purity fidelity measures separability of 
$\rho^{\rm M}(t)$. One expects that fidelity is the most restrictive quantity 
of the three - $\rho(0)$ and $\rho^{\rm M}(t)$ must be similar for $\Fn$ to be 
high. For $\Fr$ to be high, only the reduced density matrices 
$\rho_{\rm c}(0)$ and $\rho_{{\rm c}}^{\rm M}(t)$ must be similar, and finally, 
the purity fidelity $\Fp$ is high if only $\rho^{\rm M}(t)$ factorizes. 
Fidelity is the strongest criterion for stability. 

\subsubsection{Inequality between fidelity, reduced fidelity and echo purity} 

Actually, one can prove the following inequality for an arbitrary {\em pure} state $\ket{\psi}$ and an arbitrary 
{\em pure product} state $\ket{\phi_{\rm c};\phi_{\rm e}}$~\cite{Znidaric:03,Prosen:03corr},
\begin{equation}
\left|\braket{\phi_{\rm c};\phi_{\rm e}}{\psi} \right|^4 \le \left| 
\bracket{\phi_{\rm c}}{\rho_{\rm c}}{\phi_{\rm c}} \right|^2 \le \trc{\rho_{\rm c}^2},
\label{eq:ineqgen}
\end{equation}
where $\rho_{\rm c}:=\tre{\ket{\psi}\bra{\psi}}$.
\par
{\bf Proof.} Uhlmann's theorem~\cite{Uhlmann:76}, {\it i.e.} noncontractivity of the fidelity, states that tracing over an 
arbitrary subsystem can not decrease the fidelity,
\begin{equation}
\tr{\lbrack \ket{\phi_{\rm c};\phi_{\rm e}}\bra{\phi_{\rm c};\phi_{\rm e}} \ket{\psi}\bra{\psi} \rbrack} \le 
\tr{\lbrack \ket{\phi_{\rm c}}\bra{\phi_{\rm c}} \rho_{\rm c} \rbrack}. 
\end{equation}
Then, squaring and applying the Cauchy-Schwartz inequality 
$|\tr{\lbrack A^\dagger B \rbrack}|^2 \le \tr{\lbrack AA^\dagger\rbrack}\tr{\lbrack BB^\dagger\rbrack}$ we immediately 
obtain the wanted inequality (\ref{eq:ineqgen}).
\par
The rightmost quantity in the inequality $I=\tr{\lbrack \rho_{\rm c}^2
\rbrack}$ is nothing but the purity of state $\ket{\psi}$ and so does
not depend on $\ket{\phi_{\rm c};\phi_{\rm e}}$. One can think of
inequality (\ref{eq:ineqgen}) as giving us a {\em lower bound} on
purity. An interesting question for instance is, which state
$\ket{\phi_{\rm c};\phi_{\rm e}}$ optimizes this bound for a given
$\ket{\psi}$, {\it i.e.} what is the maximal attainable overlap
$\left|\braket{\phi_{\rm c};\phi_{\rm e}}{\psi} \right|^4$ (fidelity)
for a given purity. The rightmost inequality is optimized if we choose
$\ket{\phi_{\rm c}}$ to be the eigenstate of the reduced density
matrix $\rho_{\rm c}$ corresponding to its largest eigenvalue
$\lambda_1$, $\rho_{\rm c}\ket{\phi_{\rm c}}=\lambda_1 \ket{\phi_{\rm
c}}$. To optimize the left part of the inequality, we have to choose
$\ket{\phi_{\rm e}}$ to be the eigenstate of $\rho_{\rm
e}:=\trc{\rho}$ corresponding to the same largest eigenvalue
$\lambda_1$, $\rho_{\rm e}\ket{\phi_{\rm e}}=\lambda_1 \ket{\phi_{\rm
e}}$. The two reduced matrices $\rho_{\rm e}$ and $\rho_{\rm c}$ have
the same eigenvalues~\cite{Araki:70}, $\lambda_1 \ge \lambda_2 \ge \ldots
\ge\lambda_{{\mathcal N}_{\rm c}}$. For such choice of
$\ket{\phi_{\rm c};\phi_{\rm e}}$ the left inequality is actually an equality, 
$\left|\braket{\phi_{\rm c};\phi_{\rm e}}{\psi} \right|^4 = \left| 
\bracket{\phi_{\rm c}}{\rho_{\rm c}}{\phi_{\rm c}} \right|^2=\lambda_1^2$ 
and the right inequality is
\begin{equation}
\lambda_1^2 \le \tr{\lbrack \rho_{\rm c}^2 \rbrack}=\sum_{j=1}^{{\mathcal N}_{\rm c}}{\lambda_j^2},
\end{equation}
with equality iff $\lambda_1=1$. If the largest eigenvalue is close to one, 
$\lambda_1=1-\eps$, the purity will be $I =(1-\eps)^2+{\mathcal O}(\eps^2) \sim 1-2\eps$ 
and the difference between the purity and the overlap will be of the {\em second order} in $\eps$, 
$I-\left|\braket{\phi_{\rm c};\phi_{\rm e}}{\psi} \right|^4\sim \eps^2$. Therefore, for high purity 
the optimal choice of $\ket{\phi_{\rm c};\psi_{\rm s}}$ gives a sharp lower bound, {\it i.e.} its deviation 
from $I$ is of second order in the deviation of $I$ from unity.
\par
For our purpose of studying stability to perturbations, a special case of the general 
inequality (\ref{eq:ineqgen}) is especially interesting. Namely, taking for $\ket{\psi}$ 
the state after the echo evolution $\Md \ket{\psi(0)}$ and for a product state $\ket{\phi_{\rm c};\phi_{\rm e}}$ 
the initial state $\ket{\psi(0)}$ (\ref{eq:psiprod}), we obtain  
\begin{equation}
F^2(t)\le F_{\rm R}^2(t) \le F_{\rm P}(t).
\label{eq:ineq}
\end{equation}
As a consequence of this inequality we find: If fidelity is high, reduced 
fidelity and the echo purity are also high. In the case of perturbations with 
zero time average, the fidelity freezes and from the inequality we can conclude 
that the reduced fidelity and the echo purity will display a similar behavior.

\subsubsection{Uncoupled Unperturbed Dynamics}
A special, but very important, case arrises if the unperturbed dynamics $U_0$ 
represents two uncoupled systems, so we have 
\begin{equation}
U_0=U_{\rm c}\otimes U_{\rm e}.
\label{eq:U0_uncoupled}
\end{equation}
This is a frequent situation if the coupling with the environment is 
``unwanted'', so that our ideal evolution $U_0$ is uncoupled. Under these circumstances the reduced fidelity $\Fr$ and the echo purity  $\Fp$ have especially nice 
forms. 
\par
The reduced fidelity (\ref{eq:Frdef}) can be rewritten as
\begin{equation}
\Fr=\trc{\rho_{\rm c}(0) \rho_{\rm c}^{\rm M}(t)}=\trc{\rho_{\rm c}(t) \rho^\eps_{\rm c}(t)},
\end{equation}
where $\rho_{\rm c}(t)$ is the unperturbed state of the central system and $\rho^\eps_{\rm
c}(t):=\tre{U_{\eps}(t)\rho(0)U^\dagger_{\eps}(t)}$ the corresponding state obtained by perturbed 
evolution. Whereas for a general unperturbed evolution the reduced fidelity was an overlap of the initial 
state with an echo state, for a factorized unperturbed evolution it can also be interpreted as the overlap 
of the (reduced) unperturbed state at time $t$ with a perturbed state at time $t$, similarly as for fidelity.
\par
Echo purity can also be simplified for uncoupled unperturbed evolution. As $U_0$ it factorizes 
 we can bring it out of the innermost trace in the definition of echo purity and use the cyclic 
property of the trace, finally arriving at
\begin{equation}
\Fp=\trc{\{ \rho^{\rm M}_{\rm c}(t) \}^2}=\trc{\{ \rho^\eps_{\rm c}(t)\}^2}=I(t).
\end{equation}
Echo purity is therefore equal to the purity of the forward evolution. The general inequality gives 
in this case 
\begin{equation}
F^2(t) \le F^2_{\rm R}(t) \le I(t),
\end{equation}
and so the fidelity and the reduced fidelity give a lower bound on the decay of purity. Because the purity is 
frequently used in studies of decoherence this connection is especially appealing.
\par
In most of our theoretical derivations regarding the purity fidelity we shall assume a general unperturbed evolution, 
but one should keep in mind that the results immediately carry over to purity in the case of uncoupled unperturbed 
dynamics. Also a large part of our numerical demonstration in next two sections will be done on systems with an 
uncoupled unperturbed dynamics as this is usually the more interesting case. 
%%%%%%%%%%%%%5
\subsubsection{Linear Response Expansion}
\label{sec:LRE}
We proceed with the linear response expansion of reduced fidelity (\ref{eq:Frdef}) and echo purity 
 (\ref{eq:Fpdef}). We use the notation $\rho_{\rm c}:=\ket{\psi_{\rm c}(0)}\bra{\psi_{\rm c}(0)}$ 
for a pure initial density matrix for the central system and $\rho_{\rm e}:=\ket{\psi_{\rm e}(0)}\bra{\psi_{\rm e}(0)}$ 
for the environment. For explicit calculations it is convenient to use an orthogonal basis
$\ket{j;\nu}$, $j=1,\ldots,{\mathcal N}_{\rm c}$, $\nu=1,\ldots,{\mathcal N}_{\rm e}$, with the convention 
that the first basis state $\ket{1;1}:=\ket{\psi_{\rm c};\psi_{\rm e}}$ is the initial 
state. Then, inserting the second order approximation to the echo operator 
(\ref{eq:echoop2}) into the definitions, and keeping only quantities to second order in
$\eps$, we have up to ${\mathcal O}(\eps^4)$ (since third orders exactly vanish)
\begin{eqnarray}
1-\Fn&=&\left( \frac{\eps}{\hbar} \right)^2 \left\{ \bracket{1;1}{\Sigma^2(t)}{1;1}-\bracket{1;1}{\Sigma(t)}{1;1}^2 \right\} \nonumber\\
1-\Fr&=&\left( \frac{\eps}{\hbar} \right)^2 \left\{ \bracket{1;1}{\Sigma^2(t)}{1;1}-\sum_{\nu=1}^{{\mathcal N}_{\rm e}}{ \left| \bracket{1;\nu}{\Sigma(t)}{1;1}\right|^2} \right\} \\
1-\Fp&=&2\left( \frac{\eps}{\hbar} \right)^2 \left\{ \bracket{1;1}{\Sigma^2(t)}{1;1}-\sum_{\nu=1}^{{\mathcal N}_{\rm e}}{ \left| \bracket{1;\nu}{\Sigma(t)}{1;1}\right|^2} -\sum_{j=2}^{{\mathcal N}_{\rm c}}{ \left| \bracket{j;1}{\Sigma(t)}{1;1}\right|^2} \right\} \nonumber.
\label{eq:LRecho11}
\end{eqnarray}
Writing the expectation value in the initial product state as usual,
$\ave{\bullet}=\tr{\lbrack (\rho_{\rm c}\otimes \rho_{\rm
e})\bullet\rbrack}$,
we can rewrite the linear response result in basis free form
\begin{eqnarray}
1-\Fn&=&\left( \frac{\eps}{\hbar} \right)^2 \<\Sigma(t)(\mathbbm{1}\otimes\mathbbm{1}-\rho_{\rm c}\otimes\rho_{\rm e})\Sigma(t)\> \nonumber\\
1-\Fr&=&\left( \frac{\eps}{\hbar} \right)^2 \<\Sigma(t)(\mathbbm{1}-\rho_{\rm c})\otimes\mathbbm{1}\Sigma(t)\> \nonumber\\
1-\Fp&=&2\left( \frac{\eps}{\hbar} \right)^2 \<\Sigma(t)(\mathbbm{1}-\rho_{\rm c})\otimes(\mathbbm{1}-\rho_{\rm e})\Sigma(t)\>.
\label{eq:LRecho}
\end{eqnarray}
The linear response expansion of course also satisfies
the general inequality (\ref{eq:ineq}). The difference between $\Fr$
and $\Fn$ as well as between $\Fp$ and $\Fn$ results from {\em off-diagonal}
matrix elements of operator $\Sigma(t)$. One may compare these results to 
time-independent perturbative expansions valid up to quantum Zeno time 
$t_{\rm Z}$~\cite{Kubler:73,Kim:96}. Depending on the growth of the linear response
terms with time we shall again have two general categories, that of
mixing dynamics and that of regular dynamics.

In the general linear response expressions above we have not assumed anything
on the particular form of perturbation $V$ and unperturbed system $H_0$. 
It may be however, interesting to study separately the special case where the 
unperturbed system in uncoupled and the perturbation is the coupling. This we 
shall do for different situations in later sections.

\section{Quantum echo-dynamics: Non-integrable (chaotic) case}
\label{Q2}

\subsection{Fidelity and dynamical correlations}
\label{Q2D}

In the previous section we have elaborated in detail on different general
properties of echo dynamics. At this point we specialize our interest to
the case of systems which produce maximal possible dynamical disorder.
For systems which possess a well defined classical limit this means that the
latter should be non-integrable and fully chaotic. For other systems, we
assume that the system's stationary properties are well described by
Gaussian or circular ensembles of random matrices and hence fall into the 
general category of quantum chaos.
Yet, we postpone the discussion of the random matrix formulation of
echo dynamics to the next section and here concentrate on individual physical
dynamical systems which possess the property of dynamical mixing, either in 
the semiclassical or the thermodynamic limit. Again, we use the
Born expansion (\ref{eq:born}) and express fidelity in terms of dynamical 
correlations.

At first we assume that a well defined classical limit exists and that the
classical system has the strong ergodic property of {\em mixing} such that the
correlation function of the perturbation $V$ decays sufficiently fast; 
this typically (but not necessarily) corresponds to globally chaotic classical 
motion. However, reference to underlying classical dynamics is not strictly 
necessary as the corresponding quantum mixing can sometimes be established in 
thermodynamic limit without taking the classical limit. For the application of 
echo-dynamics in such a situation see Refs.~\cite{Prosen:02,Prosen:02spin}.

Due to ergodicity and mixing we can assume that averages of any dynamical 
observables over a specific initial state can be replaced by a full Hilbert 
space average, $\aave{\bullet}=\tr(\bullet)/N$, at least after times longer 
than a certain relaxation time-scale $t_{\rm E}$
\begin{equation}
\bra{\Psi_0(t)}A\ket{\Psi_0(t)} \approx \aave{A},\quad {\rm for} \quad |t| \ge t_{\rm E}.
\label{eq:ergavg}
\end{equation}
For minimal-uncertainty wave packet initial state with typical phase space 
diameter $\sqrt{\hbar}$ and exponentially unstable classical dynamics with 
characteristic exponent $\lambda$, the time scale $t_{\rm E}$ can be estimated 
as $t_{\rm E} \sim \log(1/\hbar)/\lambda$ and is sometimes known as the 
{\em Ehrenfest time}. This is the time needed for an initially localized 
wavepacket to spread effectively over the accessible phase 
space~\cite{Berman:78}. For other types of initial states this time is in 
general even shorter. For example, for a {\em random} initial state, 
Eq.~(\ref{eq:ergavg}) is satisfied for {\em any} $t$, by definition of a 
random state (\ref{eq:Avgtrace}).

Therefore, for any initial state (e.g. in the worst case for a 
minimal-uncertainty wave packet) one obtains identical results for $F(t)$ for 
sufficiently long times
\footnote{The exception might be systems with non-ergodic quantum
 behavior, for example exhibiting dynamical localization.}, 
{\it i.e.} longer than $t_{\rm E}$. The state averaged quantum correlation function 
is homogeneous in time, {\it i.e.} $C(t,t')=C(t-t')$, so we simplify the second order 
linear response formula (\ref{eq:fidlr}) for fidelity
\begin{equation}
F(t)=1-\frac{\eps^2}{\hbar^2}\left\{t C(0) + 2\int_0^t \dd t'{(t-t')C(t')} \right\} + 
\OO(\eps^4).
\label{eq:LRmixing}
\end{equation}
If the decay of the correlation function $C(t)$ is sufficiently fast, namely if its integral 
converges on a certain characteristic {\em mixing time} scale $t_{\rm mix}$, meaning that 
$C(t)$ should in general decay faster than $\OO(t^{-2})$, then the above 
formula can be further simplified. For times $t \gg t_{\rm mix}$ we can neglect the 
second term under the summation in (\ref{eq:LRmixing}) and obtain a linear fidelity decay in 
time $t$ (in the linear 
response)
\begin{equation}
F(t) = 1 - 2 (\eps/\hbar)^2 \sigma t,
\label{eq:linearmixing}
\end{equation}
with the transport coefficient $\sigma$ being
\begin{equation}
\sigma = \int_0^\infty \dd t C(t) =
\lim_{t\to\infty}{\frac{\ave{\Sigma^2(t)}-\ave{\Sigma(t)}^2}{2t}}.
\label{eq:sigmadef}
\end{equation}
Note that 
$\sigma$ has a well defined classical limit obtained from the classical correlation function and 
in the semiclassical limit this classical $\sigma_{\rm cl}$ will agree with the quantum one.
\par
We can make a stronger statement in a non-linear-response regime if we make an additional 
assumption on the factorization of higher order time-correlations, namely assuming 
$n-$point mixing \cite{Arnold:68}. This implies that $2m$-point correlation 
$\ave{V(t_1)\cdots V(t_{2m})}$ 
is appreciably different from zero for $t_{2m}-t_1 \to \infty$ only if all (ordered) time 
indices $\{t_k,k=1\ldots 2m\}$ are {\em paired} with the time differences 
within each pair, $t_{2k}-t_{2k-1}$, being of the order or less than 
$t_{\rm mix}$. Then we can make a further reduction, namely if $t\gg m t_{\rm mix}$ the terms 
in the expansion of the fidelity amplitude $f(t)$ (\ref{eq:fidech},\ref{eq:born}) are
\begin{figure}[ht!]
\centerline{\includegraphics[width=\figw\textwidth]{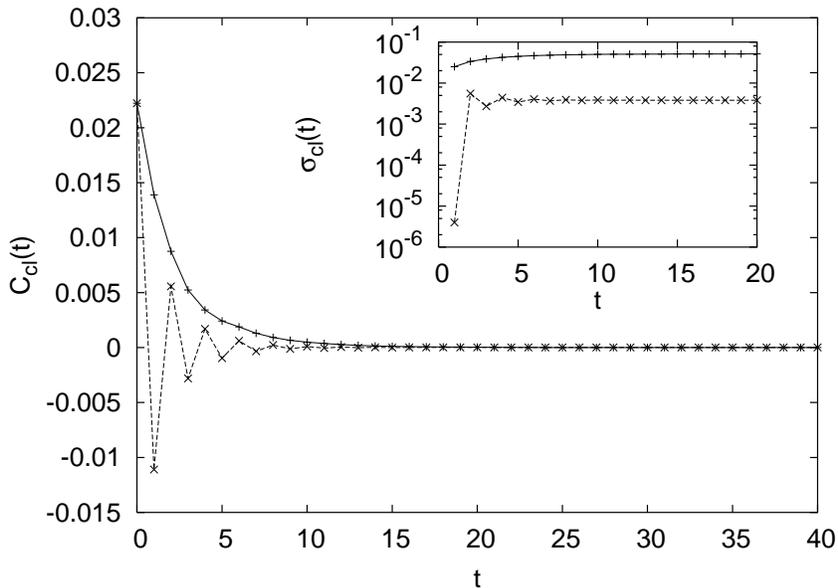}}
\caption{The classical correlation function of perturbation $V$ 
(\ref{eq:classv}) 
for chaotic kicked top and $\gamma=\pi/6$ (top solid curve) and $\gamma=\pi/2$ (bottom broken 
curve). The finite time integrated correlation function is shown in the inset.}
\label{fig:classcor}
\end{figure}
\begin{eqnarray}
%\!\!\!\!\!\!\!\!\!\!\
&\TT&\!\!\!\int_0^t \dd t_1\dd t_2 \cdots \dd t_{2m}\!
\ave{V(t_1) V(t_2) \cdots V(t_{2m})}
\rightarrow \nonumber \\
\rightarrow &\TT&\!\!\!\int_0^t \dd t_1\dd t_2 \cdots \dd t_{2m}\!
\ave{V(t_1) V(t_2)}\cdots\ave{V(t_{2m-1}) V(t_{2m})}
\rightarrow
\frac{(2m)!}{m!2^m}(2\sigma t)^m.
\label{eq:factor}
\end{eqnarray}
The fidelity amplitude is therefore $f(t)=\exp{(-\eps^2 \sigma t/\hbar^2)}$ and the fidelity is
\begin{equation}
F(t)=\exp{(-t/\tau_{\rm m})}, \qquad \qquad \tau_{\rm m}=\frac{\hbar^2}{2 \eps^2 \sigma_{\rm cl}},
\label{eq:Fnmixing}
\end{equation}
where $\tau_{\rm m}=\OO(\eps^{-2})$ is the time scale for decay and the 
subscript ``m'' stands for mixing dynamics. If the system has a classical 
limit, one may take also a classical limit of the transport coefficient 
$\sigma_{\rm cl}$, so the decay time-scale $\tau_{\rm m}$ can be computed from 
classical mechanics. We should stress again that the above result 
(\ref{eq:Fnmixing}) has been derived under the assumption of true quantum 
mixing which can be justified only in the limit $N\rightarrow\infty$, e.g. 
either in the semiclassical or the thermodynamic limit. Thus for the true 
quantum-mixing dynamics the fidelity will decay exponentially. 

The same result can also be derived using the standard Fermi golden rule 
interpreting fidelity as the transition probability and estimating the square 
of perturbation matrix elements in terms of classical correlation 
functions~\cite{Jacquod:01,Cerruti:02}. This regime of exponential fidelity 
decay is often refereed to as a {\em Fermi golden rule regime}. It is 
consistent with the effective treatment of fidelity as a Fourier transform of 
the local density of states of the eigenstates of $H_\eps$ expressed in the
eigenbasis of $H_0$, or vice versa. Within the standard random matrix 
assumptions corresponding to classical chaos, the local density of states has a 
Lorentzian form %TGin this so-called ``perturbative'' regime 
and this corresponds to an exponential decay (\ref{eq:Fnmixing}).

\begin{figure}[h]
\centerline{\includegraphics[width=\figw\textwidth]{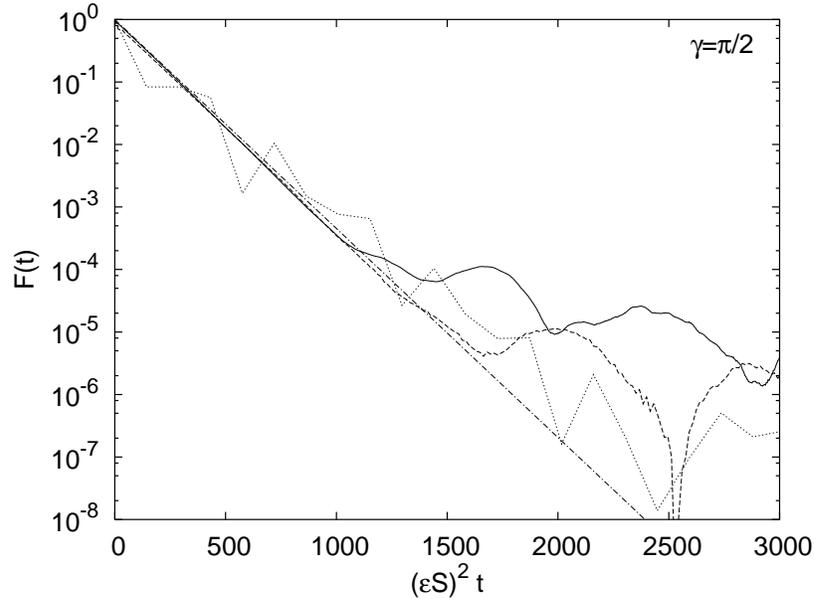}}
\caption{Quantum fidelity decay in the chaotic regime for
  $\gamma=\pi/2$ and three different perturbation strengths
  $\eps=5\times 10^{-4}$, $1\times 10^{-3}$ and $3\times 10^{-3}$
  (solid, dashed and dotted curves, respectively) is shown. The chain
  line gives theoretical decay (\ref{eq:Fnmixing}) with the classically calculated $\sigma$ seen 
in Fig.~\ref{fig:classcor}.}
\label{fig:exppi2}
\end{figure}

\begin{figure}[h]
\centerline{\includegraphics[width=\figw\textwidth]{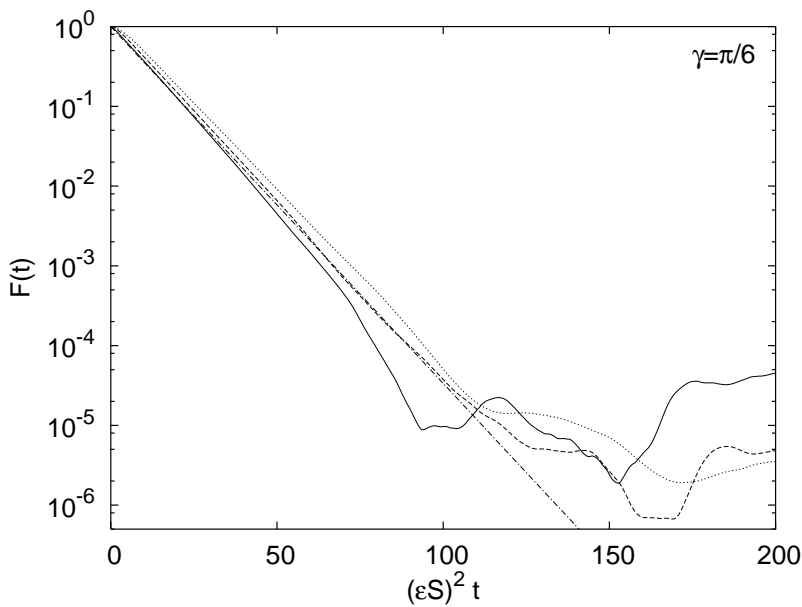}}
\caption{Similar figure as~\ref{fig:exppi2}, only for $\gamma=\pi/6$ and 
  perturbation strengths $\eps=1\times 10^{-4}$, $2\times 10^{-4}$ and 
  $3\times 10^{-4}$ (solid, dashed and dotted curves, respectively).}
\label{fig:exppi6}
\end{figure}

To numerically check the above exponential decay, we shall use the kicked top 
(\ref{eq:KTdef}) with parameter $\alpha=30$, giving totally chaotic classical 
dynamics. As argued before, one can calculate the transport coefficient 
$\sigma$ (\ref{eq:sigmadef}) by using the classical correlation function of the 
perturbation (\ref{eq:KTV}),
\begin{equation}
V_{\rm cl}=v=\frac{1}{2}z^2.
\label{eq:classv}
\end{equation}
We consider two different values of kicked top parameter $\gamma$,
namely $\gamma=\pi/2$ and $\gamma=\pi/6$. The classical correlation
functions is shown in Fig.~\ref{fig:classcor}. The correlation
function (obtained by averaging over $10^5$ initial conditions on a
sphere) is shown in the main frame. The correlation functions have
qualitatively different decay towards zero for the two chosen $\gamma$'s. In 
the inset the convergence of the classical $\sigma$ (\ref{eq:sigmadef}) is 
shown. One can see that the mixing time is $t_{\rm mix} \approx 5$. The values 
of $\sigma_{\rm cl}$ are $\sigma_{\rm cl}=0.00385$ for $\gamma=\pi/2$ and 
$\sigma_{\rm cl}=0.0515$ for $\gamma=\pi/6$. 
These values are used to calculate the theoretical decay of fidelity 
$F(t)=\exp{(-\eps^2 S^2 2 \sigma_{\rm cl} t)}$ according to 
Eq.~(\ref{eq:Fnmixing}) which is compared with numerical simulations in the
Figs.~\ref{fig:exppi2} and \ref{fig:exppi6}. We used averaging over the whole 
Hilbert space, and checked that due to ergodicity there was no difference for 
large $S$ if we choose a fixed initial state, say a coherent state. As fidelity 
will decay only until it reaches its finite size fluctuating value 
$\bar{F}$ (\ref{eq:Fnavgavg}) we choose a large $S=4000$ in order to be able 
to check the exponential decay over as many orders of magnitude as possible. In 
Fig.~\ref{fig:exppi2} the decay of quantum fidelity is shown for 
$\gamma=\pi/2$. The agreement with theory is excellent. Note that the largest 
$\eps$ shown corresponds to $\tau_{\rm m} \approx 1$ so that the condition for 
$n-$point mixing $t\gg t_{\rm mix}$ is no longer strictly satisfied, hence we 
observe some oscillations around the theoretical curve. However, overall 
agreement with the theory is still good due to the oscillatory nature of the 
correlation decay (see Fig.~\ref{fig:classcor}) fulfilling the factorization 
assumption (\ref{eq:factor}) on average.

In Fig.~\ref{fig:exppi6} for $\gamma=\pi/6$ a similar decay can be seen. In 
both cases fidelity starts to fluctuate around $\bar{F}$ calculated in the 
Section~\ref{sec:timeaveraged} for sufficiently long times $t_\infty$ (see appendix \ref{sect:appTA}).

\subsubsection{Long time behavior}
\label{sec:longtime}

\begin{figure}[h]
\centerline{\includegraphics[width=\figw\textwidth]{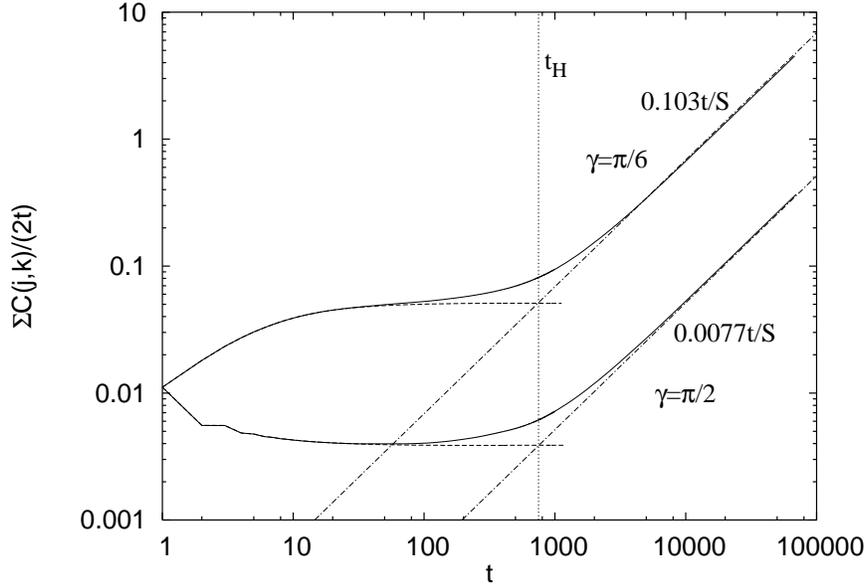}}
\caption{The finite time quantum correlation sum 
  $\sigma(t)=\sum_{j,k=0}^{t-1} {C(j,k)/2t}$ (solid curves) together with the 
  corresponding classical sum 
  $\sigma_{\rm cl}(t) = \sum_{j,k=0}^{t-1}{C_{\rm cl}(j,k)}/2t$ (dashed curves 
  saturating at $\sigma_{\rm cl}$ and ending at $t \sim 1000$) is shown for the 
  chaotic kicked top. Quantum data are for a full trace $\rho=\mathbbm{1}/N$ 
  with $S=1500$. Upper curves are for $\gamma=\pi/6$ while lower curves are for 
  $\gamma=\pi/2$. Chain lines are best fits for asymptotic linear functions 
  corresponding to $\bar{C}t/2=0.0077t/S$ for $\gamma=\pi/2$ and $0.103 t/S$ 
  for $\gamma=\pi/6$.}
\label{fig:cinf}
\end{figure}

So far, we have assumed that the quantum correlation function $C(t)$ decays to 
zero and its integral (or sum, for discrete-time dynamical systems) converges 
to $\sigma$. However, for a system with a Hilbert space of finite dimension 
$N$, the correlation function asymptotically does not decay but has a 
non-vanishing plateau $\oC$, similar to the finite asymptotic value of
fidelity $\bar{F}$. This will cause the double correlation integral to grow, 
asymptotically, quadratically with time. Because this plateau $\oC$ is small, 
the quadratic growth will overtake linear growth $2\sigma_{\rm cl} t$ only for 
large times. The time averaged correlation function 
$C(t,t')$ (\ref{eq:CFdef}) can be calculated assuming a nondegenerate 
unperturbed spectrum $E_k$ as
\begin{equation}
\oC=\lim_{t \to \infty}{\frac{1}{t^2}\int_0^t \dd t' \int_0^t \dd t'' {C(t',t'')}}=
\sum_k{\rho_{kk} (V_{kk})^2-\left( \sum_k {\rho_{kk} V_{kk} }\right)^2},
\label{eq:Cbar_def}
\end{equation}
where $\rho_{kk}$ are diagonal matrix elements of the initial density matrix $\rho(0)$ and $V_{kk}$ are 
diagonal matrix 
elements of the perturbation $V$ in the eigenbasis of the unperturbed propagator $U_0$. One can see that 
$\oC$ depends only on the diagonal matrix elements,
 in fact it is equal to the variance of the diagonal matrix elements. Since the classical 
system is ergodic and mixing, we shall use a version of the {\em quantum chaos 
conjecture}~\cite{Feingold:86,Wilkinson:87,Feingold:89,Prosen:93,Prosen:94} saying that $V_{mn}$ are 
independent Gaussian random variables with a variance given by the Fourier transformation $S(\omega)$ 
(divided by $N$) of the corresponding classical correlation function $C_{\rm cl}(t)$ at frequency 
$\omega=(E_m-E_n)/\hbar$. On the diagonal we have 
$\omega=0$ and an additional factor of $2$
due to the random matrix measure of the diagonal elements. 
%(see {it e.g.} Haake's book~\cite{Haake:91} or Mehta's~\cite{Mehta:91}). 
Using $2 \sigma_{\rm cl} t=
\int_0^t \dd t' \int_0^t \dd t'' C(t',t'')=S(0) t$ we can write
\begin{equation}
\bar{C} = \frac{2 S(0)}{N}=\frac{4\sigma_{\rm cl}}{N}.
\label{eq:Cbar}
\end{equation}
Because of ergodicity, for large $N$, $\bar{C}$ does not depend on the statistical operator $\rho$ used in the 
definition of the correlation function, provided we do not take some non generic state like a single eigenstate 
$\ket{E_k}$ for instance. 
If $U_0$ has symmetries, so that its Hilbert space is split into $s$ components of sizes $N_j$, the average 
$\bar{C}$ will be different on different subspaces, $\bar{C}_j=4\sigma_{\rm cl}/N_j$. Averaging over all invariant 
subspaces then gives 
\begin{equation}
\bar{C}=\frac{4s \sigma_{\rm cl}}{{N}},
\label{eq:Cbar_s}
\end{equation}
so that $\bar{C}$ is increased by a factor $s$ compared to the situation with only a single desymmetrized subspace. 
The fidelity decay will start to be dominated by the average plateau (\ref{eq:Cbar}) at time $t_{\rm H}$ when 
the quadratic growth takes over, $\bar{C} t_{\rm H}^2 \approx 2\sigma_{\rm cl} t_{\rm H}$,
\begin{equation}
t_{\rm H}=\frac{1}{2}N \propto \hbar^{-d},
\label{eq:np_def}
\end{equation}
which is nothing but the (dimensionless) {\em Heisenberg time} associated with 
the inverse (quasi)energy density. 
Note that for autonomous systems where energy is a conserved quantity, 
Heisenberg time scales differently $t_{\rm H} \propto \hbar^{1-d}$.
Again, if one has $s$ invariant subspaces, the Heisenberg time is reduced 
$t_{\rm H}=N/(2s)$. 
\par
In Fig.~\ref{fig:cinf} we show numerical calculation of the correlation integral (sum) for the chaotic kicked top at 
$\alpha=30$. We compare 
the classical correlation sum (the same data as in Fig.~\ref{fig:classcor}) and quantum correlation sum. 
One can nicely see the 
crossover from linear growth of quantum correlation sum 
$2 \sigma_{\rm cl} t$ for small times $t<t_{\rm H}$, 
to the asymptotic 
quadratic growth due to correlation plateau $\oC$. In addition, numerically fitted asymptotic growth 
$0.103 t/S$ and $0.0077t/S$ 
nicely agree with formula for $\oC$, using $N=S$ and classical values of transport coefficients 
$\sigma_{\rm cl}=0.0515$ and 
$0.00385$ for $\gamma=\pi/6$ and $\gamma=\pi/2$, respectively.
\par
\begin{figure}[h]
\centerline{\includegraphics[height=80mm,angle=-90]{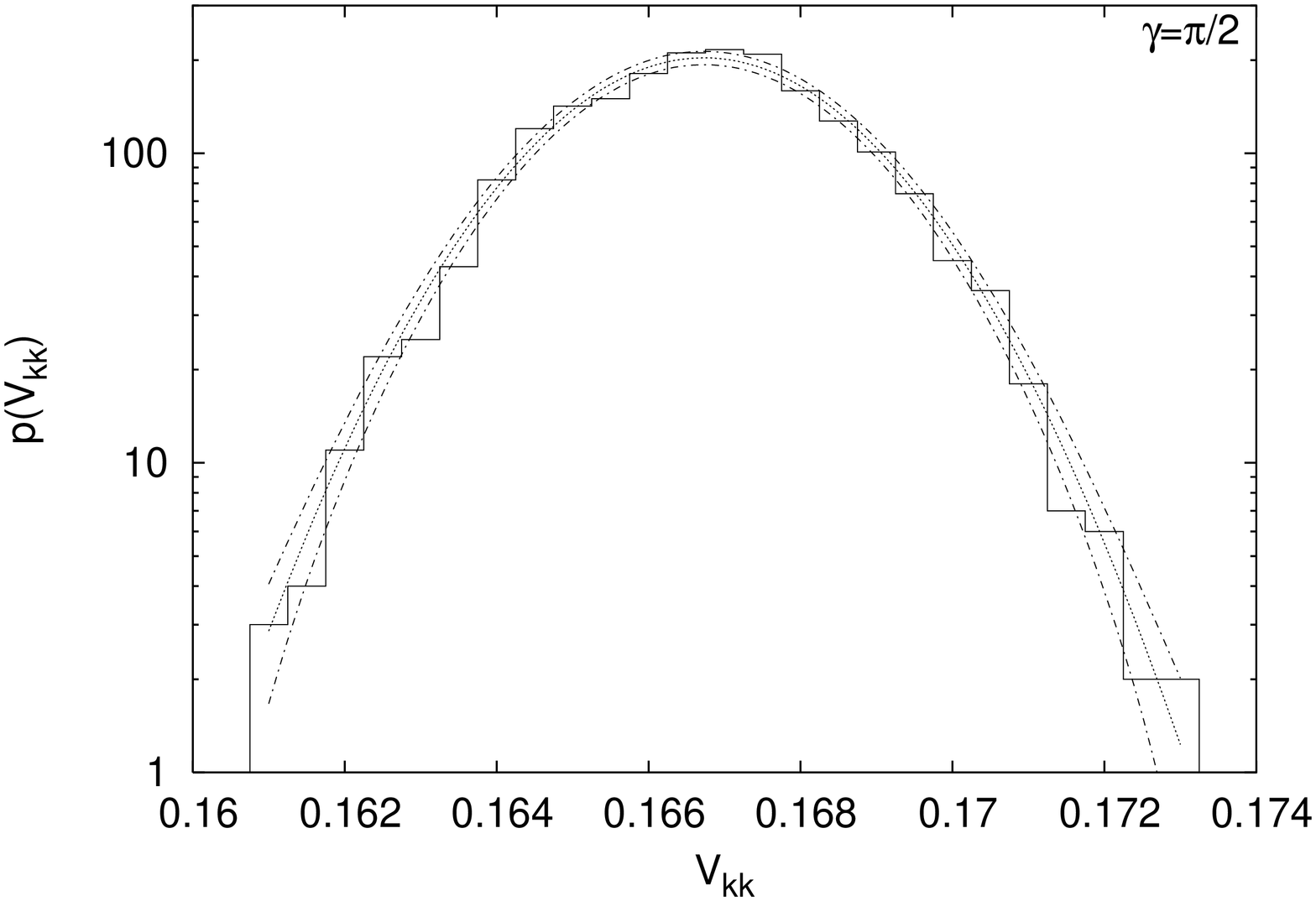}
\includegraphics[height=80mm,angle=-90]{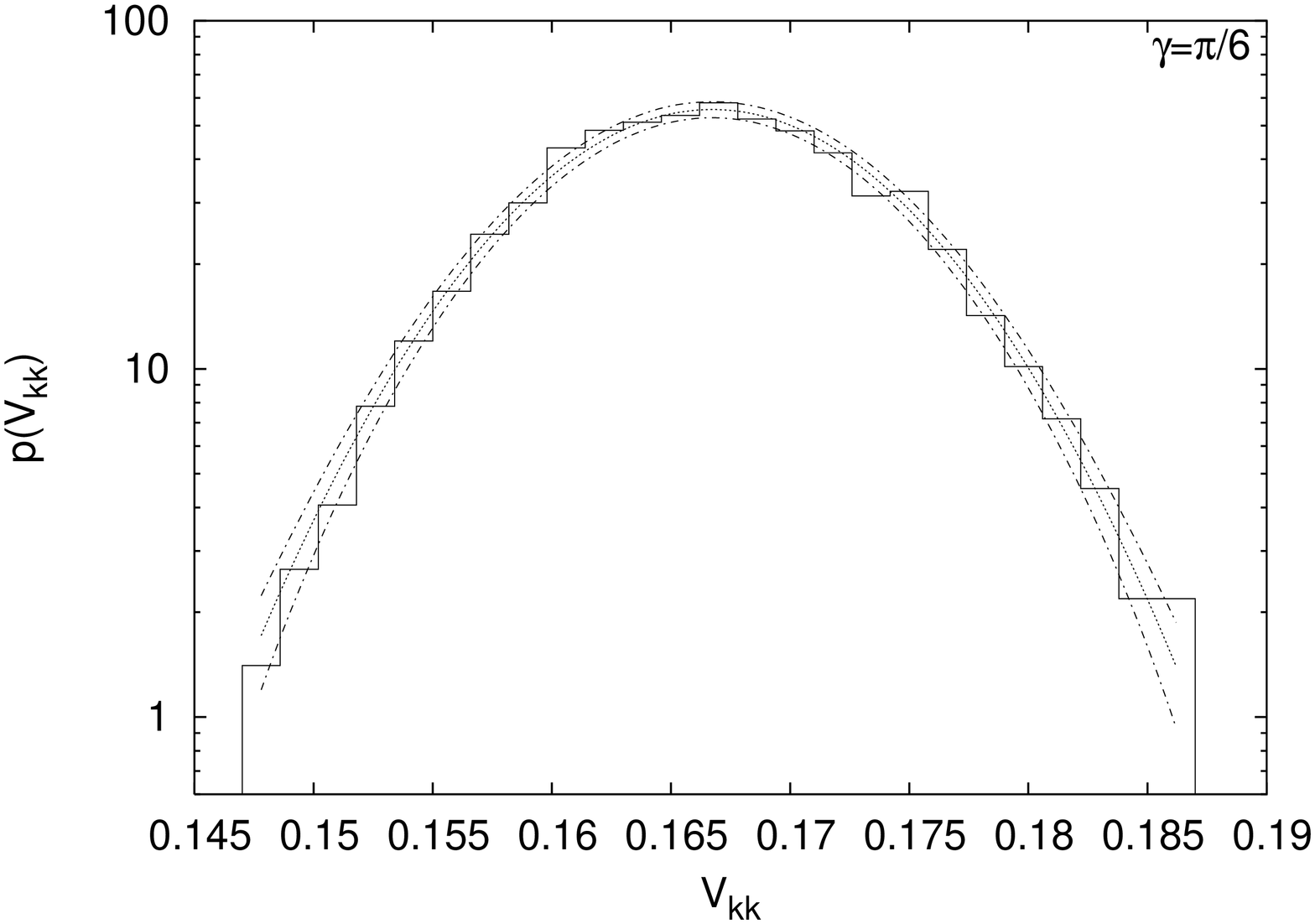}}
\caption{Histogram of the normalized distribution of the diagonal
  matrix elements $V_{kk}$ for the chaotic kicked top and $S=4000$ on
  OE subspace (\ref{eq:KTsubspaces}). The dotted line is the
  theoretical Gaussian distribution with the second moment $\oC$ and
  the two chain lines are expected $\sqrt{N_i}$ statistical deviations
  if there are $N_i$ elements in the $i-$th bin. Note the different
  $x$-ranges in two figures due to different $\sigma_{\rm cl}$ for the
  two chosen $\gamma$.}
\label{fig:Vkk}
\end{figure}
\begin{figure}[ht!]
\centerline{\includegraphics[width=\figw\textwidth]{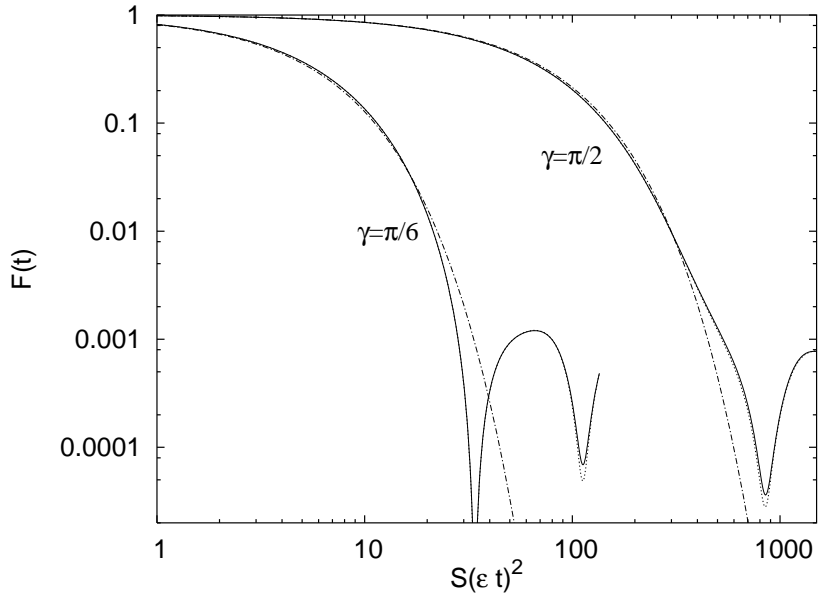}}
\caption{Quantum fidelity decay for $\eps<\eps_{\rm p}$ in the chaotic regime. 
  For $\gamma=\pi/2$ data for $\eps=1\cdot 10^{-6}$ (solid curve) and 
  $5\cdot 10^{-6}$ (dotted curve) are shown. For $\gamma=\pi/6$, 
  $\eps=3\cdot 10^{-7}$ (solid) and $1\cdot 10^{-6}$ (dotted) are shown. Note 
  that for both $\gamma$ the curves for both $\eps$ practically overlap. The 
  chain curves are theoretical predictions (\ref{eq:gaussianmixing}) with 
  classically computed $\sigma_{\rm cl}$.}
\label{fig:j15kpert}
\end{figure}

For times $t>t_{\rm H}$, and provided $\eps$ is sufficiently small, 
the correlation sum will grow
quadratically and the linear response fidelity reads
\begin{equation}
F(t)=1-\frac{\eps^2}{\hbar^2} \frac{4\sigma_{\rm cl}}{N} t^2.
\end{equation}
To derive the decay of fidelity beyond the linear response regime one
needs higher order moments of diagonal elements of perturbation
$V$. If we use the leading order BCH expression of the echo operator (\ref{eq:BCH}),
neglecting the term $\Gamma(t)\eps^2$, we have the fidelity amplitude 
$f(t)=\sum_k{\exp{(-\ii V_{kk} \eps t/\hbar)}}/{N}$, 
where we choose an ergodic average $\rho=\mathbbm{1}/{N}$. In the limit 
${N}\to \infty$ we can replace the sum with an integral over the probability 
distribution of diagonal matrix elements $p(V_{kk})=p(V)$,
\begin{equation}
f(t)=\int{\!{\rm d}V\,p(V)\exp{(-\ii V \eps t /\hbar)}}.
\label{eq:fn_fourier}
\end{equation}
For long times the fidelity amplitude is therefore a Fourier transformation of the distribution of diagonal 
matrix elements. For classically mixing systems the distribution is conjectured to be Gaussian with the second 
moment equal to $\oC$ (\ref{eq:Cbar}). This is confirmed by numerical data in Fig.~\ref{fig:Vkk}. The mean 
value of diagonal matrix elements is perturbation specific and is for our choice of the perturbation (\ref{eq:KTV}) equal to 
$\sum_k{V_{kk}}/(2S+1)=(2S+1)(S+1)/12S^2$. From the figure we can see that the distribution is indeed Gaussian 
with the variance agreeing with the theoretically predicted $\oC=4\sigma_{\rm cl}/S$. The Fourier transformation 
of a Gaussian is readily calculated and we get a Gaussian fidelity decay
\begin{equation}
F(t)=\exp{\left(-(t/\tau_{\rm p})^2 \right)},\qquad \tau_{\rm p}=\sqrt{\frac{{N}}{4 \sigma_{\rm cl}}} 
\frac{\hbar}{\eps}.
\label{eq:gaussianmixing}
\end{equation}
In order to see a Gaussian fidelity decay for mixing systems the perturbation 
strength $\eps$ must be sufficiently small. Otherwise, the fidelity will decay 
exponentially (\ref{eq:Fnmixing}) to its fluctuating plateau $\bar{F}$ before 
time $t_{\rm H}$ when the Gaussian decay starts. Requiring that the time-scale 
of exponential decay $\tau_{\rm m}$ is smaller than $t_{\rm H}={N}/2$ gives the 
critical perturbation strength $\eps_{\rm p}$,
\begin{equation}
\eps_{\rm p}=\frac{\hbar}{\sqrt{\sigma_{\rm cl}{N}}}.
\label{eq:deltap}
\end{equation}
For $\eps < \eps_{\rm p}$ we will have a Gaussian decay 
(\ref{eq:gaussianmixing}), otherwise the decay starts as an exponential 
(\ref{eq:Fnmixing}) and goes over to a Gaussian at the Heisenberg time 
$t_{\rm H}$ (see Section~\ref{R} for a uniform theory within a random matrix 
model). If the decay reaches the plateau $\overline{F}$ before or around 
$t_{\rm H}$, the decay remains purely exponential (for details see 
Section~\ref{sec:timescales}). Again we illustrate the predicted Gaussian decay for the chaotic kicked top 
with $\alpha=30$ and $S=1500$ and a full trace average over the Hilbert space. 
The results of numerical simulation, together with the theory are shown in 
Fig.~\ref{fig:j15kpert}.

The regime of Gaussian decay is sometimes referred to as the perturbative 
regime~\cite{Cerruti:02,Jacquod:01} because it can be derived using 
perturbation theory in lowest order. Writing the eigenenergies in first order
$E_k^\eps=E_k+V_{kk}\, \eps$ and the overlap matrix in zeroth order
$O_{kl}=\delta_{kl}+\OO(\eps)$ (\ref{eq:fnP}), one arrives at the Fourier 
transform formula (\ref{eq:fn_fourier}).

\subsection{Vanishing time averaged perturbation and fidelity freeze}

\label{sec:chfreeze}

Here we shall assume that the perturbation is of particular form, namely it can
be written as a time derivative of some observable $W$ (\ref{eq:Wdef}).
We shall show that a simple semi-classical theory can be developed which 
describes anomalously slow fidelity decay (fidelity freeze) in such a case.
Other types of perturbations with vanishing time-average will be considered 
within the random matrix framework in Section~\ref{Q2QF}.

The presentation here follows Ref.~\cite{Prosen:05}. Based on the general 
discussion of Section~\ref{sec:freeze}, we write a semiclassical expression for 
the fidelity below the time scale $t_2=\OO(\eps^{-1})$, which shall be 
specified later, and above the Ehrenfest time-scale 
$t_{\rm E}$, $t_{\rm E} < t < t_2$, as
\begin{equation}
F(t) \approx F_{\rm plat} = 
\left|\ave{\exp\left(-\frac{\ii\eps}{\hbar} w\right)}_{\rm cl}
\ave{\exp\left(\frac{\ii\eps}{\hbar}W\right)}\right|^2
\label{eq:Fplatch}
\end{equation}
This equation is derived from the general Eq.~(\ref{eq:Fnplateau}) in three
steps: (i) the higher order middle factor in Eq.~(\ref{eq:Fnplateau}) can be
neglected for for $t < t_2$, 
(ii) we assume that due to {\em mixing dynamics}, 
the correlation function of $\exp(-\ii W(t)\eps/\hbar)$ can be 
factorized for $t > t_{\rm mix}$ as
$$\ave{\exp(-\ii W(t) \eps/\hbar)\exp(\ii W(0)\eps/\hbar)}\approx
\ave{\exp(-\ii W(t)\eps/\hbar)} \ave{\exp(\ii W(0)\eps/\hbar)},$$
and (iii) for $t>t_{\rm E}$ the average
$\ave{W(t)}= \aave{W}$ is approximated by the classical average
\begin{equation}
\ave{w}_{\rm cl} = \frac{\int_\Omega\dd\mu{x} w(\ve{x})}{\int_\Omega\dd\mu{x}}
\end{equation}
where $w(\ve{x})$ is the classical observable corresponding to the operator 
$W$ and $\Omega$ is the classical invariant ergodic component related to the 
initial state $\Psi$, e.g. the energy surface for $E=\bra{\Psi}H_0\ket{\Psi}$.

The plateau of fidelity can be written in a compact form for the two simplest 
extreme cases of initial states: (a) for a {\em coherent initial state} (CIS), 
where the initial state average  can be approximated by the classical 
observable evaluated at the center $\ve{x}^*$ of the wave packet 
$\ave{W}\approx w(\ve{x}^*)$, hence $|\ave{\exp(-\ii W\eps/\hbar)}|\approx 1$,
and (b) for a {\em random initial state} (RIS), where the initial state average 
is equivalent to an ergodic average, 
$\ave{\exp(-\ii W\eps/\hbar)}=\aave{\exp(-\ii W\eps/\hbar)}
\approx\ave{\exp(-\ii w\eps/\hbar)}_{\rm cl}$.
Defining a classical generating function as
\begin{equation}
G(z)=\ave{\exp(-\ii z w)}_{\rm cl}
\end{equation} 
one can compactly write
\begin{equation}
F^{\rm CIS}_{\rm plat}\approx |G(\eps/\hbar)|^2 \; , \qquad
F^{\rm RIS}_{\rm plat}\approx |G(\eps/\hbar)|^4\; . 
\label{eq:NLRP}
%\label{eq:plateaus}
\end{equation}
Note that the two plateau levels satisfying {\em universal} relation
$F^{\rm RIS}_{\rm plat}\approx (F^{\rm CIS}_{\rm plat})^2$. Curiously, the same relation is 
satisfied for the case of regular dynamics (see Section~\ref{sec:ch4Veq0}).
If the argument $z=\eps/\hbar$ is large, the analytic function $G(z)$ can be 
calculated generally by the method of stationary phase. In the simplest case of 
a single isolated stationary point $\vec{x}^*$ in $N$ dimensions
\begin{equation}
|G(z)| \asymp \left|\frac{\pi}{2z}\right|^{N/2}\left|{\rm det\,}\partial_{x_j} \partial_{x_k} 
W(\vec{x}^*)\right|^{-1/2}.
\label{eq:Gasym}
\end{equation}
This expression gives an asymptotic power law decay of the plateau height 
independent of the perturbation details. Note that for a finite phase space we will have 
oscillatory {\em diffraction corrections} to Eq.~(\ref{eq:Gasym}) due to a finite range of 
integration $\int\dd\mu$ which in turn causes an interesting situation for 
specific values of $z$, namely that by 
increasing the perturbation strength $\eps$ we can actually increase the value of the plateau. 

Next we shall consider the regime of long times $t > t_2$. 
Then the second term in the exponential of 
(\ref{eq:BCH}) dominates the first one, however even the first term may not be negligible.
Up to terms of order $\OO(t \eps^3)$ we can factorize Eq.~(\ref{eq:BCHfreeze}) as
$
M_\eps(t) \approx \exp(-\ii\frac{\eps}{\hbar}(W(t)-W(0)))
\exp(-\ii\frac{\eps^2}{2\hbar}\Sigma_R(t)).
$
When computing the expectation value $f(t) = \ave{M_\eps(t)}$
we again use the fact that in the leading semiclassical order the operator ordering is irrelevant and
that any time-correlation can be factorized,
so also the second term of $\Gamma(t)$ (\ref{eq:defGamma}) vanishes. Thus we have
\begin{equation}
F(t) \approx F_{\rm plat} \left|\ave{\exp\left(-\ii\frac{\eps^2}{2\hbar}\Sigma_R(t)\right)}\right|^2,
\quad t > t_2.
\label{eq:renfid}
\end{equation}
This result is quite intriguing. It tells us that apart from a prefactor 
$F_{\rm plat}$, the decay of fidelity with residual perturbation is formally 
the same as fidelity decay with a generic (non-residual) perturbation, 
Eqs.~(\ref{eq:fidech}) and (\ref{eq:Mformal}), when one substitutes the 
operator $V$ with $R$ and the perturbation strength $\eps$ with 
$\eps_R=\eps^2/2$. 
The fact that time-ordering is absent in Eq.~(\ref{eq:renfid}) as compared with 
(\ref{eq:Mformal}) is semiclassically
irrelevant. Thus we can directly apply the general semiclassical theory of fidelity decay described in 
previous subsection,
using a renormalized perturbation $R$ of renormalized strength $\eps_R$. 
Here we simply rewrite the key results in the 'non-Lyapunov' 
(perturbation-dependent regime), namely for $\eps_R < \hbar$. Using a classical transport rate 
$\sigma := \lim_{t\to\infty}\frac{1}{2 t}(\ave{\sigma^2_R(t)}_{\rm cl}-\ave{\sigma_R(t)}_{\rm cl}^2)$
where $\sigma_R(t)$ is a classical observable corresponding to $\Sigma_R(t)$.
We have either an exponential decay 
\begin{equation}
F(t) \approx F_{\rm plat}
\exp{\left(-\frac{\eps^4}{2\hbar^2} \sigma t\right)},\quad
t < t_{\rm H}
\label{eq:Fexp}
\end{equation}
or a (perturbative) Gaussian decay 
\begin{equation}
F(t) \approx F_{\rm plat}
\exp{\left(-\frac{\eps^4}{2\hbar^2} \sigma \frac{t^2}{t_{\rm H}}\right)},\quad
t > t_{\rm H},
\label{eq:Fgau}
\end{equation}
and the crossover again happens at the Heisenberg time $t_{\rm H}$.
This is just the time when the integrated correlation function of $R(t)$ becomes dominated by 
quantum fluctuation. Comparing the two factors in (\ref{eq:Fexp},\ref{eq:Fgau}), {\it i.e.} the 
fluctuations of two terms in (\ref{eq:BCH}), we obtain a semiclassical estimate of $t_2$
\begin{equation}
t_2 \approx {\rm min}\left\{\sqrt{\frac{t_H}{\sigma}}\frac{\kappa_{\rm cl}}{\eps},
\frac{\kappa^2_{\rm cl}}{\sigma \eps^2}\right\}, \label{eq:t2}
\end{equation}
where $\kappa_{\rm cl}$ is dispersion of classical observable corresponding to 
$W$,
\begin{equation}
\kappa_{\rm cl}^2=\ave{w^2}_{\rm cl}-\ave{w}_{\rm cl}^2.
\label{eq:kappacl}
\end{equation}
Interestingly, the exponential regime (\ref{eq:Fexp}) can only take place if 
$t_2 < t_{\rm H}$. If one wants to keep $F_{\rm plat} \sim 1$, or have 
exponential decay in the full range until $F \sim 1/{\mathcal N}$, this implies 
a condition on dimensionality: $d \ge 2$. The quantum fidelity and its plateau 
values have been expressed entirely (in the leading order in $\hbar$) in terms 
of classical quantities. While the prefactor $F_{\rm plat}$ depends on the 
details of the initial state, the exponential factors of 
(\ref{eq:Fexp},\ref{eq:Fgau}) 
{\em do not}. Yet, the freezing of fidelity is a purely quantum phenomenon. 
The corresponding {\em classical fidelity} (discussed in Section~\ref{C}) 
does not exhibit freezing. Let us illustrate our theoretical findings by 
numerical examples.

\begin{figure}
\centerline{\includegraphics[angle=-90,width=\figw\textwidth]{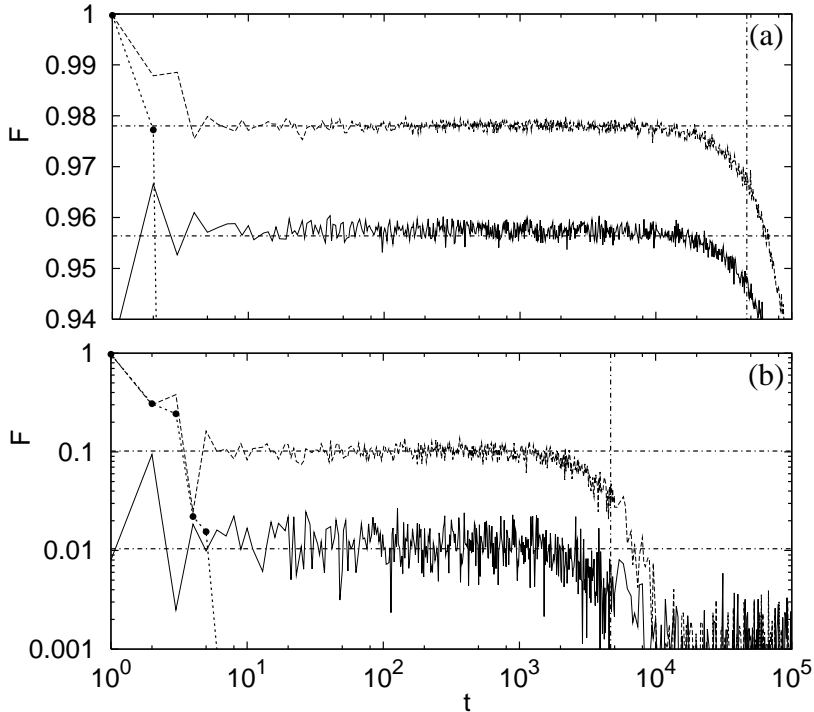}}
\caption{$F(t)$ for the kicked top, with $\eps=10^{-3}$ (a), and 
  $\eps=10^{-2}$ (b), upper curves (dashed) for CIS and the lower curves (full) 
  for RIS. Horizontal lines are theoretical plateau values (\ref{eq:NLRP}), 
  vertical lines are theoretical values of $t_2$ (\ref{eq:t2}). Points 
  represent calculation of the corresponding classical fidelity for CIS which 
  follows quantum fidelity up to the Ehrenfest ($\log\hbar$) barrier and 
  exhibits no freezing.}
\label{fig:1ktop}
\end{figure}

First we consider a quantized kicked top as an example of a one-dimensional 
system ($d=1$) with spin quantum number $S=1000$. The model is described in 
Appendix~\ref{sec:ktop}, Eqs.~(\ref{eq:KT2def},\ref{eq:W1},\ref{eq:R1}). In 
Fig.~\ref{fig:1ktop} we show that the analytical expressions in 
Eq.~(\ref{eq:NLRP}) for the plateau values agree very well with numerical 
results, not only for weak perturbation $\eps=10^{-3}$ shown in 
Fig.~\ref{fig:1ktop}(a) where linearized (linear response) expressions for the 
plateau values could be used, but also for strong perturbation $\eps=10^{-2}$ 
as shown in Fig.~\ref{fig:1ktop}(b). Integration over the sphere yields 
$G(\eps J)=
  \sqrt{\frac{\pi}{2 \eps J}} {\rm erf}(e^{\ii \pi/4} \sqrt{\eps J/2}).$
Comparing with the asymptotic general formula (\ref{eq:Gasym}) for $G(z)$
we now also find a diffractive contribution due to the oscillatory behavior of 
the complex Error function. Small (quantum) fluctuations around the theoretical 
plateau values in Fig.~\ref{fig:1ktop} lie beyond the leading order 
semiclassical description. In Fig.~\ref{fig:1ktop} we also demonstrate that 
the semiclassical formula (\ref{eq:t2}) for $t_2$ works very well.

\begin{figure}
\centerline{\includegraphics[angle=-90,width=\figw\textwidth]{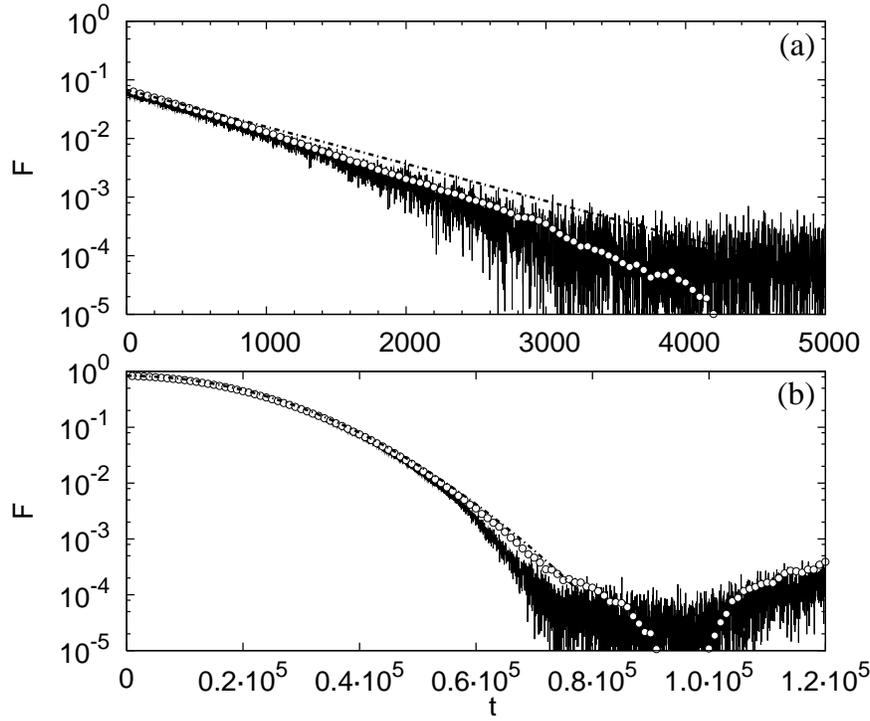}}
\caption{Long-time fidelity decay in two coupled kicked tops. For strong 
perturbation $\eps=7.5\cdot 10^{-2}$ (a) we obtain an exponential decay, and
for smaller $\eps=2\cdot 10^{-2}$ (b) we have a Gaussian decay.
Chain curves give semiclassical expressions (\ref{eq:Fexp},\ref{eq:Fgau})
using classical inputs, full curves give direct numerical simulations and
open circles give ``renormalized'' numerics in terms of renormalized
perturbation operator $R$ and perturbation strength $\eps_R=\eps^2/2$. }
\label{fig:2ktop}\end{figure}

To demonstrate the Gaussian and exponential long-time decay of fidelity 
(\ref{eq:Fexp}) with renormalized perturbation strength we look at a system of 
two ($d=2$) coupled tops described in Appendix~\ref{sec:ktop}, 
Eqs.~(\ref{eq:2KTdef}-\ref{eq:2W}). Here we work with the spin quantum number 
$S=100$. The results of the numerical simulations are shown in 
Fig.~\ref{fig:2ktop}. We show only the long-time decay, since at short times,
the behavior is qualitatively the same as for $d=1$. If the perturbation is
sufficiently strong, one obtains an exponential decay as shown in 
Fig.~\ref{fig:2ktop}(a), while for a smaller perturbation we obtain a Gaussian 
decay as shown in Fig.~\ref{fig:2ktop}(b). Numerical data have been 
successfully compared with the theory (\ref{eq:Fexp},\ref{eq:Fgau}) using 
classically calculated $\sigma=9.2 \cdot 10^{-3}$, and with the 
``renormalized'' numerics using the operator $R$. 

A similar phenomenon as quantum freeze has been observed in Ref.~\cite{Bevilaqua:04} for small perturbations consisting of a phase space displacement. 

\subsection{Composite systems}

\label{sec:chcomposite}

As for other measures of echo-dynamics for mixing and ergodic dynamical 
system we again assume a bipartite decomposition of the Hilbert space 
${\mathcal H}={\mathcal H}_{\rm c}\otimes {\mathcal H}_{\rm e}$.
For mixing dynamics the correlations decay and the linear response 
term will grow linearly with time. For large times one can argue that 
$\Sigma(t)$ should look like a random matrix and the terms giving 
the difference between the fidelity and the echo purity and the 
reduced fidelity can be estimated as
\begin{equation}
\frac{\sum_{j=2}^{{\mathcal N}_{\rm c}}{ \left| \bracket{j;1}{\Sigma(t)}{1;1}\right|^2}}
{\bracket{1;1}{\Sigma^2(t)}{1;1}}\sim \frac{\sum_{j}{|[\Sigma(t)]_{(j;1),(1;1)}|^2}}
{\sum_{j,\nu}{|[\Sigma(t)]_{(1;1),(j,\nu)}|^2}}\sim \frac{1}{{\mathcal N}_{\rm e}},
\label{eq:rm}
\end{equation}
because there are more terms in the sum for fidelity. Therefore we can
estimate the difference $\Fp-F^2(t)\sim 1/{{\mathcal N}_{\rm c}}+1/{{\mathcal
N}_{\rm e}}$ and $\Fr-\Fn \sim 1/{{\mathcal N}_{\rm c}}$. Provided both
dimensions ${\mathcal N}_{\rm c,e}$ are large and for sufficiently long
times, so the ``memory'' of the initial state is lost, we can expect the
decay of all three quantities to be the same.

Discussing linear response results (Section~\ref{sec:LRE}) in the case
of mixing dynamics we have shown that the linear decay is the same for all 
three quantities.  Similar random matrix arguments as for the linear 
response can be used also for higher order 
terms and therefore one expects that in the semiclassical limit of small 
$1/{{\mathcal N}_{\rm c}}+1/{{\mathcal N}_{\rm e}}$ we will have the same exponential decay (\ref{eq:Fnmixing})
\begin{equation}
 F_{\rm P}(t) \approx F_{\rm R}^2(t) \approx F^2(t)=\exp{(-2 t/\tau_{\rm m})},
\label{eq:three}
\end{equation}
with the decay time $\tau_{\rm m}=\hbar^2/2\eps^2\sigma_{\rm cl}$
(\ref{eq:Fnmixing}) independent of the initial state. This result is
expected to hold when $\Sigma(t)$ can be approximated with a random
matrix for large times (\ref{eq:rm}) and $V$ does not contain terms
acting on only one subspace. Such terms could cause fidelity to
decay while having no influence on purity.

\begin{figure}[h]
\centerline{\includegraphics[width=\figw\textwidth]{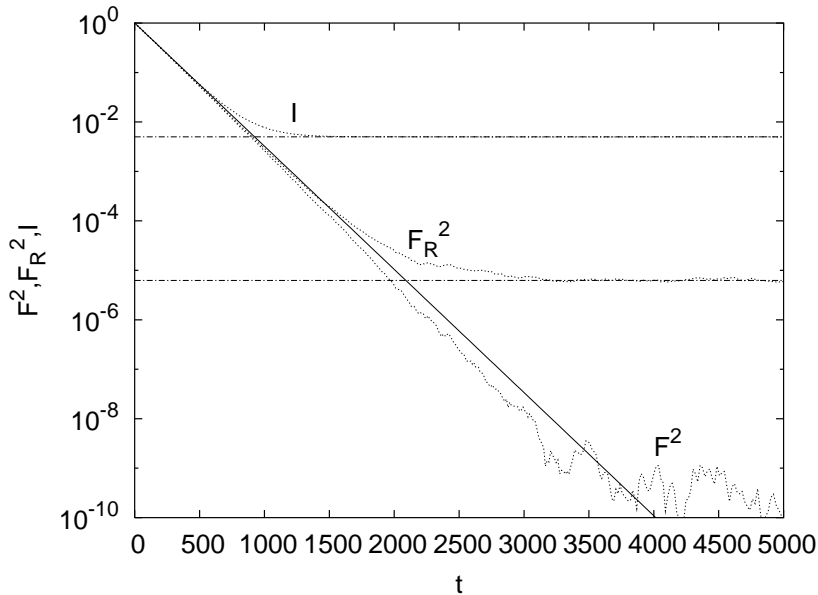}}
\caption{Decay of $F^2(t),F^2_{\rm R}(t)$ and $I(t)$ (dotted curves) in the 
  mixing regime of the double kicked top. The solid line gives the theoretical 
  exponential decay (\ref{eq:three}) with $\tau_{\rm m}$ calculated from the 
  classical $\sigma_{\rm cl}=0.056$. Horizontal chain lines give the saturation 
  values of the purity and the reduced fidelity, $1/200$ and $1/400^2$, 
  respectively.}
\label{fig:fid3030}
\end{figure}

For a numerical demonstration of this result we chose for the system of coupled
kicked tops such parameters that the corresponding classical dynamics is 
practically completely chaotic and mixing. The exact form of the perturbation 
and the parameter values are given in Appendix~\ref{sec:ktop}, 
Eq.~(\ref{eq:4KTdef}). The unperturbed system consists of two uncoupled kicked 
tops such that the coupling is due to the perturbation. In this case, echo 
purity $F_P(t)$ is the same as purity $I(t)$ of the forward evolution of the 
perturbed system. The perturbation strength and the spin size were chosen as 
$\eps=8\cdot 10^{-4}$ and $S=200$, respectively, such that 
${\mathcal N}_{\rm c,e}=2S+1=401$. The initial state was chosen as a product
of two coherent states placed at the positions
$\vartheta_{\rm c,e}^*=\pi/\sqrt{3}$, $\varphi_{\rm c,e}^*=\pi/\sqrt{2}$
on the canonical sphere, for both tops. We show in Fig.~\ref{fig:fid3030} the 
decay of the fidelity $F(t)$, the reduced fidelity $\Fr$, and the purity 
$I(t)$. Clean exponential decay is observed in all three cases, on a time scale 
$\tau_{\rm m}$ (\ref{eq:three}) given by the classical transport coefficient 
$\sigma_{\rm cl}$. We numerically calculated the classical correlation function
\begin{equation}
C_{\rm cl}(t)=\lbrack \ave{z_{\rm c}(t)z_{\rm c}(0)}_{\rm cl} \rbrack^2,
\end{equation}
where we took into account that the unperturbed dynamics is uncoupled 
and is the same for both subsystems and that $\ave{z}_{\rm cl}=0$. Taking 
only the first term $C_{\rm cl}(0)=1/9$ would give $\sigma_{\rm cl}=1/18$ 
(\ref{eq:sigmadef}) while the full sum of $C_{\rm cl}(t)$ gives a slightly 
larger value $\sigma_{\rm cl}=0.056$. Exponential decay persists 
up to the saturation value determined by the dimension of the Hilbert space.

We also wish to illustrate what happens if dimension of one of the
subspaces, say of the central system ${\mathcal N}_{\rm c}$ is not large,
but we only let the dimension of the environment ${\mathcal N}_{\rm e}$
become large. For this purpose we choose a model of a kicked Ising
spin chain for parameter values for which the model is non-integrable
and operates in the regime of quantum chaos (see Appendix~\ref{sec:KI}).

\begin{figure}
\centerline{\includegraphics[width=\figw\textwidth]{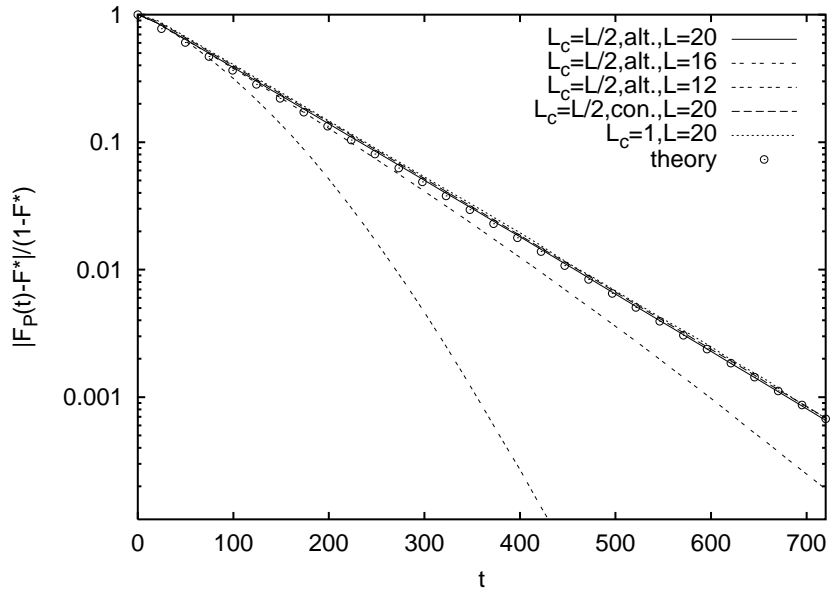}}
\caption{Decay of echo purity in kicked Ising chain in the mixing regime $h_z=1.4$ for different 
  types of division [full curve, broken dashed curves $=$ central system is 
  composed of every other spin, dashed curve $=$  central system is a
  connected half of the chain, dotted curve $=$ central system is a 
  {\em single} spin] and different sizes $L$ (indicated in the legend) and 
  $\eps'=0.01$. Theoretical decay (for $L\to\infty$) is given by sampled 
  symbols.}
\label{fig:fpe}
\end{figure}

We note that in the linear response expression for purity decay, we have
a reduced slope of purity echo increase with a factor 
$1-({\mathcal N}_{\rm e}+{\mathcal N}_{\rm c})/
({\mathcal N}_{\rm e}{\mathcal N}_{\rm c})$ with respect to $[F(t)]^2$, 
whereas for long times, the decay of purity, with the appropriate plateau
value subtracted and rescaled, is determined by the 
asymptotic decay of the echo operator, squared, {\it i.e. }
\begin{equation}
\frac{F_{\rm P}(t)-F_{\rm P}^*}{1-F_{\rm P}^*} \approx [F(t)]^2.
\label{eq:112}
\end{equation}
This is illustrated in Fig.~\ref{fig:fpe} for different cases of division
of the spin chain into the central system and the environment, and
discussed in detail in Ref.~\cite{Prosen:03corr}.

For a related work on entanglement growth in chaotic composite systems 
see~\cite{Tanaka:02,Fujisaki:03}. Some other aspects of entanglement in 
chaotic systems have been studied in Refs.~\cite{Wang:04,Scott:03,Karkuszewski:02,Ghose:04,Bandy:02,Bandy:02b,Pineda:05,Pin06,Gor06,Izrailev:04,XWang:04}.

\subsection{Semiclassical theories in terms of classical orbits}
\label{Q2S}

In the semiclassical limit (small $\hbar$) quantum fidelity can be calculated 
using the semiclassical expression for the quantum propagator in terms of 
classical orbits. Historically this was the method used by Jalabert and 
Pastawski~\cite{Jalabert:01} to derive the perturbation independent Lyapunov 
decay of quantum fidelity. They used the perturbation consisting of a uniform 
distribution of scattering potentials with a Gaussian spatial dependence. It is 
noted that in the absence of the perturbation $F(t)\equiv 1$ is recovered 
exclusively if only diagonal terms are retained, {\it i.e.} in the notation we use 
later this means orbits for which $j_\eps=j_0$, resulting in a single sum over 
orbits (instead of double). For nonzero perturbation averaging over impurities 
is argued to ``select'' only diagonal terms. Later it was shown that the same 
Lyapunov decay is found also for the classical fidelity~\cite{Veble:04}. This 
suggests that quantum Lyapunov decay is a consequence of the quantum-classical 
correspondence; see Section~\ref{C} for more details. The correspondence will 
hold until the Ehrenfest time when a wave packet is spread over a sufficiently 
large portion of Hilbert space and quantum interferences become important. We 
shall discuss in detail parameter ranges and time scales when various regimes 
of fidelity decay occur in Section~\ref{sec:timescales}, and we shall see that 
a Wigner function representation proposed in Ref.~\cite{Cucchietti:03I} 
illustrates this nicely. For a detailed discussion of the range of the Lyapunov decay in a Lorentz gas with disorder see Ref.~\cite{Cucchietti:04}. Fidelity 
decay for disordered system with diffractive scatterers has been studied
using a diagrammatic expansion in Ref.~\cite{Adamov:03}. It is worth mentioning that for sufficiently strong perturbations ($\hbar < \eps < \sqrt{\hbar}$) there are large fluctuations \cite{Silvestrov:03,Petitjean:05} and as a consequence the average fidelity decay depends on the way we average. For typical initial packets the evolution is ``hypersensitive'' to perturbations, resulting in a double exponential fidelity decay~\cite{Silvestrov:03}, $F \propto \exp{(-{\rm const}\times {\rm e}^{2\lambda t})}$. Semiclassics has been used in Refs.~\cite{Cerruti:02,Jacquod:01} to 
derive the so-called Fermi golden rule decay (\ref{eq:Fnmixing}), 
see also~\cite{Cerruti:03,Cerruti:03a}. The Fermi golden rule decay has been 
derived independently using correlation function formalism in 
Refs.~\cite{Prosen:02,Prosen:02corr}. Fermi golden rule type expressions 
involving the correlation function have also appeared as an intermediate step 
in Ref.~\cite{Jacquod:01}. Lyapunov decay in a Lorentz gas was the subject of a
numerical study in Ref.~\cite{Cucchietti:02a}. Lyapunov as well as Fermi golden 
rule regimes in a stadium billiard with disorder have been studied in 
Ref.~\cite{Cucchietti:02b}. The transition between Fermi golden rule and 
Lyapunov decay in a Bunimovich stadium has been numerically considered in 
Ref.~\cite{Wisniacki:02} while the short time decay of fidelity in the same 
system has been discussed using local density of states in 
Ref.~\cite{Wisniacki:03}. The importance of the delicate interplay between 
classical perturbation theory and the structural stability of manifolds in 
chaotic systems has been stressed in selecting the diagonal terms. The 
argumentation has been further elaborated in Ref.~\cite{Vanicek:03}, resulting 
in the so-called dephasing representation~\cite{Vanicek:04,Vanicek:05}. Using 
the dephasing representation it is possible to calculate quantum fidelity decay numerically ranging from the Fermi golden rule to the Lyapunov regime. For strong perturbations one can use dephasing
representation to investigate deviations from a simple Lyapunov decay for chaotic systems for which the
hyperbolic stretching rate is not uniform across the phase space \cite{Wang:05}, or perhaps even more interestingly, to investigate new types of perturbation dependent fidelity decay in 'weakly' chaotic systems with classically diffusive behaviour \cite{Wang:05b}.

Semiclassics can also be used to calculate purity. For chaotic systems the
exponential decay of purity has been derived in~\cite{Jacquod:04}, confirming 
earlier predictions in Refs.~\cite{Zurek:91,ZurHab93,Znidaric:03}. Chaotic time dependent oscillator is treated in Ref.~\cite{Iomin:04}.

Instead of using classical orbits in semiclassical derivations, e.g. in the
Van Vleck-Gutzwiller propagator, one can directly use the formalism of Weyl 
quantization to obtain a semiclassical expression of fidelity. Such an approach 
has been used in Ref.~\cite{Bolte:05} where the propagation of Gaussian 
wave packets is studied. Two terms are identified, the position of the center 
of the packet accounts for its translation while the metaplectic operator takes 
care of the dispersion. The metaplectic operator represents the symplectic 
transformation ({\it i.e.} canonical transformation) of the linearized motion around 
the central orbit. Independently, a similar approach has been used in 
Ref.~\cite{Combescure:05b}. 

In the present section we give a brief derivation of the semiclassical 
expression for fidelity in terms of the Wigner function of the initial state, 
the so-called dephasing representation derived in 
Refs.~\cite{Vanicek:04,Vanicek:05}. With $\veb{r},\veb{r}'$ denoting points in 
configuration space, the fidelity amplitude can be written as
\begin{equation}
f(t)=\braket{\psi_0(t)}{\psi_\eps(t)}=\int{\!\!{\rm d}\veb{r}'\, \psi_0^*(\veb{r}';t) 
\psi_\eps(\veb{r}';t)}.
\label{eq:f}
\end{equation}
If $\psi(\veb{r},0)$ is the initial state, the wave function at time $t$ is 
given by
\begin{equation}
\psi_\eps(\veb{r}';t)=\int{\!\!{\rm d}\veb{r}\, \bracket{\veb{r}'}{U_\eps(t)}{\veb{r}}\, 
\psi(\veb{r};0)}.
\label{eq:kernel}
\end{equation}
The step to semiclassics consists in replacing the quantum propagator 
$\bracket{\veb{r}'}{U_\eps(t)}{\veb{r}}$ by its semiclassical approximation 
$K_\eps^{\rm sc}(\veb{r}',\veb{r},t)$,  {\it i.e.}
$\bracket{\veb{r}'}{U_\eps(t)}{\veb{r}} \to K_\eps^{\rm sc}(\veb{r}',\veb{r},t)$. 
The first argument of $K_\eps^{\rm sc}$ will denote the end point after time $t$ 
(quantity with prime) and the second argument the initial point (without prime). 
The semiclassical Van Vleck-Gutzwiller propagator is~\cite{vanVleck:28,Gutzwiller:90},
\begin{equation}
K_\eps^{\rm sc}(\veb{r}',\veb{r},t)=\sum_{j_\eps}{\frac{1}{(2\pi \ii \hbar)^{d/2}}}
\sqrt{C_{j_\eps}} \exp{\left(\frac{\ii}{\hbar} S_{j_\eps}(\veb{r}',\veb{r},t)-\ii 
\frac{\pi}{2} m_{j_\eps} \right)},
\label{eq:K}
\end{equation}
where the sum extends over all orbits $j_\eps$ that connect points $\veb{r}$ and 
$\veb{r}'$ in time $t$,  {\it i.e.} sum over all initial momenta 
$\veb{p}_{j_\eps}$, such that the evolution for time $t$ results in 
$(\veb{r},\veb{p}_{j_\eps}) \xrightarrow{t} (\veb{r}',\veb{p}'_{j_\eps})$. 
For chaotic systems the number of contributing orbits $j_\eps$ will be very large 
already for small times and will grow exponentially with time, because the sum goes 
over all momenta, {\it i.e.} over all energies. The action is given by
\begin{equation}
S_{j_\eps}(\veb{r}',\veb{r},t)=\int_0^t{\!{\rm d}t'L_\eps(\tilde{\veb{r}}(t'),
\dot{\tilde{\veb{r}}}(t'),t')},
\end{equation}
and 
$C_{j_\eps}=|\det{(\partial^2 S_{j_\eps}/\partial \veb{r}'\partial\veb{r})}|$ 
is the absolute value of the Van Vleck determinant. Taking into account that 
$\partial S_{j_\eps}(\veb{r}',\veb{r},t)/\partial \veb{r}=-\veb{p}_{j_\eps}$, 
where $\veb{p}_{j_\eps}$ is the initial momentum, we see that $C_{j_\eps}$ is 
the Jacobian of the transformation between the initial momentum and the final 
position, 
$C_{j_\eps}=|\det{\frac{\partial \veb{p}_{j_\eps}}{\partial \veb{r}'}}|$. The
integer $m_{j_\eps}$ is a Maslov index. Writing the fidelity amplitude 
(\ref{eq:f}) in terms of $K_\eps^{\rm sc}$ and $K_0^{\rm sc}$ 
and taking the expectation value leads to an
expression involving a three-fold integral over the positions: 
$\veb{r}'$ (the position argument of the propagated wave functions at time $t$
and $\veb{r}, \tilde{\veb{r}}$ (the position arguments of the initial states
to be propagated). In addition we have a double sum over orbits $j_\eps$ of 
the perturbed dynamics, connecting 
$(\veb{r},\veb{p}_{j_\eps}) \xrightarrow{t} (\veb{r}',\veb{p}'_{j_\eps})$, 
and orbits $j_0$ of the unperturbed one, connecting $(\tilde{\veb{r}},
  \tilde{\veb{p}}_{j_0}) \xrightarrow{t} (\veb{r}',\tilde{\veb{p}}'_{j_0})$,
\begin{eqnarray}
f(t)&=&\frac{1}{(2\pi \ii \hbar)^d}\int{\!\!{\rm d}\veb{r} {\rm d}\tilde{\veb{r}}\, 
\psi^*(\tilde{\veb{r}};0)} \psi(\veb{r};0) \sum_{j_\eps,j_0}\nonumber \\
&&\times  \int{\!\!{\rm d}\veb{r}'\, \sqrt{C_{j_0} C_{j_\eps}} \exp{\left(\frac{\ii}{\hbar}
\left\{ S_{j_\eps}(\veb{r}',\veb{r},t)- S_{j_0}(\veb{r}',\tilde{\veb{r}},t)\right\}-\ii \frac{\pi}{2} 
\left\{m_{j_\eps}-m_{j_0} \right\} \right)} }.
\label{eq:fsc}
\end{eqnarray}
In the expression for fidelity $F(t)$ we have twice as many terms! The last expression is still 
rather involved but for our case of echo dynamics several simplifications are possible. One is to
use the shadowing theorem (see references in~\cite{Vanicek:04}) to identify those orbits that give the dominating contribution 
to the phase factor and the other is to transform the integration from the final position to the 
initial momentum. 

\begin{figure}[ht!]
\centerline{\includegraphics{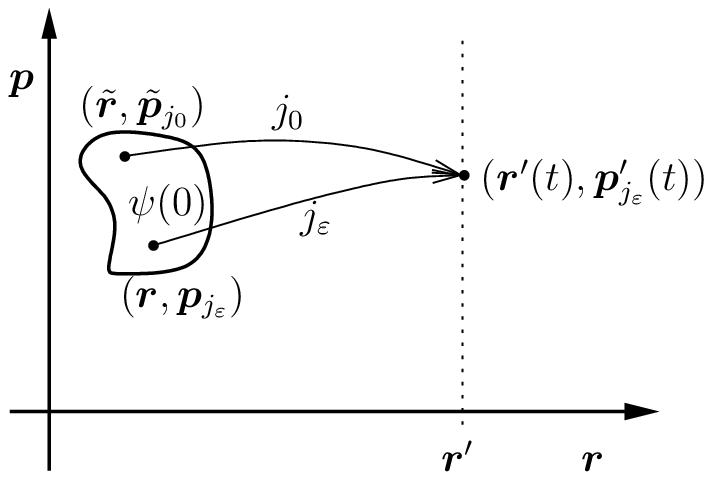}\hskip20pt\includegraphics{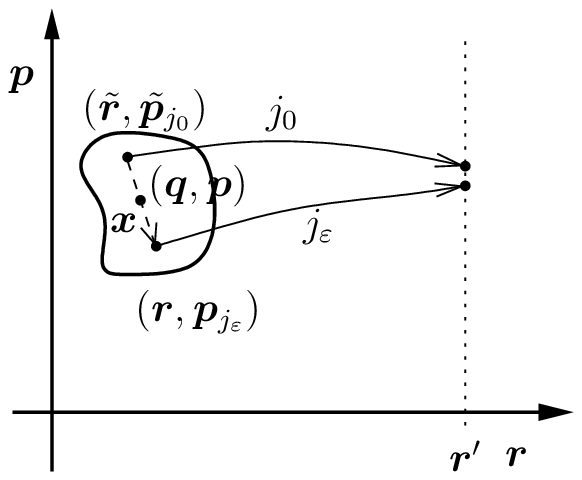}}
\caption{Orbits involved in the semiclassical calculation of the fidelity amplitude 
(\ref{eq:fsc}) after taking into account matching condition (\ref{eq:unitary}). Left 
figure denotes situation in the classical limit of exact matching and 
the right one for finite $\hbar$ where shadowing orbit pairs give major contribution.}
\label{fig:scpic}
\end{figure}
Note that the semiclassical propagator (\ref{eq:K}) does not 
satisfy a semigroup condition, {\it i.e}. 
a product of two propagators $K^{\rm sc}(t)$ for time $t$ is only approximately equal 
to a propagator for twice the time, $K^{\rm sc}(2t)$. To preserve the semigroup 
property,  {\it i.e.} preserving normalization (unitarity), a product of semiclassical 
propagators must be supplemented by the ``matching'' of momenta of two orbits occurring in 
two propagators. Formally this condition is implemented by 
making a stationary phase approximation in the integral over $\veb{r}'$, 
for details see e.g.~\cite{ChaosBook}. If the semiclassical 
limit $\hbar \to 0$ is taken prior to integration, the stationary phase condition 
$\partial S_{j_\eps}/\partial \veb{r}'-\partial S_{j_0}/\partial \veb{r}'$ has to be 
obeyed exactly. Exact matching indeed gives the decay of classical 
fidelity (see Section~\ref{C}). The matching condition at time $t$, 
\begin{equation}
\tilde{\veb{p}}'_{j_0}=\veb{p}'_{j_\eps},
\label{eq:unitary}
\end{equation}
forces the initial condition for the unperturbed orbit 
$(\tilde{\veb{r}},\tilde{\veb{p}}_{j_0})$ to be implicitly determined trough the 
matching condition for a forward perturbed evolution from 
$(\veb{r},\veb{p}_{j_\eps})$ followed by a backward unperturbed evolution. 
This suppresses one space integration over ${\rm d}\tilde{\veb{r}}$ 
in (\ref{eq:fsc}). The orbits involved in calculating this classical fidelity are 
schematically shown in the left panel of Fig.~\ref{fig:scpic}. The action difference 
$\Delta S(\veb{r},\veb{p},t)$ after forward perturbed evolution followed by backward 
unperturbed evolution is given by
\begin{equation}
\Delta S(\veb{r},\veb{p},t)=S_{j_\eps}(\veb{r}',\veb{r},t)- S_{j_0}(\veb{r}',\tilde{\veb{r}},t)=
\int_0^t{\!{\rm d}t'L_\eps(\veb{r}(t'),\dot{\veb{r}}(t'))}+\int_t^0{\!{\rm d}t'L_0(\veb{r}(t'),
\dot{\veb{r}}(t'))},
\label{eq:dS}
\end{equation}
where the initial condition for the second integral with $L_0$ is the final orbit position 
of the first integral. This difference will, in general, be large for chaotic 
systems, because perturbed and unperturbed orbits will explore vastly different 
regions of phase space. For sufficiently large time the action difference will 
depend only on the chaotic properties of the system and
not on the perturbation strength. This results in the perturbation independent 
decay of classical fidelity.

For finite $\hbar$,  {\it i.e.} first evaluating the integral and then taking the 
semiclassical limit, the above matching condition (\ref{eq:unitary}) needs to 
hold only approximately. Approximate continuity of momenta for finite 
$\hbar$ means that we have to consider also almost continuous orbits 
for which discontinuity $\tilde{\veb{p}}'_{j_0}- \veb{p}'_{j_\eps}$ is 
sufficiently small. This is shown in the right panel of Fig.~\ref{fig:scpic}.
Of course, the approximate continuity is automatically taken care of if we perform 
the stationary phase integration properly. Yet here we would like to identify 
the dominating contribution by physical arguments, namely using the shadowing theorem. 
It states for chaotic systems that, while 
two orbits starting from the same initial condition will generally exponentially diverge, 
if evolved with two slightly different evolutions, we can find exponentially close initial 
condition for one evolution such that its orbit will closely follow (shadow) the other one. 
Because the two initial conditions~\footnote{In our case there are actually two 
exponentially close ``final'' conditions at time $t$.} are exponentially 
close they should in fact be summed over 
(for finite $\hbar$) in the formula for the fidelity amplitude $f(t)$ (\ref{eq:fsc}). 
As the orbit and its shadow, a 'shadowing pair', closely follow each other, their 
action difference $\Delta S$ will be 
small,  {\it i.e.} $\Delta S \sim \mathcal{O}(\eps)$ whereas one has 
$\Delta S \sim \mathcal{O}(1)$ for non-shadowing pairs.  Therefore shadowing 
pairs cause the largest contribution to $f(t)$. It is thus sufficient to take into 
account shadowing pairs only. To arrive at the final expression we change
the integration variables ${\rm d}\veb{r}$ and ${\rm d}\tilde{\veb{r}}$ in (\ref{eq:fsc}) 
to their difference $\veb{x}=\veb{r}-\tilde{\veb{r}}$ and their mean 
$\bar{\veb{q}}=(\veb{r}+\tilde{\veb{r}})/2$. As the major contribution to 
$f(t)$ results from shadowing pairs, due to their 
small $\Delta S$ the difference $\veb{x}$ will be small and we can expand $\Delta S$ around 
$\bar{\veb{q}}$ to linear order in $\veb{x}$,
\begin{equation}
S_{j_\eps}(\veb{r}',\bar{\veb{q}}+\veb{x}/2,t)- S_{j_0}(\veb{r}',\bar{\veb{q}}-\veb{x}/2,t)=
S_{j_\eps}(\veb{r}',\bar{\veb{q}},t)- S_{j_0}(\veb{r}',\bar{\veb{q}},t)-\veb{x}\bar{\veb{p}}+ \cdots.
\label{eq:dsx}
\end{equation}
where we denoted the mean initial momentum by 
$\bar{\veb{p}}=(\veb{p}_{j_\eps}+\tilde{\veb{p}}_{j_0})/2$.
This expansion does not depend on the initial state being localized. Next, we 
transform the integration over the final positions $\veb{r}'$ to an integration 
over initial mean momenta $\bar{\veb{p}}$. Together with the original Jacobian prefactor $\sqrt{C_{j_0} C_{j_\eps}}$ this gives $\sqrt{|\det{\frac{\partial \veb{p}_{j_\eps}}{\partial \bar{\veb{p}}}}|
|\det{\frac{\partial \tilde{\veb{p}}_{j_0}}{\partial \bar{\veb{p}}}}|}$ which though is close to one for shadowing orbits. For shadowing orbits Maslov indices are equal, $m_{j_\eps}=m_{j_0}$. We thus obtain
\begin{eqnarray}
f(t)=\frac{1}{(2\pi \ii \hbar)^d}\int && \!\!{\rm d}\bar{\veb{q}}{\rm d}\bar{\veb{p}}
{\rm d}\veb{x}\, \psi^*(\bar{\veb{q}}-\frac{\veb{x}}{2};0) \psi(\bar{\veb{q}}+
\frac{\veb{x}}{2};0)\times \nonumber \\
&&\times \exp{\left(\frac{\ii}{\hbar} \left\{ S_{j_\eps}(\veb{r}',\bar{\veb{q}},t)- 
S_{j_0}(\veb{r}',\bar{\veb{q}},t) \right\}  \right)} \exp{\left(-\frac{\ii}{\hbar}\bar{\veb{p}}
\veb{x} \right)}.
\label{eq:f1}
\end{eqnarray}
The action difference 
$S_{j_\eps}(\veb{r}',\bar{\veb{q}},t)- S_{j_0}(\veb{r}',\bar{\veb{q}},t)$ 
is calculated by using forward perturbed orbit $j_\eps$ followed by an unperturbed 
backward evolution along shadowing orbit $j_0$, where the initial condition is such that 
$(\bar{\veb{q}},\bar{\veb{p}})=(\veb{r}+\tilde{\veb{r}},\veb{p}_{j_\eps}+\tilde{\veb{p}}_{j_0})/2$, 
the so-called center chord in Weyl formalism~\cite{Almeida:98}. By transforming 
to initial momenta $\bar{\veb{p}}$ we have eliminated one summation over 
momenta in Eq.~(\ref{eq:fsc}). Therefore, we should sum over all initial conditions 
with the same center chord $(\bar{\veb{q}},\bar{\veb{p}})$. As shadowing orbits are 
very close to each other there will be typically only one orbit having a given 
center chord and furthermore, to lowest order, we can assume that both orbits are 
actually the same for the purpose of calculating the action difference. For this 
single orbit we can take the initial condition to be  
the center chord $(\bar{\veb{q}},\bar{\veb{p}})$. Combining all these simplifications for
shadowing orbits and noting that $L_\eps-L_0=-\eps V$, 
if $V$ is the perturbing potential, we find 
$S_{j_\eps}(\veb{r}',\bar{\veb{q}},t)- S_{j_0}(\veb{r}',\bar{\veb{q}},t) 
\approx \int_0^t{\!{\rm d}t' V(\bar{\veb{q}}(t'),\bar{\veb{p}}(t'),t')}$. 
We recognize the Wigner function for the state $\psi(\veb{r};0)$ in Eq.~(\ref{eq:f1}),
\begin{equation}
W_{\rho}(\bar{\veb{q}},\bar{\veb{p}})=\frac{1}{(2\pi\hbar)^d}\int{\!\!{\rm d}\veb{x}\, 
\psi^*(\bar{\veb{q}}-\frac{\veb{x}}{2};0) \psi(\bar{\veb{q}}+\frac{\veb{x}}{2};0) 
\exp{\left( -\frac{\ii}{\hbar} \bar{\veb{p}}\,\veb{x}\right)}},
\end{equation}
and we finally get
\begin{equation}
f(t)=\int{\!\!{\rm d}\bar{\veb{q}}{\rm d}\bar{\veb{p}}\, W_{\rho}(\bar{\veb{q}},\bar{\veb{p}}) 
\exp{\left(-\frac{\ii}{\hbar} \eps \int_0^t{\!\!{\rm d}t' V(\bar{\veb{q}}(t'),
\bar{\veb{p}}(t'),t')} \right)} }.
\label{eq:fDP}
\end{equation} 
The final result (\ref{eq:fDP}) is the so-called dephasing 
representation~\cite{Vanicek:04,Vanicek:05}. It has a very appealing form: 
fidelity is given as the Wigner function average of the phases 
due to action differences. The importance of this expression is that it is 
valid from the Lyapunov to the Fermi golden rule regime. It can also in general describe fidelity decay in regular systems, e.g., Gaussian decay or even decay after plateau in quantum freeze~\cite{Vanicek:06}. In particular it is very handy for numerical evaluation of quantum fidelity decay in various non-generic situations or for complicated initial states. It should be also useful for systems with many degrees of freedom due to favorable scaling of its running time as compared to exact quantum calculations. 

The statistics of action differences has been discussed in Ref.~\cite{Vanicek:04b}. 
While in practice the numerical evaluation of the dephasing representation 
(\ref{eq:fDP}) turns out to reproduce the exact quantum fidelity decay in chaotic, regular as well as in mixed systems very well~\cite{Vanicek:04,Vanicek:05}, there are nevertheless some limitations. For instance, it is unable to describe the perturbative Gaussian decay for small 
perturbations which occurs due to finite size effects~\cite{Vanicek:03}. 

The exponential term in the expression for $f(t)$ (\ref{eq:fDP}) is very reminiscent of 
our approximate quantum echo operator 
$M_\eps(t)\approx \exp{(-\ii \eps \Sigma(t)/\hbar)}$ (\ref{eq:BCH}) 
where $\Sigma(t)$ is the integral of the perturbation (\ref{eq:defsig}). 
Indeed, the semiclassical fidelity amplitude could be 
obtained directly, using the Weyl-Wigner quantization~\cite{Almeida:98},  {\it i.e}. 
replacing $M_\eps(t)$ by its Weyl symbol and the density matrix $\rho$ 
by its Wigner function $W_{\rho}$ we recover the dephasing representation.

\section{Random matrix theory of echo dynamics}
\label{R}

%Title of section: Random Matrix theory of echo dynamics}

From the previous section we have a surprising dichotomy. We have two slightly
different approaches which yield for weak and very weak perturbation
respectively the Fermi Golden rule and the perturbative regime. Both results
are based on perturbation expansions, and there should be a unified theory to
describe these regimes.

As random matrix models have been extremely successful in describing a wide
field of phenomena in physics ranging from elasticity to particle
physics~\cite{Bro81,Guhr:98} associated in some sense with chaos,
it seems natural to attempt a formulation of echo dynamics in this
framework. A number of papers have appeared on this
% subject~\cite{Cerruti:03,Cerruti:03a,GPS04,StoSch04b,StoSch04,Frahm:04,GSW05,
%  GKPSSZ05,Heiner06}.
subject~\cite{Cerruti:03,Cerruti:03a,Frahm:04}.
We shall basically follow the lines of~\cite{GPS04}, which uses
the linear response approximation expressed in terms of correlation integrals,
very similar to our reasoning in Section~\ref{DF}. This method provides a
uniform approximation encompassing both regimes. It proved successful in
explaining two independent experiments (see Section~\ref{E}). Some exact
results for the random matrix theory (RMT) model used are also
available~\cite{StoSch04b,StoSch04,GKPSSZ05,Heiner06}.
We give the outlay of the model and derive a linear response formula for
fidelity decay extending its validity to long times in a form valid strictly
in the weak perturbation limit. We compare this result to new more detailed
calculations or the kicked top and to the exact RMT solution. We then present
a considerable body of new material extending this model to purity decay, and
with modified perturbations we discuss various situations where quantum freeze
can occur~\cite{GKPSSZ05,Heiner06,Pin06}.

The RMT model we present, allows to describe
fidelity decay in a chaotic system, under a static global perturbation;
that model could be extended to treat also noisy perturbations, but at least
in the case of uncorrelated noise, a direct statistical treatment is more
adequate (see Section~\ref{sec:noisypert}).
Chaoticity is meant to justify choosing the unperturbed system from one of
the Gaussian invariant ensembles~\cite{CVG80,BGS84,LS92}.
With the word ``global'' we mean that in the eigenbasis of the unperturbed
Hamiltonian, the perturbation matrix is not sparse. Banded matrices may occur,
as long as their bandwidth is so large, that a full matrix would yield similar
results. We shall mainly concentrate on this case, because it is most
amenable to analytic treatment. Where other Hamiltonians are used we shall
point this out and explain why it is necessary.

We consider a perturbed Hamiltonian in a form, typical for
RMT (see e.g. \cite{SimAlt93,Dietz:96}):
\begin{equation}
H_{\eps'} = \cos(\eps ')\; H_0 + \sin(\eps ')\; H_1 \: ,
\label{R:Hcos}\end{equation}
where $H_0$ and $H_1$ are chosen from one of Cartan's classical
ensembles~\cite{Cartan:35,Guhr:98}. %TG characterized by the index $\beta$.
This scheme has the advantage that the perturbation does not change the level
density of the Hamiltonian, and thus
%TG does not make necessary the normalization of time in
avoids the need to normalize time in the
echo dynamics, as discussed at the end of Section~\ref{sec:conslaws}.
We will generally be interested in situations where $\eps'$ scales as
$1/\sqrt{N}$, where $N$ denotes the dimension of the Hamiltonian matrices.
In this case, the matrix elements of the perturbation couple a finite number of
neighboring eigenstates of the unperturbed system, largely independent on $N$.
This assumption allows to describe all regimes, from $\eps'=0$ up to the Fermi
golden rule regime, and beyond. For large $N$, we may therefore linearize the
trigonometric functions in Eq.~(\ref{R:Hcos}). It is then convenient, to
fix the average level spacing of $H_0$ to be one in the center of the spectrum,
and to require that the off-diagonal matrix elements of $V= H_1/\sqrt{N}$ have
unit variance such that
\begin{equation}
H_\eps = H_0 + \eps\; V \; ,
\label{R:Heps}\end{equation}
where $\eps = \sqrt{N}\eps'$.
It is easy to check that corrections to the Heisenberg time are of
order $\mathcal{O}(1/N)$. The matrix
$H_0$ can have special properties (for example in Section~\ref{Q2PE} on
echo purity), but typically it will be derived from a random matrix taken from
the classical ensembles. We may use different ensembles for $H_0$ and $V$.
In many cases, the ensemble of perturbations is invariant under the
transformations that diagonalize $H_0$. We can then choose $H_0$ to be
diagonal with a spectrum $\{E^0_j\}$ with given spectral statistics.
In this situation we can unfold the spectrum that defines
$H_0$, to have average level density one along the entire spectrum, or
we can restrict our considerations to the center of the spectrum. This
restricts us to situations, where the spectral density may be assumed
constant over the energy spread of the initial state. Other cases could be
important, but have, to our knowledge, so far not been considered in RMT.

The one parameter family $H_\eps$ defines echo dynamics as discussed in
Section~\ref{D}. In what follows, the eigenbasis of $H_0$ will be the only
preferred basis except in Section~\ref{Q2PE}, where we deal with entangled
subsystems. Unless stated otherwise, we consider initial states to be
random, but of finite span in the spectrum of $H_0$. Eigenstates of $H_0$ are
one limiting case and random states with maximal spectral span the other.

The spectral span of the initial state and the spreading width
of the $H_0$-eigenstates in the eigenbasis of $H_\eps$, determine the only
relevant time scales. They should be compared to the Heisenberg time, which
has been fixed to $t_H=1$, by unfolding the spectrum of $H_0$. In the limit
$N\to\infty$, the Zeno time (of order $t_H/N$; see Section~\ref{sec:zeno})
plays no role.

Here, we are interested in the decay of fidelity or some other measure
of echo dynamics. The main results cover essentially the range from the
perturbative up to the Fermi golden rule regime~\cite{Jacquod:01,Cerruti:02}.
The analysis of the quantum freeze and an exact analytical result for the
random matrix model will provide additional and/or different regimes.
The Lyapunov regime~\cite{Jalabert:01,Jacquod:01} as
well as the particular behavior of coherent states are certainly not within
the scope of RMT.

We shall find that in many situations the fidelity amplitude is self
averaging; see Eq.~(\ref{Q2LR:F}). Therefore we mainly concentrate on the
fidelity amplitude, and do not bother with the more complicated averages for
fidelity itself.

This section is organized as follows:
The linear response approximation and the exact analytic
treatment of the fidelity amplitude are discussed in Sections~\ref{Q2LR},
and~\ref{Q2SS}, respectively. The quantum freeze case is studied in
Section~\ref{Q2QF}. Composite systems (two coupled subsystems) are
considered in Sections~\ref{Q2PE} and~\ref{Q2PD}.

\subsection{\label{Q2LR} Linear response theory}

Recall that the echo operator $M_\eps(t)$ can be written to second order
in $\eps$ as~\cite{Prosen:02corr,GPS04}
\begin{equation}
M_\eps(t) = \Ione -\rmi\, 2\pi\eps\int_0^t\d t' \; \tilde V(t')
 - (2\pi\eps)^2 \int_0^t\d t'\int_0^{t'}\d t''\; \tilde V(t')\,
  \tilde V(t'') + \mathcal{O}(\eps_0^3) \; .
\label{Q2LR:defM}\end{equation}
To obtain the fidelity amplitude we have to compute the average
of $M_\eps(t)$ with respect to both, the ensemble defining the random
perturbation $V$, and the one defining the spectrum $\{E^0_\alpha\}$ of
$H_0$. Provided that $H_0$ and $V$ are statistically independent, 
the linear term in $\eps$ averages to zero.
The quadratic term is determined by the two-point time correlation function
\begin{equation}
\la [\tilde V(\tau)\,\tilde V(\tau')]_{\nu,\nu'}\ra
= \sum_\mu \la V_{\nu,\mu} \; V_{\mu,\nu'}\ra\;
  \la e^{2\pi \rmi [(E_\nu - E_\mu)\tau+(E_\mu-E_\nu')\tau']} \ra
= \delta_{\nu,\nu'}\;\left\{ \frac{2}{\beta_V}+\delta(\tau-\tau')
 - b_2(\tau-\tau')\right\} \; .
\label{R:dav}\end{equation}
The middle expression contains two separate averages, the first is taken over
the ensemble of $V$ and the second over the spectral ensemble of the
unperturbed Hamiltonian. The average over the matrix elements of $V$ yields
delta functions that allow to rewrite the spectral average in terms of the
two-point correlation function of the spectrum; as the spectral ensemble is
invariant under rearrangements of the eigenenergies, the result is independent
of the index $\nu$. The spectral two-point correlation function is then
translational invariant and the spectral form factor $b_2$~\cite{Mehta:91} is
conveniently introduced. The constant $\beta_V$ depends on the classical
ensemble from which the perturbation was taken. It can take the values 1,2,
and 4 for the orthogonal, unitary, and symplectic ensembles, respectively.

Inserting this result into Eq.~(\ref{eq:fidech}) for the fidelity
amplitude we get \cite{GPS04}
\begin{equation}
\la f_\eps(t)\ra
 = 1-(2\pi \eps)^2\left[ t^2/\beta_V + t/2 - \int_0^t \d\tau'\int_0^{\tau'}
  \d\tau\; b_2(\tau) \right] + \mathcal{O}(\eps^4) \;.
\label{R:rmt-lr}\end{equation} 
Any stationary ensemble from which $H_0$ may be chosen, yields a particular
two-point function $b_2$. The correlation integrals over $b_2$ for the GOE and
the GUE are discussed in~\cite{GPS04}. To demonstrate the effect of spectral
correlations by contrast, we occasionally use a random level sequence. In this
case $b_2(t)=0$ and the last term in~(\ref{R:rmt-lr}) vanishes. Typically
(at least in the case of the classical ensembles), spectral correlations lead
to a positive $b_2$, such that the fidelity decay will be slowed down.

The result~(\ref{R:rmt-lr}) shows two remarkable features: The first is that
the linear and the quadratic term in $t$ scale both with $\eps^2$
The second
is about the two possibly different ensembles used for the perturbation and
for $H_0$. The characteristics of $V$ affect only the prefactor of the
$t^2$-term, while the characteristics of $H_0$ affect only the two-point form
factor $b_2$.

In experiments or numerical simulations averaging over $H_0$ may be
unnecessary, due to the self-averaging properties of fidelity. For this to be
effective, we need an initial state which covers a sufficiently large
number of eigenstates of $H_0$. The two-point form factor $b_2$ can then be
obtained by averaging over energy or frequency intervals (spectral average).

As an expansion in time, Eq.~(\ref{R:rmt-lr})
contains the leading terms for the perturbative
as well as the Fermi golden rule results~\cite{Peres:84,Jacquod:01,Cerruti:02}.
Both are exponentials of the corresponding terms. It is then tempting to simply
exponentiate the entire $\eps^2$-term to obtain
\begin{equation}
\la f_\eps(t)\ra = \exp\left[-(2\pi \eps)^2 \left( t^2/\beta_V + t/2
  - \int_0^t\d\tau'\int_0^{\tau'}\d\tau\; b_2(\tau)\;\right)\right]
\label{Q2LR:expLR} \end{equation}
This expression will prove to be extremely accurate for perturbation strengths
up to the Fermi golden rule regime. Some justification for the exponentiation
is given in~\cite{Prosen:03ptps}. While exponentiation in the perturbative
regime is trivially justified, our result shows that for times $t\ll t_H$, we
always need the linear term in $t$ to obtain the correct answer. In
experiments the interplay of both terms has been proven to be
important~\cite{SGSS05,SSGS05,GSW05}; see also Section~\ref{ES}. On the
other hand, for stronger perturbations fidelity has decayed before Heisenberg
time to fluctuation levels or to levels where our approximation fails.
Comparison with the exact result~\cite{StoSch04b,StoSch04}
(Section~\ref{Q2SS}) will show that the exponentiation allows to extend the
linear response result from a validity of $\la f(t)\ra \approx 1$ to a validity
range of $\la f(t)\ra \gtrsim 0.1$

Note that the pure linear response result~(\ref{R:rmt-lr}) is probably all we
need for quantum information purposes, as processes with fidelity less than $1-\eta$,
where $\eta \sim 10^{-4}$, are not amenable to
quantum error correction schemes~\cite{Nielsen:01}. For considerations about
stability in echo dynamics, on the other hand, the exponentiated
formula~(\ref{Q2LR:expLR}) gives a clear and simple expression. The exact
treatment will show where to expect additional effects, but experiments at
this time are still limited to
$\la f(t)\ra \gtrsim 0.1$~\cite{SGSS05,SSGS05,GSW05}.

Fidelity can be calculated in the linear response approximation along the
same lines as above. One obtains~\cite{GPS04}:
\begin{equation}
\la F_\eps(t)\ra = \la |f_\eps(t)|^2\ra = \la f_\eps(t)\ra^2 + (2\pi\eps)^2\;
  (2/\beta_V)\; {\it ipr}\; t^2\; +\mathcal{O}(\eps^4) \; .
\label{Q2LR:F}\end{equation}
Here ${\it ipr} = \sum_{\nu} |\braket{E_\nu}{\Psi}|^4$ 
indicates the {\em inverse participation ratio} of the initial state
expanded in the eigenbasis of $H_0$. This equation displays two extreme
effects: On the one hand it shows the self-averaging properties of this
system. For states with a large spectral span in $H_0$ the correction term
that marks the difference between $\la F_\eps(t)\ra$  and
$|\la f_\eps(t)\ra|^2$ goes to zero as the
inverse participation ratio becomes small ($\sim 1/N$), see also Eq.~(\ref{eq:Avg4order}) for the difference between the average fidelity amplitude and the average fidelity. 
On the other hand, for an eigenstate of $H_0$, ${\it ipr}=1$, and hence
the quadratic term in
Eq.~(\ref{Q2LR:F}) disappears. Moreover, the correlations cancel the
linear term after the Heisenberg time. Thus, we find that after Heisenberg
time the decay stops for an $H_0$ taken from a GUE and continues only
logarithmically for a GOE~\cite{GPS04}. That situation is very similar to the
quantum freeze considered in Section~\ref{Q2QF}.

Situations involving states with a small spectral span have been analyzed in
Ref.~\cite{HKCG04}. There fidelity decay has been computed for two perturbed
Hamiltonians $H_\eps$ and $H_{-\eps}$, with perturbations of opposite sign.
Under these circumstances, it has been found that the evolving states have
maximal overlap for different evolution times. Note that by construction, the
Heisenberg times are the same for both evolutions, in distinction to the cases
we discuss in the Sections~\ref{sec:conslaws} and~\ref{ES}.

\begin{figure}
\centerline{\input{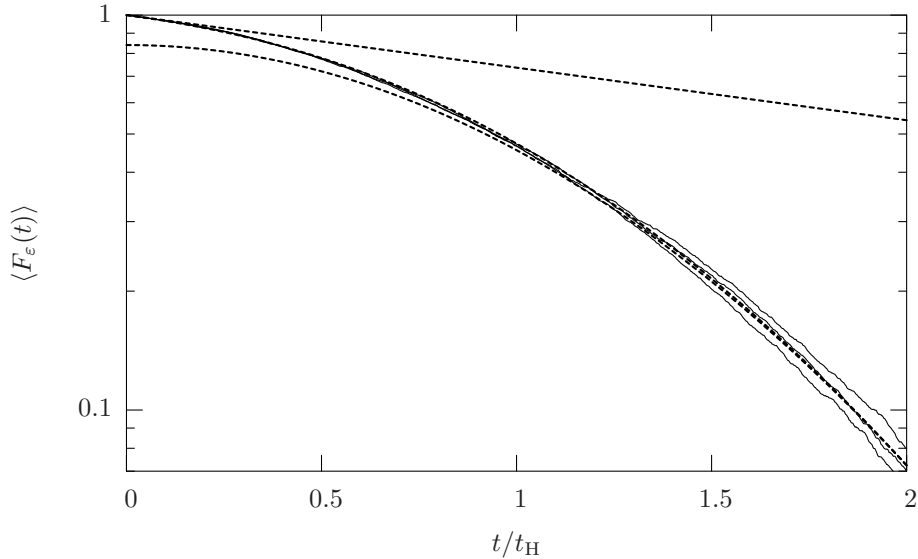}}
\caption{Simulation of fidelity decay for a dynamical model (the desymmetrized
kicked rotor) in the cross-over regime. The three thin solid lines show the
result of the simulation (for three different initial conditions), the dashed
line the corresponding exponentiated linear response approximation. An
additional dashed line shows the exponential part, another one the Gaussian
part of that approximation (for details see text). The Heisenberg time is
$t_{\rm H}=1000$.}
\label{RMT-KT1}\end{figure}

We illustrate the above results with the help of dynamical models,
the kicked rotor and the kicked top (see Appendix~\ref{sec:ktop}).
We show cases, where the time-reversal symmetry is conserved (by both the
unperturbed Hamiltonian $H_0$ and the perturbation $V$), and other cases,
where the time-reversal symmetry is broken (again by both, $H_0$ and $V$). The
former corresponds to a GOE spectrum, perturbed by a GOE matrix, the latter
to a GUE spectrum, perturbed by a GUE matrix.

In Fig.~\ref{RMT-KT1} we show fidelity decay in the cross-over regime. 
The numerical calculations have been carried out by D.~F.~Martinez~\cite{DFM:privcomm}.
The quantum propagator for the kicked rotor is constructed in position space,
such that the Floquet matrix is symmetric~\cite{TwoSch03}.
However, we changed the kick potential from $\cos\varphi$ to
$\cos\varphi - \sin\, 2\varphi$ in order to break the reflection symmetry.
That assures that the time reversal symmetry is the only remaining symmetry
in the system. The perturbation is implemented by a small increment of
the kicking strength $K$, which has been chosen in the regime of complete
chaos $K \approx 10$. The perturbation strength $2\pi\eps\approx 0.554$ has
been computed
from the integrated classical correlation function, $\sigma_{\rm cl}$
as explained in Section~\ref{Q2D}. Thus,
in the comparison of theory and numerical experiment, there is no free
parameter. We plot three curves, for different initial states. Those were
coherent (Gaussian) wave packets located at arbitrarily chosen positions
on the $p= 1/2$ axis in phase space. In this way, the initial states have
real coefficients in the position basis.
%\tgcomment{The Heisenberg time can be easily increased up to $t_H= 10^4$ or
%even $t_H= 10^5$, if necessary.}
The additional dashed lines are just to guide the eye, as to the validity of
the Fermi golden rule and the perturbative expressions, respectively.

\begin{figure}[t!]
\centerline{\includegraphics[angle=-90,width=\figw\textwidth]{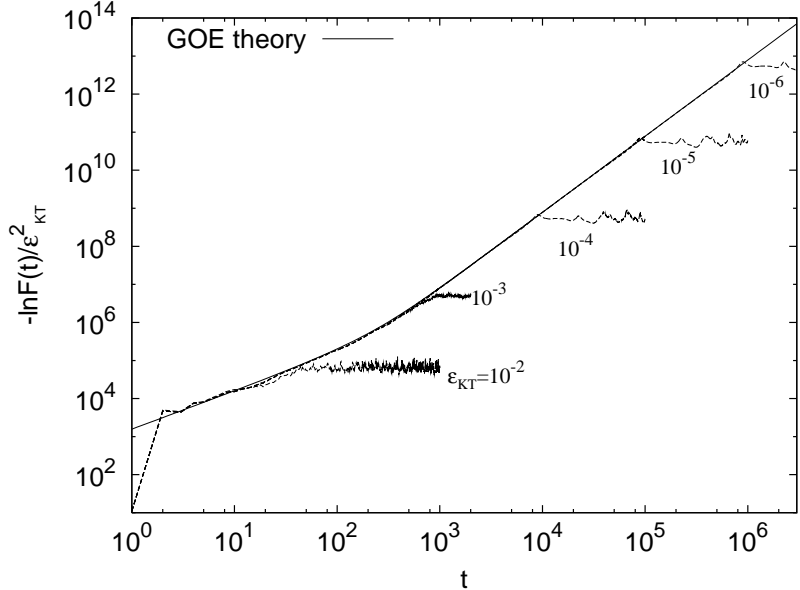}}
\caption{The scaled logarithm of the fidelity as a function of time for
a time-reversal invariant kicked top. The quantity is chosen such that
different perturbation strengths $\eps$ give the same theoretical
curve (solid line), based on the Eq.~(\ref{Q2LR:expLR}).
The crossover between the exponential (Fermi golden rule regime) and
Gaussian decay (perturbative regime) occurs around the Heisenberg time
$t_{\rm H} = 400$. The numerical data for different perturbation strengths are
shown by dashed curves.}
\label{fig:scalingGOE}\end{figure}

\begin{figure}[ht]
\centerline{\includegraphics[angle=-90,width=\figw\textwidth]{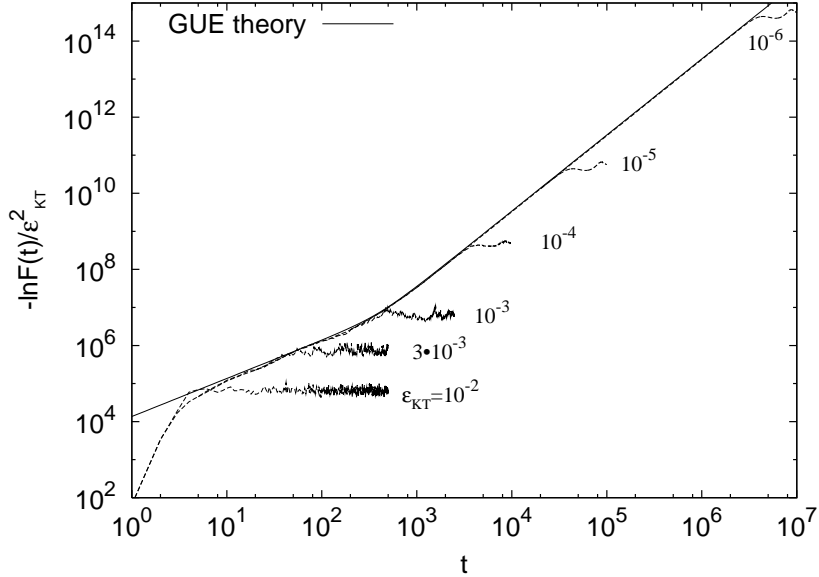}}
\caption{The scaled logarithm of the fidelity as a function of time for
a kicked top, with broken time-reversal invariance (similar graph, as in
Fig.~\ref{fig:scalingGOE}). The theoretical curve (solid line) is based on
the Eq.~(\ref{Q2LR:expLR}), with $\beta_V=2$ and
$b_2(t)$ describing GUE level fluctuations. The Heisenberg time is $t_H=400$.
The numerical data for different perturbation strengths are shown by dashed
curves.}
\label{fig:scalingGUE}\end{figure}

To check explicitly the scaling in the perturbation strength we finally
perform simulations with the kicked top (Appendix~\ref{sec:ktop}). Numerical 
results for time-reversal invariant kicked top with $S=400$ 
(propagator~(\ref{eq:KTdef})) and perturbations ranging over a whole set of 
kicked top perturbation strengths $\eps_{\rm KT} = 10^{-6} $ to 
$\eps_{\rm KT} = 10^{-2}$ are shown in Fig.~\ref{fig:scalingGOE}. The RMT 
perturbation parameter $\eps$ is given in terms of ``physical'' kicked top 
perturbation $\eps_{\rm KT}$ (written simply as $\eps$ in propagators in 
Appendix~\ref{sec:ktop}) as 
$2\pi \eps=S \eps_{\rm KT} \sqrt{2 \sigma_{\rm cl} \mathcal{N}}$, resulting 
in $2\pi \eps$ ranging from $7.9$ to $7.9 \cdot 10^{-4}$ for the shown 
$\eps_{\rm KT}$, see also Section~\ref{sec:ktopRMTfreeze} for further 
details on obtaining $\eps$ from $\eps_{\rm KT}$. Plotting 
$-{\eps_{\rm KT}}^{-2}\, \ln F_\eps(t)$ versus $t$ on a
double-log scale has the consequence that all curves should follow a single
line given by Eq.~(\ref{Q2LR:expLR}). However, the numerical results
deviate at %TG the same line except for
the very end and the very beginning.
Differences for long times are due to the scaling, since the saturation value
of fidelity does actually not depend on $\eps$.  At short times the individual
properties of the dynamical system show up, and short periodic orbits give the
expected non-universal contributions. Figure~\ref{fig:scalingGOE} shows that
for most perturbations considered, the exponentiated linear response expression
fits very well and explains the transition from linear to quadratic behavior.
The strongest perturbation is the exception. Here saturation sets in before
non-generic effects have died out, without leaving any space for generic RMT
behavior.

To complete the picture, we show in Fig.~\ref{fig:scalingGUE} a similar
calculation for a kicked top which breaks the time-reversal symmetry
(Appendix~\ref{sec:ktop}, Eq.~(\ref{eq:ktopguefreeze})) and spin size $S=200$. 
It corresponds to the random matrix
model~(\ref{R:Heps}), where both parts, $H_0$ and $V$ are chosen from a GUE.
Again we see essentially the same features as in Fig.~\ref{fig:scalingGOE}.
The comparison with the exponentiated linear response formula shows
similar agreement as for the GOE case.

\subsection{\label{Q2SS} Supersymmetry calculation for the fidelity
 amplitude}
% Supersymmetric Ansatz by St{\" o}ckmann

The exponentiated linear response formula~(\ref{Q2LR:expLR}) agrees very well 
with dynamical models as we have seen above, and as it has been
reported in the literature~\cite{GPS04,Haug:05}. It also agrees with 
experiments~\cite{SGSS05,SSGS05,GSW05} on which we shall report in 
Section~\ref{ES}.
Nevertheless it is quite clear, that this approach is not justified for
perturbations stronger than one, even if we omit the fact, that the
exponentiation is only heuristically justified. In the last twenty years many
problems in RMT have been solved exactly, and indeed
recently St\" ockmann and Sch\" afer~\cite{StoSch04b,StoSch04} have solved
the model given in Eq.~(\ref{R:Heps}) exactly, for GOE or GUE matrices,
in the limit of infinite dimensions.

More specifically, they choose $H_0$ and $V$ independently but both either
from the GOE or the GUE, and compute the fidelity amplitude $\la f_\eps(t)\ra$
with the help of supersymmetry techniques. A detailed discussion would not
be adequate in this review, and we refer to~\cite{VWZ85} for an introduction to
the techniques used and to the original paper~\cite{StoSch04b} for details.
They obtain
\begin{equation}
\la f_\eps(t)\ra = \frac{1}{t}\;\int_0^{{\rm min}(t,1)}\d u\;
  (1+t-2u)\; \e^{- (2\pi\eps)^2\, (1+t-2u)\, t/2}
\label{Q2SS:GUE}\end{equation}
for the GUE case and
\begin{align}
\la f_\eps(t)\ra &= 2\int_{{\rm max}(0,t-1)}^t\d u\int_0^u\d v\;
  \frac{(t-u)(1-t+u)\, v\, ( (2u+1)\, t -t^2+v^2)}
     {(t^2-v^2)^2\,\sqrt{(u^2-v^2)((u+1)^2-v^2)}}\notag\\
&\qquad\times \e^{- (2\pi\eps)^2\, [ (2u+1)\, t -t^2+v^2]/2} \; .
\label{Q2SS:GOE}\end{align}
for the GOE case. These solutions are valid for arbitrary but fixed
perturbation strength, in the limit $N\to\infty$.

\begin{figure}
\includegraphics[width=7cm]{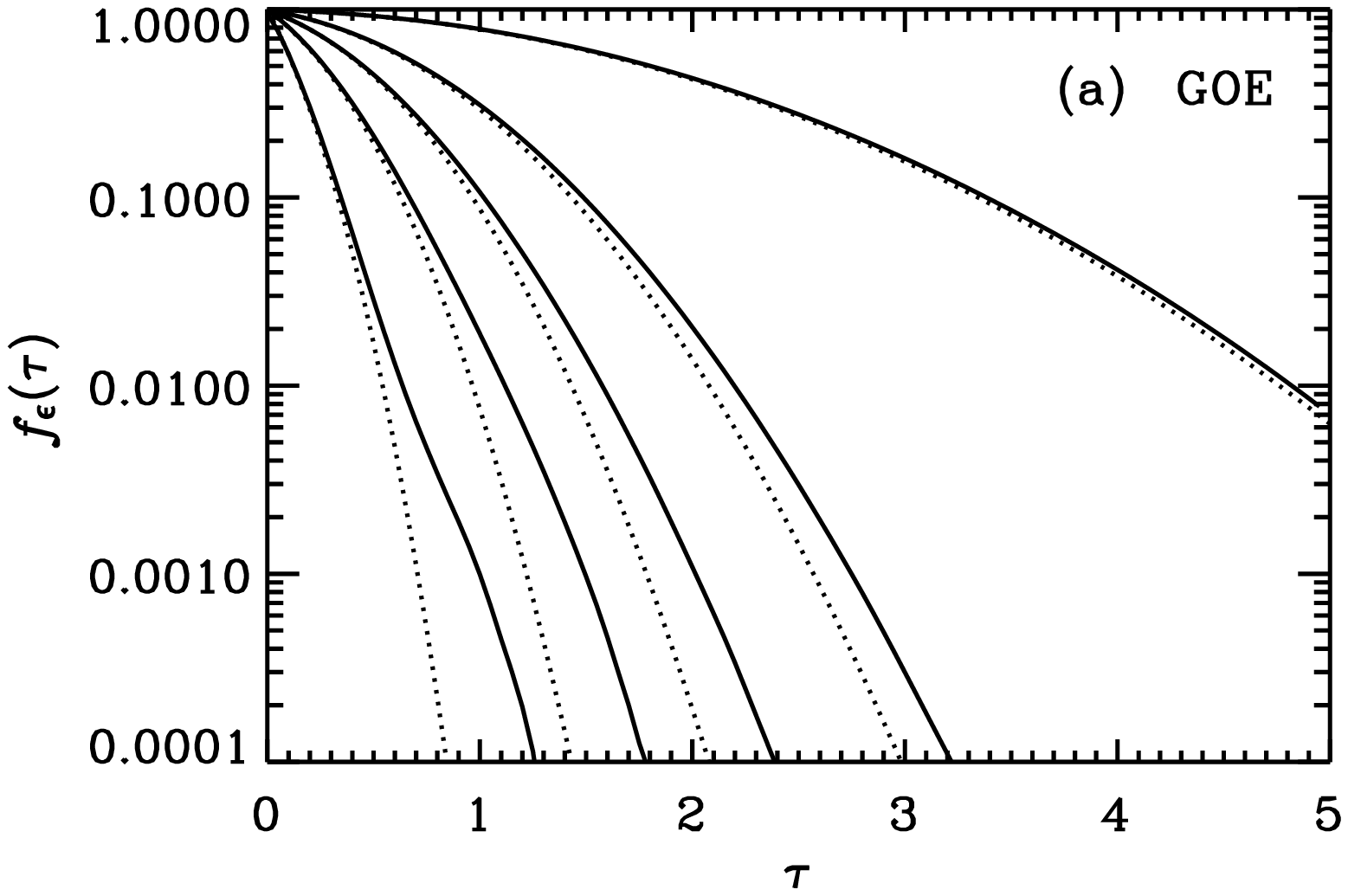}\qquad
\includegraphics[width=7cm]{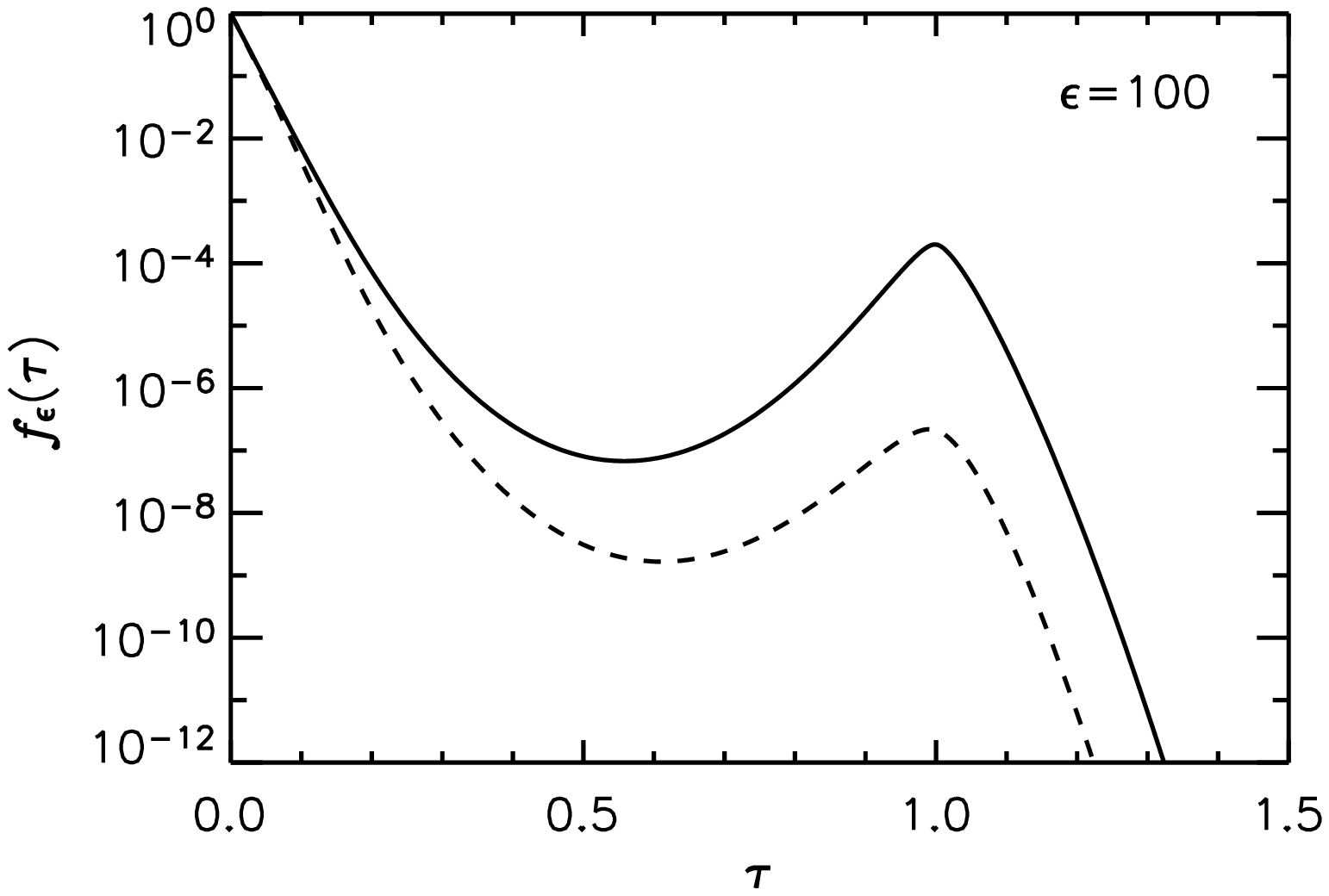}
\caption{Fidelity amplitude decay for the random matrix model, defined in
Eq.~(\ref{R:Heps}) (taken from~\cite{StoSch04b}).
Part (a) shows $\la f_\eps(t)\ra$ for the GOE case, as obtained
from the exact expression, Eq.~(\ref{Q2SS:GOE}), (solid lines), together with
the same quantity, as obtained from the exponentiated linear response result,
Eq~(\ref{Q2LR:expLR}) (dashed lines). The perturbation strength has been set
to the following values: $(2\pi\, \eps)^2= 0.2,\, 1\, , 2\, , 4\, $ and $10$.
Part~(b) shows $\la f_\eps(t)\ra$ with $(2\pi\, \eps)^2= 100$ for the GUE case
(solid line) and the GOE case (dashed line), as obtained from the exact
expressions, Eqs.~(\ref{Q2SS:GUE}) and (\ref{Q2SS:GOE}), respectively.}
\label{Q2SS:f:StoSch}\end{figure}

In Fig.~\ref{Q2SS:f:StoSch}, we reproduce two graphs from~\cite{StoSch04b}.
The left one, compares the exact and the exponentiated linear response result
for $\la f_\eps(t)\ra$ for the GOE case. For large perturbations we see there
a qualitative difference in the shape of fidelity decay as a shoulder is
forming in the exact results. For even stronger perturbations depicted on the
right hand side, this becomes notorious as a revival appears at Heisenberg
time.  Yet, the revival is noticeable only for very small fidelities of the
order of $10^{-4}$ for the GUE and  $10^{-6}$ for the GOE. In all cases
agreement with the exponentiated linear response formula is limited to
$\la f(t)\ra \gtrsim 0.1$. It is thus adequate for most applications, and
indeed it was difficult to come up with a dynamical model which can show the
revival.

\begin{figure}
\centerline{\includegraphics[width=\figw\textwidth]{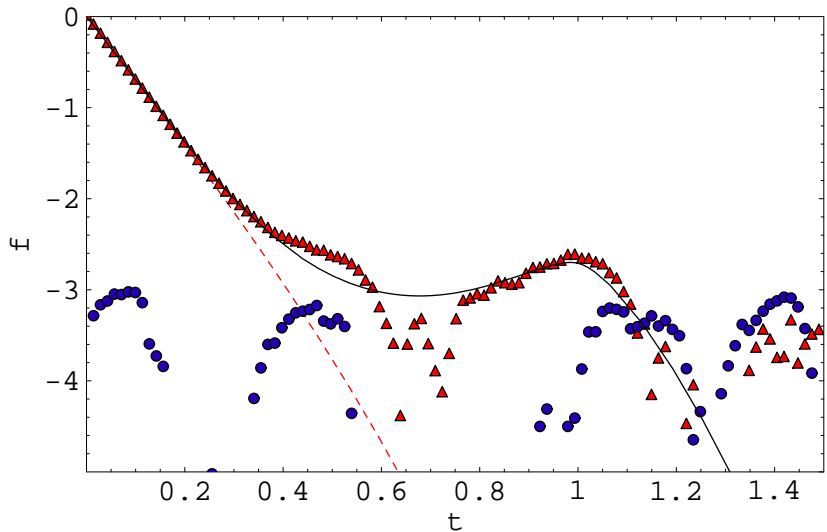}}
\caption{The decay of the fidelity amplitude in a dynamical system, in the
case of broken time-reversal symmetry, in a regime where the partial revival
is observable (taken from Ref.~\cite{Pin05}). More precisely,
$\log_{10} \la f(t)\ra$ is plotted. The triangles show the real part
of the fidelity amplitude as obtained from the numerical simulation; the
circles show the imaginary part, which goes to zero as $1/\sqrt{N}$ 
due to state 
averaging. The perturbation strength is
$(2\pi\eps)^2= 31.78$. The solid line shows the exact theoretical result,
Eq.~(\ref{Q2SS:GUE}), the dashed line shows the exponentiated linear
response result, Eq.~(\ref{Q2LR:expLR}).}
\label{Q2SS:f:pineda}\end{figure}

Naturally the random matrix model does not capture the Lyapunov
regime, which depends on semi-classical properties. Yet it is not obvious that
such a regime always exists. For example for a kicked spin chain we have never
discovered this regime, and it may well be that there is no semi-classical
regime for this system. This dynamical model has been used~\cite{Pin05}, to
illustrate the partial revival,
as shown in Fig.~\ref{Q2SS:f:StoSch}. The authors used a multiply kicked
Ising spin chain in a Hilbert space spanned by 20 spins, and averaged
only over a few initial conditions. The aim was to obtain the partial revival
with as little averaging as possible, relying on the self-averaging properties
of the fidelity amplitude. It has been checked that the model shows random
matrix behavior as far as its spectral statistics is concerned. The
result for the decay of the fidelity amplitude is reproduced in
Fig.~\ref{Q2SS:f:pineda}.  Yet it will probably be difficult to see this
effect in an experiment, and we do not know how it may appear in a dynamical
system which does display Lyapunov decay at times short compared to the
Heisenberg time. %\tgcomment{Why worry about Lyapunov decay?}

The partial revival of the fidelity amplitude occurs not only in the case of
a full GOE (GUE) perturbation as discussed here, but also in the case of a
perturbation with vanishing diagonal elements (see Section~\ref{Q2QF}, below). 
In~\cite{StoSch04} the partial revival has been explained with an analogy to
the Debye-Waller factor. A direct, semiclassical and hence dynamical
explanation could be obtained by periodic orbit expansions similar 
to~\cite{Mue04,Mue05} (for general chaotic systems) or~\cite{GnuAlt04} (for 
a quantum graph model).

\subsection{\label{Q2QF} Quantum freeze}

In Section~\ref{Q2LR}, we found that after the Heisenberg time, fidelity
decay is essentially Gaussian [{\it cf.} discussion below
Eq.~(\ref{Q2LR:expLR})]. It is determined by the diagonal
elements, {\it i.e.} the time average of the perturbation (in the interaction
picture). If this term is zero or very small, we should see a considerable
slowing down of fidelity decay.
That such a possibility exists in principle  was first noted
in~\cite{Prosen:03} for integrable systems (see
Section~\ref{sec:ch4Veq0}) and in~\cite{Prosen:05} for chaotic systems, as
described in Section~\ref{sec:chfreeze}. However, in both cases the
perturbation was chosen to depend in a very particular way on the unperturbed
Hamiltonian, with the effect that fidelity became largely independent
of the dynamics in the unperturbed system. We shall discuss that particular
choice in the context of RMT at the end of this subsection.

Here we shall mainly consider situations where the perturbation is
some RMT matrix with zero diagonal. In the notation of
Section~\ref{sec:freeze}, it means that the off diagonal elements will form
the residual interaction $V_{\rm res}=V$. The different options to implement
such a scenario have been discussed in Section~\ref{sec:freeze}. We
concentrate here on the case, where $H_0$ is chosen from a GOE and $V$ is an
antisymmetric hermitian random Gaussian matrix. This case is of interest,
because it is the only random matrix model, so far, for which an exact
analytical result has been obtained~\cite{GKPSSZ05,Heiner06}. We shall show these
below, yet we shall mainly concentrate on an extension of the previous
linear response treatment, as well as an approach based on second order
(time independent) perturbation theory. These approaches give more insight
into the mechanism, and they are more easily generalized to other interesting
cases. Indeed we shall again find a plateau in fidelity decay, which will be
flat if $H_0$ is chosen from a GUE and almost flat (up to logarithmic
corrections) if it is chosen from the GOE.
This plateau begins at Heisenberg time, and we are able to describe the
decay to the plateau and the plateau itself in terms of the linear response
results. The finally ensuing %TG Gaussian
decay is obtained from second order perturbation theory. A brief description
of some of these results can be found in~\cite{GKPSSZ05}.

\subsubsection{Linear response approximation}

In linear response, we have seen that the expectation value
$\la V_{ij}\, V_{kl}\ra$ is the essential ingredient for the calculation of
fidelity.  We can compute this average with equal ease for the following three
ensembles of perturbations:
(a) If $V$ is taken from a GOE with deleted diagonal, {\it i.e.}
$\la V_{ij}\, V_{kl}\ra = (\delta_{ik}\delta_{jl} + \delta_{il}\delta_{jk})
(1-\delta_{ij})$, then we find
\begin{equation}
\la V_{ik}\, V_{kj}\ra = (\delta_{ik}\delta_{kj} + \delta_{ij})(1-\delta_{ik})
 = \delta_{ij}\, (1-\delta_{ik}) \;.
\end{equation}
(b) If $V$ is taken from a GUE with deleted diagonal, {\it i.e.}
$\la V_{ij}\, V_{kl}\ra = \delta_{il}\delta_{jk}\, (1-\delta_{ij})$ then
we just obtain the same result for $\la V_{ik} V_{kj}\ra $. Thus, in both cases, we obtain:
\begin{align}
\la \tilde V(\tau)\, \tilde V(\tau')\ra_{ij} &= \sum_k
  \e^{2\pi\rmi\, (E_i-E_k)\, \tau}\; \e^{2\pi\rmi\, (E_k-E_j)\, \tau'} \;
  \delta_{ij}\, (1-\delta_{ik})
 = \delta_{ij}\sum_{k\ne i} \e^{2\pi\rmi\, (E_i-E_k)\, (\tau -\tau')} \; .
\end{align}
(c) If $V$ is taken from an ensemble of imaginary antisymmetric matrices we may
write
\begin{equation}
V = \rmi\; A \qquad
\la A_{ij}\, A_{kl}\ra = \delta_{ik}\delta_{jl} - \delta_{il}\delta_{kj} \; .
\end{equation}
This yields
\begin{align}
\la \tilde V(\tau)\, \tilde V(\tau')\ra_{ij} &= - \sum_k
  \e^{2\pi\rmi\, (E_i-E_k)\, \tau}\; \e^{2\pi\rmi\, (E_k-E_j)\, \tau'} \;
  (\delta_{ik}\delta_{kj} - \delta_{ij})
 = \delta_{ij}\sum_{k\ne i} \e^{2\pi\rmi\, (E_i-E_k)\, (\tau -\tau')} \; .
\end{align}
In all three cases, we obtain the same result, which is in fact quite similar to
the cases without deleted diagonal considered in Section~\ref{Q2LR}. If
we write the sum of exponentials in terms of the spectral two-point correlation
function (as it has been done in Eq.~(\ref{R:rmt-lr}), we obtain
\begin{equation}
\la f_\eps(t)\ra = 1- (2\pi \eps)^2
  \left[ t/2-\int_0^t d\tau'\int_0^{\tau'} d\tau\; b_2(\tau)\right]
  + \mathcal{O}(\eps^4) \; .
\label{R:rmt-lr-fr}\end{equation}
Comparing this expression to Eq.~(\ref{R:rmt-lr}), we note that the
$t^2$-term is missing. This has the peculiar consequence that the
characteristics of $V$ [{\it i.e.} whether we consider case (a), (b), or (c)]
have no effect on the linear response result. It solely depends on the
spectral statistics of $H_0$, encoded in the two-point form factor $b_2$.
If we deal with an $H_0$ taken from a GUE then the decay will stop at
Heisenberg time, while for one selected from a GOE we will have a very slow
decay, determined by the logarithmic behavior of $b_2$. We can get explicit 
formulae for the freeze after Heisenberg time. Evaluating the integral in 
Eq.~(\ref{R:rmt-lr-fr}) for GOE and GUE cases, we get
for times $t > t_{\rm H}$,
\begin{align}
\ave{f_\eps(t)} &= 1-(2\pi \eps)^2\frac{\log{(2t)}+2}{12}+\mathcal{O}(\eps^4) 
  \quad :\quad {\rm GOE} \notag \\
\ave{f_\eps(t)} &= 1-(2\pi \eps)^2\frac{1}{6}+\mathcal{O}(\eps^4)
  \quad :\quad {\rm GUE} \; .
\label{eq:rmt_plateau}
\end{align}
We can see the time independence for systems without time-reversal symmetry, 
while for time-reversal invariant systems freezing is not perfect as the 
plateau value decreases logarithmically, {\it i.e.} very slowly with time. 
These predictions are illustrated and compared to
simulations for a dynamical model in Figs.~\ref{fig:ktopgoefreeze} 
and~\ref{fig:ktopguefreeze} in Section~\ref{sec:ktopRMTfreeze}. 

\begin{figure}
\centerline{\input{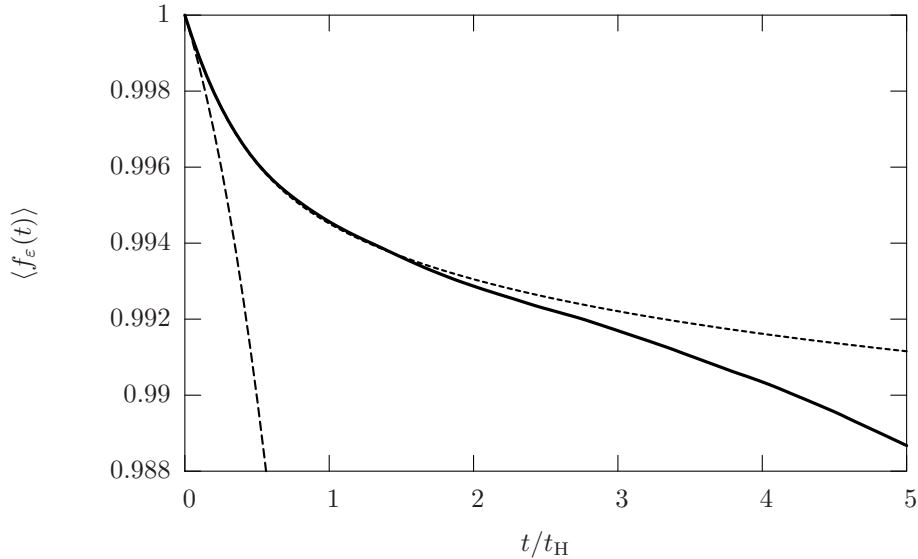}}
\caption{Numerical simulation of fidelity amplitude decay for a GOE
spectrum perturbed with a GOE matrix with deleted diagonal. The perturbation
strength was $\eps= 0.025$ (for further details, see text). The thick solid
line shows the result of the simulation, the short dashed line shows the
linear response result, Eq.~(\ref{R:rmt-lr-fr}). The long dashed line
shows the linear response result for an ordinary GOE perturbation.}
\label{Q2QF:f1}\end{figure}

In Appendix~\ref{app:HBS} we show that the Born series can be computed to
fourth order and estimates of higher orders can be given in the limit of large
$N$. This is done both for case (a) and case (c), and again the results agree.
By consequence, any difference between the two kinds of perturbations must be
of strictly non-perturbative type. That is also confirmed by numerics and thus
in Fig.~\ref{Q2QF:f1}, we only show  case (a), {\it i. e.} we give,
for a GOE Hamiltonian perturbed by a GOE with deleted
diagonal, a comparison of the above formula with a Monte Carlo simulation of
the random matrix model. Here we used just a linear rescaling of the
unperturbed spectrum to obtain unit average level spacing in the center.
Therefore we used rather large matrices of dimension $N=400$, and random
initial states with small spectral span (over the 20 central eigenstates
of $H_0$). With an ensemble of $n_{\rm run}=2000$, we obtained for the
fidelity amplitude the thick solid curve. The linear response approximation,
Eq.~(\ref{R:rmt-lr-fr}) is plotted by a thin short-dashed line. The
agreement is rather good up to $2\, t_{\rm H}$. From then on, we find
increasing deviations. This point of departure from the linear response
approximation can be moved towards larger times, by decreasing the perturbation
strength further. A strict plateau cannot be observed due to the remaining 
logarithmic decay of the linear response result. The great gain in fidelity
can be appreciated by comparing to the decay of the fidelity amplitude for an
ordinary GOE perturbation (long-dashed line).

Note that this derivation does not depend in any way on the time
averaged perturbation being identically zero. Rather we use the fact that
diagonal and off diagonal elements enter independently, the former influencing
the $t^2$-term in Eq.~(\ref{R:rmt-lr}) only. We can therefore equally well
deal with diagonal elements which are simply suppressed by a constant
factor. That factor would lead to a corresponding suppression of the above
mentioned quadratic term. This opens a new range of applications which will
have to be considered in the future.

\subsubsection{\label{Q2PT} Beyond fidelity freeze}

Higher order terms in $\eps$ will cause an end of the freeze, and we shall
use time-independent perturbation theory to describe this phenomenon. Recall 
that in the case of a standard perturbation such considerations lead to 
quadratic decay setting in around Heisenberg time~\cite{Jacquod:01,Cerruti:02}.
It is precisely the absence of this term in the present situation that is 
responsible for the freeze. Here, we consider the case, where the
second order term is needed, e.g. due to the
unusually small or vanishing diagonal elements in the perturbation matrix.
%TG A case of particular physical relevance is the perturbation of a GOE matrix
%TG by an anti-symmetric imaginary random Gaussian matrix, which describes a
%TG perturbation which breaks the time-reversal symmetry~\cite{GKPSSZ05}.

In a formally exact manner, the fidelity amplitude can be written in terms of
the eigenvalues of the perturbed and the unperturbed system, and the orthogonal
transformation between their respective eigenbases (\ref{eq:fnP}):
\begin{equation}
f(t) = 
  \sum_{\alpha\beta} \Psi^*_\alpha \Psi_\beta
  \e^{2\pi\rmi\, E_\alpha\, t}\; O_{\alpha\beta}\;
  \e^{-2\pi\rmi\, \tilde E_\beta\, t}\; O_{\alpha\beta}\; \; .
\end{equation}
We assume random initial states $\Psi$ with well defined local density of
states (in the eigenbasis of $H_0$). The spectrum of $H_0$ is denoted by
$\{ E_\alpha\}$, while the spectrum of $H_\eps$ is denoted by
$\{ \tilde E_\alpha\}$. For times which are long compared to the Heisenberg
time, we may neglect phases from energies of different states due to phase
randomization. This amounts to approximate the above sum by its diagonal
contribution $\alpha =\beta$ as
\begin{equation}
f(t)= \sum_\alpha |\Psi_\alpha|^2\; O_{\alpha\alpha}^2\;
  \e^{-2\pi\rmi\, \Delta_\alpha\, t}\qquad
\Delta_\alpha = E_\alpha- \tilde E_\alpha \; .
\label{RG:longt}\end{equation}
Finally, we may assume a sufficiently small perturbation, such that
$O_{\alpha\alpha}^2$ may be set equal to one.
\footnote{A more phenomenological assumption of statistically independent
eigenvectors and eigenvalues, which would allow to pull the average
$\la O_{\alpha\alpha}^2\ra$ out of the sum, does not prove very useful, if
compared to the exact analytical result~\cite{GKPSSZ05}.} This leads to
\begin{equation}
f(t) \approx \sum_\alpha |\Psi_\alpha|^2\;
  \lla \e^{-2\pi\rmi\, \Delta_\alpha\, t}\rra \; .
\label{RG:longt2}\end{equation}
In second order perturbation theory, the level shifts $\Delta_\alpha$ are
given by
\begin{equation}
\Delta_\alpha = \eps V_{\alpha\alpha} + \eps^2
  \sum_{\gamma\ne\alpha} \frac{|V_{\alpha\gamma}|^2}{E_\alpha -E_\gamma}
 + {\mathcal O}(\eps^3) \; .
\label{Q2PT:shifts}\end{equation}
The first order term, containing the diagonal elements of $V$, is statistically
independent from the second order term. As a consequence, the ensemble average
inside the sum factorizes into a purely Gaussian decay and a
factor which corresponds to the quantum freeze situation
\begin{equation}
f(t) \approx \sum_\alpha |\Psi_\alpha|^2\;
  \lla\e^{-2\pi\rmi\, \eps V_{\alpha\alpha}\, t}\rra\;
  \la S_\tau(E_\alpha)\ra \qquad
 S_\tau(E_\alpha)= \left. \e^{-\rmi\tau\, \sum_{\gamma > 1}
  |V_{1\gamma}|^2/(E_1 -E_\gamma)} \right|_{E_1= E_\alpha} \; ,
\label{Q2PT:genfreez}\end{equation}
where $\tau= 2\pi\, \eps^2\, t$.
The above construction allows to treat the energy argument in
$\la S_\tau(E)\ra$ as
a free variable. For the following, we discard the exponential which
contains the diagonal matrix elements of $V$ (it usually factors out, anyway).
In addition we assume that in the eigenbasis of $H_0$, the initial state has
non-zero coefficients only in a narrow region in the center of the spectrum.
That allows to remove the sum, and consider $\la S_\tau(E)\ra$ at the
energy $E=0$.
More general cases can be considered, but require to replace the sum over
$\alpha$ by a convolution-type integral. For $E\ne 0$,  $\la S_\tau(E)\ra$
can be reduced to $\la S_\tau(0)\ra$, by employing a principal value
integral to express the smooth part of the sum in the exponent.

We thus consider initial states, such that
$\la f_\eps(t)\ra \approx \la S_\tau(0)\ra$. Then, the decay of the
fidelity amplitude only depends on the spectral correlations of the ensembles
considered. It involves the level curvature
$\kappa= \sum_{\gamma >1}^N |V_{1\gamma}|^2/E_\gamma - E_1 $,
which has been studied in some detail in the literature (see
Ref.~\cite{Oppen:94,Oppen:95,FyoSom95} and references therein). Using a closed 
form of the distribution of $\kappa$ it is a simple exercise to compute 
$\la f_\eps(t)\ra$:
\begin{equation}
{\mathcal P}(\kappa) = \frac{C_\beta}
  {(\kappa^2 + \pi^2)^{1+\beta/2}} \quad : \quad
\la S_\tau(0)\ra = C_\beta\int\d \kappa\;
 \frac{\cos\, \tau\kappa}{(\kappa^2 + \pi^2)^{1+\beta/2}}\qquad
\tau= 2\pi\eps^2\, t \; ,
\label{Q2PT:curvdist}\end{equation}
where $C_\beta$ assures proper normalization of $\mathcal{P}(\kappa)$.
The cosine integral in Eq.~(\ref{Q2PT:curvdist}) gives
\begin{equation}
\la f_\eps(t)\ra \approx \la S_\tau(0)\ra
 = \begin{cases} \e^{-\pi\tau}\; (1+ \pi\tau) &: \beta= 2\\
  \pi\tau\; K_1(\pi\tau) &: \beta= 1\\
  \e^{-\pi\tau} &: \beta= 0 \end{cases}\; ,
\label{Q2PT:fres}\end{equation}
where $K_1(z)$ is the modified Bessel function~\cite{AbrSte70}. The resulting 
$f_\eps(t)$ gives the decay of fidelity (within the second order perturbative 
approximation) after the freeze plateau. Note that the decay time scales as 
$1/\eps^2$ as opposed to $1/\eps$ in the absence of the freeze. In all cases,
we assume $H_0$ to be diagonal. The case $\beta =2$ describes a GUE spectrum
perturbed by a GUE matrix $V$, with deleted diagonal. It means that the moduli
squared $|V_{1\gamma}|^2$ in Eq.~(\ref{Q2PT:genfreez}) have an exponential
distribution. The case $\beta =1$ describes a GOE spectrum perturbed by a GOE
matrix, with deleted diagonal. In that case, the moduli squared
$|V_{1\gamma}|^2$ have a Porter-Thomas distribution~\cite{PorTho56}. This
perturbation is equivalent to an antisymmetric imaginary random Gaussian
matrix. This perturbation has been treated exactly using 
supersymmetry techniques~\cite{GKPSSZ05,Heiner06}. The last case $\beta =0$ 
describes a spectrum with uncorrelated eigenvalues, while $V$ is chosen as in 
the $\beta =1$ case. That case has not been considered in~\cite{FyoSom95}. It 
fits rather accidentally to the formula~(\ref{Q2PT:curvdist}) for
$\mathcal{P}(\kappa)$, when setting $\beta =0$. We include that case, in order
to get a more complete overview on the effects of level repulsion on the decay
of the fidelity amplitude in the quantum freeze situation. The fact that a
spectrum without correlations gives indeed an exponential decay, can be
easily checked by direct evaluation of the separable average~\cite{G99}.
Furthermore, one finds that this result only depends on the average
of the matrix element $|V_{1\gamma}|^2$, and not on its distribution.

\begin{figure}
\centerline{\input{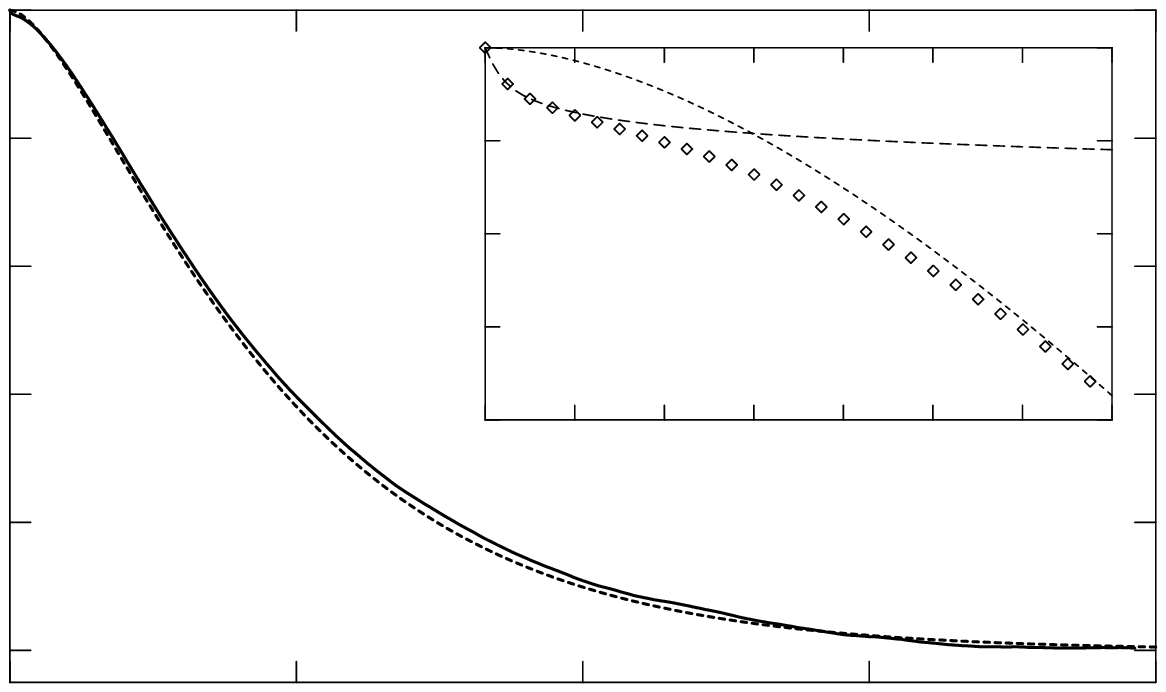}}
\caption{Numerical simulation of the fidelity amplitude decay for a GOE
spectrum perturbed with a GOE matrix with deleted diagonal. The perturbation
strength and all other parameters were chosen as in Fig.~\ref{Q2QF:f1}.
The thick solid line shows the result of the simulation, the short dashed line
shows the result from perturbation theory, Eq.~(\ref{Q2PT:fres}) with
$\beta=1$.
%TG Note that the time-axis is scaled to coincide with $\tau$ in that
%TG equation.
In the inset, we show the transition from linear response (long dashed line)
to time independent perturbation theory (short dashed line). The numerical
data are plotted with diamonds.}
\label{Q2PT:f01}\end{figure}

In Fig.~\ref{Q2PT:f01} we show the decay of the fidelity amplitude for a GOE
spectrum, perturbed by a GOE matrix with deleted diagonal. The perturbation
strength $\eps=0.025$ and the other parameters are exactly the same as in
Fig.~\ref{Q2QF:f1}; but here we show the long time behavior (main graph). The
figure shows that our perturbation theory provides a very accurate description
of the data. In the inset we focus on smaller times, where the transition
from the linear response regime to the perturbative regime can be observed.
For the GUE case ($\beta =2$) and the Poisson case ($\beta =0$) we obtain
similar agreement.

\paragraph*{The effect of spectral correlations (level repulsion)}
In view of Eq.~(\ref{Q2PT:shifts}), it is clear that level repulsion plays a
crucial role for the behavior of the fidelity amplitude. The degree of level repulsion determines whether the level curvature
$\kappa= \sum_{\gamma >1}^N |V_{1\gamma}|^2/E_\gamma$,
has a finite second moment, or not. If it has, we may expand $S_\tau(0)$ in
angular brackets in Eq.~(\ref{Q2PT:genfreez}) up to the quadratic term, and
average term by term. That would give
\begin{equation}
f_0(E,t) \sim 1- \frac{\tau^2}{2}\; \la \kappa^2\ra \; .
\label{Q2PT:fexpand}\end{equation}
It is not difficult to see that $\la \kappa^2\ra$ is finite for the GUE
spectrum, only. For a GOE spectrum, or a Poisson spectrum (without
correlations) that average diverges. Equation~(\ref{Q2PT:fexpand}) will be
used below to consider the case of a GUE spectrum perturbed by a GOE matrix
with deleted diagonal.

The "small" time behavior of the fidelity amplitude can be obtained from the
exact results, Eq.~(\ref{Q2PT:fres}). Here, small time really means small
$\tau$; the time $t$ should still be much larger than the Heisenberg time.
The corresponding asymptotic expansions yield to lowest order
\begin{equation}
\la f_\eps(t)\ra \sim \begin{cases}
  1 - \pi^2\tau^2/2 &: \text{GUE} \\
  1 - \frac{1}{2}\left( -\ln(\pi\tau) +\frac{1}{2} +\ln 2 -\gamma\right)\;
     \pi^2\tau^2 &: \text{GOE}\\
  1 - \pi\tau &: \text{Poisson}
\end{cases} \; ,
\label{Q2PT:fasy}\end{equation}
where $\gamma$ is Euler's gamma constant~\cite{AbrSte70}.
These expansions demonstrate that in the GOE and the Poisson case, the second
moment of the level curvature $\kappa$ do not exist.

\begin{figure}
\centerline{\input{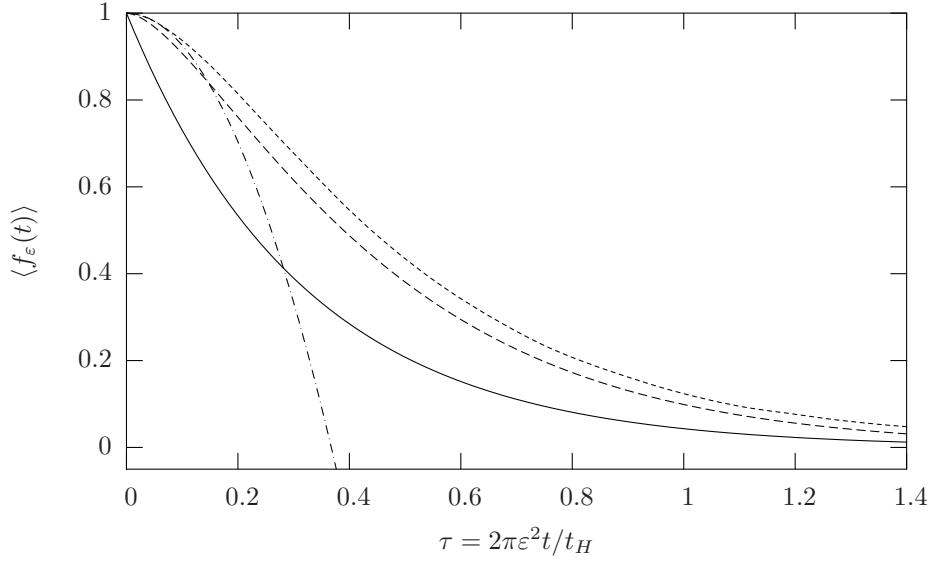}}
\caption{Numerical simulation of fidelity amplitude decay for different
random spectra, perturbed with a GOE matrix with deleted diagonal. For the
GOE (long-dashed line) and the Poisson spectrum (solid line) we show the
analytical results of Eq.~(\ref{Q2PT:fres}). For the GUE spectrum we show
a numerical simulation %TG with $\eps=10^{-3}$
(short-dashed line) and the leading order quadratic behavior (dash-dotted
line). For additional details, see text.}
\label{Q2PT:f04}\end{figure}

In Fig.~\ref{Q2PT:f04}, we illustrate the effect of level correlations on the
decay of the fidelity amplitude, in the presence of a GOE perturbation (with
deleted diagonal). For the GOE and the uncorrelated spectrum, the curves shown
are just the analytical results given in Eq.~(\ref{Q2PT:fres}). For the
GUE spectrum, there is no complete analytical result available (note that it is
a rather artificial situation), and we therefore show an actually quite
accurate\footnote{We expect this simulation to reproduce the result of the
perturbation theory exactly, within the finite linewidth in the figure. In
particular, we checked that further decrease of $\eps$ or increase of $N$ (the
dimension of the Hamiltonian matrix) gives no noticeable change on the scales
of the figure.}
numerical simulation. The Poisson spectrum has no correlations at all, and by
consequence the decay of the fidelity amplitude is the fastest. From $\tau=0$
it starts of linearly. Next comes the GOE spectrum, with linear level
repulsion, and therefore much slower decay $\la f(t)\ra$. The GUE spectrum
gives leads to the slowest decay of the fidelity amplitude. It is the only
case, where the $\la f(t)\ra$ behaves quadratically for small $\tau$. The
corresponding coefficient will be calculated below; the resulting
quadratic decay is plotted with a dash-dotted line.

As mentioned above, a GUE spectrum perturbed by a GOE matrix (with deleted
diagonal), is a rather artificial construction.
%TG Here, it is considered just for the sake of the argument, {\it i.e.}
%TG to show the influence of the level repulsion on fidelity
%TG decay in the quantum-freeze situation.
Even though, there is no complete analytical result available (in terms of the
perturbation theory, developed above), we may still obtain the leading order
behavior at small times. This is due to the fact that for the GUE spectrum,
the level curvature has a finite second moment
\begin{equation}
\la \kappa^2\ra = \la |V_{1\gamma}|^4\ra\; \left. \lla \sum_{\gamma=2}^N
  (E_1 - E_\gamma)^{-2}\rra \right|_{E_1=0} \; ,
\end{equation}
where $\la |V_{1\gamma}|^4\ra = 2\, (3)$ for a GUE (GOE) perturbation.
By comparison of Eq.~(\ref{Q2PT:fexpand}) with the
$\beta=2$ case in Eq.~(\ref{Q2PT:fasy}) we find that
$\la f_\eps(t)\ra \sim 1 - 3\, \pi^2\,\tau^2/4$ (dash-dotted line in
Fig.~\ref{Q2PT:f04}).

\subsubsection{\label{Q2CP} Commutator perturbation}

% {\bf TP: I have modified definition of matrix W below such that it really
% corresponds to the commutator. Thomas Gorin please check! I think
% it does not modify further analysis.}

Let us take a perturbation which is proportional to the commutator of a random 
GOE matrix $W$ with $H_0$ (in order to remain Hermitean, the perturbation 
must be imaginary). 
\begin{equation}
H= H_0 + \eps\; V \qquad V_{kl}= \ii W_{kl}\; (E_k-E_l) \; .
\label{Q2QF:RMMdef}\end{equation}
In the interaction picture, the perturbation $V$ is equivalent to the 
time-derivative of $W$, as detailed in Section~\ref{sec:freeze}.
Here, we use a banded perturbation matrix $W_{kl}$, and consequently
also banded $V_{kl}$, because otherwise the limit of
infinite matrix dimension $N\to\infty$ will be ill defined. It is then
convenient to assume that the basis states are arranged in such a way that
$E_i < E_j$ for $i<j$ (for technical convenience we also assume a
non-degenerate spectrum).

\begin{figure}
\mbox{}\qquad\qquad\input{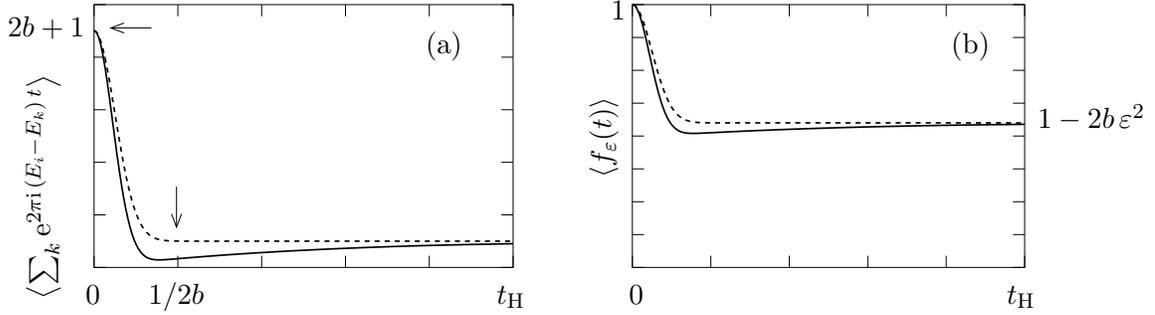}
%%%%%%%%%%%%%%%%%%%%%%%%%%%%%%%
% exported from xfig with 70% %
%%%%%%%%%%%%%%%%%%%%%%%%%%%%%%%
\caption{Schematic behavior of the fidelity amplitude [panel (b)] and the
two-point correlations [panel (a)], which are the determining quantities for
that behavior (see text for details). For the purpose of illustration, we
assume a small band width. The solid lines show the results for a GOE spectrum,
the dashed lines that for an uncorrelated (Poisson) spectrum.}
\label{Q2CP:fscheme}\end{figure}

\paragraph*{Linear response}
The crucial quantity is the average
\begin{equation}
\sum_k \la \tilde V_{ik}(\tau)\, \tilde V_{kj}(\tau')\ra = \sum_k
  \e^{2\pi\rmi\, (E_i-E_k)\, \tau}\; \e^{2\pi\rmi\, (E_k-E_j)\, \tau'}\;
  \la V_{ik}\, V_{kj}\ra \; .
\end{equation}
Since $W$ is hermitian, we obtain under the above assumptions:
\begin{equation}
\la \tilde V(\tau)\, \tilde V(\tau')\ra_{ij}= \delta_{ij}\; \sum_k\;
  (E_i-E_k)^2
  \e^{2\pi\rmi\, (E_i-E_k)\, (\tau-\tau')}\; \la |W_{ij}|^2\ra \qquad
 \la |W_{ij}|^2\ra = \theta\big (\, b- |i-j|\, \big ) \; ,
\end{equation}
where $\theta(j)= 1$ for $j\ge 0$ and $\theta(j)= 0$ otherwise.
That is the simplest case of a banded matrix, but more realistic band profiles
can be treated along the same lines. Inserting this into the
Eq.~(\ref{Q2LR:defM}) for the echo operator in the linear response
approximation, we get
\begin{align}
\la M_\eps(t)\ra_{ij} &= \delta_{ij}\left[
  1 + \eps^2 \sum_{k= \max(1,i-b)}^{\min(N,i+b)} \big (\,
  \e^{2\pi\rmi\, (E_i-E_k)\, t} - 2\pi\rmi\, (E_i-E_k)\, t - 1\, \big )
  \right] \notag\\
&= \delta_{ij}\big [\, 1 - \eps^2\; \mathcal{C}(t)\, \big ]\; .
\label{Q2CP:band}\end{align}
For an initial state which stays away from the border of the spectrum,
the linear term averages to zero (as a spectral average). In the limit of
large band-width, the remaining sum of exponential phases tend to an
expression involving the two-point form factor, similar to the case of
Eq.~(\ref{R:dav}):
\begin{equation}
  \sum_{k= \max(1,i-b)}^{\min(N,i+b)} \e^{2\pi\rmi\, (E_i-E_k)\, t}
  \to \delta(t) - b_2(t) +1 \; .
\end{equation}
With that information in mind, we plot in
Fig.~\ref{Q2CP:fscheme} the schematic behavior of that sum
for finite band width $b$; panel (a). The solid line
shows the result for a GOE spectrum, the dashed line for the case $b_2(t)=0$
(uncorrelated levels).  The time $t= t_H/(2b)$ may be considered as the
Ehrenfest time of the system. So we may say that
the correlation integral $\mathcal{C}(t)$ increases from zero to some value
between $2b$ and $2b+1$, within a time which is of the oder
of the Ehrenfest time. Panel (b) of Fig.~\ref{Q2CP:fscheme} shows the
schematic behavior of the fidelity amplitude; its very fast decay to the
practically constant value of $f_{\rm plateau}= 1- 2b\, \eps^2$. This shows
that a commutator perturbation yields a behavior of the fidelity amplitude
which is to a large extent insensitive to the particular dynamics of the
unperturbed system. This is in line with the ``surprising'' similarities
of quantum freeze between integrable and chaotic
systems~\cite{Prosen:03,Prosen:05}. For perturbations which have a vanishing
diagonal due to other reasons, the situation is different, as we
have seen above.

\paragraph*{Time independent perturbation theory}
As the diagonal of the perturbation $V$ is zero, we must
evaluate the eigenenergies of $H$ in second order perturbation theory, just as
in Eq.~(\ref{Q2PT:shifts}). For a perturbation of the
form~(\ref{Q2QF:RMMdef}), the level shifts are given by
\begin{equation}
\Delta_\alpha= \tilde E_\alpha-E_\alpha = \eps^2\sum_{\gamma\ne\alpha}
    |W_{\alpha\beta}|^2\, (E_\alpha-E_\gamma) \; .
\label{Q2QF:shifts}\end{equation}
For this expression to be valid, the shifts in the
eigenenergies must be small compared to the average level distance (which
is set equal to one). For definiteness, let us consider the banded random
matrix model, Eq.~(\ref{Q2CP:band}), as used in the linear response
calculation. Then we obtain, from Eq.~(\ref{Q2PT:genfreez}):
\begin{equation}
\la f_\eps(t)\ra \approx \sum_\alpha |\Psi_\alpha|^2\;
  \lla e^{\rmi\tau\, \Delta_\alpha}\rra \qquad
\Delta_\alpha = \sum_{1 \le |\gamma-\alpha| \le b}
  |W_{\alpha\beta}|^2\, (E_\alpha-E_\gamma) \; ,
\end{equation}
where $\tau=2\pi\, \eps^2\, t$. In this equation, we have assumed that the
initial state stays away from the matrix borders. Then, as long as the
level density remains constant, the phase average
$\la e^{\rmi\tau\, \Delta_\alpha}\ra$ is translational invariant, and we
obtain
\begin{equation}
\la f_\eps(t)\ra \approx \lla e^{\rmi\tau\, \Delta_\alpha}\rra
 = 1 - \frac{\tau^2}{2}\; \la \Delta_\alpha^2\ra + \mathcal{O}(\tau^4) \; .
\label{Q2CP:2pres}\end{equation}
In distinction to the perturbations with deleted diagonal, here the second
moment of the level shifts usually exist. We find:
\begin{equation}
\la \Delta_\alpha^2\ra \approx 2\;
  \la |W_{\alpha\beta}|^4\ra \left( \int_0^b\d x\; x^2 + b^2/2.\right)
 = 2\; \la |W_{\alpha\beta}|^4\ra \left( \frac{b^3}{3} + \frac{b^2}{2}
  + \mathcal{O}(b) \right) \; .
\end{equation}
The sum over $\gamma$ is here approximated with an integral. Here, we added
the next leading term of the Euler-Maclaurin expansion (it amounts to the
inversion of the familiar trapezoidal integration rule)~\cite{AbrSte70}.
Note that the $b^2$-term also depends on the spectral
correlations. Thus, the expansion above is correct only in the case of an
uncorrelated (Poisson) spectrum. For a GOE spectrum, the $b^2$-term must be
modified.

\begin{figure}
\makebox[0pt][l]{\includegraphics[scale=1.1]{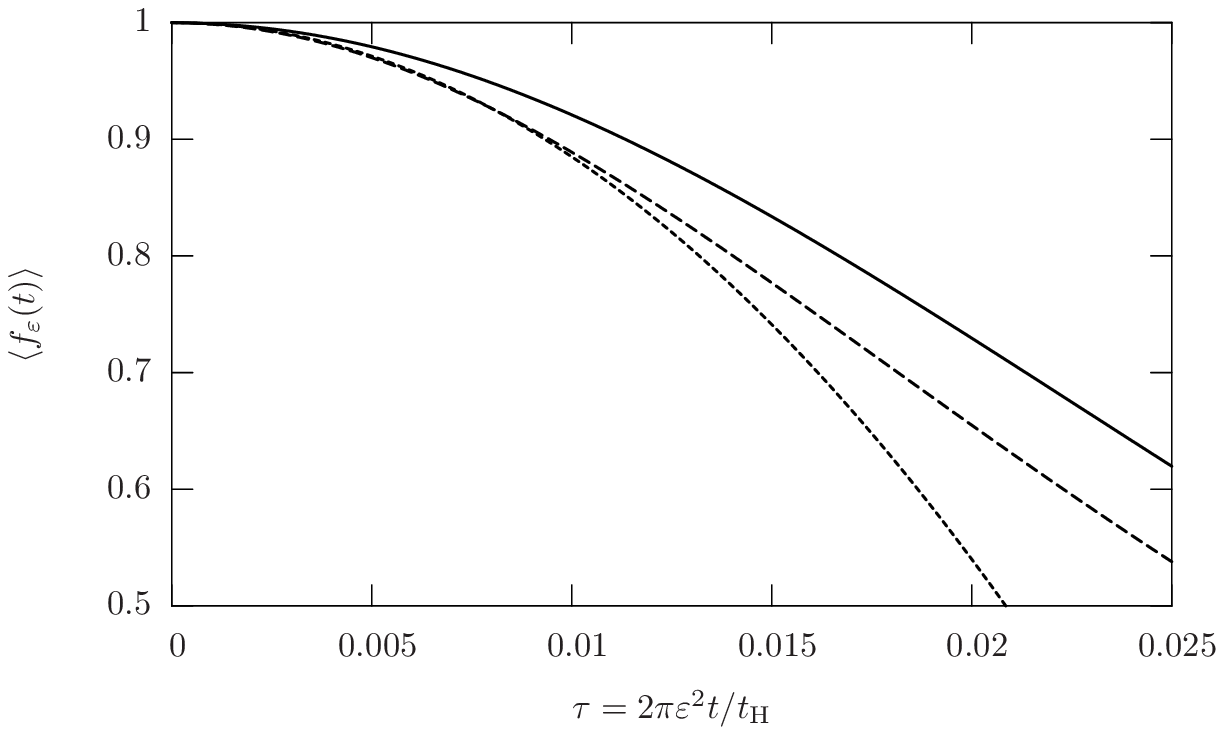}}\hspace{50pt}
\makebox[0pt][l]{\raisebox{45pt}{\includegraphics[scale=0.7]{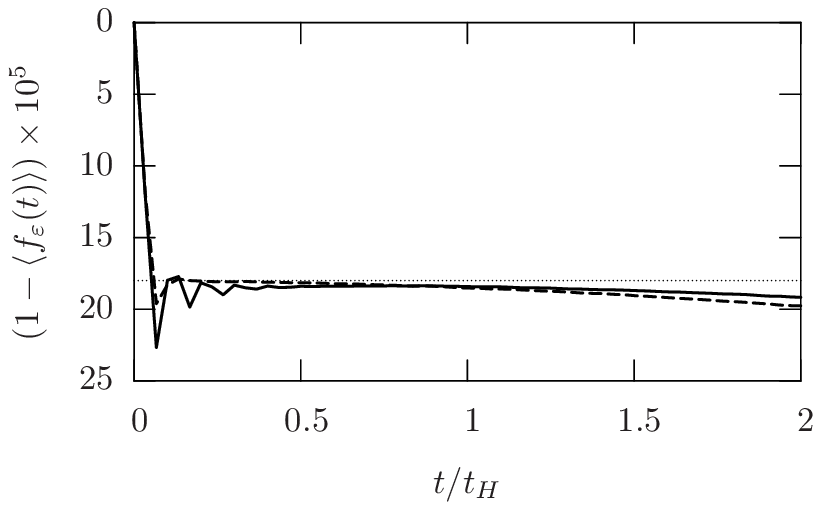}}}
\caption{The fidelity amplitude in the case of a banded commutator 
  perturbation. The case of a GOE spectrum (solid lines) and a Poisson spectrum 
  (long dashed lines). The theoretical expectations based on linear response 
  ($f_{\rm plateau}$; dotted line) and second order perturbation theory
  (short dashed lines). The main figure shows the long time behavior, where we 
  scaled time to coincide with $\tau$, as used in Eq.~(\ref{Q2CP:2pres}). The
  insert shows the short time behavior for times of the order of the Heisenberg 
  time. For details on the parameters used in the simulation, see text.}
\label{Q2CP:f01}\end{figure}

In Fig.~\ref{Q2CP:f01} we show the behavior of the fidelity amplitude for
a banded commutator perturbation. We performed two simulations, one with a
GOE spectrum and another one with a Poisson (uncorrelated) spectrum. We use
a small band-width $b=10$ to observe finite size effects in the band-width.
The perturbation strength was $\eps= 0.003$, and the dimension of the matrices
$N=100$. We performed an ensemble average over $n_{\rm run}=40\, 000$
realizations, and an average over $n_{\rm tr}=50$ initial states, which were
eigenstates of the unperturbed system, located in the center of the spectrum.

The main figure shows the long time behavior for both simulations,
(GOE spectrum: solid line, Poisson spectrum: long dashed line) together with
the theoretical expectation, Eq.~(\ref{Q2CP:2pres}). We find good
agreement in the Poisson case, while the correlations in the GOE spectrum lead
to a slower decay. As explained above, we expect the difference between
correlated and uncorrelated spectra to disappear in the limit of large
band-width. The insert shows the short time behavior for times
of the order of the Heisenberg time. We compare the simulations with the
plateau value $f_{\rm plateau}$ found from the linear response approximation.
We find good agreement, and we recognize a weak influence of the two-point
form factor, as explained in Fig.~\ref{Q2CP:fscheme}.

\subsubsection{RMT freeze in dynamical systems}
\label{sec:ktopRMTfreeze}

We shall demonstrate the theory for fidelity freeze in RMT by a dynamical model. 
We again choose the kicked top. As the numerical results for a commutator type 
perturbations have already been presented (Figs.~\ref{fig:1ktop} 
and~\ref{fig:2ktop}), we shall focus on perturbations whith vanishing diagonal
elements, either due to symmetries or due to setting them to zero by hand, see 
also general discussion in Section~\ref{sec:freeze}. 
 
\begin{figure}[ht]
\centerline{\includegraphics[angle=-90,width=\figw\textwidth]{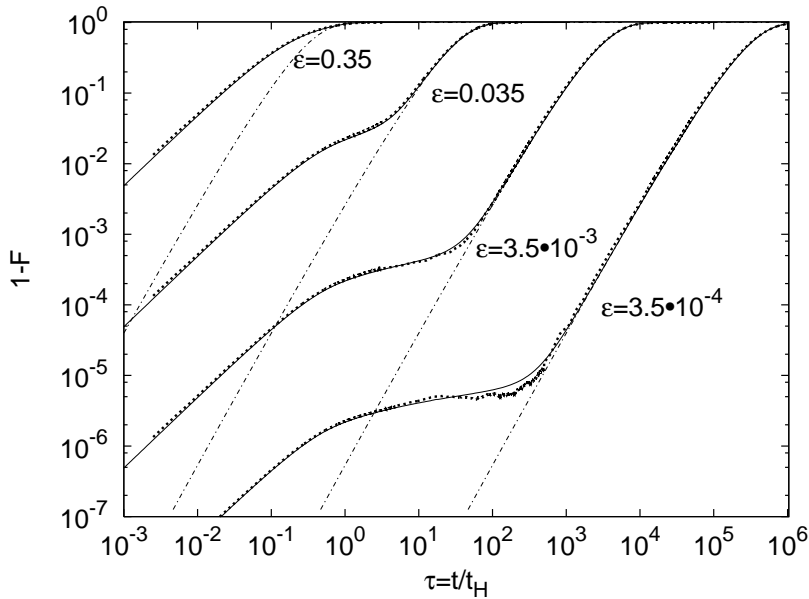}}
\caption{Log-freeze in a kicked top, GOE case, the same data as 
in~\cite{GKPSSZ05}. The solid line shows the exact result obtained by 
supersymmetry~\cite{GKPSSZ05}, the chain line shows the long-time approximation 
Eq.~(\ref{eq:long}), the thick dotted lines show numerical simulations with 
the kicked top. At small times, $1-F$ increases linearly as given by the
standard Fermi golden rule decay (\ref{eq:short}). Around the Heisenberg time 
$t_{\rm H}=400$ this regime ends and the freeze begins. The logarithmic behavior ends 
where the asymptotic decay of the freeze begins. This is well described by the 
long-time approximation.}
\label{fig:ktopgoefreeze}
\end{figure}

First we are going to discuss unperturbed system with time reversal symmetry, 
therefore corresponding to GOE theory (COE for a kicked system). For this we take 
a symmetrized chaotic kicked top $U_\eps^{\rm sym}$ given in 
Appendix~\ref{sec:models}, Eq.~(\ref{eq:ktopGOEsym}). Because the perturbation 
for $U_\eps^{\rm sym}$ is complex antisymmetric (in the eigenbasis of 
unperturbed $U_0^{\rm sym}$) we expect to find quantum freeze. 
The symmetrization, {\it i.e.}, splitting operator $P$ in $U_\eps^{\rm sym}$ into 
two parts, is essential. Without symmetrization there is no freeze and 
fidelity decays on much shorter time scale. For instance, for parameters used 
in Fig.~\ref{fig:ktopgoefreeze} fidelity for an unsymmetrized dynamics decays 
on a time scale which is approximately $1000$ times shorter! In order to 
compare RMT with the numerics we have to re-introduce physical units. The 
Heisenberg time is $\mathcal{N}=2S$, if $S$ is the spin size, and the second 
moment of matrix elements of the perturbation is 
$\ave{|V_{j k\to j}|^2}=2\sigma_{\rm cl}/\mathcal{N}$, where $\sigma_{\rm cl}$ 
is an integral of the classical correlation function (see Section~\ref{Q2D}). 
For times smaller than the Heisenberg time fidelity will decrease linearly, 
according to (\ref{eq:linearmixing}),
\begin{equation}
1-F\approx(\eps_{\rm KT} S)^2 2\sigma_{\rm cl} t,
\label{eq:short}
\end{equation}
where we temporarily use $\eps_{\rm KT}$ for kicked top perturbation strength 
(denoted simply by $\eps$ in Appendix~\ref{sec:ktop}) to distinguish it from 
RMT perturbation strength $\eps$, which is given as 
$2 \pi \eps=\eps_{\rm KT}\sqrt{4 \sigma_{\rm cl} S^3}$. For intermediate times, 
when we have a log-freeze (\ref{eq:rmt_plateau}), we have instead
\begin{equation}
1-F\approx \eps_{\rm KT}^2 S^3 4\sigma_{\rm cl} \frac{\ln{(t/2S)}+2+\ln{2}}{6}
 \; .
\label{eq:middle}
\end{equation}
For large times we can 
use the approximate theoretical prediction (\ref{Q2PT:fres}) noting that in 
physical units we have $\tau=\eps_{\rm KT}^2 S^2 \sigma_{\rm cl} t/\pi$ and
thus
\begin{equation}
F=\eps_{\rm KT}^4 S^4 \sigma_{\rm cl}^2 t^2 
  [ K_1(\eps_{\rm KT}^2 S^2 \sigma_{\rm cl} t)]^2 \; .
\label{eq:long}
\end{equation}
In Fig.~\ref{fig:ktopgoefreeze} we compare numerical results with the 
theoretical prediction~(\ref{eq:long}) as well as with the exact result 
from~\cite{GKPSSZ05}, all for $S=200$ and different 
$\eps_{\rm KT}=10^{-6},\ldots,10^{-3}$. We have averaged over $400$ independent 
kicked top realizations, taking one random initial state for each. Independent 
kicked top realizations are obtained by drawing the parameter $\alpha$ in
Eq.~(\ref{eq:ktopGOEsym}) at random from a Gaussian probability distribution 
with the center at $\alpha=30$ and unit variance. To ensure statistical 
independence the standard deviation divided by the number of realizations
must be large as compared to the so-called level collision 
time which scales as $\sim 1/\sqrt{\mathcal{N}}$ and is determined by the level velocity and the mean level spacing. The averaging over independent spectra is here absolutely 
essential to reproduce the log-freeze. The agreement with the log-freeze 
(\ref{eq:middle}) holds only up to the time determined by the smallest level 
spacing and therefore scales as $\sim \sqrt{m}$, if $m$ is the number of 
ensembles used in the averaging. The asymptotic result (\ref{eq:long}) is
similarly sensitive to averaging.

\begin{figure}[ht]
\centerline{\includegraphics[angle=-90,width=\figw\textwidth]{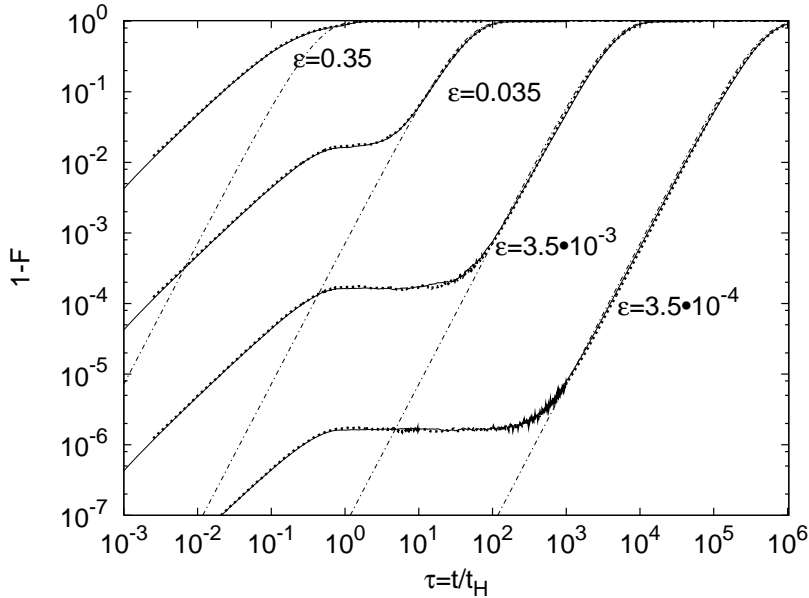}}
\caption{Freeze in a kicked top, for the GUE case. The diagonal elements are 
  put to zero by hand. The thick dashed lines show the numerical simulations, 
  the solid lines correspond to the exact RMT result from~\cite{GKPSSZ05}, and
  the chain lines show the asymptotic result~(\ref{eq:GUElong}). The plateau 
  during freeze is constant in the GUE case, as described by 
  Eq.~(\ref{eq:GUEmiddle}).}
\label{fig:ktopguefreeze}
\end{figure}

Next we shall give an example for a system without time-reversal symmetry, 
belonging to the CUE. The one step propagator is given by 
Eq.~(\ref{eq:ktopguefreeze}). For the numerical simulations, we use exactly
the same paramters as in the GOE case; this refers to the spin magnitude 
$S=200$, the random distribution for $\alpha$, the initial states, and the 
perturbation strengths. In order to obtain freeze we set diagonal elements of 
the perturbation to zero by hand. The short time decay does not depend on the 
symmetry class and it is the same as for the GOE case (\ref{eq:short}). The 
plateau during quantum freeze is for the GUE case independent of time,
\begin{equation}
1-F=\eps_{\rm KT}^2 S^3 4\sigma_{\rm cl}\frac{1}{3}\; .
\label{eq:GUEmiddle}
\end{equation}
On the other hand, Eq.~(\ref{Q2PT:fres}) yields for the long time decay
\begin{equation}
F=(1+\eps_{\rm KT}^2 S^2 \sigma_{\rm cl} t)^2 \exp{(-2\eps_{\rm KT}^2 S^2 \sigma_{\rm cl} t)}.
\label{eq:GUElong}
\end{equation}
In Fig.~\ref{fig:ktopguefreeze} we again find nice agreement between
numerical simulations and theory.

\subsection{\label{Q2PE} Echo purity}

The random matrix model developed in~\cite{GPS04} and described above can be
generalized to cover the evolution of entanglement following the lines of
Refs.~\cite{GS02b,GS03}. The new essential ingredient is to write the
Hilbert space $\mathcal{H}$ as a tensor product of two Hilbert spaces
$\mathcal{H}= \mathcal{H}_{\rm c}\otimes\mathcal{H}_{\rm e}$, which we will for
convenience call ``central system'' and ``environment''. For the dimension
$N$ of $\mathcal{H}$, we get: $N= n_{\rm c}\, n_{\rm e}$, where $n_{\rm c}$ and
$n_{\rm e}$ are the dimensions of $\mathcal{H}_{\rm c}$ and
$\mathcal{H}_{\rm e}$, respectively.

We first consider an unperturbed Hamiltonian $H_0\in$ GOE, which couples the
two subspaces randomly and strongly, while the perturbation $V$ will be drawn
independently from the same ensemble. The entire argument that follows will
go through analogously for $H_0\in$ GUE. Echo purity, {\it i.e.} the purity
of an initial state evolved with the echo operator, is defined in 
Section~\ref{embf}, Eq.~(\ref{eq:Fpdef}).
To calculate that quantity we need a representation
of the echo operator $\tilde M_\eps(t)$ in a product basis with respect to the
two factor spaces. Some orthogonal transformation will yield this as
\begin{equation}
\tilde M_\eps(t)= O\; M_\eps(t)\; O^T \; .
\end{equation}
As $H_0$ and $V$ are chosen from orthogonally invariant ensembles, the
orthogonal transformation $O$ will be random and independent of the product
spaces $\mathcal{H}_{\rm c}$ and $\mathcal{H}_{\rm e}$. We shall average over
it using the invariant measure of the orthogonal group as in~\cite{GS02b,GS03}.
Since purity is of second order in the density matrix the expressions become
involved and we have to digress to some formal considerations before actually
working out the linear response approximation.

\paragraph*{Purity form}
Since purity is defined as a trace of the square of a density operator, its
treatment in RMT usually leads to a rather messy jungle of
indices~\cite{GS02b,GS03}. To avoid that, we define a linear form
$p[\, .\,]$ on the space of operators in $\mathcal{H}^{\otimes2}$.
To this end we require that for any tensor product of operators
$\varrho_1,\varrho_2$ acting on $\mathcal{H}$ the purity functional evaluates
to
\begin{equation}
p[\varrho_1 \otimes \varrho_2] = {\rm tr}_{\rm c} \big (\,
  {\rm tr}_{\rm e}\, \rho_1\; {\rm tr}_{\rm e}\, \rho_2\, \big) \; .
\label{Q2PE:psym}\end{equation}
Due to linearity this suffices for a complete definition of the purity
form. The purity itself, reads: $I(t)= p[\varrho\otimes\varrho]$.
For that expression to be physically meaningful, $\varrho$ should be a
density matrix. However, the following algebraic manipulations will
eventually involve operators, which are not density operators.
%This expression is probably only physically meaningful in this case
%but the more general form $p[.\/, .\/]$ allows to perform some
%algebra with auxiliary matrices which may not be density matrices.

For $H= H_0 + \eps\; V$ and $\rho^M(t)$ denoting the initial state $\rho(0)$
propagated with the echo operator $M(t)= U_0^\dagger(t)\; U(t)$, the echo
purity may be written as
\begin{equation}
F_P(t)= p[\rho^M(t)\otimes \rho^M(t)] \; .
\end{equation}
In order to compute the echo purity in linear response approximation, we
use the corresponding expression for the echo operator,
Eq.~(\ref{Q2LR:defM}) in the eigenbasis of $H_0$. Due to the bilinearity
of the tensor product, we obtain for the
linear response expression of the echo operator in the product basis
\begin{align}
&\tilde M(t) \approx \Ione -2\pi\rmi\, \eps\; \tilde I - (2\pi\eps)^2\;
  \tilde J \label{Q2PE:echoop}\\
\tilde I &= O\; I\; O^T \qquad I_{\alpha\beta}= \int_0^t\d\tau\;
  \e^{2\pi\rmi\, E_\alpha\,\tau}\; V_{\alpha\beta}\;
  \e^{-2\pi\rmi\, E_\beta\,\tau} \\
\tilde J &= O\; J\; O^T \qquad
J_{\alpha\beta}= \sum_\xi
  \int_0^t\d\tau\int_0^\tau\d\tau'\; \e^{2\pi\rmi\, E_\alpha\,\tau}\;
  V_{\alpha\xi}\; \e^{-2\pi\rmi\, E_\xi\, (\tau-\tau')}\; V_{\xi\beta}\;
  \e^{-2\pi\rmi\, E_\beta\, \tau'}\; ,
\label{Q2PE:IuJ}\end{align}
where $O^T\, H_0\, O = {\rm diag}(E_\mu)$.
As mentioned above, $V$ will be drawn from a GOE, while the orthogonal matrix
$O$ is taken from the orthogonal group provided with the invariant Haar
measure. We shall see
that in linear response approximation, echo purity depends on the spectrum of
$H_0$ only via its two-point correlations, which are conveniently expressed in
terms of the two-point form factor $b_2$. All three components are assumed to
be statistically independent.
\footnote{This assumption is crucial and should be carefully checked, in
practice.}

The starting point is thus more general than what we have considered in
Section~\ref{embf},
inasmuch as we allow for non-separable (with respect to
$\mathcal{H}_{\rm c}$ and $\mathcal{H}_{\rm e}$) and even non-pure
(in $\mathcal{H}$) initial states. While we treat the technical details in
Appendix~\ref{app:EP}, we state here, that the echo purity is of the general
form
\begin{equation}
F_P(t) \approx p[\rho(0)\otimes\rho(0)] - 2\; (2\pi\eps)^2\;
  \mathcal{C}_P(t)\; ,
\label{Q2PE:LR1}\end{equation}
where the correlation integral $\mathcal{C}_P(t)$ is independent of $\eps$.

In the case of a pure and separable initial state $\rho(0) =
|\psi_c(0)\ra \, \la\psi_c(0)| \otimes |\psi_e(0)\ra \, \la\psi_e(0)|$ we
may use the orthogonal invariance of $H_0$ and $V$ under arbitrary orthogonal
transformations, and the invariance of purity under orthogonal transformations
in the factor spaces, to obtain the following compact expression:
\begin{equation}
\mathcal{C}_P(t)= 2\, \left(1 - \frac{n_{\rm c}+n_{\rm e}-2}{N+2}\right)\;
  \left( t^2 + t/2 - \int_0^t\d\tau'\int_0^{\tau'}\d\tau\; \la b_2(E,\, \tau)\ra
  \right) \; .
\label{Q2PE:LR2}\end{equation}
Note that the average over the random orthogonal transformation, implies that
the decay is influenced by the spectral correlations from everywhere in the
spectrum. For that reason we use the notation $\la b_2(E,\, \tau)\ra$ which
means a global spectral average of the two-point form factor over the whole
spectrum. The decay of the echo purity is of the same form as the fidelity
amplitude decay, but faster by a factor of
$4\, [1- (n_{\rm c}+n_{\rm e}-2)/(N+2)]$. This fact
is consistent with the inequality~(\ref{eq:ineqgen}) which states that the
square of fidelity must be smaller or equal to purity; for our RMT model in
linear response approximation the equality holds exactly in
the limit $n_{\rm c},n_{\rm e}\to\infty$.

These results, in particular Eqs.~(\ref{Q2PE:LR1}) and (\ref{Q2PE:LR2}), apply
to the quantum simulation of a dynamical model (the kicked Ising spin chain,
Appendix~\ref{sec:KI}) considered in Section~\ref{sec:chcomposite}. While those
considerations were restricted to the Fermi golden rule regime (exponential
decay), the present results cover the whole range from the Fermi golden rule
to the perturbative regime. Being obtained from linear response theory, their
validity is naturally restricted to time regions where the echo operator is
sufficiently close to the identity. However, we expect that this defect can
be cured by phenomenological exponentiation of the linear response result,
similar to the fidelity amplitude case. If in addition the possibly quite large
limit values of the echo purity are taken into account 
(as described in Section~\ref{sec:chcomposite}), we expect
to obtain an accurate general theoretical description of echo purity
decay. Numerical simulations to check these ideas are considered, but have not
been performed, so far. See however, Fig.~\ref{fig:fpe}, which presents
some cases for relatively small Hilbert space dimensions, where the echo-purity
deviates very clearly from an exponential decay. These deviations indicate the
presence of an additional term, quadratic in time, which would be in line with
Eq.~(\ref{Q2PE:LR2}). Unfortunately, the presence of a number of discrete
symmetries in the dynamical system used, makes a quantitative comparison
exceedingly difficult.

Some special cases could be considered. In many dynamical models, such as
for the Jaynes-Cummings model with counter-rotating terms~\cite{Prosen:03evol}
the perturbation may only act in one of the two subsystems. As on the
other hand the eigenstates of $H_0$ (when transformed to a product basis) imply
a complicated orthogonal transformation, we still expect the above result to
provide a reasonable description of the decay of echo purity.

One special case of fundamental importance will be treated in
the following Section~\ref{Q2PD}. There we shall assume $H_0$ to be
separable, the perturbation providing the only coupling.

\subsection{\label{Q2PD} Purity decay}

The results we developed
in the previous subsection (and in particular in Appendix~\ref{app:EP}) can
readily be modified to obtain an excellent model for the decay of purity in
a pure forward time evolution. To that end, consider a
Hamiltonian $H_0$ which does not establish any interaction
between the two subsystems, such that it can be written as
$H_0= h_{\rm c} \otimes \Ione + \Ione \otimes h_{\rm env}$, where $h_{\rm c}$
and $h_{\rm env}$ act on $\mathcal{H}_{\rm c}$ and $\mathcal{H}_{\rm e}$,
respectively. Clearly the purity of any density matrix for the central system
will not be affected by a time evolution with this Hamiltonian. Thus
entanglement and purity decay will be entirely due to the forward time
evolution.
A simulation with a corresponding dynamical model is discussed in
Section~\ref{sec:chcomposite}, Fig.~\ref{fig:fid3030}. While the focus there
is on the semiclassical behavior which involves quantum mechanically strong
perturbations to reach the Fermi golden rule regime, we consider decay times
of the order of the Heisenberg time. As before, the results here will
therefore cover the whole range from the Fermi golden rule to the perturbative
regime.

Assuming that the perturbation $V$ as well as the
Hamiltonians $h_{\rm c}$ and $h_{\rm env}$ are taken from appropriate random
matrix ensembles, we can still apply our results obtained for echo purity.
%However, due to the additivity of the eigenenergies of $h_{\rm c}$ and
%$h_{\rm env}$ in the unperturbed system, its spectrum will be essentially a
%random uncorrelated sequence of levels. 
However, at the level of echo dynamics, we make an additional approximation.
Since the additivity of the eigenenergies of $h_{\rm c}$ and $h_{\rm env}$ in 
the unperturbed system tend to destroy spectral correlations, anyway, we
replace the spectrum of $H_0$ by a sequence of uncorrelated random levels.
We approximate the purity decay $I(t)$ of a separable $H_0$ with the decay of 
the echo purity $F_P(t)$ of a simpler $H_0$, which is no longer separable.
In this way, we get for the growth of entanglement as measured by the purity 
of an evolving state with initial density matrix $\varrho_0$
\begin{equation}
\la I(t)\ra \approx \la F_P(t)\ra 
 = I(0) - 2\; (2\pi\eps)^2\; \mathcal{C}_P(t) + \mathcal{O}(\eps^4)\; ,
\end{equation}
with a simplified correlation integral due to the absence of correlations in
the spectrum of $H_0$ (see Appendix~\ref{appEPD}).
For the case of pure initial states we thus get the following
linear response result:
\begin{equation}
\la I(t)\ra \approx 1- 4\; (2\pi\eps)^2\left\{ t^2\; \big [\,
  1 - {\rm Ipr}_2\, \varrho_0\, \big ] + \frac{t}{2}\; \big [\,
  1 - \frac{n_{\rm c}+n_{\rm e}- 1 - {\rm Ipr}_2\,\varrho_0}{N}\, \big ]
  \right\} + \mathcal{O}(\eps^4)\; ,
\label{Q2PD:LRres}\end{equation}
where ${\rm Ipr}_2= {\rm Ipr}(\, {\rm tr}_{\rm e}\, \varrho_0\, )
 + {\rm Ipr}(\, {\rm tr}_{\rm c}\, \varrho_0\, )
 - {\rm Ipr}\, \varrho_0$. Here, we have generalized the inverse participation
ratio to density matrices by defining:
${\rm Ipr}\, \varrho = {\rm tr}\, (\, {\rm diag}\, \varrho\, )^2$, where
${\rm diag}\, \varrho$ is the diagonal part of the density matrix $\varrho$.
In the case of pure states, $\varrho= |\psi\ra\, \la\psi|$, this definition
reduces to the familiar definition of the inverse participation ratio. Note
that even if $\varrho_0$ is assumed pure in this equation, the partial traces
of $\varrho_0$ are generally not.

\begin{figure}
\centerline{\input{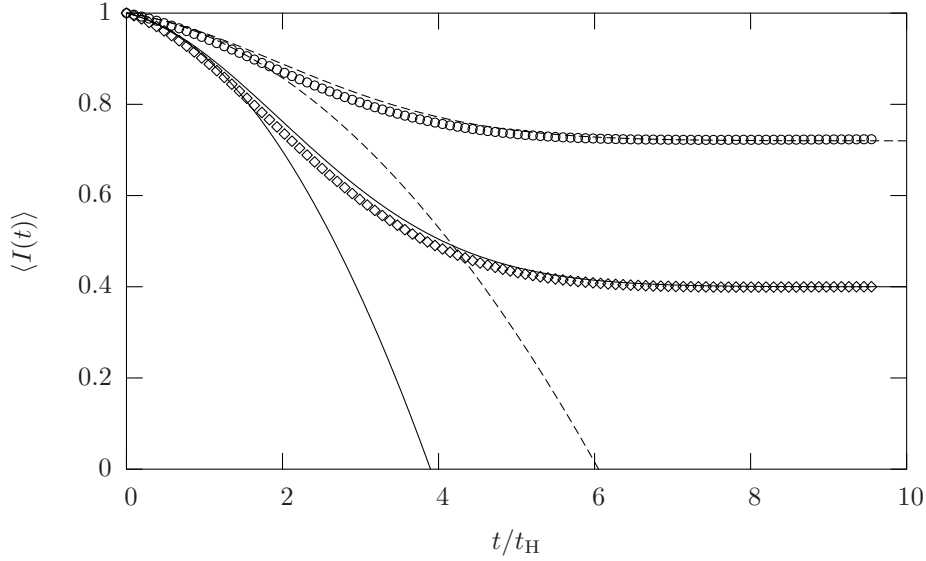}}
\caption{Purity decay
%Decay of the echo purity (here, it is identical to purity) 
in a random matrix simulation. The initial
state is a tensor product of two random states in the respective factor spaces.
The simulations are shown by diamonds ($n_{\rm c}=n_{\rm e}=10$) and by circles
($n_{\rm c}=2, n_{\rm e}=50$). The corresponding theoretical curves show that
bare linear response result, Eq.~(\ref{Q2PD:LRres}) together with its
exponentiated version for the case $n_{\rm c}=n_{\rm e}=10$ (solid lines) and
the case $n_{\rm c}=2, n_{\rm e}=50$ (dashed line). The exponentiated version
contains a phenomenological treatment of the asymptotic value (for details,
see text).}
\label{Q2PE:f:puran}\end{figure}

\begin{figure}
\centerline{\input{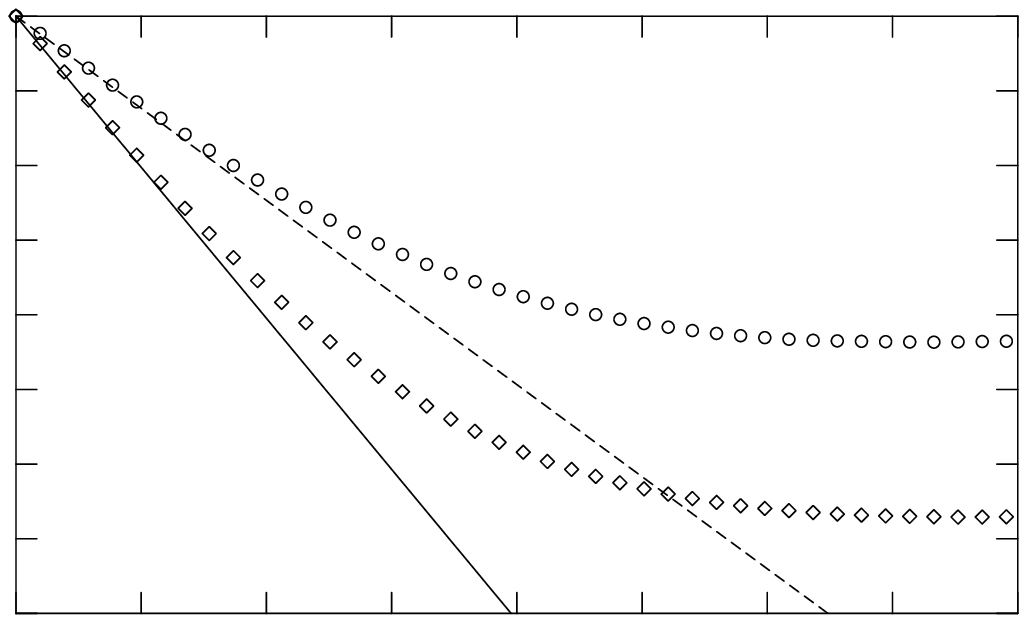}}
\caption{Purity decay
%Decay of the echo purity (here, it is identical to purity) 
in a random matrix simulation. The initial state is a product state
from the product basis of eigenststes of the uncoupled Hamiltonian with an 
energy located at the center of the spectrum. The simulations are shown by
diamonds ($n_{\rm c}=n_{\rm e}=10$) and by circles
($n_{\rm c}=2, n_{\rm e}=50$). The corresponding theoretical curves show the
plain linear response result, Eq.~(\ref{Q2PD:LRres:bas}), by a solid line
($n_{\rm c}=n_{\rm e}=10$) and a dashed line ($n_{\rm c}=2, n_{\rm e}=50$),
respectively.}
\label{Q2PE:f:purity}\end{figure}

Equation~(\ref{Q2PD:LRres}) can be tested by random matrix simulations. To
this end we assume the initial state to be pure and separable, $I(0)=1$. We
choose rather small dimensions of the Hilbert spaces involved, and compare in
Figs.~\ref{Q2PE:f:puran} and \ref{Q2PE:f:purity} the
cases $n_{\rm c}=n_{\rm e}=10$ (diamonds) and $n_{\rm c}=2, n_{\rm e}=50$
(circles). In those figures we show the average purity $\la I(t)\ra$ (it 
equals the average echo purity $\la F_P(t)\ra$ in the present
model), as a function of time, restricting ourselves to rather small
perturbations $\eps= 0.025$. In Fig.~\ref{Q2PE:f:puran} we use
initial states which are
products of random states in the respective subsystems. In that case,
${\rm Ipr}_2\, \varrho_0 = 3/(n_{\rm c}+2) + 3/(n_{\rm e}+2) -
  9/((n_{\rm c}+2)(n_{\rm e}+2))$. Here, the perturbation strength is
actually strong enough to observe the cross-over from linear to quadratic
decay. For each case, $n_{\rm c}=n_{\rm e}=10$ (solid lines) and
$n_{\rm c}=2, n_{\rm e}=50$ (dashed lines), we plot two theoretical curves,
the behavior according to Eq.~(\ref{Q2PD:LRres}) and its exponentiated
version.  For the latter we also took into account the finite limit values of
the purity at large times, in the way as described in
Section~\ref{sec:chcomposite}, Eq.~(\ref{eq:112})
\begin{equation}
\la I(t)\ra \approx F_P^* + (1-F_P^*)\;
\exp\big [\, 2\, \mathcal{C}_P(t)\, (1-F_P^*)^{-1}\, \big ] \; .
\end{equation}
In the absence of a theoretical prediction for the limit value $F_P^*$, we
fitted it to the numerical simulation. There are noticeable differences 
between the numerical simulations and the analytical predictions. We believe
these differences to be due to the particular form of the level density for
$H_0$ and the fact that we use random initial states with maximal spectral 
span. Since $H_0$ is the sum of two random spectra with constant level
density, the level density of $H_0$ has a triangular shape, whereas in the
analytical work the level density is assumed to be constant.

Figure~\ref{Q2PE:f:purity} shows the purity decay for initial states which are
eigenstates of $H_0$. In that case, the quadratic term in
Eq.~(\ref{Q2PD:LRres}) cancels, and one obtains
\begin{equation}
\la I(t)\ra = 1- 2\; (2\pi\eps)^2\; t\;
  \left( 1 - \frac{n_{\rm c}+n_{\rm e}-2}{N}\right) + \mathcal{O}(\eps^4)\; .
\label{Q2PD:LRres:bas}\end{equation}
We find that the linear response result is indeed valid, also for that case.
At small times the theoretical curves converge to the numerical results.
However, even when the saturation value of purity is taken into account (as
done in Fig.~\ref{Q2PE:f:puran}), the exponentiated linear response formula
(not plotted) does not correctly describe the numerical simulations. Note 
that due to the absence of the $t^2$-term the saturation values are much 
higher.

In~\cite{Pin06} a similar result is obtained
for an even  more involved situation relevant for quantum information.

\section{Quantum echo-dynamics: Integrable case}
\label{Q1}

We shall now turn our attention to integrable dynamics. 
The system can actually also have a mixed phase space, provided the initial 
state is located in the integrable part of phase space so that we can locally 
define action-angle variables used in the following theoretical derivations. 
We shall call such systems {\em regular}. Throughout this section we shall 
assume for simplicity that the perturbation is time 
independent~\footnote{This of course does not mean that the perturbation 
$\tilde{V}(t)$ in the interaction picture is time independent.} in the 
Schr\" odinger picture, {\it i.e.} $V(t)=V$.

The action-angle variables used in the theoretical derivations are introduced 
in Section~\ref{sec:ch4action}. Afterwards, we discuss fidelity decay for 
perturbations with non-zero time average (Section~\ref{sec:ch4Vneq0}) and then 
for perturbations with zero time average, resulting in the so-called fidelity 
freeze (Section~\ref{sec:ch4Veq0}). In Section~\ref{sec:compInt} measures for 
composite systems are discussed, in particular purity. 
Section~\ref{sec:ch4mixed} discusses fidelity decay in systems with mixed phase 
space, for cases where the purely regular theory exposed in the previous 
subsections is not sufficient.   

If the dynamics is integrable it is useful to define a {\em time-averaged perturbation} $\oV$ (\ref{eq:timeaverage}),
\begin{equation}
 \oV=\lim_{t\to \infty}{\frac{1}{t}\int_{0}^{t}{\!\!dt'\tilde{V}(t')}}=\lim_{t\to\infty}{\frac{\Sigma(t)}{t}}.
\label{eq:Vbardef}
\end{equation}
For sufficiently small $\hbar$  the above integral converges for some classical 
averaging time $t_{\rm ave}$. Observe that $\oV$ is by construction a 
constant of motion, $[\oV,U_0]=0$. For chaotic dynamics $\oV$ goes to zero in the semiclassical limit. For regular dynamics on the other hand there can be two different situations depending on the operator $\oV$: (i) the typical case where $\oV \neq 0$ and (ii) the special case where 
$\oV=0$, which gives rise to an entirely different behavior known as quantum freeze~\cite{Prosen:03}, for a general discussion of freeze see Section~\ref{sec:freeze}. If $\oV$ is nonzero the double integral of
the correlation function $C(t',t'')$ (\ref{eq:CFdef}) will also grow quadratically with time and
so one can also define an {\em time-averaged correlation function} $\oC$,
\begin{equation}
\oC=\lim_{t\to \infty}\frac{1}{t^2}\int_{0}^{t}{\int_0^t{\!\! dt''dt' C(t'',t')}}.
\label{eq:Cbarreg}
\end{equation}
$\oC$ is the plateau value of the correlation function for 
$t\to \infty$ (or the value around which the correlation function oscillates).
Correlation functions in regular systems will typically have a plateau due to the 
existence of conserved quantities. Equivalently $\oC$ can be written in 
terms of the average perturbation operator $\oV$,
\begin{equation}
\oC=\< \oV^2 \>-\< \oV\>^2\; ,
\label{eq:oC}
\end{equation}
{\it i.e.}, $\oC$ is the variance of the time averaged perturbation. For sufficiently 
small $\hbar$, the quantum correlation function can be replaced by its 
classical counterpart, {\it i.e.}, the quantum operator $V$ can be replaced by the
classical oservable $v$, such that $\oC$ is given by classical dynamics.
The time scale on which $\oC$ converges is again $t_{\rm ave}$. Physically,
this is  the time in which the classical correlation functions decay to their 
asymptotics. The linear response expression for fidelity (\ref{eq:fidlr}) can 
now be written as
\begin{equation}
F(t)=1-\frac{\eps^2}{\hbar^2} \oC t^2+\cdots \; ,
\label{eq:reglr}
\end{equation} 
valid for $t>t_{\rm ave}$.
The time scale for fidelity decay scales as $\sim 1/\eps$. Here, the quadratic
decay is not a consequence of trivial conservation laws, as discussed in 
Section~\ref{sec:conslaws}, but of the regular dynamics with non-decaying 
correlations. Also it must not be confused with the so-called Zeno decay, 
Section~\ref{sec:zeno}, which is also quadratic but happens on a very short 
time-scale on which the correlation function does not yet decay, regardless of 
the dynamics~\cite{zeno:77,zeno:80}. For a numerical comparison of Zeno 
time-scale and our quadratic decay; see Fig.~\ref{fig:2ktopd}. 

Of course one should keep in mind that the behavior of the correlation function 
(\ref{eq:CFdef}) depends on the dynamics as well as on the perturbation $V$ 
itself. For instance, even for integrable dynamics the correlation function can 
decay to zero for a sufficiently ``random'' perturbation. In such a case the 
fidelity decay in the linear response regime will be linear even for regular 
dynamics. Indeed, exponential decay of fidelity has been found in 
Ref.~\cite{Emerson:02} also in a regular regime if a random transformation of 
the basis is made before applying the perturbation (making the perturbation 
effectively random). Note that such a random perturbation has no classical 
limit, nevertheless it could be important for quantum computing where typical 
errors acting on individual qubits may not have a classical 
limit~\cite{rossini:04b} either. The application of perturbations which have 
no classical limit can result in a situation where the presence of symmetries 
can cause the fidelity decay to be exponential and slower than in the absence 
of symmetries~\cite{Weinstein:05b}. A more detailed account of fidelity studies 
in the context of quantum computing is given in Section~\ref{sec:AppQI}.

Passing to action-angle operators will 
turn out to make derivations easier and will furthermore enable us to use classical 
action-angle variables in the leading semiclassical order, thereby expressing 
quantum fidelity in terms of classical quantities, even though in some cases 
quantum and classical fidelity may behave quite differently. Therefore, before 
proceeding with the evaluation of $F(t)$ beyond the linear response let us have a look at the quantum 
action-angle operator formalism. 

\subsection{Action-angle operators}

\label{sec:ch4action}
Since we assume the classical system to be integrable (in the case of near-integrable systems at least locally, by KAM theorem) we
can employ action-angle variables, $\{j_k,\theta_k,k=1\ldots d\}$, in a system with $d$ 
degrees of freedom. In the present section, dealing with the regular regime 
we shall use lowercase letters to denote {\em classical variables} and capital letters to 
denote the corresponding {\em quantum operators}. For instance, the quantum Hamiltonian 
will be given as $H(\veb{J},\veb{\Theta})$ whereas its classical limit will be written 
as $h(\veb{j},\veb{\theta})$.

As our unperturbed Hamiltonian is integrable, it is a function of
actions only, {\it i.e.} $h_0=h_0(\veb{j})$. 
The solution of the classical equations of motion is simply
\begin{equation}
\veb{j}(t) = \veb{j}, \qquad \veb{\theta}(t) = \veb{\theta} + \veb{\omega}(\veb{j})t \pmod{2\pi},
\label{eq:classeq}
\end{equation}
with a dimensionless frequency vector $\veb{\omega}(\veb{j}) = \partial h_0(\veb{j})/\partial\veb{j}$.
In Section~\ref{sec:ch4Veq0} it will come handy to expand the
classical limit $v(\veb{j},\theta)$ of a perturbation operator $V$ into a Fourier series,
\begin{equation}
v(\veb{j},\veb{\theta}) = \sum_{\veb{m}\in\Z^d} v_{\veb{m}}(\veb{j}) 
e^{\ii\veb{m}\cdot\veb{\theta}},
\label{eq:fourier}
\end{equation}
where the multi-index $\veb{m}$ has $d$ components. The classical limit of the 
time-averaged perturbation $\bar{V}$ is $\bar{v} = v_{\veb{0}}(\veb{j})$, 
{\it i.e.} only the zeroth Fourier mode of the perturbation survives time averaging.

In quantum mechanics, one quantizes the action-angle variables using the 
EBK procedure (see {\it e.g.} Ref.~\cite{Berry:77}) where one defines the 
action (momentum) operators $\veb{J}$
and angle operators $\exp(\ii\veb{m}\cdot\veb{\Theta})$ satisfying the canonical
commutation relations, 
$[J_k,\exp(\ii\veb{m}\cdot\veb{\Theta})] = \hbar m_k \exp(\ii\veb{m}\cdot\veb{\Theta}), k = 1,\ldots,d$.
As the action operators are mutually commuting they have a common eigenbasis 
$\ket{\veb{n}}$ labeled by the $d$-tuple of quantum numbers 
$\veb{n}=(n_1,\ldots,n_d)$,
\begin{equation}
\veb{J}\ket{\veb{n}} = \veb{J}_{\veb{n}}\ket{\veb{n}}.
\label{eq:Jeigen}
\end{equation}
Here $\veb{J_n}=\hbar \{\veb{n} + \veb{\alpha}\}$ is an eigenvalue of 
operator $\veb{J}$ in an eigenstate $\ket{\veb{n}}$ and 
$0 \le \alpha_k \le 1$ are the Maslov indices which are irrelevant for 
the leading order semiclassical approximation we shall use. It follows that 
the angle operators act as shifts, $\exp(\ii \veb{m}\cdot\veb{\Theta})\ket{\veb{n}} = \ket{\veb{n}+\veb{m}}$.
The Heisenberg equations of motion can be solved in the leading 
semiclassical order by using classical equations of motion, {\it i.e.} replacing quantum $H_0(\hbar \veb{n})-H_0(\hbar(\veb{n}-\veb{m}))$ with its classical limit $\veb{m}\cdot \veb{\omega}(\hbar \veb{n})$ and disregarding the operator ordering,
\begin{eqnarray}
\veb{J}(t) &=& {\rm e}^{\ii H_0  t/\hbar}\veb{J} {\rm e}^{-\ii H_0  t/\hbar} = \veb{J},\nonumber\\
{\rm e}^{\ii\veb{m}\cdot\veb{\Theta}(t)} 
&=& {\rm e}^{\ii H_0  t/\hbar}{\rm e}^{\ii\veb{m}\cdot\veb{\Theta}} {\rm e}^{-\ii H_0  t/\hbar} \cong 
{\rm e}^{\ii\veb{m}\cdot\veb{\omega}(\veb{J})t} {\rm e}^{\ii\veb{m}\cdot\veb{\Theta}},
\end{eqnarray}
in terms of the frequency operator $\veb{\omega}(\veb{J})$. Throughout 
this paper we use the symbol $\cong$ for 
'semiclassically equal', {\it i.e}. asymptotically equal to leading order 
in $\hbar$.
Similarly, time evolution of any other observable is obtained to 
leading order by substitution of classical with quantal action-angle 
variables. For instance the perturbation $\tilde{V}(t)$ (\ref{eq:fourier}) is
\begin{equation}
\tilde{V}(t) = {\rm e}^{\ii H_0  t/\hbar}V {\rm e}^{-\ii H_0  t/\hbar} \cong
\sum_{\veb{m}} v_{\veb{m}}(\veb{J}) 
{\rm e}^{\ii\veb{m}\cdot\veb{\omega}(\veb{J})t}{\rm e}^{\ii\veb{m}\cdot\veb{\Theta}}.
\label{eq:Vtquant}
\end{equation}

\subsection{Perturbations with non-zero time average}

\label{sec:ch4Vneq0}
In this section we assume $\oV \neq 0$. 
We have seen (\ref{eq:BCH}) that the echo operator for sufficiently 
small $\eps$ can be written as
\begin{equation}
M_\eps(t)\approx\exp\left\{-\frac{\ii}{\hbar}\left(
\Sigma(t)\eps + \frac{1}{2}\Gamma(t)\eps^2\right)\right\},
\label{eq:ch4echo}
\end{equation}
where $\Gamma(t)$ (\ref{eq:defGamma}) grows at most linearly with $t$. 
If $\oV\neq 0$, the first term in the 
exponential will grow as $\Sigma(t)= \oV t$ for times longer than the classical averaging 
time $t_{\rm ave}$ (\ref{eq:Vbardef}). Therefore, provided $\eps<1$, the term involving 
$\Gamma(t)$ will be $1/\eps$ times smaller than the first one 
and can be neglected. Replacing $\Sigma(t)$ with $\oV t$ we write 
the fidelity amplitude as
\begin{equation}
f(t)=\ave{{\rm e}^{-\ii \eps t \oV/\hbar}},\qquad t \gg t_{\rm ave}.
\label{eq:ch4fvbar}
\end{equation}   
As the average perturbation operator $\oV$ is diagonal in the eigenbasis 
of actions, $\oV\; \ket{\veb{n}}=\oV_\veb{n} \ket{\veb{n}}$, the above expectation value is especially 
simple in the action eigenbasis $\ket{\veb{n}}$,
\begin{equation}
f(t)=\sum_{\veb{n}}{\exp{\left(-\ii \eps t \oV_\veb{n}/\hbar \right)} 
D_\rho(\hbar \veb{n})},\qquad D_\rho(\hbar \veb{n})=\bracket{\veb{n}}{\rho}{\veb{n}}.
\label{eq:LDOS}
\end{equation}
Now we make a leading order semiclassical approximation by 
replacing quantum operator $\oV$ by its classical limit $\bar{v}(\veb{j})$ and replacing 
the sum over quantum numbers $\veb{n}$ with the integral over classical 
actions $\veb{j}$ (note small-cap letter). By denoting with $d_\rho(\veb{j})$ 
the classical limit of $D_\rho(\hbar \veb{n})$, we arrive at the 
fidelity amplitude,
\begin{equation}
f(t) \cong \hbar^{-d}\int{\!{\rm d}^d \veb{j} \exp{\left(-\ii 
\frac{\eps}{\hbar} t \bar{v}(\veb{j})\right)} \,d_\rho(\veb{j})}.
\label{eq:fnASI}
\end{equation}  
This equation will serve us as a starting point for all calculations of fidelity decay in
regular systems for non-residual perturbations. The replacement of the sum with the 
action space integral (ASI) is 
valid up to such classically long times $t_{\rm a}$, that the variation of the argument 
in the exponential across one Planck cell is small,
\begin{equation}
t_{\rm a}=\frac{1}{|\partial_{\veb{j}}\bar{v}|\eps}\sim \hbar^0/\eps.
\label{eq:regulart*}
\end{equation}
Subsequently we shall see that the fidelity decays in times shorter than this, and thus 
the ASI approximation is justified. The ASI representation 
(\ref{eq:fnASI}) will be now used to evaluate the fidelity for different 
initial states $d_\rho(\veb{j})$.

\subsubsection{Coherent initial states}

First we take the initial state to be a coherent 
state $\ket{\veb{j}^*,\veb{\theta}^*}$ centered at action and angle 
position $(\veb{j}^*,\veb{\theta}^*)$. Expansion coefficients of a general 
coherent state (a localized wave packet) can be written as
\begin{equation}
\braket{\veb{n}}{{\veb{j}^*,\veb{\theta}^*}} = 
\left(\frac{\hbar}{\pi}\right)^{d/4}
\!\!\!\left|\det\Lambda\right|^{1/4}
\exp\left\{-\frac{1}{2\hbar}(\veb{J}_{\veb{n}} - \veb{j}^*)\cdot\Lambda(\veb{J}_{\veb{n}}-
\veb{j}^*) - 
\ii\veb{n}\cdot\veb{\theta}^*\right\},
\label{eq:coh}
\end{equation}
where $\Lambda$ is a positive symmetric $d\times d$ matrix of squeezing 
parameters. 
The classical limit of $D_\rho$ (\ref{eq:LDOS}) is therefore
\begin{equation}
d_\rho(\veb{j}) = (\hbar/\pi)^{d/2}\left|\det\Lambda\right|^{1/2}
\exp(-(\veb{j}-\veb{j}^*)\cdot\Lambda (\veb{j}-\veb{j}^*)/\hbar).
\label{eq:drhocoh}
\end{equation}
The ASI (\ref{eq:fnASI}) for the fidelity amplitude can now be evaluated 
by the stationary phase method with the result~\cite{Prosen:02corr},
\begin{equation}
F(t)=\exp{\left\{ -(t/\tau_{\rm r})^2\right\}},\qquad  \tau_{\rm r}=
\frac{1}{\eps}\sqrt{\frac{2\hbar}{\bar{\veb{v}}'\cdot\Lambda^{-1}\bar{\veb{v}}'}},
\label{eq:Fnregcoh}
\end{equation}
where the derivative of the average perturbation is
\begin{equation}
\bar{\veb{v}}' = \frac{\partial\bar{v}(\veb{j}^*)}{\partial\veb{j}},
\label{eq:vbarc}
\end{equation}
and is evaluated at the position of the initial packet $\veb{j}^*$.
Comparing the fidelity (\ref{eq:Fnregcoh}) with the linear response 
formula (\ref{eq:reglr}) we see that the average correlation function 
for a coherent initial state, 
$\bar{C}=\frac{1}{2}\hbar(\bar{\veb{v}}'\cdot\Lambda^{-1}\bar{\veb{v}}')$, 
is proportional to $\hbar$ because the size of the packet scales with $\hbar$.

We thus find a Gaussian decay of fidelity for coherent initial states in regular systems. This has 
very simple physical interpretation. As the fidelity is an overlap of the 
initial state with the state obtained after an echo, {\it i.e.} after propagation 
with $M_\eps(t)=\exp{(-\ii \bar{V}(\veb{J})\eps t/\hbar)}$, we can 
see that the effective Hamiltonian for this evolution is  
$\bar{V}(\veb{J})$. It depends only on action variables and therefore 
its classical limit $\bar{v}(\veb{j})$ generates a very simple classical 
evolution. Only the frequencies of tori are changed by the amount 
$\Delta \veb{\omega}=\eps \bar{\veb{v}}'$, while the shapes of tori do not change. 
This change in frequency
causes the ``echo'' packet $M_\eps(t) \ket{\psi(0)}$ to move {\em
  ballistically} away from its initial position and as a consequence
fidelity decays. The functional form of this decay is directly connected 
with the shape of the initial packet. For coherent initial states it is
Gaussian while for other
forms of localized initial packets it will be correspondingly different 
but with the same dependence of the regular decay time $\tau_{\rm r}$ on the 
ballistic
separation ``speed'' $\bar{\veb{v}}'$ and perturbation strength $\eps$. 
Note that the decay of classical fidelity is more complicated, see Section~\ref{C}.

In one dimensional systems ($d=1$) another phenomenon will be
observable. After long times the echo packet will make a whole
revolution around the torus causing the fidelity to be large
again. This will happen after the so-called beating time $t_{\rm b}$ 
determined by the condition~\footnote{In the presence of symmetries 
the beating time can be shorter than the one given by Eq.~(\ref{eq:tb}).} 
$\eps \bar{v}'(j^*) t_{\rm b}=2\pi$,
\begin{equation}
t_{\rm b}=\frac{2\pi}{\bar{v}' \eps}.
\label{eq:tb}
\end{equation}
This beating phenomena is particular to one dimensional systems as in 
general the incommensurability of frequencies will suppress the revivals 
of fidelity in more than one degree of freedom systems ({\it i.e.} $t_{\rm b}$ 
for derivative of $\bar{v}$ in each action direction would have to be the 
same), for a numerical example of a two degree of freedom system see 
Ref.~\cite{Prosen:02corr}. Strong revivals of fidelity have been discussed also 
in Ref.~\cite{Sanka:03}. In case of commensurate frequencies one can have 
revivals also in many-dimensional systems, for an example of $d=2$ dimensional 
Jaynes-Cummings model see Refs.~\cite{Angelo:01,Prosen:03corr}. In 
Ref.~\cite{Combescure:05a} quantum fidelity has been exactly calculated for a
system with the Hamiltonian $H_\eps=p^2/2+g(t)q^2/2+\eps/q^2$, with an
arbitrary periodic function $g(t)$, and it has been shown that there are 
perfect periodic recurrences of fidelity.  

\begin{figure}[ht]
\centerline{\includegraphics[width=\figw\textwidth]{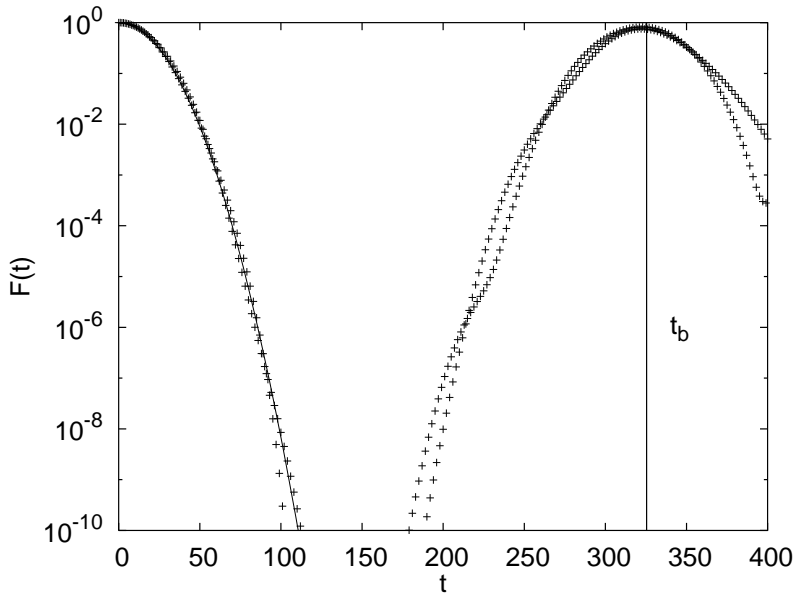}}
\caption{Quantum fidelity for a coherent initial state of the regular kicked 
  top (\ref{eq:KTdef}) [from~\cite{Prosen:02corr}]. The solid line shows
  the predicted Gaussian decay (\ref{eq:ch4regcohtheo}), the point symbols 
  ``+'' show the numerical results. The vertical line indicates the theoretical 
  beating time (\ref{eq:tb}) which is particular to one degree of freedom 
  systems.}
\label{fig:ch4regcoh}
\end{figure}

To illustrate the above theory for fidelity decay we compare it in 
Fig.~\ref{fig:ch4regcoh} with the results of numerical 
simulation. The model we choose is a kicked top with a unit-step 
propagator given in Eq.~(\ref{eq:KTdef}), using $\gamma=\pi/2$ and small $\alpha=0.1$ 
giving regular dynamics. Evaluation of the theoretical 
formula (\ref{eq:Fnregcoh}) in the limit of vanishing $\alpha$ 
gives for this particular perturbation the theoretical fidelity decay
\begin{equation}
F(t)=\exp{\left(- \eps^2 S t^2 y^2(1-y^2)/8 \right)},
\label{eq:ch4regcohtheo}
\end{equation}
if $y$ is the position of the initial coherent state on a sphere. Nice 
agreement of numerics ($S=100$ and $\eps=0.0025$) with the 
theory can be seen as well as fidelity recurrence at the predicted time 
$t_{\rm b}$. 

\subsubsection{Random initial states}

\label{sec:ch4RIS}
In this section we shall consider random initial states, that is pure states 
whose expansion coefficients are independent Gaussian random numbers.
% in the basis of the unperturbed system. 
%TG The basis in which we draw this random coefficients is for spin systems 
%TG the eigenbasis of $S_{\rm z}$ operator. 
In the semiclassical limit, {\it i.e.} for large Hilbert space 
dimension $\mathcal{N}$, the expectation value of an echo operator is self 
averaging, meaning that the fidelity for one particular random state is
equal to the fidelity averaged over the whole Hilbert space. Therefore, we 
calculate $f(t)$ via Eq.~(\ref{eq:fidech}) choosing the initial state as 
the density matrix $\rho=\mathbbm{1}/\mathcal{N}$. The expansion coefficients 
take the form $D_\rho(\veb{J})=1/\mathcal{N}$ (\ref{eq:LDOS}) and the 
corresponding classical quantity is
\begin{equation}
d_\rho(\veb{j})=\frac{(2\pi\hbar)^d}{\mathcal{V}},
\label{eq:ch4dRIS}
\end{equation}
where we calculated the dimension of the Hilbert space $\mathcal{N}$ using 
the Thomas-Fermi rule. For short times the decay of fidelity is again 
quadratic (\ref{eq:reglr}), albeit with a different $\oC$ than for 
coherent states. For longer times the ASI (\ref{eq:fnASI}) giving $f(t)$ 
can be calculated using a stationary phase method if the phase 
$\eps \bar{v} t/\hbar$ of the echo operator changes fast enough. This 
is valid for $t > \hbar/\eps$. Note on the other hand, that there is 
also an upper limit for ASI determined by $t_{\rm a}\sim 1/\eps$ 
(\ref{eq:regulart*}). In contrast to the coherent initial state, we can 
now have more than one stationary point. If we have $p$ points, 
$\veb{j}_\eta, \eta=1,\ldots,p$, where the phase is stationary, 
$\partial\bar{v}(\veb{j}_\eta)/\partial\veb{j}=\veb{0}$, the ASI
gives~\cite{Prosen:02corr}
\begin{equation}
f(t) = \frac{(2\pi)^{3d/2}}{\mathcal{V}}\left|\frac{\hbar}{t\eps}\right|^{d/2}\sum_{\eta=1}^p
\frac{\exp\{-\ii t\bar{v}(\veb{j}_\eta)\eps/\hbar - \ii \nu_\eta \}}{|\det 
\bar{\veb{V}}_\eta|^{1/2}},
\label{eq:regpower}
\end{equation}
where $\{ \bar{\veb{V}}_\eta \}_{kl} = \partial^2 \bar{v}(\veb{j}_\eta)/\partial j_k\partial j_l$ 
is a matrix of second
derivatives at the stationary point $\veb{j}_\eta$, and $\nu_\eta = \pi(m_+ - m_-)/4$
where $m_{\pm}$ are the numbers of positive/negative eigenvalues of the matrix
$\bar{\veb{V}}_\eta$. In above derivation we also assumed that phase 
space is infinite. In a finite phase space we shall have diffractive 
oscillatory
corrections in the stationary phase formula, see
numerical results below or Ref.~\cite{Prosen:03}. It is interesting to 
note the
power-law dependence on time and perturbation strength. In the simple case 
of one stationary point, fidelity will asymptotically decay as
\begin{equation}
F(t) \asymp \left[\hbar/(t\eps)\right]^d,
\label{eq:ch4fris}
\end{equation}
where the sign $\asymp$ will denote ``in the asymptotic limit'' throughout this
section. An interesting, though still largely unexplored, question is what happens in 
the thermodynamic limit $d\to \infty$? From the above formula we see that 
with increasing dimensionality $d$ of a system the decay gets faster. 
This allows for a possible crossover to a Gaussian decay when approaching 
the thermodynamic limit. Such behavior has been observed in a class of 
kicked spin chains~\cite{Prosen:02}. Agreement with Gaussian or expeonetial decay beyond 
linear response is frequently observed also
for finite $d$, {\it e.g.} in a spin model of quantum
computation~\cite{Prosen:01}.

\begin{figure}[ht]
\centerline{\includegraphics[width=\figw\textwidth]{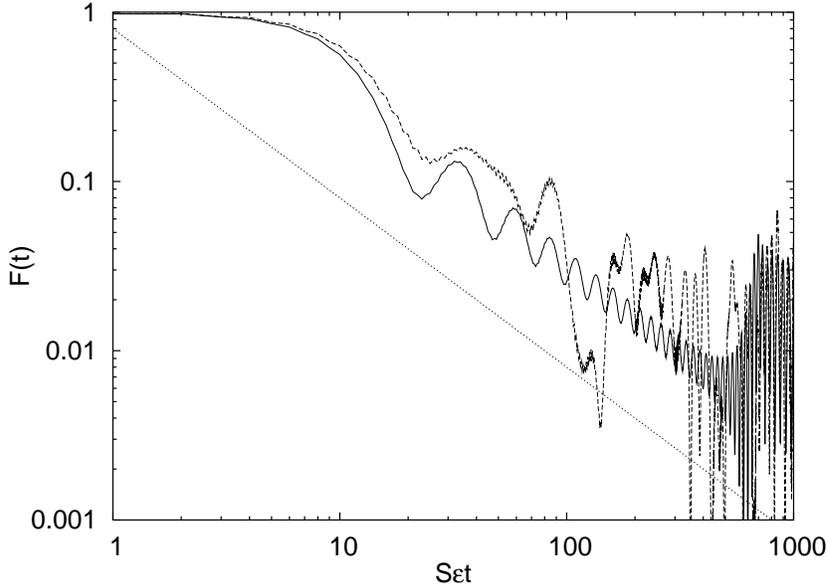}}
\caption{Power law decay of fidelity for the regular kicked top (\ref{eq:KTdef})
[from~\cite{Prosen:02corr}] and random initial state. The solid curve gives the results of 
a numerical simulation for $\rho=\mathbbm{1}/\mathcal{N}$, the dotted line gives a 
predicted asymptotic decay $\propto t^{-1}$, and the wiggling dashed 
curve represents the numerics for a single random state, $\mathcal{N}=100$.}
\label{fig:ch4analit}
\end{figure}
In Fig.~\ref{fig:ch4analit} we show results of numerical simulation 
for the same regular kicked top used for coherent initial states (\ref{eq:KTdef}). Note the 
oscillating decay due to finite phase space.

\subsubsection{Average Fidelity for Coherent States}

\label{sec:ch4avgF}
One might be also interested in fidelity averaged over the position of the initial coherent state,
\begin{equation}
\ave{F(t)}_\veb{j}=\frac{(2\pi)^d}{\mathcal{V}}\int{\!{\rm d}^d \veb{j} F(t,\veb{j})}.
\label{eq:CISavg}
\end{equation}
The asymptotic decay of average fidelity will be dominated by regions in the action space where 
the decay $F(t,\veb{j})$ is the slowest. Denoting all $\veb{j}$ dependent terms by a 
non-negative scalar function $g(\veb{j})$, the fidelity decay for a single coherent state can be 
written as $F(t,\veb{j})=\exp{(-\eps^2 t^2 g(\veb{j})/\hbar)}$ 
(\ref{eq:Fnregcoh}). For large $\eps^2 t^2/\hbar$ the main contribution 
to the average will come from regions around zeros of $g(\veb{j})$, where the 
fidelity decay is slow. In general there can be many zeros due to divergences in $\tau_{\rm r}$ 
(\ref{eq:Fnregcoh}), but for simplicity let us assume there is a single 
zero at $\veb{j}_*$ of order $\eta$. The asymptotic decay can then be 
calculated and scales as~\cite{Znidaric:04}
\begin{equation}
\ave{F(t)}_\veb{j} \asymp \left( \frac{\hbar}{\eps^2 t^2}\right)^{d/\eta}.
\label{eq:Fasim}
\end{equation}
The asymptotic decay is therefore algebraic with the power depending on the number of degrees of
freedom $d$ and the order $\eta$ of the zero. For infinite phase space $\eta$ can only be
an even number, whereas for a finite space $\eta$ can also be odd. Note 
that for regular systems the fidelity averaged over the position of 
coherent states $\ave{F(t)}_\veb{j}$ is not necessarily equal to the 
fidelity for a random initial state (\ref{eq:ch4fris})~\footnote{For chaotic systems 
the two averages of course agree.}. Whereas for fidelity decay of 
a random initial state zeros of $\bar{v}(\veb{j})$ matter, for 
$\ave{F(t)}_\veb{j}$ singularities of $\tau_{\rm r}$ (\ref{eq:Fnregcoh}) 
determine asymptotic decay ({\it e.g.} zeros of $\bar{\veb{v}}'$).

\begin{figure}[ht!]
\centerline{\includegraphics[width=\figw\textwidth]{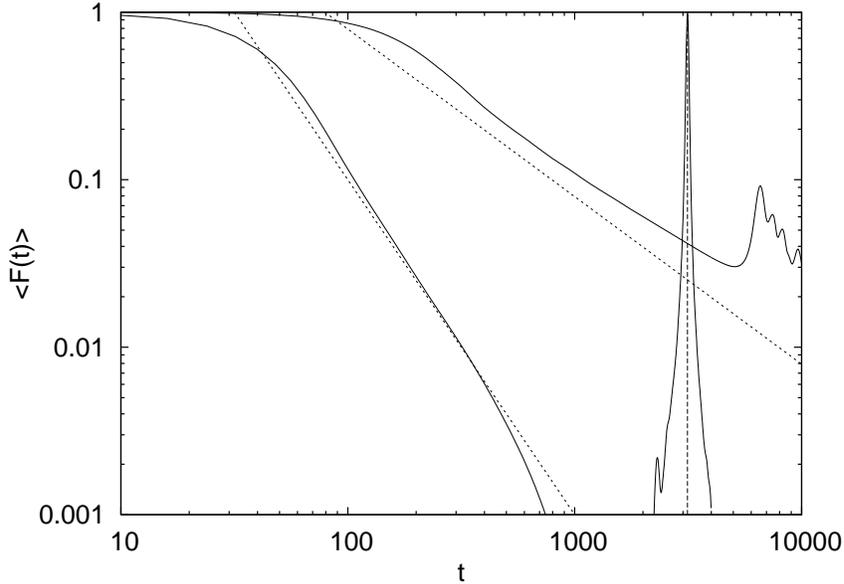}}
\caption{Average fidelity decay $\ave{F(t)}_\veb{j}$ for a regular 
kicked top [from~\cite{Znidaric:04}]. For two different perturbations, Eq.~(\ref{eq:KTdef}) 
and Eq.~(\ref{eq:KT1def}), different asymptotic power law is obtained. Top set of curves is for a 
perturbation in $\alpha$ (\ref{eq:KTdef}) and bottom is for a perturbation 
in the rotation angle (\ref{eq:KT1def}). Full curves in both cases represent 
numerics while dotted lines are theoretical predictions (\ref{eq:avgZ}) and 
(\ref{eq:avgY}) without free fitting parameters.}
\label{fig:ch4avg}
\end{figure}
To illustrate the above points we calculated the average fidelity decay for a
kicked top and two different perturbations. In the first case we take 
the system already used for coherent states (\ref{eq:KTdef}), for which 
$g(j)=j^2(1-j^2)/8$ (\ref{eq:ch4regcohtheo}). We have three zeros, 
with the asymptotic decay given by the one of highest order, $\eta=2$, 
causing asymptotic decay,
\begin{equation}
\ave{F(t)}_{\veb{j}} \asymp \frac{\sqrt{2\pi}}{\eps t \sqrt{S}}.
\label{eq:avgZ}
\end{equation}
As a second example we take the same regular kicked top system and 
parameters, only the perturbation is now in rotation angle (\ref{eq:KT1def}), whereas before it
was in the twist parameter $\alpha$. The function $g$ is in this case $g(j)=(1-j^2)/2$. 
As opposed to the previous case, we now have only two zeros of order $\eta=1$ at $j_*=\pm 1$. 
The asymptotic decay is therefore
\begin{equation}
\ave{F(t)}_{\veb{j}} \asymp \frac{1}{\eps^2 t^2 S}.
\label{eq:avgY}
\end{equation}
In Fig.~\ref{fig:ch4avg} the results of a numerical simulation are compared 
with theory. Good agreement is observed with the asymptotic predictions 
(\ref{eq:avgZ}) and (\ref{eq:avgY}). Strong revival of fidelity for the case 
of perturbation in the rotation angle (\ref{eq:KT1def}) seen in the figure 
is particular to this perturbation because the beating time $t_{\rm b}$ 
(\ref{eq:tb}) does not depend on $j$. Generally such revivals are absent 
from $\ave{F(t)}_{\veb{j}}$. We can therefore see that the asymptotic decay 
of average fidelity $\ave{F(t)}_{\veb{j}}$ is algebraic with the power 
being {\em system dependent}.

\subsection{Perturbations with zero time average}

\label{sec:ch4Veq0}
Above we considered general perturbations, for which the 
double correlation integral (\ref{eq:fidlr}) in regular systems was growing 
quadratically with time, {\it i.e.} time averaged perturbation $\oV$ is nonzero. 
In the present section we discuss the situation 
when this correlation integral does not grow with time (asymptotically for 
large times), that is the time averaged
perturbation is zero, $\oV \equiv 0$. This are the so called {\em residual}
perturbations because $V=\Vres$. Most of the results have been published in 
Ref.~\cite{Prosen:03}. We shall heavily rely on general derivations presented in
Section~\ref{sec:freeze}. As the short time fidelity 
is given by the operator $\Sigma(t)$ (\ref{eq:SW}) and its norm does 
not grow with time, the fidelity will {\em freeze} at a constant value called
the plateau (\ref{eq:Fnplateau}). After a sufficiently long time it will again start to decay 
due to the operator $\Gamma(t)$ (\ref{eq:GR}). The main tool for theoretical derivations 
will be similar as for non-residual perturbations, that is using action-angle variables and 
evaluating resulting semiclassical ASI using stationary phase method. 

\subsubsection{Coherent initial states}
\label{sec:freezeCIS}

Similarly as we expanded the perturbation $V$ (\ref{eq:fourier}) into a 
Fourier series we can also expand the operator $W(t)$ which is needed to calculate the 
freeze plateau (\ref{eq:Fnplateau}). We write it in 
terms of action-angle operators, $W=W(\veb{J},\veb{\Theta})$ and then 
write the corresponding classical quantity $w(\veb{j},\veb{\theta})$ as,
\begin{equation}
w(\veb{j},\veb{\theta}) = \sum_{\veb{m}\neq\veb{0}}w_{\veb{m}}(\veb{j})
{\rm e}^{\ii\veb{m}\cdot\veb{\theta}}.
\label{eq:fourierW}
\end{equation}
The expansion coefficients $w_\veb{m}$ are easily expressed in terms of $v_\veb{m}$,
\begin{equation}
w_\veb{m}=-\ii \frac{v_\veb{m}}{\veb{m}\cdot\veb{\omega}}.
\label{eq:ch4wm}
\end{equation}
The time dependent value of the plateau (\ref{eq:Fnplateau}) is now in the leading semiclassical order
\begin{equation}
f_{\rm plat}(t) \cong \ave{\exp{\left(-\ii \eps w(\veb{J},\veb{\Theta}+
\veb{\omega}t)/\hbar\right)}\exp{\left(\ii \eps w(\veb{J},\veb{\Theta})/\hbar \right)}}.
\end{equation}
For sufficiently large time, say $t>t_1$, the 
phase $\veb{\omega}(\veb{J})t$ in the argument of the first exponential 
function will change over the initial state by more than $2\pi$. In this
case the averaging over initial states will yield the same result as 
averaging over time. Therefore, for $t>t_1$ we can replace the time dependent 
$f_{\rm plat}(t)$ with its time average,
\begin{equation}
f_{\rm plat} \cong \lim_{t \to \infty}{\frac{1}{t}\int_{0}^t{\!dt'f(t')}}.
\label{eq:ch4favg}
\end{equation}
But averaging over time is equal to averaging over angle, so we can write 
the expression for fidelity plateau as
\begin{equation}
f_{\rm plat} \cong \ave{\exp{\left( \ii \frac{\eps}{\hbar} w(\veb{J},
\veb{\Theta}) \right)} \int{\frac{{\rm d}^d\veb{x}}{(2\pi)^d} 
\exp{\left( -\ii\frac{\eps}{\hbar} w(\veb{J},\veb{x}) \right)} } }.
\label{eq:fplatreg}
\end{equation}
This is a general expression for $f_{\rm plat}$ valid for any initial state. 
Now we shall assume the initial state to be a coherent state.

Let us first estimate the time $t_1$ after which the above averaging is 
justified. We demand $\veb{m}\Delta\veb{\omega}t_1 > 2\pi$, where 
$\Delta\veb{\omega}$ is a frequency change over the packet and $\veb{m}$ 
is the mode number of certain Fourier mode (\ref{eq:fourierW}) $w_{\veb{m}}$. 
The frequency change over the initial packet is $\Delta \veb{\omega} \approx \Omega \Delta\veb{j}$, 
where $\Omega_{jk}=\partial \omega_j(\veb{j}^*)/\partial j_k$ and $\Delta \veb{j}$ is the size of 
the initial packet. For coherent states (\ref{eq:drhocoh}) it is 
$\Delta\veb{j}\approx \sqrt{\hbar/\Lambda}$ so that we get an estimate of $t_1$ for 
coherent initial states, $t_1 \approx \sqrt{\Lambda}/(\veb{m}\Omega \sqrt{\hbar})$. 
A more rigorous derivation~\cite{Prosen:03} shows that the Fourier mode $w_{\veb{m}}$ 
decays with a Gaussian envelope on the time scale $t_1$,
\begin{equation}
t_1 = \frac{1}{\sqrt{\frac{\hbar}{4}\left(\veb{m}\cdot \Omega \Lambda^{-1}
\Omega^{\rm T}\veb{m}\right)}}
\propto \hbar^{-1/2}.
\label{eq:t1}
\end{equation}
Therefore, all time dependent terms in $f_{\rm plat}(t)$ decay with a 
Gaussian envelope on the above time scale $t_1$. After $t_1$, but before 
$t_2$ when $\Gamma(t)$ becomes important, the fidelity is constant and 
we have a fidelity freeze. We proceed to evaluate the plateau value of this freeze 
$f_{\rm plat}$ (\ref{eq:fplatreg}) 
for coherent initial states. We shall use the fact that the expectation 
value of an arbitrary quantity is to leading order equal to this quantity 
evaluated at the position of the packet, 
$\ave{\exp(-(\ii\eps/\hbar)g(\veb{J},\veb{\Theta}))} \cong 
\exp(-(\ii\eps/\hbar)g(\veb{j}^*,\veb{\theta}^*))$, 
for sufficiently smooth function $g$, provided that the size of the wave-packet
$\sim\sqrt{\hbar}$ is smaller than the oscillation scale of the
exponential $\sim \hbar/\eps$, {\it i.e.} provided $\eps \ll \hbar^{1/2}$.
Then the squared modulus of $f_{\rm plat}$ (\ref{eq:fplatreg}) reads as
\begin{equation}
F_{\rm plat}^{\rm CIS} \cong \frac{1}{(2\pi)^{2d}}
\left|\int {\rm d}^d\veb{\theta} \exp\left(-\frac{\ii\eps}{\hbar}w(\veb{j}^*,
\veb{\theta})\right)\right|^2,
\label{eq:plateauCS}
\end{equation}
where superscript CIS is abbreviation for coherent initial state, {\it i.e.} localized 
wave packet. This equation is the expression for the plateau for arbitrary residual perturbation. 
The integral over angles can not be done in general but if $\eps<\hbar$ then only the 
linear response expression for the plateau is needed. 
Expanding $F_{\rm plat}^{\rm CIS}$ we obtain
\begin{equation}
1-F_{\rm plat}^{\rm CIS}=\frac{\eps^2}{\hbar^2}\nu_{\rm CIS},\qquad \nu_{\rm CIS}= 
\sum_{\veb{m}\neq\veb{0}}{|w_\veb{m}(\veb{j}^*)|^2}.
\label{eq:nuCIS}
\end{equation}
The integral in Eq.~(\ref{eq:plateauCS}) can be done 
analytically if the perturbation $w(\veb{j},\veb{\theta})$ 
(\ref{eq:fourierW}) has a {\em single} nonzero Fourier mode, say 
$w_{\pm\veb{m_0}}$. In that case the integral gives
\begin{equation}
F_{\rm plat}^{\rm CIS}=J_0^2\left(2\frac{\eps}{\hbar} |w_{\veb{m}_0}(\veb{j}^*)|\right),
\label{eq:ch4J0}
\end{equation}
were $J_0$ is the zero order Bessel function. The plateau for a single mode residual perturbation 
is therefore given by a Bessel function.

\begin{figure}[h!]
\centerline{\includegraphics[width=\figw\textwidth]{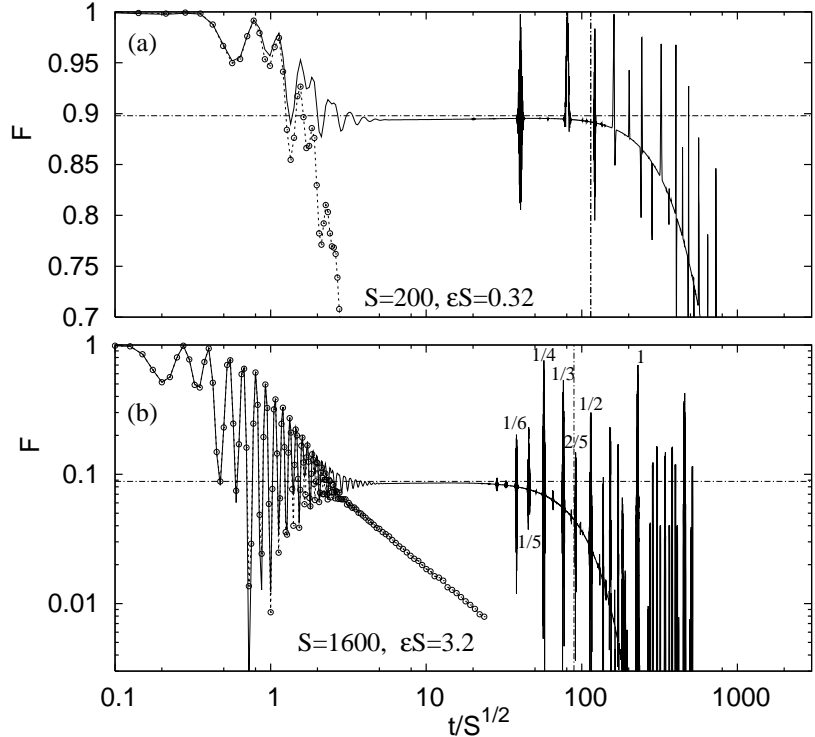}}
\caption{Fidelity freeze for coherent initial states in a regular kicked 
  top (\ref{eq:KT3def}) [from~\cite{Prosen:03}]. Symbols connected with dashed 
  lines denote the corresponding classical fidelity while the solid line
  represents a quantum simulation. After the initial decay, when the quantum 
  fidelity agrees with the classical one (until $t_1$), quantum fidelity 
  freezes at the plateau (horizontal chain line, Eq.~(\ref{eq:ch4J0})). 
This plateau lasts long (note the log scale) until at $t_2$ (\ref{eq:t2coh}), denoted by vertical 
chain line, decay starts again. In (b) we also indicate 
fractional $k/p$ resonances with 
$k/p$ marked on the figure.} 
\label{fig:ch4shortcoh}
\end{figure}

\begin{figure}[h!]
\centerline{\includegraphics[width=\figw\textwidth]{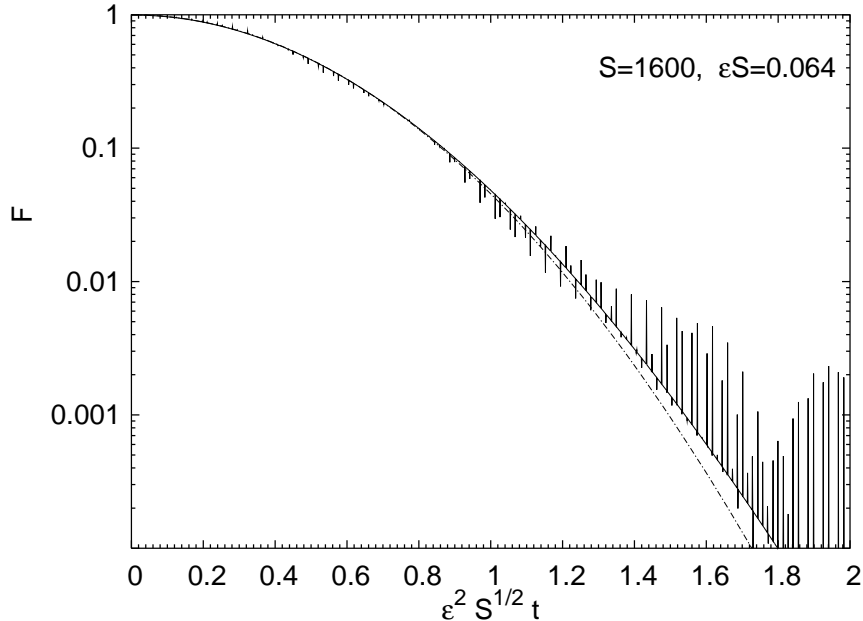}}
\caption{Long time ballistic decay after the freeze plateau for the regular 
  kicked top (\ref{eq:KT3def}) with 
coherent initial state and residual perturbation [from~\cite{Prosen:03}]. Chain curves indicate 
theoretical Gaussian decay (\ref{eq:reslong}) with analytically computed 
coefficients. The decay is the same as for non-residual perturbations but with a "renormalized" 
perturbation strength $\eps^2/2$ and perturbation $\bar{r}$ instead of $\bar{v}$.}
\label{fig:longcoh}
\end{figure}

As a simple illustration of the above theory we again take the kicked top 
with a one-step propagator given in Eq.~(\ref{eq:KT3def}). Using classical 
$w(j,\theta)$ (\ref{eq:wclas}) and a single-mode formula (\ref{eq:ch4J0}) we get
\begin{equation}
F_{\rm plat}^{\rm CIS}=J_0^2\left(\eps S \frac{\sqrt{1-j^{*2}}}{2\sin{(\alpha j^*/2)}} \right).
\label{eq:bessel}
\end{equation}
Comparison of theory with numerics is shown in Fig.~\ref{fig:ch4shortcoh}.
We can see that up to time $t_1$ (\ref{eq:t1}) quantum fidelity follows 
the classical one (circles in Fig.~\ref{fig:ch4shortcoh}). After $t_1$ 
the correspondence breaks as the classical fidelity decays as a power law, 
while quantum fidelity freezes. For details about classical fidelity see Section~\ref{C}. 
The value of the plateau 
agrees with the linear response formula (\ref{eq:nuCIS}) for weak perturbations and with the full 
expression (\ref{eq:bessel}) for strong perturbation. Vertical chain lines 
show theoretical values of $t_2$ (calculated later), which is the time 
when the plateau ends. Also, at certain times  quantum fidelity exhibits 
resonances or strong revivals. These ``spikes'' occurring at 
regular intervals can be seen also in a long time decay of fidelity in 
Fig.~\ref{fig:longcoh}. These are called the echo resonances 
and are particular to one-dimensional systems and localized initial packets, 
so that the variation of the frequency derivative 
$\omega'={\rm d}\omega(j)/{\rm d}j$ over the wave packet is small. Due to the
discreteness of quantum energies, at certain times a constructive 
interference occurs, resulting in a revival of fidelity. These echo resonances must not 
be confused with the revivals of fidelity due to the beating phenomenon (\ref{eq:tb}) which is 
classical, whereas echo resonances are a quantum phenomenon. The time of occurrence of a 
resonance $t_{\rm res}$ is given by
\begin{equation}
t_{\rm res}=\frac{2\pi}{\hbar \omega'},
\label{eq:ch4tr}
\end{equation}
or at fractional multiples of basic $t_{\rm res}$. Echo resonances 
have been described and explained in detail in Ref.~\cite{Prosen:03} and they can 
also be observed 
in the numerical results shown in Refs.~\cite{Sanka:03} and~\cite{Weinstein:05}. For a more 
mathematically oriented derivation of the condition under which they occur see 
also Ref.~\cite{Combescure:05b}. The same resonant times are also found for 
revivals of the wave-packet after just 
forward evolution studied in Refs.~\cite{Braun:96} and~\cite{Leichtle:96}, 
see also the review~\cite{Robinett:04}. For the quantum kicked rotor in a 
regime of quantum resonance similar phenomenon as freezing in regular systems 
can be observed due to the formal similarity of propagator~\cite{Wimberger:06}. 
It is important 
to realize that freeze is pure quantum phenomenon; classical fidelity does not exhibit freeze 
(see figure \ref{fig:ch4shortcoh}) even though the perturbation is exactly the same 
as in the quantum case!
 
The operator $\Sigma(t)$ determines the plateau. Long time decay
on the other hand will be dictated by the operator $\Sigma_{\rm R}(t)$ (\ref{eq:SR_def}). For
long times, when this decay will take place, we can define a time averaged
operator $R$ (\ref{eq:GR}),
\begin{equation}
\bar{R}=\lim_{t \to \infty}{ \frac{\Sigma_{\rm R}(t)}{t} }=
\lim_{t \to \infty}{ \frac{1}{t} \int_0^t{\!dt' R(t')} },
\label{eq:Rbar}
\end{equation}
and approximate $\Gamma(t) \approx \bar{R} t$. This approximation is justified, because the
fidelity decay will happen on a long time scale $\sim 1/\eps^2$,
whereas the average of $R$ converges in a much shorter classical 
averaging time $t_{\rm ave}$. For $R$ the semiclassical limit $r$ can be calculated 
using the Poisson brackets instead of commutators, $r=-\{ w,dw/dt\}$. When 
we average $r$ over time, only the zeroth Fourier mode survives resulting in
\begin{equation}
\Gamma(\veb{J})/t \cong \bar{r}(\veb{J}),\qquad \bar{r}(\veb{j}) = 
-\sum_{\veb{m}\neq\veb{0}}{\veb{m}
\cdot\partial_\veb{j}\left\{|w_\veb{m}(\veb{j})|^2 \, \veb{m}\cdot\veb{\omega}(\veb{j})\right\}}.
\label{eq:oGw}
\end{equation}
Provided $\eps/\hbar$ is sufficiently small ({\it i.e.} the plateau is high) the echo 
operator can be for long times approximated by $M_\eps\cong 
\exp{(-\ii \eps^2 \bar{R} t/2\hbar)}$. Long time decay of fidelity can then be calculated by 
evaluating the ASI in analogy to the procedure developed for non-residual perturbations 
(\ref{eq:fnASI}), replacing the perturbation $\eps$ with 
$\eps^2/2$ and $\bar{v}$ with $\bar{r}$. Using formula 
$F(t)=\exp{(-(t/\tau_{\rm r})^2)}$ (\ref{eq:Fnregcoh}), we obtain 
long-time fidelity decay for coherent states and residual perturbations,
\begin{equation}
F(t) \cong \exp{\left\{ -(t/\tau_{\rm rr})^2\right\}} ,\qquad \tau_{\rm rr}=
\frac{1}{\eps^2}\sqrt{\frac{8\hbar}{\bar{\veb{r}}'\cdot\Lambda^{-1}\bar{\veb{r}}'}}.
\label{eq:reslong}
\end{equation}
The semiclassical value of $\bar{r}$ is given in Eq.~(\ref{eq:oGw}) and its
derivative is $\bar{\veb{r}}'=\partial_\veb{j}\bar{r}$. The decay time
scales as $\tau_{\rm rr}\sim \hbar^{1/2} \eps^{-2}$ and is thus
smaller than the upper limit $t_{\rm a}\sim \hbar^0 \eps^{-2}$ of
the validity of the ASI approximation. Remember that the above formula is 
valid only if the plateau is close to $F=1$, such that the term $\Sigma(t)$ 
can be neglected. 
For such small $\eps$ the crossover time $t_2$ from the plateau to the 
long time decay can be estimated by comparing the linear response plateau 
(\ref{eq:nuCIS}) with the long time decay (\ref{eq:reslong}), resulting in 
$t_2\approx \tau_{\rm rr} \eps \sqrt{\nu_{\rm CIS}}/\hbar\sim\hbar^{-1/2}\eps^{-1}$. 
For stronger perturbations, namely up to $\eps \sim \sqrt{\hbar}$ time $t_2$ can be 
estimated as $\tau_{\rm rr}$. We therefore have
\begin{equation}
t_2 = \min\{1,\frac{\eps}{\hbar}\nu_{\rm CIS}^{1/2}\} \tau_{\rm rr} = 
\min\{{\rm const}\ \hbar^{1/2}\eps^{-2},{\rm const}\ \hbar^{-1/2}\eps^{-1}\}.
\label{eq:t2coh}
\end{equation}
The theoretical prediction for the departure point $t_2$ (\ref{eq:t2coh}) of 
fidelity from the plateau is plotted as a vertical chain line in 
Fig.~\ref{fig:ch4shortcoh}. In Fig.~\ref{fig:longcoh} we show results of 
long time numerical simulation for the kicked top (\ref{eq:KT3def}). The 
agreement with theoretical Gaussian decay (\ref{eq:reslong}) is good. 

\subsubsection{Freeze in a harmonic oscillator}

\label{sec:freezeHO}
The time $t_1$ (\ref{eq:t1}) when the plateau starts can diverge for systems with vanishing frequency 
derivatives $\Omega$. Such is the case for instance in harmonic oscillator where 
$\Omega=0$. One can show that 
in this case the plateau is not constant but oscillates and is in the linear response 
regime by factor $1/\hbar$ {\em higher} than in systems with $\Omega \neq 0$. 
The full expression for the plateau for the harmonic oscillator and coherent initial state 
is~\cite{Znidaric:04}
\begin{equation}
F_{\rm plat}^{\rm CIS}=\exp{\left(-\frac{\eps^2}{\hbar}\nu_{\rm har} \right)},
\label{eq:ch4harplat}
\end{equation}
where $\nu_{\rm har}$ is some time-dependent coefficient. This expression must be compared 
with the Bessel function (\ref{eq:ch4J0}) for systems with nonzero $\Omega$. 
Because Bessel functions decay asymptotically in an algebraic way, the plateau for large 
perturbations is {\em smaller} in harmonic oscillators than in general systems 
(note that for small perturbations it is {\em larger}). As opposed to general situation, 
for harmonic oscillator classical fidelity agrees with the quantum one despite the freeze. 
Because of inequality (\ref{eq:ineq}) 
there is also freeze in reduced fidelity and purity. The plateau in reduced fidelity scales 
the same as for fidelity (\ref{eq:ch4harplat}) whereas is does not depend on $\hbar$ for purity. 
The long time decay of fidelity or of reduced fidelity and purity after the plateau ends does 
not show any peculiarities for the harmonic oscillator. For further details as well as for 
a numerical example in Jaynes-Cummings system see Ref.~\cite{Znidaric:04}.

\subsubsection{Rotational orbits}

So far we have always talked about the operator $\oV$ being either zero or 
not, resulting in a qualitatively different decay of quantum fidelity. 
Similar behavior as for $\oV=0$ might occur for particular localized initial 
states provided the expectation value of $\oV$ and its powers are zero, 
$\ave{\oV^n}\equiv 0$, even if on the operator level we have $\oV\neq 0$. Such 
situation might occur in regular systems having 
rotational orbits in classical phase space. For these we may have 
perturbations which have $\ave{\bar{v}}_{\rm class}=0$ and correspondingly quantum freeze occurs. 
Note that for classical fidelity only the functional 
dependence will change from a Gaussian to a power-law as one puts the initial 
coherent state in the
rotational part of phase space, in both cases depending on $\eps t$~\cite{Benenti:03}.
Quantum fidelity on the other hand decays as a Gaussian on a time scale $\tau_{\rm r}\sim 1/\eps$ in 
the case of $\ave{V}\neq 0$, while it displays a freeze and later decays on a much longer time 
scale $\tau_{\rm rr} \sim 1/\eps^2$ for $\ave{V}=0$. Freeze of quantum fidelity and correspondingly 
short correspondence between quantum and classical fidelity for an initial 
state put in the rotational part of phase space has been observed 
in Ref.~\cite{Sanka:03}.

Interesting behavior might emerge at borders 
between regions with $\ave{\oV^n}=0$ and $\ave{\oV^n}\neq 0$, for instance close 
to separatrices that separate rotational parts of a phase space from the rest. 
The decay of quantum fidelity at such a border has been studied 
in Ref.~\cite{Weinstein:05}. While far away in the rotational part of phase space 
the freeze has been observed, even though on the operator level one has 
$\oV\neq 0$, interesting and non-trivial behavior has been seen in the 
transition region. There fidelity decays as a power law, depending on the 
product $\eps t$, {\it i.e.} on a time scale $\tau \sim 1/\eps$. Such 
power-law decay is actually not a complete surprise, as the average fidelity 
(Section~\ref{sec:ch4avgF}) also decays as a power-law due to regions of 
phase space with diverging time-scale $\tau_{\rm r}$ of Gaussian decay 
(\ref{eq:Fnregcoh}). Still, exact theoretical understanding of the fidelity 
decay close to separatrices is lacking. 
 
\subsubsection{Random initial states}

Similarly as for non-residual perturbations (Section~\ref{sec:ch4RIS}) we again 
assume a finite Hilbert space of sufficient size, such that the difference 
between fidelity for one specific random state and an average over all 
states can be neglected. We shall replace quantum 
averages with classical averages, 
$\ave{g(\veb{J},\veb{\Theta})}\cong \frac{1}{\mathcal{V}}\int{\!{\rm d}^d\veb{j}{\rm d}^d\veb{\theta} 
\, g(\veb{j},\veb{\theta})}$. First we calculate the plateau. 
Calculating the plateau for coherent initial states, we argued that 
after time $t_1$ (\ref{eq:t1}) the fidelity is time-independent and we 
have a plateau. For random initial states the same happens, only the time 
$t_1$ is different. One can use the same argumentation as for coherent 
states, taking into account that the effective size $\Delta\veb{j}$ of 
random states is $\Delta \veb{j}\sim \mathcal{O}(1)$, resulting in 
\begin{equation}
t_1 \sim \frac{1}{|\veb{m}\Omega|}\sim \eps^0 \hbar^0.
\end{equation}
Compared to coherent states, $t_1$ is now independent of $\hbar$ and is therefore smaller.
After averaging time $t_1$ we can use the general time averaged expression for the value 
of the plateau (\ref{eq:fplatreg}), arriving at
\begin{equation}
f_{\rm plat}^{\rm RIS}\cong\frac{(2\pi)^d}{\mathcal{V}}\int{\!{\rm d}^d 
\veb{j} \left| \int\!\frac{{\rm d}^d\veb{\theta}}{(2\pi)^d} 
\exp\left(-\frac{\ii\eps}{\hbar}w(\veb{j},\veb{\theta})\right) \right|^2},
\end{equation}
where RIS  stands for random initial state. Interestingly, 
the plateau for a random initial state is the {\em
average} plateau for a coherent state squared, where averaging is done
over the position of the initial coherent state. If we denote the plateau 
for a coherent initial state centered at $\veb{j}^*$ by $F_{\rm plat}^{\rm CIS}
(\veb{j}^*)$ (\ref{eq:plateauCS}), then the plateau for a random initial 
state  $F_{\rm plat}^{\rm RIS}$ is
\begin{equation}
F_{\rm plat}^{\rm RIS}\cong\left| \frac{(2\pi)^d}{\mathcal{V}}\int{\!{\rm d}^d \veb{j} 
F_{\rm plat}^{\rm CIS}(\veb{j})}\, \right|^2.
\label{eq:RISplat}
\end{equation}
One word of caution is necessary though concerning the above formula. The 
$w(\veb{j},\veb{\theta})$ (\ref{eq:ch4wm}) needed in the calculation of 
$F_{\rm plat}^{\rm CIS}(\veb{j}^*)$ (\ref{eq:plateauCS}) diverges at points in phase space where 
$\veb{m}\cdot\veb{\omega}(\veb{j})=0$. For a coherent initial state this was no 
real problem as divergence would occur
only if we placed the initial packet at such a point. For random initial
state though, there is an average over the entire action space in the
plateau formula (\ref{eq:RISplat}) and if there is a single diverging point 
somewhere in the phase space it will cause divergence. 
The solution to this is fairly simple. If we compare the semiclassical formula for 
$w_\veb{m}$ (\ref{eq:ch4wm}) with its quantum version for matrix elements 
$W_{jk}$ (\ref{eq:Wmatrix}), we see that there are no divergences in 
$W_{jk}$ as the matrix element $V_{jk}$ is for residual perturbations 
zero precisely for those $j,k$ which would give divergence due to 
vanishing $E_j-E_k$. Remembering that the integral in the formula for $F_{\rm plat}^{\rm RIS}$ 
over $\veb{j}$ is actually
an approximation for a sum over $\hbar\veb{n}$ the remedy 
is to retain the original sum over the eigenvalues of the action 
operator excluding all diverging terms. The formula for the plateau 
in the case of such divergences is therefore
\begin{equation}
F_{\rm plat}^{\rm RIS}\cong \left| \frac{1}{\mathcal{N}}\sum_{\veb{n}}^{\veb{m}\cdot
\veb{\omega}(\hbar\veb{n})\neq 0}{F_{\rm plat}^{\rm CIS}(\hbar\veb{n})}\, \right|^2.
\label{eq:RISsing}
\end{equation}
In the summation over $\veb{n}$ we exclude all terms in which, for 
any constituent Fourier mode $\veb{m}$, we would have $\veb{m}\cdot\veb{\omega}(\hbar\veb{n})= 0$. 
To summarize, if there are no points with $\veb{m}\cdot\veb{\omega}=0$ in phase space on can use 
the integral formula (\ref{eq:RISplat}), 
otherwise a summation formula (\ref{eq:RISsing}) must be used. 
\begin{figure}[h!]
\centerline{\includegraphics[width=\figw\textwidth]{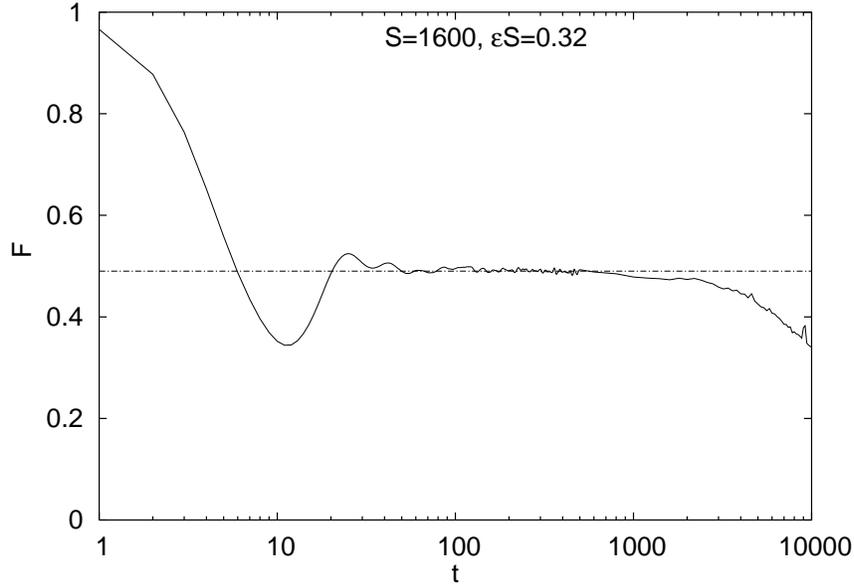}}
\caption{Fidelity plateau for a random initial state for the regular kicked top (\ref{eq:KT3def}) 
[from~\cite{Prosen:03}]. The chain line shows the theoretical value of the plateau as 
computed from Eq.~(\ref{eq:RISsing}). Due to near resonant terms with 
small $\veb{m}\cdot\veb{\omega}(\veb{j})$ the plateau is lower than in Fig.~\ref{fig:ch4shortcoh}, 
even though the perturbation is of the same strength.}
\label{fig:shortRanSing}
\end{figure}

To illustrate the summation formula for the plateau, we use the same kicked 
top as for coherent states (\ref{eq:KT3def}). As one can see in 
Fig.~\ref{fig:shortRanSing} the agreement with numerics is very good. Compared 
to Fig.~\ref{fig:ch4shortcoh}(a) for coherent state (where the same perturbation strength is used) 
the plateau is here much lower due to near-diverging terms. 
For a numerical example of a system with no singularities, where the integral 
formula (\ref{eq:RISplat}) can be used, see Ref.~\cite{Prosen:03}.

For long times, after the plateau ends at $t_2\sim 1/\eps$, the fidelity 
again starts to decrease due to increasing $\Gamma(t)$. The calculation 
using ASI is analogous to the case of non-residual perturbations, Eq.~(\ref{eq:regpower}), 
one just has to replace $\bar{v}$ with $\bar{r}$ and 
$\eps$ with $\eps^2/2$. 
We shall only give the result,
\begin{equation}
F(t)\asymp \left(\frac{t_{\rm ran}}{t} \right)^d,\qquad t_{\rm ran} = {\rm const}\times 
\frac{\hbar}{\eps^2}.
\label{eq:RISasymp}
\end{equation}
For comparison with numerical simulation see Ref.~\cite{Prosen:03}.

Finally, we make a short comment on the behavior of average fidelity 
(averaged over positions of coherent states) for residual perturbations. 
For non-residual perturbations we have seen in Section~\ref{sec:ch4avgF} 
that the asymptotic decay is determined by the divergences of the decay 
time $\tau_{\rm r}$. For residual perturbation the situation is very similar. 
During the plateau, until time $t_2$, averaging of the plateaus for 
coherent states gives the plateau for random states. Thus in 
the average fidelity one still has a plateau of order $1-F\sim (\eps/\hbar)^2$ (in the 
linear response). After the plateau, we have to average 
Gaussian decay (\ref{eq:reslong}), where in general divergences of 
$\tau_{\rm rr}$ will cause a power-law asymptotic decay. The power will 
depend on the nature of the divergence and therefore on the perturbation The only 
difference to general perturbations is that the 
decay depends on the parameter $\eps^2 t$ and not on $\eps t$ as 
for non-residual perturbations. Note though that some researchers~\cite{Jacquod:03} have 
predicted a universal $t^{-3d/2}$ decay of average fidelity for 
residual perturbations, that is with the power depending only on the 
dimension $d$ and not on perturbation specifics. The reason for the 
discrepancy between these results is presently unclear.

\subsection{Composite systems and entanglement}

\label{sec:compInt}
In case of composite systems one might not be interested in the influence of 
perturbations on the whole system but just on some ``central'' subsystem of interest. Stability of a 
subsystem is quantified by reduced fidelity $\Fr$ (\ref{eq:Frdef}) and by echo purity $\Fp$ 
(\ref{eq:Fpdef}). We shall mostly discuss coherent initial states, {\it i.e.} a direct product 
of localized packets for central subsystem and environment, for which semiclassical 
treatment is the most simple. At the end we shall also comment on other possible choices of 
initial states.

The calculation of the reduced fidelity is quite analogous to the one for fidelity 
(\ref{eq:Fnregcoh}). We again employ a classical time averaged perturbation 
$\bar{v}(\mathbf{j})$, where the action vector $\mathbf{j}$ has $d_{\rm c}$ components 
$\mathbf{j}^{\rm c}$ for the central system and $d_{\rm e}$ components $\mathbf{j}^{\rm e}$ 
for the environment, $\mathbf{j}=(\mathbf{j}^{\rm c},\mathbf{j}^{\rm e})$, if $d_{\rm c}$ and 
$d_{\rm e}$ are the number of degrees of freedom (DOF) of the central subsystem and environment, respectively. 
The shape of the initial packet is given by squeezing matrices $\Lambda_{\rm c}$ and $\Lambda_{\rm e}$ 
for the central subsystem and the environment, respectively. The derivatives of the 
classical $\bar{v}$ with respect to actions can be written as
\begin{equation}
\bar{\veb{v}}'=(\bar{\veb{v}}'_{\rm c},\bar{\veb{v}}'_{\rm e}),\qquad \bar{\veb{v}}'_{\rm c}=
\frac{\partial \bar{\veb{v}}(\veb{j}^*)}{\partial \veb{j}^{\rm c}},\qquad \bar{\veb{v}}'_{\rm e}=
\frac{\partial \bar{\veb{v}}(\veb{j}^*)}{\partial \veb{j}^{\rm e}}.
\end{equation}
Performing Gaussian integrals in action space gives reduced fidelity~\cite{Znidaric:03}
\begin{equation}
\Fr=\exp{\left(-\frac{\eps^2}{\hbar^2} \bar{C}_{\rm R} t^2 \right)},\qquad \bar{C}_{\rm R}=
\frac{1}{2}\hbar \left( \bar{\veb{v}}_{\rm c}'\Lambda^{-1}_{\rm c}\bar{\veb{v}}'_{\rm c} \right).
\label{eq:Frexact}
\end{equation} 
A single parameter $\bar{C}_{\rm R}$ determines the reduced fidelity decay. This must be compared 
with $\bar{C}=\frac{1}{2}\hbar \left( \bar{\veb{v}}_{\rm c}'\Lambda^{-1}_{\rm c}\bar{\veb{v}}'_{\rm c}
+ \bar{\veb{v}}_{\rm e}'\Lambda^{-1}_{\rm e}\bar{\veb{v}}'_{\rm e} \right)$ (\ref{eq:Fnregcoh}) which 
governs fidelity decay. Compared to fidelity, reduced fidelity depends understandably only on 
perturbation derivatives with respect to actions of the central system. But note that the average 
coupling $\bar{v}$ depends also on the dynamics of the environment and so $\bar{C}_{\rm R}$ itself 
does depend on the environment.

The derivation of echo purity (or in special case of uncoupled unperturbed dynamics of purity) is a 
little more involved. We have to calculate averages of the form
\begin{eqnarray}
\Fp\cong \hbar^{-2d} \int{\!\! d\veb{j}\, d\tilde{\veb{j}}\, \exp{\left(-\ii \frac{\eps t}{\hbar} 
\Phi\right)} d_\rho(\veb{j}) d_\rho(\tilde{\veb{j}})}, \nonumber \\
\Phi=\bar{v}(\veb{j}^{\rm c},\veb{j}^{\rm e})-\bar{v}(\tilde{\veb{j}}^{\rm c},\veb{j}^{\rm e})+
\bar{v}(\tilde{\veb{j}}^{\rm c},\tilde{\veb{j}}^{\rm e})-\bar{v}(\veb{j}^{\rm c},\tilde{\veb{j}}^{\rm e}),
\label{eq:Iint}
\end{eqnarray}
with $d_\rho(\veb{j})$ being the initial density (\ref{eq:drhocoh}) of the wave packet. When 
expanding $\bar{v}(\mathbf{j})$ around the position $\mathbf{j}_*$ of the initial packet, the 
constant term, the linear terms as well as the diagonal quadratic terms cancel exactly, regardless 
of the position of the initial packet. The lowest non-vanishing contribution comes from non-diagonal 
quadratic term
\begin{equation}
-\ii \frac{\eps t}{\hbar} \left[ (\veb{j}^{\rm c}-\tilde{\veb{j}}^{\rm
 c})\cdot\bar{v}_{\rm ce}''(\veb{j}_*)(\veb{j}^{\rm e}-\tilde{\veb{j}}^{\rm e}) \right],
\end{equation}
where $\bar{v}''_{\rm ce}$ is a $d_{\rm c} \times d_{\rm e}$ matrix 
of mixed second derivatives of $\bar{v}$ evaluated at the position of the initial packet,
\begin{equation}
\left( \bar{v}''_{\rm ce} \right)_{kl}=\frac{\partial^2 \bar{v}}{\partial (\veb{j}^{\rm c})_k 
\partial(\veb{j}^{\rm e})_l}.
\label{eq:vce}
\end{equation}
After performing integrations we get the final result (see Ref.~\cite{Znidaric:05} for details)
\begin{equation}
\Fp=\frac{1}{\sqrt{\det{(\mathbbm{1}+(\eps t)^2 u)}}},\quad u=\Lambda_{\rm c}^{-1} 
\bar{v}''_{\rm ce} \Lambda_{\rm e}^{-1} \bar{v}''_{\rm ec},
\label{eq:Imain}
\end{equation}
where $u$ is a $d_{\rm c}\times d_{\rm c}$ matrix involving 
$\bar{v}''_{\rm ce}$ and its transpose $\bar{v}''_{\rm ec}$. Note that the 
matrix $u$ is a classical quantity (independent of $\hbar$) that depends 
only on the observable $\bar{v}$ and on the position of the initial packet. 
For small $\eps t$ we can 
expand the determinant and we get initial quadratic decay in the linear response regime 
(special case of Eq.~(\ref{eq:LRecho})), 
\begin{equation}
\Fp=1-\frac{1}{2}(\eps t)^2 \tr{}{u}+\cdots.
\label{eq:FpLR}
\end{equation}
The most prominent feature of the expression (\ref{eq:Imain}) for the echo purity 
(or purity) decay for initial product wave packets is its 
$\hbar$ independence. In the linear response calculation this 
$\hbar$-independence has been theoretically predicted in Ref.~\cite{Prosen:03evol} as 
well as numerically confirmed~\cite{Znidaric:03}. We see that the scaling of the decay time 
$\tau_{\rm P}$ of $\Fp$ is $\tau_{\rm P} \sim 1/\eps$. This means that echo purity will 
decay on a very long time scale and so localized wave packets are universal 
pointer states~\cite{Zurek:81}, {\it i.e.} states robust against decoherence~\cite{Zurek:91}. 
  For large times we use the fact that $\det(\mathbbm{1} + z u)$ is
a polynomial in $z$ of order $r={\rm rank\,}(u)$ and find asymptotic power law decay
$\Fp \asymp {\rm const}\,(\eps t)^{-r}$.
Note that the rank of $u$ is bounded by the smallest of the subspace dimensions, {\it i.e.} 
$1\le r \le {\rm min}\{d_{\rm c},d_{\rm e}\}$, 
since the definition (\ref{eq:Fpdef}) is symmetric with respect to 
interchanging the roles of the subspaces ``c'' and ``e''.
Let us give two simple examples: (i) For $d_{\rm c}=1$ and for {\em any} $d_{\rm e}$ we shall always 
have asymptotic power law decay with $r=1$. If a {\em single} DOF of 
the subsystem ``c''  is coupled with {\em all} DOF of the subsystem ``e'', {\it e.g.} 
$\bar{v}=j^{\rm c} \otimes (j^{\rm e1}+j^{\rm e2}+\cdots)$, then 
$|\bar{v}''|^2 \propto d_{\rm e}$ and we have 
$\Fp \asymp 1/(\eps t \sqrt{d_{\rm e}})$; (ii) Let us 
consider a multidimensional system where the matrix $u$ is of rank one so it
can be written as a direct product of 
two vectors, $u=\veb{x} \otimes \veb{y}$. The determinant occurring in $\Fp$ is 
$\det{(\mathbbm{1}+(\eps t)^2u)}=1+(\eps t)^2 \veb{x}\cdot \veb{y}$. 
Such is the case for instance if we have a coupling of the same strength 
between all pairs of DOF. The dot product is then 
$\veb{x}\cdot \veb{y} \propto d_{\rm c} d_{\rm e}$ and we find 
$\Fp \asymp 1/(\eps t \sqrt{d_{\rm c} d_{\rm e}})$. The power of 
the algebraic decay is independent of both $d_{\rm c}$ and $d_{\rm e}$. The bottom line is, that 
the power law of the asymptotic decay depends on the perturbation. 
In Refs.~\cite{Jacquod:04,Jacquod:04erratum} the author predicted $\sim t^{-d}$ decay of purity for 
intermediate times and universal $\sim t^{-2d}$ decay for large times. While the intermediate decay 
is in accord with our theory, its long time decay is not. For numerical examples where 
asymptotic decay of purity exhibits different powers see Ref.~\cite{Znidaric:05}.
\begin{figure}[h]
\centerline{\includegraphics[width=\figw\textwidth]{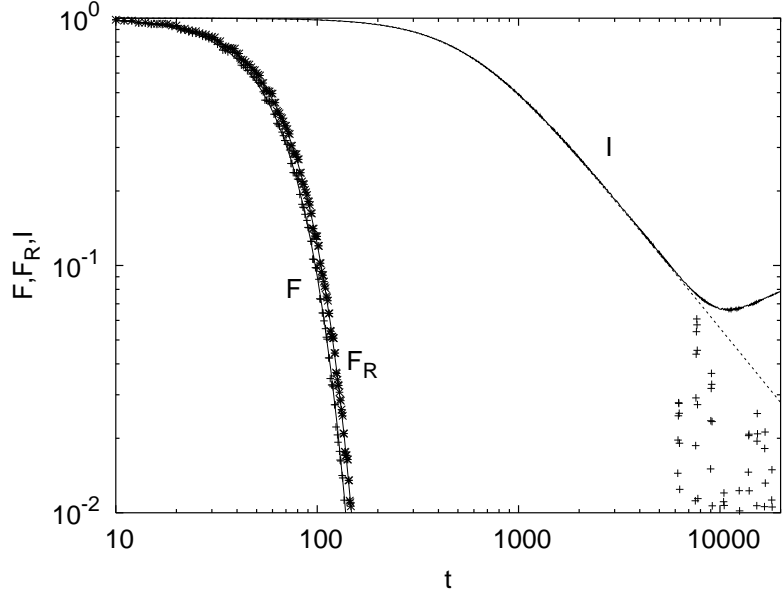}}
\caption{Decay of fidelity $F(t)$, reduced fidelity $F_{\rm R}(t)$ and purity 
  $I(t)$ for integrable dynamics and localized initial packets 
  [from~\cite{Znidaric:04}]. The theoretical Gaussian decay for the fidelity 
  (\ref{eq:Fnregcoh}) and the reduced fidelity (\ref{eq:Frexact}) overlaps with 
  the numerics (symbols). Purity decays asymptotically as a power law
(\ref{eq:Imain}) and also nicely agrees with the numerics (full curve), until a 
finite size saturation level is reached. Note that purity decays on a much longer time scale
than fidelity and reduced fidelity due to independence of its decay time on the small parameter $\hbar$.}
\label{fig:fidJ10000}
\end{figure}

To demonstrate the theory of reduced fidelity and echo purity decay we again take two coupled 
kicked tops, using a regular propagator in Eq.~(\ref{eq:3KTdef}). We take $S=100$, 
coupling $\eps=0.01$ and coherent initial state. As the unperturbed dynamics is uncoupled $\Fp$ 
is actually equal to purity $I(t)$. The results of the numerical simulation are plotted in 
Fig.~\ref{fig:fidJ10000} together with the theoretical predictions (\ref{eq:Frexact}) for 
reduced fidelity and for purity (\ref{eq:Imain}). No fitting is involved.

In the above paragraph we have discussed coherent initial states. What about other choices? We 
shall consider two possibilities: (i) superposition of two coherent states - so called cat states 
and (ii) random initial states. In both cases we assume that the unperturbed evolution is uncoupled, 
so echo purity equals purity. Let us first describe the decay of purity $I(t)$ for cat states,
\begin{equation}
\ket{\psi(0)}=\frac{1}{\sqrt{2}} \left(\ket{\psi_{\rm c1}(0)}+\ket{\psi_{\rm c2}(0)} 
\right) \otimes \ket{\psi_{\rm e}(0)},
\label{eq:cat}
\end{equation}
where the initial positions of the wave packets $\psi_{\rm c1}(0)$, $\psi_{\rm c2}(0)$ and
 $\psi_{\rm e}(0)$ are $\veb{j}^{\rm *c1}$, $\veb{j}^{\rm *c2}$ and $\veb{j}^{\rm *e}$, 
respectively. Assuming that the initial packets of the central subsystem are orthogonal the
initial reduced density matrix is
\begin{equation}
\rho_{\rm c}(0)=\frac{1}{2}\left( \ket{\psi_{\rm c1}(0)}\bra{\psi_{\rm c1}(0)} +
 \ket{\psi_{\rm c2}(0)}\bra{\psi_{\rm c2}(0)} +
\ket{\psi_{\rm c2}(0)}\bra{\psi_{\rm c1}(0)} +
\ket{\psi_{\rm c1}(0)}\bra{\psi_{\rm c2}(0)} 
\right),
\label{eq:rho0}
\end{equation}
and represents coherent superposition of two macroscopically separated packets
 -- superposition of ``dead and alive cat'' \cite{Cat:35} -- see also a popular 
account in Ref.~\cite{Zurek:91}. 
What we would like to illustrate is decoherence. Decoherence is a process where 
off diagonal matrix elements of the initial reduced density matrix dissappear due to the coupling with the
environment, so that the resulting reduced density matrix $\rho_{\rm c}^{\rm mix}(\tau_{\rm dec})$ 
is a statistical mixture of two diagonal contributions only,
\begin{equation}
\rho_{\rm c}^{\rm mix}(\tau_{\rm dec})=\frac{1}{2}\left( 
\ket{\psi_{\rm c1}(\tau_{\rm dec})}\bra{\psi_{\rm c1}(\tau_{\rm dec})} +
\ket{\psi_{\rm c2}(\tau_{\rm dec})}\bra{\psi_{\rm c2}(\tau_{\rm dec})}
\right),
\label{eq:rhodec}
\end{equation}  
with $\tau_{\rm dec}$ being decoherence time. Here we already used the fact that purity
for individual localized packets decays on a long $\hbar$ independent time scale 
(\ref{eq:Imain}) and
because we expect $\tau_{\rm dec}$ to be much shorter, we can use propagated packets as pointer states.
During the process of decoherence purity will decay from $I(0)=1$ for the initial 
reduced density matrix $\rho_{\rm c}(0)$ to $I(\tau_{\rm dec})=1/2$ for decohered matrix
$\rho_{\rm c}^{\rm mix}(\tau_{\rm dec})$. Using similar techniques as for the calculation of 
echo purity for individual coherent states (\ref{eq:Imain}), one can also calculate 
purity for cat states, the result being~\cite{Znidaric:05b}
\begin{equation}
I(t)=\frac{1}{4}(I_1(t)+I_2(t)+2F_{\rm e}(t)),
\label{eq:Icat}
\end{equation}
where $I_{1,2}(t)$ are purities (\ref{eq:Imain}) for individual constituent coherent states, 
that is for $\psi_{\rm c1,c2}(0)\otimes \psi_{\rm e}(0)$, while $F_{\rm e}(t)$ is a 
cross-correlation quantity similar to reduced fidelity. Explicitly, it is given by
\begin{eqnarray}
F_{\rm e}(t)=\exp{\left(-t^2/\tau_{\rm dec}^2 \right)},\qquad 
\tau_{\rm dec}= \sqrt{\hbar /C_{\rm e}}/\eps  \nonumber \\
C_{\rm e}=\frac{1}{2}\left( \bar{v}'_{\rm e}(\veb{j}^{\rm *c1})-\bar{v}'_{\rm e}(\veb{j}^{\rm *c2}) 
\right) \Lambda_{\rm e}^{-1} \left( \bar{v}'_{\rm e}(\veb{j}^{\rm *c1})-
\bar{v}'_{\rm e}(\veb{j}^{\rm *c2}) \right).
\label{eq:Fe}
\end{eqnarray}
Decoherence for regular systems can now be understood from the very different time scales on
which $I_{1,2}$ and $F_{\rm e}$ decay. $F_{\rm e}(t)$ decays fast as a Gaussian on an $\hbar$-dependent time scale while $I_{1,2}(t)$ decay slowly on an 
$\hbar$-independent time scale. At the point when $F_{\rm e}(t)$ becomes negligibly small, purity will be 
simply $I(t)\approx (I_1(t)+I_2(t))/4$ and therefore approximately equal to $1/2$. The decay of 
$F_{\rm e}(t)$ therefore signifies decoherence with the decoherence time scale $\tau_{\rm dec}$ being 
simply equal to the decay time of $F_{\rm e}(t)$ (\ref{eq:Fe}). On the other hand, the decay of 
$I_{1,2}(t)$ is a signal of relaxation of individual packets to equilibrium. In the simple case 
when $\bar{v}$ has coupling terms between all pairs of degrees of freedom one can see~\cite{Znidaric:05b} 
that $C_{\rm e}$ scales as $C_{\rm e} \sim d_{\rm c} d_{\rm e} (\veb{j}^{\rm *c1}-\veb{j}^{\rm *c2})^2$ 
and therefore the decoherence time scales as 
$\tau_{\rm dec}\sim \sqrt{\hbar/(d_{\rm c} d_{\rm e})}/(\eps |\veb{j}^{\rm *c1}-\veb{j}^{\rm *c2}|)$. 
It therefore goes to zero in any of the following four limits: 
the semiclassical limit $\hbar \to 0$, the 
thermodynamic limit of large environment and/or large central subsystem, 
$d_{\rm c,e}\to \infty$, the 
limit of macroscopic superpositions, {\it i.e.} of large packet separation, 
$|\veb{j}^{\rm *c1}-\veb{j}^{\rm *c2}| \to \infty$ and finally the limit of strong coupling. 
Note that these are exactly the limits one needs 
to explain the lack of cat states in classical macroscopic world of large objects! One should note that 
irreversible decoherence is possible only in the thermodynamic limit. For a finite integrable
system there will be partial revivals of purity at large times. This revivals get less prominent and
happen at larger times with increasing dimensionality $d$ or decreasing $\hbar$. Also, due to high
sensitivity of systems to perturbation, reversing the process of decoherence will be in practice
close to impossible. Therefore, for all practical purposes one can observe decoherence also in a 
sufficiently large but finite regular system. For the theory presented here (\ref{eq:Icat}) we needed
that the average perturbation $\oV$ converged, {\it i.e.} time is larger than the classical 
averaging time $t_{\rm ave}$. But for sufficiently separated packets decoherence 
time $\tau_{\rm dec}$ will eventually get very small~\cite{Braun:01,Strunz:03,Strunz:03a}, 
even smaller than $t_{\rm ave}$. We can circumvent this difficulty by noting that 
for very short times one can instead of the time averaged $\oV$ use just an instantaneous 
time independent $V(0)$. Then our theory can again be used~\cite{Znidaric:05b}. 
The theory developed here therefore applies to truly macroscopic superpositions of 
localized wave packets (for which $\tau_{\rm dec}$ is very small) regardless 
whether the dynamics are regular or chaotic.    

\begin{figure}[h]
\centerline{\includegraphics[angle=-90,width=\figw\textwidth]{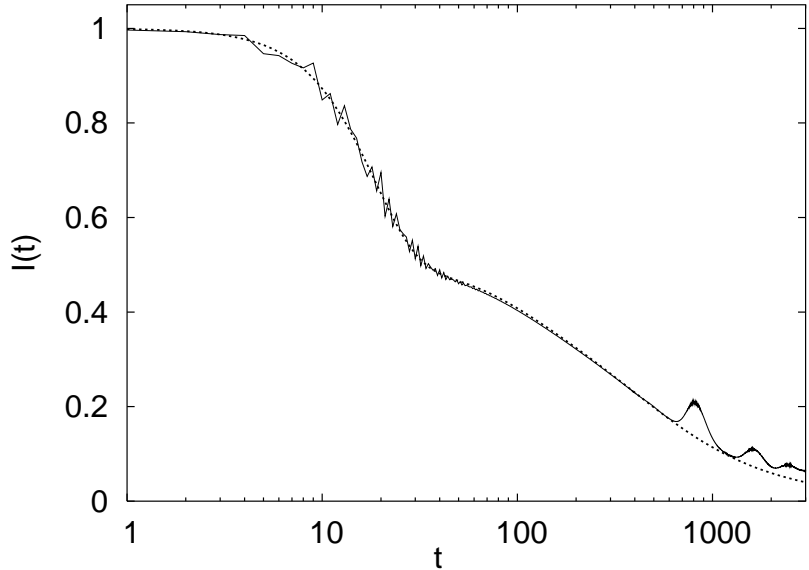}}
\caption{Decay of purity $I(t)$ (\ref{eq:Icat}) in a regular system for an
  initial cat state (\ref{eq:cat}) [from~\cite{Znidaric:05b}]. The dotted 
  curve shows the theoretical prediction (\ref{eq:Icat}) with no fit 
  parameters, and the full curve shows numerics. The initial decay of purity 
  until $\tau_{\rm dec}\sim 30$ is given by a fast Gaussian decay of
  $F_{\rm e}(t)$ (\ref{eq:Fe}). At the end of decoherence purity is
  $I(\tau_{\rm dec})=1/2$, then the relaxation of individual packets begins, 
  visible as a slow power-law decay of purities $I_{1,2}(t)$ (\ref{eq:Imain}).}
\label{fig:Icat}
\end{figure}

Let us now demonstrate the decay of purity for cat states (\ref{eq:Icat}) by a 
numerical experiment. We take a system of two coupled anharmonic oscillators 
with $\hbar=1/500$ and coupling $\eps=0.01$, for details see 
Ref.~\cite{Znidaric:05b}. The results are shown in Fig.~\ref{fig:Icat}, 
where one can nicely see two different regimes of purity decay. In the first 
one fast Gaussian decoherence happens until purity reaches $I=1/2$ at 
$\tau_{\rm dec}\sim 30$, and in the second one a slow algebraic relaxation 
takes place.
 
In the context of quantum information perhaps more interesting than coherent states are random 
initial states. For regular dynamics echo purity will initially decay quadratically, but on $\hbar$ 
dependent scale that is typically much shorter than for coherent initial states. Echo purity for 
random states in a spin chain (a system having no classical limit) has been studied 
in Ref.~\cite{Prosen:02spin}. In the linear response regime a quadratic decay has been observed 
and interestingly the full functional form of the decay approached a Gaussian in the 
thermodynamical limit. Quadratic decay of concurrence, a quantity describing entanglement 
for mixed states, has been observed for an integrable spin chain~\cite{Pineda:05}. 
In Ref.~\cite{Rafal:04} the initial entanglement production rate (time derivative of purity) 
as well as its asymptotic saturation value have been studied for random state averaging and 
for coherent state averaging. In the integrable regime the initial entanglement production rate 
is much larger for random state averages than for coherent state averages due to a different 
$\hbar$ dependence of the corresponding correlation functions. Decoherence in a system of two $\frac{1}{2}$ spins coupled to a bath composed of a number of spin $\frac{1}{2}$ particles has been studied in Ref.~\cite{Lages:05}. It has been numerically found that decoherence is faster when the internal bath dynamics is chaotic than when it is regular. Interestingly, in a certain regime of parameters decoherence decreases by increasing coupling strength to the bath. 

An interesting question is to what extent the decay of e.g. purity can be described classically.  
The classical analog of decoherence has been considered in 
Refs.~\cite{Wehrl:78,Gong:03,Gong:03pra,Gong:03pra2}. 
Using a perturbative approach the influence of the type of the perturbation and of the dynamics on 
the quantum-classical correspondence were explored, see also Ref.~\cite{Angelo:04} for a study of 
quantum-classical correspondence of entanglement in coupled harmonic oscillators.

We shall not discuss the decay of reduced fidelity 
or of echo purity in the case of freeze. Based on the general inequality between the 
three quantities (\ref{eq:ineq}), we can immediately state that reduced fidelity and echo 
purity will also exhibit freeze. For an example of purity freeze in a harmonic oscillator 
see Ref.~\cite{Znidaric:04} and also Section~\ref{sec:freezeHO}. 

\subsection{Mixed phase space}
\label{sec:ch4mixed}

For classically mixed situation, with
coexisting regular and chaotic regions in phase space,
much less is known than for purely regular (Section~\ref{Q1}) or chaotic 
(Section~\ref{Q2}) dynamics. 
If a localized packet is started completely within a region of phase space 
having regular/chaotic structure one can use the appropriate theory for 
quantum fidelity. One should be cautious 
though as there can be different time-scales involved that are not present 
in purely regular or chaotic situations. The 
behavior of fidelity in the border region between regular and chaotic 
components might be particularly tricky. Two effects can be important at such a border: (i) classical 
dynamics in border region is ``sticky'', {\it i.e.} diffusive, causing a slow power-law decay of 
various classical quantities~\cite{Karney:83,Chirikov:84,Ruffo:96}. 
Here the value of $\hbar$ might be very important for the impact this 
classical phenomenon has on quantum fidelity decay. On the other hand, there 
are also pure quantum effects that will influence fidelity decay, (ii) 
the transition between $\ave{\oV}\neq0$ for regular region (in typical 
situation) and $\ave{\oV}\to 0$ in chaotic region occurs at the border. 
We have seen that for regular dynamics the fidelity typically 
decays as a power-law at points with vanishing average perturbation 
(actually vanishing $\bar{v}'$), for instance around separatrices. Similarly, 
average fidelity (\ref{eq:Fasim}) also has a power-law decay. 
Therefore, both effects suggest a power-law decay of quantum fidelity at 
the border between chaotic and regular regions of phase space. The fidelity 
decay in such region has been numerically studied in Ref.~\cite{Weinstein:02}. 
They were able to fit a two-parameter power-law formula to the numerically 
obtained fidelity decay, although the results were less than convincing 
as the power-law covered less than one order of magnitude. To numerically evaluate quantum fidelity for systems with mixed classical phase space on can use the 
dephasing representation. In Refs.~\cite{Vanicek:04,Vanicek:05} it has been 
shown that the dephasing representation works well in such situations. Other 
numerical results are reported in Ref.~\cite{Haug:05}.

Another point worth mentioning is the actual transition from regular to 
chaotic behavior as some parameter of the system is varied. Due to finite 
$\hbar$ the transition will generally not happen at the same parameter value 
as for classical dynamics. In particular, in Ref.~\cite{Wang:04} the parameter 
region of weak chaoticity in a saw-tooth map has been studied and deviations 
from the Fermi golden rule and Lyapunov regime of quantum fidelity decay have 
been observed. The authors studied fidelity decay around the transition point 
between the Fermi golden rule and Lyapunov regime, occurring at 
$\eps\sim\eps_{\rm r}\propto\hbar$. By increasing $\eps/\hbar$ the deviation 
from the Fermi golden rule starts first, after which the decay rate of quantum 
fidelity approaches the classical Lyapunov exponent in an oscillatory way. The 
reason for this transitional behavior are strong fluctuations in the fidelity 
amplitude, causing deviations between $\vert\ave{f}\vert^2$ and $\ave{F}$.

\section{Classical echo-dynamics}
\label{C}

In this section we shall define the corresponding classical quantity, 
characterizing classical Loschmidt echoes, namely the {\em classical fidelity}. 
One can use a strategy which is completely analogous to the quantum case to 
derive a theoretical description of the decay of the classical fidelity. The 
results found here may be of interest for themselves, or can serve as a 
reference for the quantum results. For example, within the time scale of the 
order of Ehrenfest time ($\sim \log\hbar$ for chaotic systems), classical 
fidelity and quantum fidelity should strictly agree.

Classical fidelity has been considered in several papers. 
In~\cite{Prosen:02corr} a definition and a linear response treatment has been 
given. Benenti and Casati \cite{Benenti:02} considered it in the context of 
quantum classical correspondence in a diffusive system with dynamical 
localization. Eckhardt~\cite{Eckhardt:03} treated its generalization to 
stochastic flows, whereas Benenti et al.~\cite{Benenti:03,Benenti:03b} 
considered the asymptotic decay of the classical fidelity, both in the case of 
regular and chaotic dynamics. In Refs.~\cite{Veble:04,Veble:05} a theory has 
been developed for the short-time regime of classical fidelity decay and its 
relation to the Lyapunov spectrum.

The classical fidelity $F_{\rm cl}(t)$ can be defined as
\begin{equation}
F_{\rm cl}(t)=\int_{\Omega}\dd{\ve x} \rho_\eps({\ve x},t) \rho_0({\ve
x},t), \label{fido}
\end{equation}
where the integral is extended over the phase space, and
\begin{equation}
\rho_0({\ve x},t)=\op U_0^t \rho({\ve x},0), 
\quad \rho_\eps({\ve x},t)=\op U_\eps^t \rho({\ve x},0)
\end{equation}
give the classical Liouvillean evolution for time $t$ of the initial 
phase space density $\rho({\ve x},0)$. In order to have $F_{\rm cl}(t)=1$, we have
to require that the phase space density is square normalized
$\int \dd{\ve x}\rho^2({\ve x},0)=1$.
We note that classical dynamics of $L^2$ phase space densities is unitary due to 
phase volume conservation, {\em i.e.} the Liouville theorem.
Therefore, we can also build a classical echo picture and form a unitary
{\em classical echo propagator} that composes perturbed forward evolution with the 
unperturbed backward
evolution which is also unitary and is given by
\beq
\op U_{\rm E}(t)=\op U^\dagger_0(t)\,\op U_\eps (t) \label{eq:Md}.
\eeq
Writing the phase space density after an echo as
\beq
\rho_{\rm E}(\ve{x},t) = U_{\rm E}(t) \rho_0(\ve{x},0)
\eeq
we rewrite the classical Loschmidt echo as
\beq
F_{\rm cl}(t) = 
\int_{\Omega}\dd{\ve x} \rho_{\rm E}({\ve x},t) \rho_0({\ve
x},0). \label{fidoE}
\eeq
The above definitions (\ref{fido},\ref{fidoE}) can be shown to correspond to
the classical limit of quantum fidelity, if written in terms of Wigner 
functions \cite{Prosen:02corr,Karkuszewski:02}. In the ideal case of a perfect 
echo ($\eps=0$), fidelity does not decay, $F_{\rm cl}(t)=1$. However, due to 
chaotic dynamics, when $\eps\ne 0$ the classical echo decay sets in after a 
time scale
\begin{equation}
t_\eps\sim\frac{1}{\lambda}
\ln\left(\frac{\nu}{\eps}\right),
\end{equation}
required to amplify the perturbation up to the size $\nu$ of the
initial distribution. Thus, for $t\gg t_\eps$ the recovery of
the initial distribution via the imperfect time-reversal procedure
fails, and the fidelity decay is determined by the decay of
correlations for a system which evolves forward in time according
to the Hamiltonians $H_0$ (up to time $t$) and $H_\eps$ (from
time $t$ to time $2t$). This is conceptually similar to the
``practical'' irreversibility of chaotic dynamics: Due to the
exponential instability, any amount of numerical error in computer
simulations rapidly erases the memory about the initial distribution. 
In the present case, the coarse-graining which leads to irreversibility is 
not due to round-off errors but to a perturbation in the Hamiltonian.

Another somehow related concept named quantum-classical fidelity has 
recently been introduced and studied \cite{horvat:06}, which measures 
$L^2$ distance between the Wigner function and the corresponding classical 
phase space density as a function of time. It is a very suitable tool for 
precise 
definitons of various Ehrenfest 
time-scales and a detailed study of quantum-classical correspondence. 
Quantum-classical fidelity can also be interpreted as a ``classical'' 
fidelity if the Plack constant $\hbar$ is treated as a perturbation.

\subsection{Short time decay of classical fidelity}

The analysis presented below follows Refs.~\cite{Prosen:02corr,Veble:04}.
The propagation of classical densities in phase space is governed
by the unitary Liouville evolution $\op{U}_\eps(t)$
\beq
\frac{\dd}{\dd t} \op U_\eps(t)={\op L}_{H_\eps(\ve{x},t)} 
\op U_\eps(t)
\label{eq:dUdt}
\eeq
where
${\op  L}_{\! A(\ve{x},t)} =  
\left(\ve \nabla\! A(\ve{x},t)\right)\cdot\mat J \ve \nabla,$ 
$A$ is any observable, and
\beq
H_\eps(\ve{x},t) = H_0 (\ve{x},t) + \eps V(\ve{x},t),
\label{eq:ham}
\eeq
is a generally time-dependent family of classical 
Hamiltonians with perturbation
parameter $\lambda$. The matrix $\mat J$ is the usual symplectic unit.
Similarly,
$
\dd\op U^\dagger_\eps(t)/\dd t= - \op U^\dagger_\eps (t) 
{\op  L}_{H_\eps(\ve{x},t)}.
$
Using Eqs.~(\ref{eq:dUdt},\ref{eq:ham},\ref{eq:Md}) and writing 
$\op{U}_\eps(t) = \op{U}_0(t) \op{U}_{\rm E}(t)$ we get
\beq
\frac{\dd}{\dd t} \op U_{\rm E} (t)= \left\{ \op U_0^\dagger(t)
{\op L}_{\eps V(\ve{x},t)}\op U_0(t)\right\} \op U_{\rm E} (t). 
\label{eq:timeder}
\eeq
Classical dynamics have the nice property that the evolution
is governed by characteristics that are simply the classical phase
space trajectories, so the action of the evolution operator on any
phase space density is given as
$
\op U_0(t)~\rho=\rho \circ \ve \phi_t^{-1},
$
where ${\ve \phi}_t^{-1}$ denotes the backward (unperturbed) phase space flow
 from time $t$ to time $0$. Similarly, the backward evolution is given by
$
\op U_0^\dagger(t)~\rho=\rho \circ \ve \phi_t,
$
where $\phi_t$ represents the forward phase space flow
from time $0$ to time $t$. Here and in the following we assume the 
dynamics to start at time $0$.

We note that echo-dynamics (\ref{eq:Md}) can be treated as
Liouvillean dynamics in the {\em interaction picture}, since
\beqa
&&\left\{ \op U_0^\dagger(t) { \op L}_{A(\ve{x},t)} \op U_0 (t) \rho \right\}(\ve x)= \\&&
=\op U_0^\dagger(t)\left(\ve \nabla_{\ve{x}}  A(\ve x,t)\right) \cdot \mat J \ve \nabla_{\ve x}
 \rho\left(\ve \phi_t^{-1}(\ve x)\right)=\nonumber\\
&&
=
\left(\ve\nabla_{\ve{\phi}_t(\ve{x})} A (\ve \phi_t(\ve x),t)\right)
\cdot \mat J \ve \nabla_{\ve{\phi}_t(\ve{x})} \rho\left(\ve \phi_t^{-1}\left(\ve \phi_t(\ve x)\right)\right)
=\nonumber\\
&&
=
\left(\ve\nabla_{\ve{x}} A (\ve \phi_t(\ve x),t)\right)
\cdot \mat J \ve \nabla_{\ve{x}} \rho(\ve x)
=\left\{ { \op L}_{A\left(\ve \phi_t(\ve x),t\right)} ~\rho\right\}({\ve x}). \nonumber
\eeqa
In the last line we used the invariance of the Poisson bracket under the flow. 
This extends Eq.~(\ref{eq:timeder}) to the form (\ref{eq:dUdt})
\beq
\frac{\dd}{\dd t} \op U_{\rm E} (t)= 
{ \op L}_{H_{\rm E}(\ve x,t)}~\op U_{\rm E} (t) 
\eeq
where the {\em echo Hamiltonian} is given by 
\beq
H_{\rm E}({\ve x},t) = \eps V\left(\ve \phi_t(\ve x),t \right). \label{eq:Hf}
\eeq
The function $H_{\rm E}$ is nothing but the perturbation
part $\eps V$ of the original Hamiltonian, which, however, is evaluated
 at the point that is 
obtained by forward propagation with the unperturbed original Hamiltonian. 
It is important to stress that this 
is not a perturbative result but an exact expression. Also, even if the 
original Hamiltonian system was time independent, the echo dynamics
obtains an explicitly time dependent form.
Trajectories of the echo-flow are given by Hamilton equations
\beq
\dot {\ve x}= \mat J~ \ve \nabla H_{\rm E} (\ve x,t). 
\label{eq:hameq}
\eeq
At this point we limit our discussion to time independent
Hamiltonians and perturbations. The slightly more general case of periodically 
driven systems reducible to symplectic maps shall be discussed later.

Inserting (\ref{eq:Hf}) into Eq.~(\ref{eq:hameq}) yields
\beq
\dot {\ve x} = 
\eps \mat J \ve \nabla_{\ve x} V(\ve \phi_t(\ve x)) =
\eps \mat J {\mat M_t^T(\ve x)} (\ve \nabla V)(\ve \phi_t(\ve x)) \label{eq:dyn}.
\eeq
Here we have introduced the {\em stability matrix} $\mat M_t (\ve x)$,
$[\mat M_t(\ve x)]_{i,j} = \partial_j [\ve\phi_t(\ve x)]_i$.

\subsubsection{Linear response regime for Lipschitz continuous initial 
  density}

Let us first discuss the classical echo dynamics in the 
{\em linear response regime}, {\em i.e.} in the case where the total 
{\em echo displacement}
\begin{equation}
\Delta {\ve x}(t) = \int_0^t \dot{\ve{x}} \dd t' = 
\eps \int_0^t \dd t' \mat J {\mat M_{t'}^T(\ve x)} (\ve \nabla V)(\ve \phi_{t'}(\ve x)) 
\end{equation}
is small compared to all other classical phase space scales.
Writing classical fidelity in the symmetric representation we immediately arrive at the 
simple linear response result
\begin{eqnarray}
F_{\rm cl}(t) 
&=& \int_{\Omega}\dd{\ve x} \rho(\ve{x} + \Delta\ve{x}(t)/2)\rho(\ve{x}-\Delta\ve{x}(t)/2) \nonumber \\
&=& 1 - \frac{1}{4}\eps^2 \dd t'\int_\Omega\dd\ve{x} \left( \nabla\rho(\ve{x})\cdot
\int_0^t\dd t'\mat J {\mat M_{t'}^T(\ve x)} (\ve \nabla V)(\ve \phi_{t'}(\ve x))\right)^2
\label{eq:clLRf}
\end{eqnarray}
Now it's easy to specialize the above formula to the cases of chaotic and regular dynamics.
In the case of chaotic dynamics, the stability matrix ${\mat M}_t(\ve{x})$ will after a
short time $t$ expand any initial vector with exponential $\exp(\lambda_{\rm max} t)$ 
where $\lambda_{\rm max}$ is the maximal Lyapunov exponent of the flow.
Since the time integral of an exponential is still an exponential, 
the expression in the big
bracket on the RHS of Eq.~(\ref{eq:clLRf}) can be written as 
$2C[\rho]\exp(\lambda_{\rm max} t)$ where $C[\rho]$ is a constant depending 
only on the initial density.
Hence
\begin{equation}
F^{\rm ch}_{\rm cl}(t) = 1 - \eps^2 C[\rho]^2 \exp(2\lambda_{\rm max}|t|) + \OO(\eps^4).
\label{eq:clLRfch}
\end{equation}
For sufficiently strong perturbations, $\eps>\hbar$, and typical initial packets quantum fidelity decays in a similar way~\cite{Silvestrov:03}. On the other hand, for regular dynamics, the stability matrix is of parabolic 
type (all eigenvalues $1$ but having defective Jordan canonical form) so
it always stretches an arbitrary initial vector linearly in time and
\begin{equation}
F^{\rm reg}_{\rm cl}(t) = 1 - \eps^2 C'[\rho]^2 t^2 + \OO(\eps^4)
\label{eq:clLRfreg}
\end{equation}
where $C'[\rho]$ is another constant depending on the initial density $\rho$ only.

However, please note that in deriving the expression (\ref{eq:clLRf}) 
we have assumed that the
initial density $\rho(\ve{x})$ was {\em smooth} and {\em differentiable}.
In fact, the above result still holds for general Lipschitz continuous 
densities \cite{Veble:05}.

\subsubsection{Linear response regime for discontinuous initial density}

It is interesting though, that the result is quite different for discontinuous 
initial densities, like for example characteristic functions on sets in phase space
which may be of considerable interest in classical statistical mechanics.
In such a general case it is useful to rewrite the classical fidelity as
\begin{equation}
F_{\rm cl}(t) = 1 - 
\frac{1}{2}\int_\Omega [\rho(\ve{x}) - \rho(\ve{x}+\Delta\ve{x}(t))]^2
\end{equation}
This representation is a trivial consequence of the square normalizability of phase-space
density and Liouville theorem for the echodynamics $\ve{x}+\Delta\ve{x}(t)$.
Now it is easy to see that for Lipschitz continuous densities the previous
results (\ref{eq:clLRfch},\ref{eq:clLRfreg}) follow, while for {\em discontinuous}
densities the main contribution to the integral, for small $\eps$, {\em i.e.} 
for small $|\Delta\ve{x}|$, comes from the set of phase space points 
$\ve{x}$ where $\rho(\ve{x})$ is discontinuous, where the contribution is proportional to
discontinuity gap times $|\Delta\ve{x}|$, namely
\begin{equation}
F_{\rm cl}(t) = 1 - \int_\Omega \dd\ve{x} D_\rho(\ve{x})|\Delta\ve{x}(t)|
\end{equation}
where $D_\rho(\ve{x})$ is a certain {\em distribution} taking account for the 
discontinuity gaps of the density $\rho(\ve{x})$. So, the main distinction from the 
continuous case is that the fidelity drop is 
proportional to the length of the echo displacement $|\Delta\ve{x}|$ instead of
its square.
Since the latter asymptotically expands as $e^{\lambda_{\rm max}|t|}$ or $|t|$
for chaotic and regular dynamics, respectively, we obtain a universal linear 
response formulae for fidelity decay of discontinuous densities
\begin{eqnarray*}
F^{\rm ch,dis}_{\rm cl}(t) &=& 
1 - |\eps|C''[\rho]\exp(\lambda_{\rm max}t) + \OO(\eps^2),\\
F^{\rm reg,dis}_{\rm cl}(t) &=&
1 - |\eps|C'''[\rho]|t| + \OO(\eps^2),
\end{eqnarray*}
where $C''[\rho]$ and $C'''[\rho]$ are constants depending only on the
details of the initial density $\rho$.

\subsubsection{General (multi-)Lyapunov decay for chaotic few body systems}

From now on we assume that the flow $\ve{\phi}_t$ has very strong ergodic
properties, {\em e.g.} it is uniformly hyperbolic and Anosov, and consider
to what extend can classical echo dynamics be explored in order to understand
classical fidelity decay beyond the linear response approximation.
To understand the dynamics (\ref{eq:dyn}) we need to explore the properties of 
$\mat M_t$. We start by writing the matrix
\begin{equation}
\mat M_t^T(\ve x) \mat M_t(\ve x)=
\sum_j e^{2 \lambda_j t} d_j^2(\ve x,t)
 \ve v_j(\ve x,t)\otimes \ve v_j(\ve x,t)
\end{equation}
expressed in terms of orthonormal 
eigenvectors $\ve v_j(\ve x,t)$ and eigenvalues
$d_j^2(\ve x,t) \exp(2 \lambda_j t)$. After the ergodic time $t_{\rm e}$
necessary for the echo trajectory to explore the available region of phase 
space, Oseledec theorem \cite{Oseledec:68} guarantees 
that the eigenvectors of this matrix converge to Lyapunov eigenvectors being
independent of time, while $d_j(\ve x,t)$ grow slower than exponentially, 
so the leading exponential growth defines the Lyapunov exponents $\lambda_j$.
Similarly, the matrix
\begin{equation}
\mat M_t(\ve x) \mat M_t^T(\ve x) =\sum_j e^{2 \lambda_j t} c_j^2(\ve x_t,t)
 \ve u_j(\ve x_t,t)\otimes \ve u_j(\ve x_t,t), 
\end{equation}
where $\ve x_t=\ve\phi_t(\ve x)$, has the same eigenvalues 
[$c^2_j(\ve x_t,t) \equiv d^2_j(\ve x,t)$], and its eigenvectors depend on the 
final point $\ve x_t$ only,
as the matrix in question can be related to the backward evolution. 
The vectors $\{\ve u_j(\ve x_t)\}$, $\{\ve v_j(\ve{x})\}$, 
constitute left, right, part, respectively, of the {\em singular value decomposition} of 
$\mat M_t(\ve{x})$, so we write for $t \gg t_{\rm e}$
\beq
\mat M_t(\ve x)=
 \sum_{j=1}^N \exp(\lambda_j t) 
~\ve e_j(\ve \phi_t(\ve x))
\otimes  \ve f_j(\ve x) 
\label{eq:Mt}
\eeq
assuming that the limits 
$\ve e_j(\ve x) = \lim_{t\to\infty} c_j(\ve x,t)\ve u_j (\ve x,t)$,
$\ve f_j(\ve x) = \lim_{t\to\infty} d _j(\ve x,t) \ve v_j(\ve x,t)$
exist. Rewriting Eq.~(\ref{eq:dyn}) by means of Eq.~(\ref{eq:Mt})
we obtain
\beq
\dot {\ve x}=
\eps \sum_{j=1}^N \exp(\lambda_j t) 
~W_j(\ve \phi_t(\ve x))~
 \ve h_j(\ve x) 
\label{eq:dyn1}
\eeq
where $\ve h_j(\ve x)=\mat J\ve f_j(\ve x)$, and introducing new observables
\beq
W_j(\ve x)=
\ve e_j(\ve x)\cdot \ve \nabla V(\ve x).
\eeq

Note that for general hyperbolic systems the signs of the vector fields 
$\ve e_j(\ve x)$, $\ve f_j(\ve x)$ may actually not be unique, so they 
rigorously exist only as {\em director} fields, {\em i.e.} rank one tensor
fields. But since from a physics point of view this represents only a 
technical difficulty \cite{Veble:05}, we shall ignore the problem in the 
following discussions.

For small perturbations the echo trajectories
remain close to the initial point $\ve x(0)$ for times large in comparison to the
internal dynamics of the system ($t_{\rm e}$, Lyapunov times, decay of correlations, etc), 
and in this regime
the echo evolution can be linearly decomposed along different independent
directions $\ve h_j(\ve x(0))$ 
\beq
\ve x(t)=\ve x(0) +\sum_{j=1}^N y_j(t)
 ~\ve h_j(\ve x(0)).
\label{eq:decomp}
\eeq
For longer times, the point $\ve x(t)$ moves away from the initial point,
but the dynamics is still governed by the local unstable vectors at the 
evolved point. Therefore the decay of fidelity is governed by the
spreading of the densities along the conjugated unstable manifolds 
defined by the conjugate vector fields ${\ve h}_j(\ve x) = J {\ve f}_j(\ve x)$.

Inserting (\ref{eq:decomp}) into (\ref{eq:dyn1}) we obtain for each direction
$\ve h_j$
\beq
\dot y_j =  \eps ~ \exp(\lambda_j t) W_j(\ve \phi_t(\ve x)).
\label{eq:yyy}
\eeq
For {\em stable} directions with $\lambda_j<0$, after a certain time, the 
variable $y_j$ becomes a constant of the order $\eps$.

For {\em unstable} directions with $\lambda_j>0$, we
introduce a new variable $z_j$ as $y_j = \eps \exp(\lambda_j t) z_j$ and 
rewrite the above equation as
\beq
\dot z_j + \lambda_j z_j = W_j(\ve \phi_t(\ve x)).
\label{eq:damp}
\eeq
The right hand side of this equation is simply the evolution of the
observable $W_j$ starting from a point in phase space $\ve{x}=\ve{x}(0)$. 
Due to assumed ergodicity of the flow $\ve{\phi}_t$, $W_j(\ve \phi_t(\ve x))$
has well defined and {\em stationary} statistical properties such as
averages and correlation functions.
The linear damped Eq.~(\ref{eq:damp}) is formally equivalent to a
Langevin equation (with deterministic noise) and hence its
solution $z_j(t)$ has a well defined time- and $\eps$-independent 
probability distribution $P_j(z_j)$. Its moments can be expressed 
in terms of moments and correlation functions
of the deterministic noise $W_j$, in particular $\overline{W_j}=0$. 
The analysis generalizes to the case of explicitly time-dependent $W_j$.

Going back to the original coordinate $y_j$ we obtain its distribution as
$K_j(y_j)=P_j(z_j)\dd z_j/\dd y_j$,
or
$
K_j(y_j)=P_j\left(\exp(-\lambda_j t) y_j/\eps\right)
\exp(-\lambda_j t)/\eps.
$
This probability distribution tells us how, on average, points
within some initial (small) phase space set 
of characteristic diameter $\nu$ spread along a locally well defined 
unstable Lyapunov direction $j$ and therefore represents an 
averaged kernel of the evolution of such densities along
this direction. Starting from the initial localized density $\rho_0$, of 
small width $\nu$ such that the decomposition (\ref{eq:decomp}) does
not change appreciably along $\rho_0$, echo dynamics yields for the 
densities
$
\rho_t(\ve y)=\int \dd^N \ve{y}'
\rho_0(\ve y^\prime)
\prod_j K_j(y_j-y_j^\prime).
$
For the stable directions $j$ we set $K_j(y_j) = \delta(y_j)$, as the 
shift of $y_j$ (of order $\eps$) can be neglected as compared to 
unstable directions. This also implies that the assumption
$\eps \ll \nu$ is necessary in order to get any echo at
all after not too short times. Classical fidelity (\ref{fidoE}) 
can now be written as
$F_{\rm cl}(t)=\int \dd^N\ve y ~\rho_0(\ve y) ~\rho_t(\ve y)$.
As long as the width $\nu_j$ of $\rho_0$ along the unstable 
direction $j$ is much
larger than the width of the kernel $K_j$, there is no appreciable
contribution to fidelity decay in that direction. At time
\beq
t_j=(1/\lambda_j)\log(\nu_j/(\eps\gamma_j)),
\eeq 
where $\gamma_j$ is a typical width of the distribution $P_j$, the width of
the kernel is of the order of the width of the distribution along the
chosen direction. After that time, the overlap between the two
distributions along the chosen direction
starts to decay with the same rate as the value of
the kernel in the neighborhood of $y_j=0$, which is
$\propto \exp(-\lambda_j t)$. The total overlap decays as
\beq
F_{\rm cl}(t)\approx \prod_{j;~t_j<t} \exp\left[-\lambda_j (t-t_j)\right], 
\label{eq:multilyap}
\eeq
where only those unstable directions contribute to the decay for
which $t_j>t$. As the time $t_j$ is shorter the higher the corresponding
Lyapunov exponent $\lambda_j$, fidelity will initially decay with
the largest Lyapunov exponent $\lambda_1$. In chaotic 
systems with more than two degrees of freedom we, however,
 expect to observe an increase of decay rate after the time
$t_2$, etc. Eq.~(\ref{eq:multilyap}) provides
good description for classical fidelity as long as $F_{\rm cl}(t)$ does not 
approach the
saturation value $F_\infty \sim \nu^N$ where the asymptotic decay
of classical fidelity is then given by the leading eigenvalue of 
Perron-Frobenius operator \cite{Benenti:03b}.

\begin{figure}
\centerline{\includegraphics[width=\figw\textwidth]{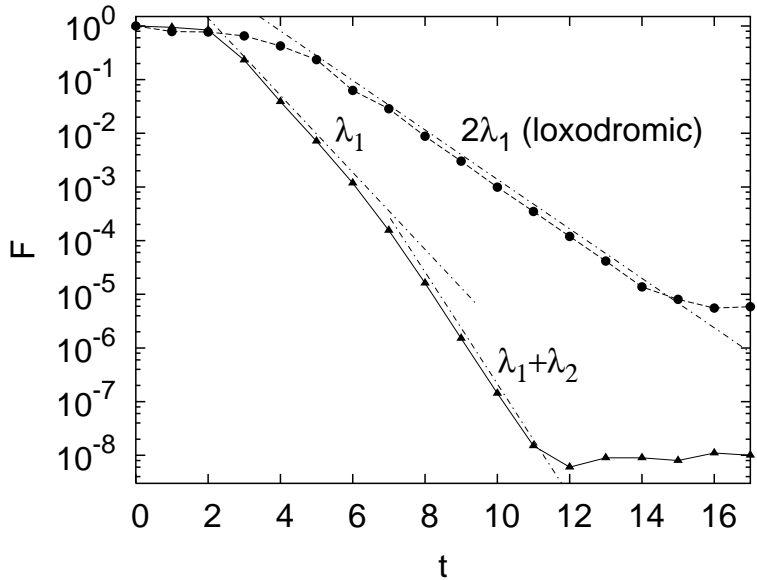}}
\caption{Decay of classical fidelity for two examples of 4D cat maps 
perturbed as explained in text [taken from \cite{Veble:04}]. 
Triangles refer to doubly-hyperbolic case where initial set was a 4-cube
$[0.1,0.11]^4$, and $\eps=2\cdot 10^{-4}$, whereas circles
refer to loxodromic case where initial set was $[0.1,0.15]^4$, and 
$\eps=3\cdot 10^{-3}$. In both cases initial density was sampled by $10^9$ points.
Chain lines give exponential decays with theoretical rates, 
$\lambda_1=1.65,\lambda_1+\lambda_2=2.40$ (doubly-hyperbolic),
and $2\lambda_1=1.06$ (loxodromic).
\label{fig:3}}
\end{figure}

The result (\ref{eq:multilyap}) does not only explain a simple Lyapunov decay 
of Loschmidt echoes as observed in many numerical experiments on two 
dimensional systems, but also predicts a cascade of decays with increasing rates given by the sums of first few Lyapunov exponent for few body chaotic systems.
For example, in the first non-trivial case of a four dimensional dynamics with two positive Lyapunov exponents one can have two distinct behaviors.
 In the simple {\em doubly hyperbolic} case of well separated individual Lyapunov exponents 
$\lambda_1 > \lambda_2$ the decay is expected to go through a cascade of 
increasing rates, first $\lambda_1$ and then $\lambda_1 + \lambda_2$, 
whereas in the {\em loxodromic} case $\lambda_1=\lambda_2$ the rate is $2\lambda_1$.
We illustrate this numerically for fourdimensional cat maps, {\em i.e.} 
4-volume preserving automorphisms on a 4-torus, 
$\ve{x}'= \mat{C} \ve{x} \pmod{1},$ $\ve x \in [0,1)^4$. The choices
$$
\mat C_{\rm d-h}=\left[
\begin{array}{rrrr}
2 &-2 &-1 & 0 \\
-2& 3 & 1 & 0 \\
-1& 2 & 2 & 1 \\
2 &-2 & 0 & 1
\end{array}\right],\
\mat C_{\rm lox}=\left[
\begin{array}{rrrr}
0 & 1 & 0 & 0 \\
0 & 1 & 1 & 0 \\
1 &-1 & 1 & 1 \\
-1&-1 &-2 & 0
\end{array}\right]
$$
are two examples representing the doubly-hyperbolic and loxodromic case. 
The matrix $\mat C_{\rm d-h}$ has the unstable eigenvalues $\approx 5.22,2.11$,
while the largest eigenvalues of $\mat C_{\rm lox}$ 
are $\approx 1.70 \exp(\pm {\rm i} 1.12)$. The perturbation for
both cases was done by applying an additional map at each time step
$\bar{x}_1=x'_1 + \eps\sin\left(2\pi x_3\right) \pmod{1}$,
$\bar{x}_{2,3,4}=x'_{2,3,4}$.
In Fig.~\ref{fig:3} we show the two types of decay which agree with
theoretical predictions.

\subsection{Asymptotic long time decay for chaotic dynamics}

However, our theory above is valid only up to a time-scale in which the
echo-dynamics spreads all over the available phase space.
After that time, one should use a different approach to understand
asymptotic (long time) decay of classical fidelity. In this regime, there 
does not yet exist any  rigorous approach to fidelity decay, 
hence we describe a heuristic approach of the paper \cite{Benenti:03b}, 
also demonstrating different regimes of classical 
fidelity decay in chaotic classical maps. We illustrate the general phenomena 
in a standard model of classical chaos, namely the {\em sawtooth map} 
which is defined by 
\begin{equation}
\overline{p}=p+F_0(\theta), \quad \overline{\theta}=\theta
+\overline{p} \pmod{2\pi}, \label{saw}
\end{equation}
where $(p,\theta)$ are conjugated action-angle variables,
$F_0=K_0(\theta -\pi)$, and the over-bars denote the variables
after one map iteration. We consider this map on the torus $0\leq
\theta < 2\pi$, $-\pi L\leq p <\pi L$, where $L$ is an integer.
For $K_0>0$ the motion is completely chaotic and diffusive, with
Lyapunov exponent given by
$\lambda=\ln\{(2+K_0+[(2+K_0)^2-4]^{1/2})/2\}$. For $K_0>1$ one
can estimate the diffusion coefficient $D$ by means of the random
phase approximation, obtaining $D\approx (\pi^2/3)K_0^2$. In order
to compute the classical fidelity, we choose to perturb the
kicking strength $K=K_0+\eps$, with $\eps\ll K_0$. In
practice, we follow the evolution of a large number of trajectories, which
are uniformly distributed inside a given phase space region of
area $A_0$ at time $t=0$. The fidelity $f(t)$ is given by the
percentage of trajectories that return back to that region after
$t$ iterations of the map (\ref{saw}) forward, followed by the
backward evolution, now with the perturbed strength $K$, in the
same time interval $t$. In order to study the approach to
equilibrium for fidelity, we consider the quantity
\begin{equation}
g(t)=(f(t)-f(\infty))/(f(0)-f(\infty)) \; .
\end{equation}
Thus, $g(t)$ drops from $1$ to $0$ as $t$ goes from $0$ to
$\infty$. We note that $f(0)=1$ while, for a chaotic system,
$f(\infty)$ is given by the ratio $A_0/A_c$, with $A_c$ the area
of the chaotic component to which the initial distribution
belongs.

\begin{figure}
\centerline{\includegraphics[width=\figw\textwidth]{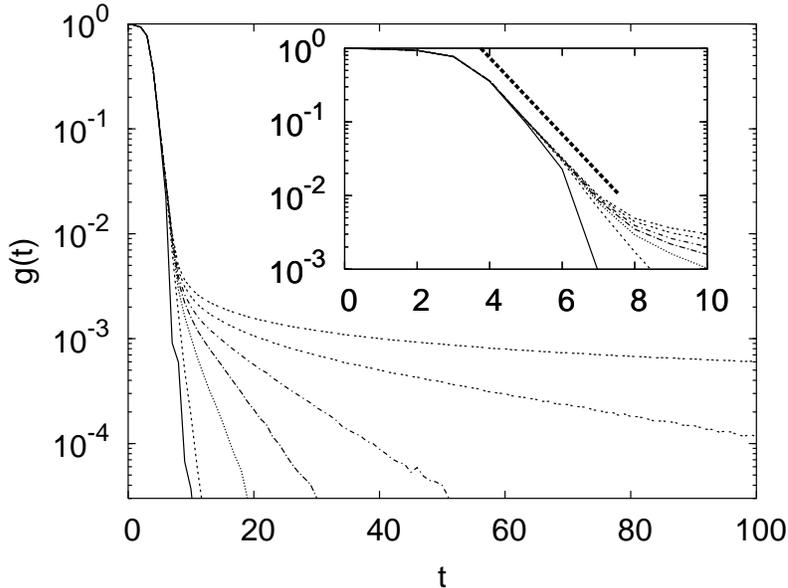}} \caption{ Decay
of the fidelity $g(t)$ for the sawtooth map with the parameters
$K_0=(\sqrt{5}+1)/2$ and $\eps=10^{-3}$ for different values
of $L=1,3,5,7,10,20,\infty$ from the fastest to the slowest
decaying curve, respectively [taken from \cite{Benenti:03b}]. The initial phase space density is
chosen as the characteristic function on the support given by the
$(q,p)\in[0,2 \pi)\times[-\pi/100,\pi/100]$. Note that between the
Lyapunov decay and the exponential asymptotic decay there is a
$\propto 1/\sqrt{t}$ decay, as expected from diffusive behavior.
Inset: magnification of the same plot for short times, with the
corresponding Lyapunov decay indicated as a thick dashed line.}
\label{fig:fidelity}
\end{figure}

\begin{figure}
\centerline{\includegraphics[width=\figw\textwidth]{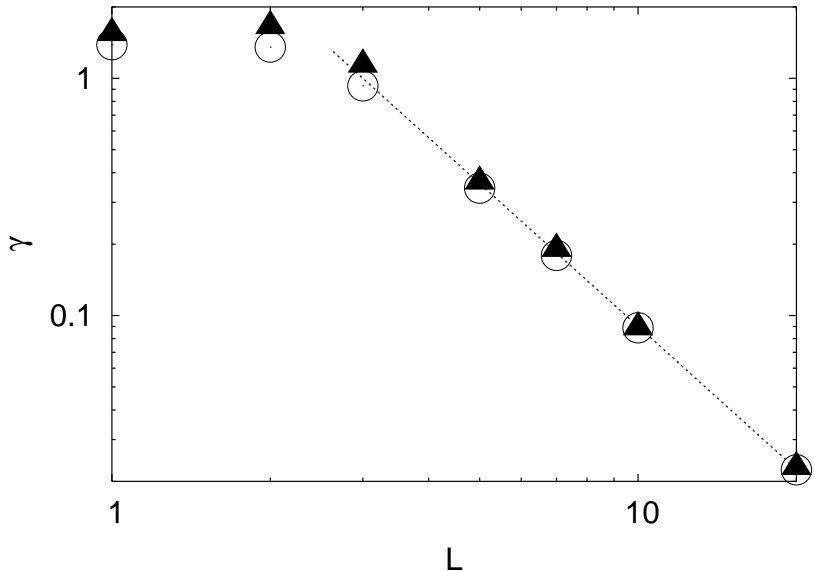}}
\caption{Asymptotic exponential decay rates of fidelity for the
sawtooth map ($K_0=(\sqrt{5}+1)/2$, $\eps=10^{-3}$) as a
function of $L$ [taken from \cite{Benenti:03b}]. The rates are extracted by fitting the tails of
the fidelity decay in the Fig.~\ref{fig:fidelity} (triangles) and
from the discretized Perron-Frobenius operator (circles). The line
denotes the $\propto 1/L^2$ behavior of the decay rates, as
predicted by the Fokker-Planck equation.}
\label{fig:decrates}
\end{figure}

The behavior of $g(t)$ is shown in Fig.~\ref{fig:fidelity}, for
$K_0=(\sqrt{5}+1)/2$ and different $L$ values. One can see that
only the short time decay is determined by the Lyapunov exponent $\lambda$,
$f(t) = \exp(-\lambda t)$. 
In this figure we demonstrate that the Lyapunov decay is followed by a 
power law  decay  $\propto 1/\sqrt{D t}$, a manifestation of deterministic
diffusion taking place on the cylindrical phase space (for large $L$) 
\cite{Benenti:02}, up to the diffusion time 
$t_D\sim L^2/D$, and then the asymptotic relaxation to equilibrium takes place
exponentially, with a decay rate $\gamma$ 
(shown in Fig.~\ref{fig:decrates}) ruled
not by the Lyapunov exponent but by the largest Ruelle-Pollicott
resonance.

These resonances have been determined for the sawtooth map by diagonalizing a 
discretized (coarse-grained) classical propagator~\cite{Benenti:03b} and 
determining the spectral gap between the modulus of the largest eigenvalue
and the unit circle.
In Fig.~\ref{fig:decrates} we
illustrate the good agreement between the asymptotic decay rate of
fidelity (extracted from the data of Fig.~\ref{fig:fidelity}) and the decay 
rate $\gamma$ as predicted by the gap in the discretized Perron-Frobenius 
spectrum. 

More generally, one can conclude that the long time asymptotics of
the (shifted) classical fidelity $g(t)$ behaves in the same way as a typical
temporal correlation function of the (forward) dynamics at time $2t$.
This is true even for systems where the decay of correlations is given by 
power laws and the spectral gap of the Perron-Frobenius
operator vanishes, {\em e.g.} for the stadium billiard \cite{Benenti:03b}.

\subsection{Asymptotic decay for regular dynamics}

As for the decay of classical fidelity in the regime of
regular classical, the theory has been developed in Ref.~\cite{Benenti:03}.
We shall not outline the details here but just repeat the main result.

The evolution of initial phase space density $\rho({\ve x},0)$ under regular 
dynamics is quite trivial, as it evolves quasi-periodically on the set of
neighboring tori as can be expressed explicitly in the action-angle
coordinates \cite{Benenti:03}. We note that a perturbation of an integrable 
system, being of KAM (Kolmogorov-Arnold-Moser) type i.e. sufficiently smooth 
in the phase space variables, can cause two main effects on the KAM tori of an 
unperturbed system: (i) either the shape of the tori changed, or (ii) the 
frequencies of the motion changed. Depending on whether one of the two effects 
is dominant, we have
two different types of decay: (i) algebraic decay $f(t) \propto t^{-d}$, where
$d$ is the number of degrees of freedom if mainly the shape of the tori 
is changed, (ii) ballistic decay, which is basically determined by the shape
of the initial density $\rho({\ve x},0)$, i.e. Gaussian if the latter is a 
Gaussian, if mainly the frequencies of the tori are changed.

\begin{figure}
\centerline{\includegraphics[width=\figw\textwidth]{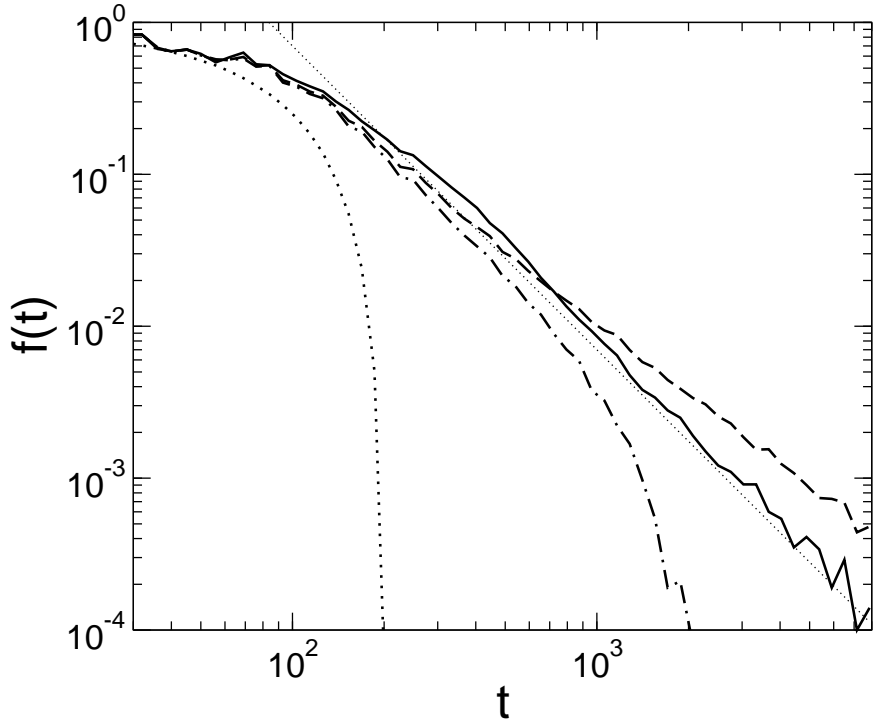}}
\caption{Fidelity decay for the rectangular billiard 
for various values of the parameter $\beta=$ 0 (full line),  
0.232 (dashed), 0.3 (dot-dashed)
and $\pi/2$ (dotted) [taken from \cite{Benenti:03}].
The parameters
of the system have been chosen as follows: $\alpha_1=(\sqrt{5}+1)/2$, 
$\alpha_2=1$. In all cases
the initial phase space density is a hyper-rectangle centered around
$I_1=1$, $I_2=1$, $\Theta_1=1$ and $\Theta_2=1$ with all sides
of length $\nu_{I_1}=\nu_{I_1}=
\nu_{\Theta_1}=\nu_{\Theta_2}=0.02$. 
The perturbation parameter is $\eps=3\times 10^{-4}$ and 
the number of trajectories $N=10^5$. The $\propto 1/t^2$ decay is 
shown as a thin dotted line.}
\label{fig:rect}
\end{figure}

In Fig.~\ref{fig:rect} we show an example of classical fidelity decay in
an integrable rectangular billiard, with the action-angle Hamiltonian ($d=2$)

\begin{equation}
H_0=\frac{\alpha_1}{2}I_1^2+\frac{\alpha_2}{2}I_2^2
\end{equation}
which is perturbed by a generic perturbation of the form
\begin{equation}
V=\cos(\beta) \cos(\Theta_1) \cos(\Theta_2)
+\sin(\beta) I_1 I_2\nonumber.
\end{equation}
Depending on the value of the parameter $\beta$, the 
perturbation mainly affects either the shape of the tori or 
their frequencies.

\subsection{Classical fidelity in many-body systems}

In systems with increasing, macroscopic number of degrees of freedom the above 
theoretical considerations can still be applied though somehow qualitatively different 
behavior is obtained. The details can be found in Ref.~\cite{Veble:05}.

The central result of that analysis is that the fidelity of a many-body system 
(in the larger $d$ limit), which is perturbed by a {\em global} perturbation 
(i.e. the one which perturbs all degrees of freedom in an approximately 
uniform way), and for which Lyapunov exponents around the maximal 
one $\lambda_{\rm max}$ are smoothly distributed with a well defined density 
in the thermodynamic limit $d\to\infty$, decays with a doubly exponential law
\begin{equation}
F_{\rm cl}(t) = \exp(-A d \eps^\beta \exp(\beta\lambda_{\rm max}t)) \; .
\end{equation}
$A$ is some system specific constant which does not depend on $d,\eps,t$.
As we have discussed already within the classical linear response 
approximation, it turns out that it is of crucial importance whether
the initial phase space density is (Lipschitz) 
continuous or not, namely we have index $\beta=1$ for
discontinuous distribution, and $\beta=2$ for Lipschitz continuous one.

\subsection{Universal decay in dynamically mixing systems which lack exponential 
sensitivity to initial conditions}

Recently, another possibility of relaxation and mixing behavior in classical dynamical
systems has been investigated \cite{CasatiProsen:99,CasatiProsen:00}, 
which does not require exponential sensitivity on initial conditions. Even 
though it may appear quite exotic at the first sight, such a situation can 
arise quite often, for example in polygon billiards of two and higher 
dimensions, hard point gases in one dimension, etc. The decay of classical 
fidelity for such situations has been addressed in Ref.~\cite{Jingua:05}. It 
has been found that under quite general dynamical conditions of mixing and 
marginally stable (parabolic) dynamics (being a consequence of the lack of 
exponential sensitivity to initial conditions), the fidelity decay exhibits a
universal form (for small perturbations $\eps$) namely
\begin{equation}
F_{\rm cl}(t) = \phi(B|\eps|^{2/5}t)
\end{equation}
where $B$ is some constant depending on details of temporal correlation decay 
of the perturbation and $\phi(x)$ is a {\em universal} function with the 
following asymptotic properties: $1 - \phi(x) \propto |x|^{5/2}$, for 
$|x| \to 0$, and $\phi(x) = \exp(-|x|)$, for $|x|\to\infty$.

\section{Time scales and the transition from regular to chaotic behavior}
\label{sec:timescales}

It is instructive to summarize all different time scales involved in the decay
of fidelity for regular and chaotic dynamics. As for general, or generic 
perturbations one does not have freeze, 
we shall limit our discussion to the situation with a 
nonzero time averaged perturbation. 

\subsection{Chaotic dynamics}

\label{sec:timechaos}
\begin{figure}[ht]
\centerline{\includegraphics[width=0.5\textwidth]{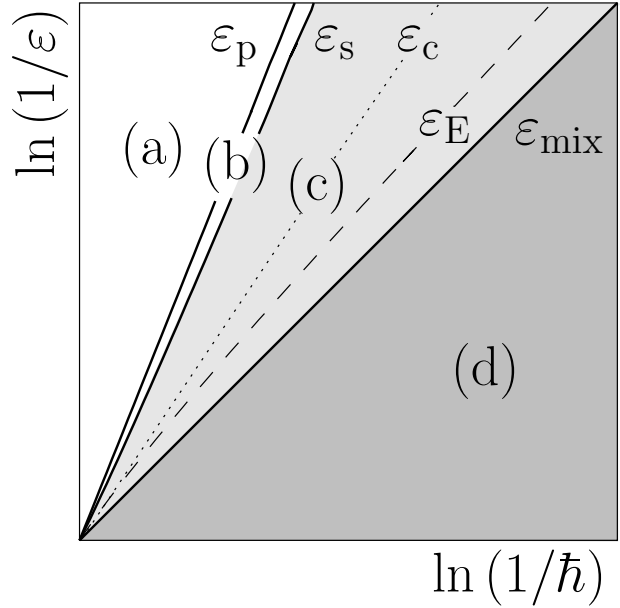}}
\caption{Schematic view of different fidelity decay regimes for mixing dynamics 
[from~\cite{Prosen:02corr}]. For instance, by increasing 
the perturbation fidelity goes through a Gaussian perturbative regime (a), then exponential 
Fermi golden rule (c) and finally for $\eps > \eps_{\rm mix}$ it decays 
almost instantly. The limits $\hbar \to 0$ and $\eps \to 0$ do not commute! 
For details see text.}
\label{fig:delta1}
\end{figure}

The exposition here closely follows the one in Ref.~\cite{Prosen:02corr}.
For chaotic dynamics there are six relevant time scales 
(even seven for coherent initial states): the (short) Zeno time, the classical 
mixing time $t_{\rm mix}$ on which correlation function decays; the quantum decay time of the fidelity 
$\tau_{\rm m}$ (\ref{eq:Fnmixing}); the Heisenberg time $t_{\rm H}$ (\ref{eq:np_def}) 
after which the system starts to ``feel'' finiteness of Hilbert space and 
effectively begins to behave as an integrable system; the decay time 
$\tau_{\rm p}$ (\ref{eq:gaussianmixing}) of perturbative Gaussian decay present 
after $t_{\rm H}$; the time $t_\infty$ when the fidelity reaches the 
finite size plateau; 
for coherent initial states we have in addition the Ehrenfest time $t_{\rm E}$ 
up to which we have quantum-classical correspondence. Depending on the 
interrelation of these time scales, {\it i.e.} depending on the 
perturbation strength $\eps$, Planck's constant $\hbar$ and the dimensionality $d$, 
we shall also have different decays of fidelity. All different regimes can be 
reached by {\it e.g.} fixing $\hbar$ and increasing $\eps$. 
Let us follow different decay regimes as we increase $\eps$ 
(shown in Fig.~\ref{fig:delta1}):
\begin{enumerate}
\item[(a)] For $\eps<\eps_{\rm p}$ we will have $t_{\rm H} < \tau_{\rm m}$. 
  This means that at the Heisenberg time, the fidelity due to exponential decay 
  (\ref{eq:Fnmixing}) will still be close to $1$, $F(t_{\rm H}) \approx 1$, and 
  we will see mainly a Gaussian decay due to finite 
Hilbert space (\ref{eq:gaussianmixing}). The critical $\eps_{\rm p}$ below which we will see 
this regime has already been calculated and is (\ref{eq:deltap})
\begin{equation}
\eps_{\rm p}=\frac{\hbar}{\sqrt{\sigma_{\rm cl}{\mathcal N}}}=\hbar^{d/2+1}\,
\frac{(2\pi)^{d/2}}{\sqrt{\sigma_{\rm cl}{\mathcal V}}}.
\label{eq:dp}
\end{equation}
For $\eps<\eps_{\rm p}$ the fidelity will have Gaussian decay with the decay time 
$\tau_{\rm p}$ (\ref{eq:gaussianmixing})
\begin{equation}
\tau_{\rm p}=\frac{\hbar^{1-d/2}}{\eps}\sqrt{\frac{{\mathcal V}}{4\sigma_{\rm cl} (2\pi)^d}}.
\label{eq:taup}
\end{equation}

\item[(b)] For $\eps_{\rm p}< \eps < \eps_{\rm s}$ we will have a crossover 
  from the initial exponential decay (\ref{eq:Fnmixing}) to the asymptotic 
  Gaussian decay (\ref{eq:gaussianmixing}) at time $t_{\rm H}$ illustrated in
  Figs.~\ref{fig:scalingGOE} and~\ref{fig:scalingGUE}. This regime occurs if 
  $\tau_{\rm m} < t_{\rm H} < t_\infty$. With increasing perturbation, 
$t_\infty$ will decrease and the upper border $\eps_{\rm s}$ is determined by the condition 
$t_\infty=t_{\rm H}$. Denoting a finite size plateau by $\Fn\sim 1/{\mathcal N}^\mu$, with 
$\mu$ lying between $1$ and $2$, depending on the initial state (see Appendix~\ref{sect:appTA}), 
we have the condition $\exp{(-(t_{\rm H}/\tau_{\rm p})^2)}=1/\mathcal{N}^\mu$ which gives
\begin{equation}
\eps_{\rm s}=\frac{\hbar}{\sqrt{\sigma_{\rm cl}{\mathcal N}}} \sqrt{\mu \ln{{\mathcal N}}}=
\eps_{\rm p} \sqrt{\mu \ln{{\mathcal N}}}.
\label{eq:ds}
\end{equation}
Further increasing the perturbation, we reach perhaps the most interesting regime, in which 
quantum fidelity can decay faster the more chaotic the systems is. In this regime the exponential 
decay persists until the plateau is reached.

\item[(c)] For $\eps_{\rm s} < \eps < \eps_{\rm mix}\, (\eps_{\rm E})$ we will have an 
exponential decay until $t_\infty$. The upper border $\eps_{\rm mix}$ is determined by the 
condition $\tau_{\rm m}=t_{\rm mix}$ which is a point where the argument leading to the 
factorization of $n-$point correlation function breaks down. For random initial states 
we get
\begin{equation}
\eps_{\rm mix}=\frac{\hbar}{\sqrt{2 \sigma_{\rm cl} t_{\rm mix}}}=
\eps_{\rm p}\sqrt{\frac{{\mathcal N}}{2t_{\rm mix}}}.
\label{eq:dmix}
\end{equation}
Note that the relative size of this window $\eps_{\rm mix}/\eps_{\rm s}=
\sqrt{{\mathcal N}/2\mu t_{\rm mix}\ln{{\mathcal N}}}$ increases both in the semiclassical 
$\hbar \to 0$ and in the thermodynamic $d \to \infty$ limit.

For coherent initial states the quantum correlation function relaxes on a slightly longer 
time scale than $t_{\rm mix}$, namely on the Ehrenfest time 
$t_{\rm E}\sim -\ln{\hbar}/\lambda$. Until $t_{\rm E}$ quantum packet 
follows the classical trajectory and afterwards interferences start to build leading 
to the breakdown of quantum-classical correspondence. Equating $\tau_{\rm m}=t_{\rm E}$ 
gives the upper border for coherent states
\begin{equation}
\eps_{\rm E}=\frac{\hbar}{\sqrt{-\ln{\hbar}}}\, \sqrt{\frac{\lambda}{2\sigma_{\rm cl}}}= 
\eps_{\rm mix}\, \sqrt{\frac{\lambda t_{\rm mix}}{-\ln{\hbar}}}.
\label{eq:de}
\end{equation}

\item[(d)] For $\eps > \eps_{\rm mix}$ the perturbation 
is so strong that the quantum fidelity decays before $t_{\rm mix}$, 
{\it i.e.} perturbed and unperturbed dynamics are essentially 
unrelated and fidelity decays almost instantly.
\end{enumerate}
For coherent initial states the upper border of regime (c) is at $\eps_{\rm E}$ 
which is smaller than the lower border $\eps_{\rm mix}$ of regime (d), which 
opens up the possibility of another regime between (c) and (d), namely for 
$\eps_{\rm E}< \eps < \eps_{\rm mix}$ the fidelity will decay within the 
Ehrenfest time which grows logarithmically with the number of contributing 
states ${\mathcal N}$ (i.e. with $\log(1/\hbar)$). In this regime the decay of 
quantum fidelity is the same as the decay of classical fidelity and can be 
explained in terms of classical Lyapunov exponents~\cite{Veble:04}. The 
relative width of this regime 
$\eps_{\rm mix}/\eps_{\rm E}=\sqrt{\ln{(1/\hbar)}/\lambda t_{\rm mix}}$ again
grows logarithmically with $1/\hbar$. In regime (d), that is for $\hbar < \eps < \sqrt{\hbar}$, quantum dynamics acctually becomes ``hypersensitive'' to perturbations~\cite{Silvestrov:03}. For typical initial packets the decay is double exponential, $F \propto \exp{(-{\rm const}\times {\rm e}^{2\lambda t})}$. Due to large fluctuations between different initial conditions for times smaller than the Ehrenfest time the average fidelity depends on the way we average. For instance, averaging fidelity one obtains exponential Lyapunov-like decay, $\ave{F(t)} \propto {\rm e}^{-\lambda_1 t}$, whereas by averaging the logarithm of the fidelity one gets a double exponential decay, $\ave{\log{F(t)}}\propto {\rm e}^{2 \lambda_{-2} t}$, see Ref.~\cite{Silvestrov:03} for details. 

Similar information as from Fig.~\ref{fig:delta1} can be gained from Fig.~{6} 
in Ref.~\cite{Cucchietti:04}.

For composite systems we have seen in Section~\ref{sec:chcomposite} that reduced 
fidelity and echo purity decay semiclassically on the same time scale as fidelity, 
the difference between time scales being 
$\sim 1/\mathcal{N}_{\rm c}+1/\mathcal{N}_{\rm e}$ if 
$\mathcal{N}_{\rm c,e}$ are dimensions of two subspaces. Therefore, time scales 
involved in the decay of 
either reduced fidelity or echo purity (purity) are exactly the same as for fidelity.

Similar regimes as for the fidelity decay were also obtained in 
Refs.~\cite{Cohen:99,Cohen:00,Cohen:00a} when studying energy spreading in parametrically
driven systems, see also Ref.~\cite{Wisniacki:02b} for a connections between fidelity and local density of states.

\subsection{Regular dynamics}

For regular dynamics the situation is much simpler. We have only three relevant time scales: 
(i) the classical averaging time $t_{\rm ave}$ in which the average
perturbation operator converges to $\bar{V}$ (\ref{eq:Vbardef}), 
(ii) the quantum fidelity decay time $\tau_{\rm r}$ and, 
(iii) the time $t_\infty$ when the fidelity reaches a fluctuating plateau due 
to the finite dimension of Hilbert space. For times smaller than $t_{\rm ave}$ decay 
is system and state specific and can not be discussed in general. After 
$t_{\rm ave}$ the fidelity decay is first quadratic in time as dictated by the linear response 
formula (\ref{eq:reglr}). The decay time $\tau_{\rm r}$ scales as $\sim \sqrt{\hbar}/\eps$ 
(\ref{eq:Fnregcoh}) for coherent initial states and as $\sim \hbar/\eps$ (\ref{eq:regpower}) 
for random initial states. Beyond linear response the functional dependence of the decay is typically a
Gaussian for coherent initial states and power law for random initial states with the before mentioned decay 
time $\tau_{\rm r}$ (by the decay time in the case of power law dependence we mean the time when the fidelity reaches a fixed value, e.g., $\frac{1}{2}$). 
For weak perturbation this decay persists until $t_{\rm infty}$ when the 
finite size effects take in and fidelity approaches a time-averaged saturation value (\ref{eq:Fnavg}).
However, for stronger perturbations and very small $\hbar$, such that the time-scale $t_{\rm a}$ (\ref{eq:regulart*}) of 
applicability of action space integral approximation
% (\ref{eq:fnANSI}) 
may be shorter than $t_\infty$, one may
observe another regime of asymptotic fidelity decay after $t_{\rm a}$ \cite{Weinstein:05}, 
which is typically again a power law (however, of different origin than for random initial states). 
Detailed semiclassical theory of this regime can be found in Ref.~\cite{Wang:06}.

The time $t_\infty$ again depends on the initial state as well as on 
the Hilbert space dimension ${\mathcal N}$. For random initial states the power law decay gets faster 
with increasing dimensionality $d$ of the system, and is conjectured to approach a Gaussian decay 
in the thermodynamic limit~\cite{Prosen:02}.  

In contrast to chaotic systems, for regular systems the decay time scale of reduced fidelity and 
of echo purity for composite systems differs from that of fidelity. For coherent initial states 
the reduced fidelity decays as a Gaussian on a time scale 
$\tau \sim \sqrt{\hbar}/\eps$ (\ref{eq:Frexact}), i.e.
with the same scaling as fidelity (\ref{eq:Fnregcoh}) but with a different prefactor. 
Echo purity (or purity) for coherent initial states on the other hand initially decays quadratically but then asymptotically goes into a power law decay, 
$F_{\rm P} \sim 1/(\eps t)^r$ (\ref{eq:Imain}), where the power $r$ depends on the 
perturbation and is bounded by $1 \le r \le {\rm min}(d_{\rm c},d_{\rm e})$ if $d_{\rm c,e}$ 
are the numbers of degrees of freedom of two subsystems. The important point though is that the whole decay does not depend on $\hbar$. For small $\hbar$ the decay of echo 
purity is therefore much slower than that of fidelity. If the initial state of the central subsystem is 
a superposition of two packets, the so-called cat state, the decay time of purity is much smaller and scales 
the same as for reduced fidelity or fidelity for a single coherent state, {\em i.e.} as 
$\tau_{\rm dec}\sim \sqrt{\hbar}/\eps$ (\ref{eq:Icat}). For random initial states echo purity and 
reduced fidelity decay on a time scale $\tau \sim \hbar/\eps$.

\subsection{Comparison, chaotic vs. regular}

\label{sec:compare}
Let us compare decay time scales of chaotic and regular systems. One might expect that quantum 
fidelity will decay faster for chaotic systems than for regular, at least such is the case for 
classical fidelity (Section~\ref{C}). As we shall see, this is not necessarily the case. Quantum fidelity
decay can be faster for regular systems! 

The fidelity decay time scales as $\sim 1/\eps$ for regular systems, while it is 
$\sim 1/\eps^2$ for mixing dynamics. This opens up an interesting possibility: it is 
possible that the fidelity decays {\em faster} for regular system than for chaotic one. 
Demanding $\tau_{\rm r}<\tau_{\rm m}$ we find that for sufficiently small 
$\eps$ one will indeed have faster fidelity decay in regular systems. This will happen for
\begin{equation}
\eps <
\left\{
\begin{array}{ll}
\eps_{\rm r}=\hbar\,\bar{C}^{1/2}/2\sigma_{\rm cl} \propto \hbar      & \hbox{random init. state} \\ 
\eps_{\rm c}=\hbar^{3/2}\sqrt{\bar{\veb{v}}'\cdot \Lambda^{-1} \bar{\veb{v}}'/8 
\sigma^2_{\rm cl}}\propto \hbar^{3/2} & \hbox{coherent init. state}
\end{array}
\right. .
\label{eq:deltacon}
\end{equation}
We explicitly wrote the result for random initial states $\eps_{\rm r}$ and coherent 
initial states $\eps_{\rm c}$ as the two have different scaling with $\hbar$. We can see 
that for random initial states $\eps_{\rm r}$ scales in the same way as $\eps_{\rm mix}$ 
and so one has faster decay of fidelity in regular systems provided 
$\eps<\eps_{\rm r} \sim \eps_{\rm mix}$. For a coherent initial state this can be satisfied 
above the perturbative border $\eps > \eps_{\rm p}$ only in more than one dimension $d > 1$. 
In one dimensional systems $\eps_{\rm p}$ and $\eps_{\rm c}$ have the same scaling with 
$\hbar$ and whether we can observe faster decay of fidelity in regular systems than in chaotic 
ones depends on the values of $\sigma_{\rm cl}$ and $\bar{\veb{v}}'$. We stress that our result 
does not contradict any of the existing findings on quantum-classical correspondence. For 
example, a growth of quantum dynamical entropies~\cite{Alicki:96,Miller:99} persists only 
up to logarithmically short Ehrenfest time 
$t_{\rm E}$, which is also the upper bound for the Lyapunov decay of quantum 
fidelity~\cite{Jalabert:01} and within which one would always find 
$F^{\rm reg}(t) > F^{\rm mix}(t)$ (for coherent states) above the perturbative border 
$\eps > \eps_{\rm p}$, whereas the theory discussed here reveals new nontrivial 
quantum phenomena with a semiclassical prediction (but not correspondence!) much beyond that
time. If we let $\hbar\to 0$ first, and then $\eps \to 0$, {\it i.e.} we keep 
$\eps \gg \eps_{\rm r,\rm c}(\hbar)$,
then we recover the result supported by classical intuition, namely that the 
regular (non-ergodic) dynamics
is more stable than the chaotic (ergodic and mixing) dynamics. On the other hand, if
we let $\eps\to 0$ first, and only after that $\hbar\to 0$, {\it i.e.} satisfying 
Eq.~(\ref{eq:deltacon}), we find somewhat counterintuitive result saying 
that chaotic (mixing) dynamics is more stable 
than the regular one. How can we understand this? Finite $\hbar$ causes that there is a lower
limit on the size of structures, see for instance figures of Wigner functions in 
Section~\ref{sec:wigner}. Therefore, for sufficiently small perturbations 
quantum fidelity can not exhibit Lyapunov decay because the latter implies 
occurrence of smaller and smaller structures as time progresses. So what comes 
into play for quantum fidelity are correlations between perturbations applied at 
different times. If correlations are small, like in a chaotic system, perturbations 
will ``add up'' in a slow random way, causing a slow decay of quantum fidelity. 

\begin{figure}[th!]
\centerline{\includegraphics[width=\figw\textwidth]{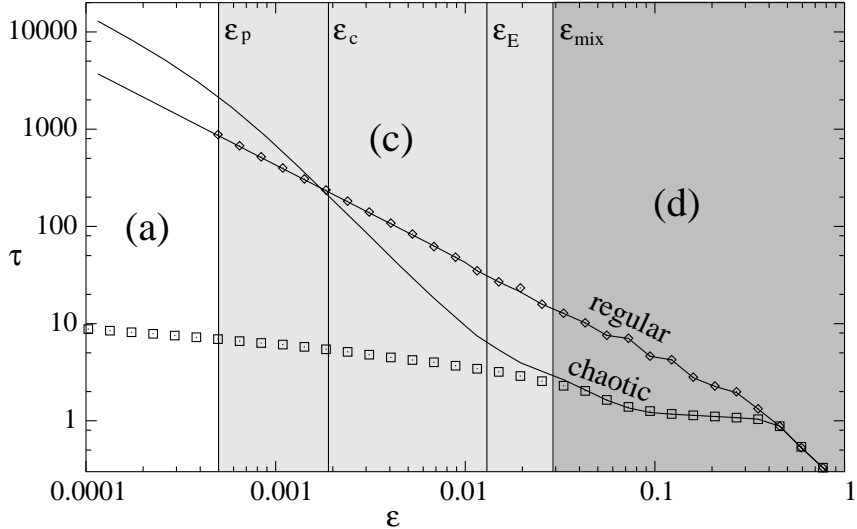}}
\caption{Numerically calculated decay time of the quantum and classical 
  fidelity for the double kicked top (\ref{eq:4KTdef}) in regular and chaotic
  regimes. The two solid lines show numerically calculated decay times of 
  quantum fidelity for chaotic and regular regimes. Symbols are decay times of 
  classical fidelity, diamonds for the regular and squares for the chaotic 
  regime. Vertical lines show the position of perturbation borders 
  (\ref{eq:borders}). The shading and the letters (a),(c) and (d) correspond 
  to the regimes described in Fig.~\ref{fig:delta1}. The Zeno regime 
  corresponds to very short times $\tau<1$ (i.e. strong perturbations 
  $\eps>0.4$).}
\label{fig:2ktopd}
\end{figure}

Let us now demonstrate the above regimes by a numerical example 
(Fig.~\ref{fig:2ktopd}). As already noted, for a one dimensional system 
($d=1$), the `surprising' behavior of the regular decay time being smaller than 
the mixing one, $\tau_{\rm r} < \tau_{\rm m}$, is for coherent initial states 
possible only around the border $\eps_{\rm p}$ (unless $\sigma_{\rm cl}$ is 
very small) where the exponential decay in the mixing regime goes over to a 
Gaussian decay due to finite ${\mathcal N}$. However, for more than one degree 
of freedom, such behavior is generally possible well above the finite size 
perturbative border $\eps_{\rm p}$. To have general situation we shall therefore 
use double kicked top ($d=2$). The propagator is given in 
Appendix~\ref{sec:ktop}, Eq.~(\ref{eq:4KTdef}). Depending on the parameters the 
corresponding classical system is chaotic or regular. As initial state we take 
a product of spin coherent states (\ref{eq:SU2coh}) 
$\ket{\vartheta,\varphi} \otimes \ket{\vartheta,\varphi}$ with 
$(\vartheta,\varphi)=(\pi/\sqrt{3},\pi/\sqrt{2})$ and $S=100$. We numerically 
calculated the dependence of the decay time $\tau$ (when fidelity reached 
$1/e \approx 0.37$) on the perturbation strength $\eps$ for chaotic and regular 
cases. We then compared these numerical data with theoretical predictions. 
Using classically calculated $\sigma=0.058$ one gets chaotic decay time 
$\tau_{\rm m}$ (\ref{eq:Fnmixing}) $\tau_{\rm m}=\frac{8.6}{\eps^2 S^2}$. 
To determine the decay time of quantum fidelity for regular situation, 
Eq.~(\ref{eq:Fnregcoh}), one needs the coefficient
$\sqrt{\bar{v}'\Lambda^{-1}\bar{v}'}$. We determined it by fitting the 
dependence of $\tau_{\rm r}$, such that 
$\tau_{\rm r}=\frac{4.5}{\eps\sqrt{S}}$, which gives 
$\sqrt{\bar{v}'\Lambda^{-1}\bar{v}'}=0.31$.
Note that the coefficient $\sqrt{\bar{v}'\Lambda^{-1}\bar{v}'}$ has been 
obtained by numerical fitting only for convenience. In principle it could be obtained from classical dynamics, but we 
would again have to resort to numerical calculations as the regular system is not
completely integrable but is rather in a mixed KAM-like regime. The values of $\sigma$ and 
$\sqrt{\bar{v}'\Lambda^{-1}\bar{v}'}$ can then 
be used to calculate various perturbation borders as discussed before. 
For our numerical values we get (\ref{eq:dp},\ref{eq:deltacon},\ref{eq:de},\ref{eq:dmix})
\begin{equation}
\eps_{\rm p}=0.0005,\quad \eps_{\rm c}=0.0019,\quad \eps_{\rm E}=0.013,\quad 
\eps_{\rm mix}=0.029.
\label{eq:borders}
\end{equation}
In addition to the decay time of quantum fidelity we also numerically computed 
the decay time of classical fidelity. All these data are shown in 
Fig.~\ref{fig:2ktopd}.  We can see that in the regular regime the
quantum and the classical fidelity agree in the whole range of
$\eps$. In the chaotic regime things are a bit more complicated. By decreasing
perturbation from $\eps=1$ we are at first in regime of very strong
perturbation where the fidelity decay happens faster than any
dynamical scale and it does not depend on whether we look at chaotic
or regular system or quantum or classical fidelity. There the fidelity decays
within the Zeno times-scale, see Section~\ref{sec:zeno}. For smaller $\eps$ 
the regular and chaotic decays start to differ. In chaotic situation the 
quantum and classical fidelity still agree. This correspondence breaks down
around $\eps_{\rm mix}$ where the quantum fidelity starts to follow
the theoretical $\tau_{\rm m}$, while the classical fidelity decay is
$\tau_{\rm clas}=\log{(0.25/\eps)}/\lambda$, with $0.25$ being a
fitting parameter (depending on the width of the initial packet) and
$\lambda=0.89$ is the Lyapunov exponent, read from
Fig.~\ref{fig:2ktop1}. For an explanation of this classical
decay see Section~\ref{C}. Incidentally, in our chaotic system the 
classical mixing time is very short, $t_{\rm mix}\sim 1$, and we see that the
correspondence breaks down already slightly before $\eps_{\rm E}$. The
quantum fidelity decay time $\tau_{\rm m}$ is valid until a
perturbative border $\eps_{\rm p}$ is reached, when finite Hilbert
space dimension effects become important and the decay times become equal to 
$\tau_{\rm p}$. Note that for $\eps<\eps_{\rm c}$ we indeed have faster 
fidelity decay for {\em chaotic} than for {\em regular} dynamics.

\subsection{Increasing chaoticity}

\label{sec:trans}
\begin{figure}[htb!]
\centerline{\includegraphics[width=95mm]{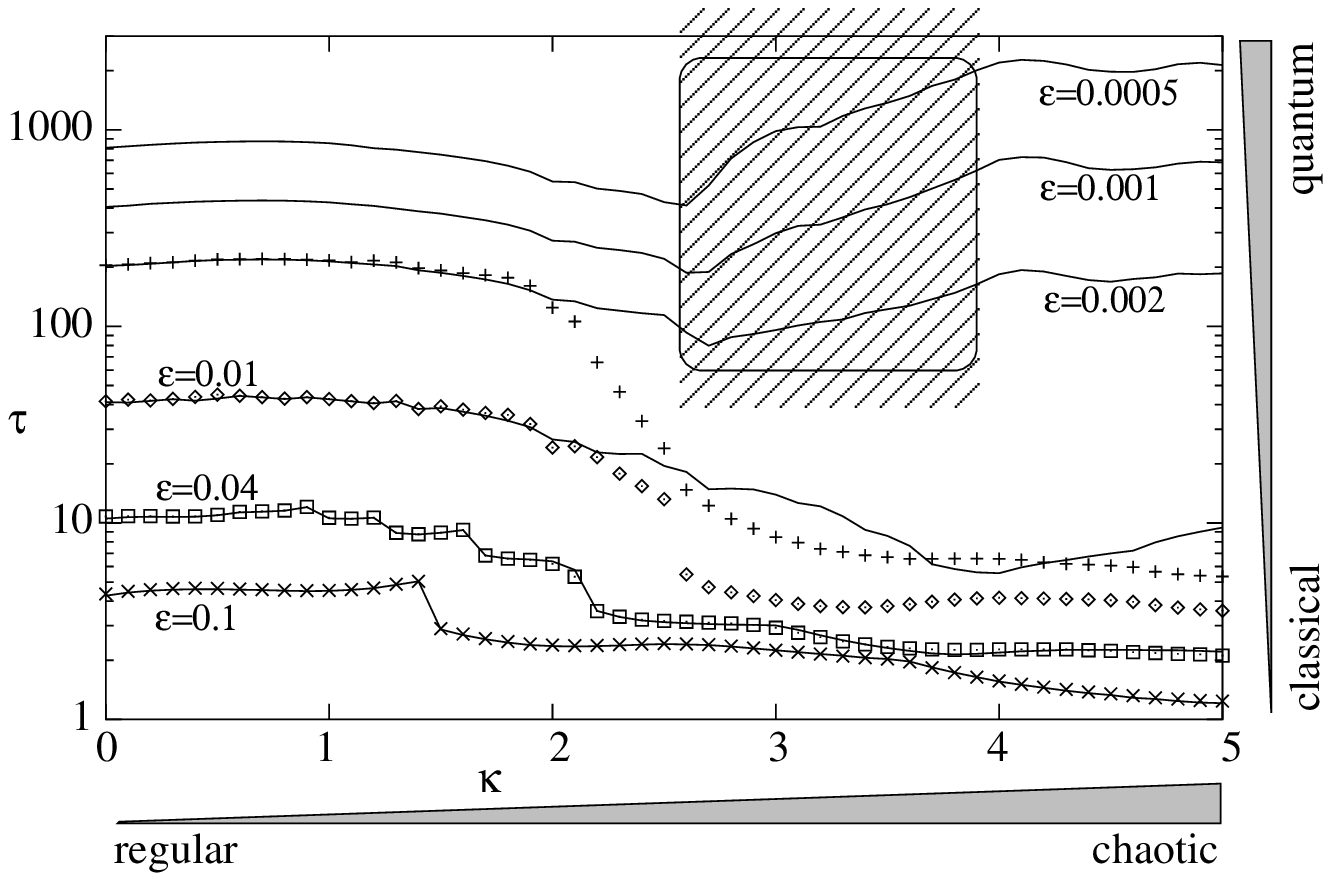}\includegraphics[width=67mm]{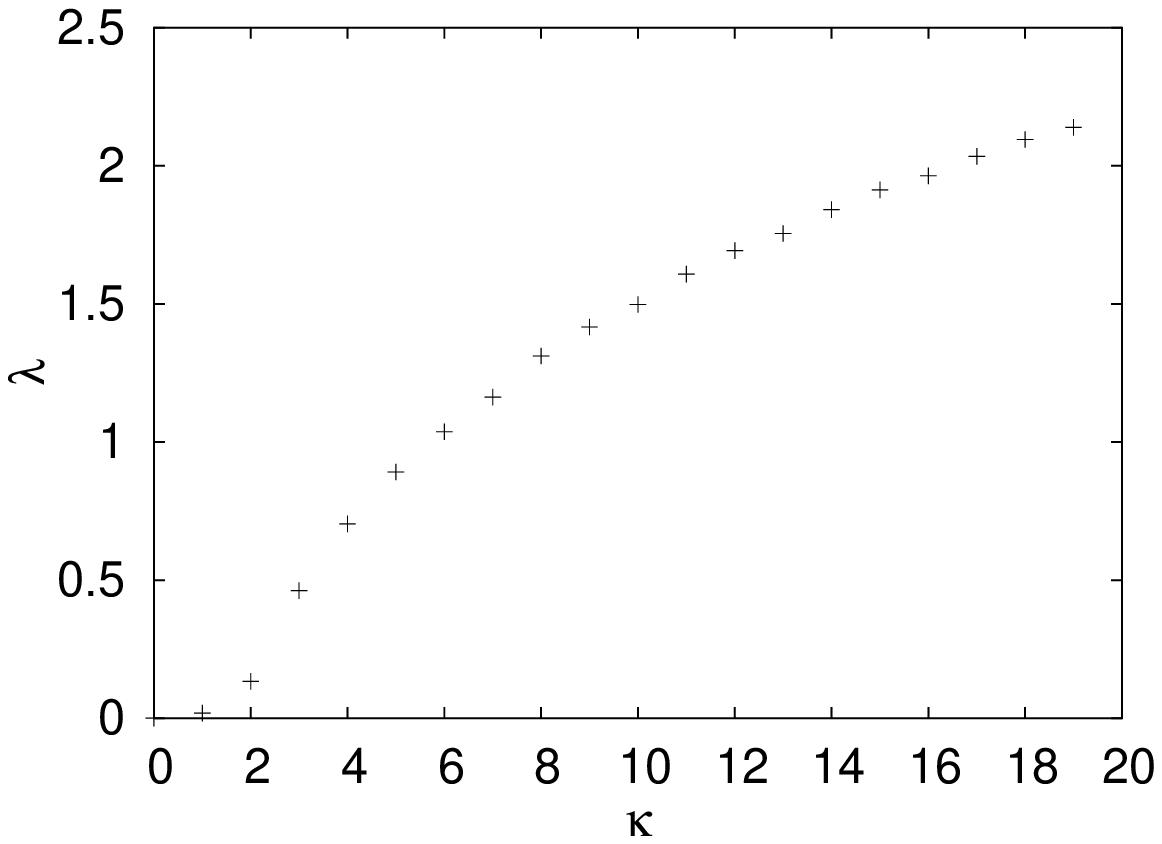}}
\caption{Numerically calculated dependence of quantum (solid lines) and 
  classical (symbols) fidelity decay times on the parameter $\kappa$ for the 
  double kicked top (\ref{eq:4KTdef}). Different curves are for different 
  perturbation strengths $\eps$. By increasing $\kappa$ the classical dynamics 
  goes from regular to chaotic, see also the right panel showing the dependence 
  of the largest Lyapunov exponent on $\kappa$. By decreasing $\eps$, on the 
  other hand, we go from the regime of quantum-classical correspondence for 
  $\eps>\eps_{\rm E}$, towards a genuinely quantum regime in which for chaotic 
  dynamics we can {\em increase} the decay time by {\em increasing} 
  chaoticity -- shaded region for the three smallest $\eps$.}
\label{fig:2ktop1}
\end{figure}

In Fig.~\ref{fig:2ktopd} we saw there is a range of perturbations for which 
quantum fidelity decay is faster for regular than for chaotic situation. 
Another interesting aspect of our correlation function formalism can also be 
seen for chaotic dynamics alone. Because the decay rate of the fidelity in a 
chaotic situation is proportional to the integral of the correlation function 
$\sigma$, a stronger chaoticity will typically result in a {\em faster decay} 
of the correlation function $C(t)$ and therefore in smaller $\sigma$, resulting 
in slower fidelity decay. This means that {\em increasing} chaoticity (of the 
classical system) will {\em increase} quantum fidelity, i.e. stabilize quantum 
dynamics. Of course, for this to be observable we have to be out of the regime 
of quantum-classical correspondence. This phenomenon is illustrated in 
Fig.~\ref{fig:2ktop1}, where we show similar decay times as in 
Fig.~\ref{fig:2ktopd}, i.e. the same system and initial condition, but this 
time depending on the parameter $\kappa$ of the double kicked 
top~(\ref{eq:4KTdef}). The parameter $\kappa$ controls the chaoticity of the 
classical dynamics, i.e. at $\kappa=1$ we are in the quasi-regular regime and 
for larger $\kappa$ we get into the chaotic regime. This can also be seen from 
the dependence of the Lyapunov exponent on $\kappa$ in the right panel of the 
figure. Data is shown for six different perturbation strengths $\eps$. 
We can see that in the regular regime ($\kappa < 2$) the classical fidelity 
agrees with the quantum one regardless of $\eps$. In the chaotic regime though, 
the agreement is present only for the two largest $\eps$ shown, where we have 
$\eps>\eps_{\rm mix}$ (\ref{eq:borders}). For $\eps<\eps_{\rm c}$ and for 
chaotic dynamics (three smallest $\eps$) we get into the non-intuitive regime 
(shaded region in Fig.~\ref{fig:2ktop1}) where the quantum fidelity will 
increase if we increase chaoticity. Note that this growth of the decay time 
stops at around $\kappa \sim 4$ because the classical mixing time 
$t_{\rm mix}$ gets so small that the transport coefficient is given by its time 
independent first term $\sigma=C(0)/2$ alone and so $\sigma$ cannot be 
decreased any further. From the figure one can also see that the decay 
time in the transition region, where the corresponding classical system has a 
mixed phase space, does not change monotonically with $\kappa$. Such a behavior
is system specific; for some additional results for the kicked top, see 
Ref.~\cite{Haug:05}. 

At the end let us make a brief comparison of decay times of purity for coherent 
initial states. In chaotic situation purity decays in the same way as fidelity, 
$\tau \sim \hbar^2/\eps^2$, so there is nothing new. 
On the other hand for regular dynamics purity decays on an $\hbar$ 
independent time scale, $\tau \sim 1/\eps$, which is therefore for 
small $\hbar$ much larger than for chaotic dynamics. 
Still, because of different dependence on perturbation strength $\eps$ 
({\em i.e.} coupling) one can have situation where purity decay in regular system 
is faster than in chaotic one. Comparing time scales of decays 
(\ref{eq:Imain},\ref{eq:three}) within the range of linear response
we see that in order to have such a situation we must have
\begin{equation}
\eps < \eps^{\rm I}_{\rm c}=\hbar^2 \sqrt{{\rm tr\,} u}/(4\sigma_{\rm cl})\propto \hbar^2,
\label{eq:ecI}
\end{equation}
where $u$ is $\hbar$-inedependent matrix of second derivatives used 
in (\ref{eq:Imain}).

\subsection{Echo measures in terms of Wigner functions}

\label{sec:wigner}
In Section~\ref{Q2S} we have derived a semiclassical expression for fidelity 
in terms of the Wigner function of the initial state (\ref{eq:fDP}). That was 
just an approximation. On the other hand, one can also express fidelity in 
terms of two Wigner function exactly. Wigner functions, or more generally Weyl 
symbols for quantum operators, have a nice property that the trace of a product 
of two operators is equal to the overlap integral of two corresponding Weyl 
symbols, $\tr{AB}=\int{W_{\rm A} W_{\rm B} {\rm d}\Omega}$, where $W_{\rm A,B}$ 
are the corresponding Weyl symbols and the integral is over the whole phase 
space.  Here we will need a special case of this equality for density 
matrices, 
\begin{equation}
\tr{(\rho_{\rm A} \rho_{\rm B})}=\int{\! W_{\rho_{\rm A}}W_{\rho_{\rm B}} {\rm d}\Omega},
\end{equation}
where $\rho_{\rm A,B}$ are now two density matrices and $W_{\rho_{\rm A,B}}$ 
the corresponding Wigner functions. For more information about Wigner functions 
for systems described 
by a canonical variables $[q,p]=\ii \hbar$ see, e.g. book~\cite{Schleich}, 
whereas for the definition of Wigner function for spin systems ({\it e.g.} our kicked top models) 
see Ref.~\cite{Agarwal:81}. Remembering that 
the fidelity can be written as a trace of the product of the initial density matrix $\rho(0)$ 
and density matrix after an echo, $\rho^{\rm M}(t)=M_\eps(t) \rho(0) M_\eps^\dagger(t)$, 
or equivalently, as a trace of two forward propagated density matrices 
$\rho^{\eps}(t)=U_\eps(t)\rho(0)U_\eps^\dagger(t)$ and 
$\rho^{0}(t)=U_0(t)\rho(0)U_0^\dagger(t)$, we have
\begin{equation}
F(t)=\tr{\rho(0)\rho^{\rm M}(t)}=\int{W_{\rho(0)} W_{\rho^{\rm M}(t)} {\rm d}\Omega}=
\int{W_{\rho^0(t)} W_{\rho^\eps(t)} {\rm d}\Omega}.
\label{eq:Fwig}
\end{equation}
Similarly, one can write the reduced fidelity (\ref{eq:Frdef}) and echo purity (\ref{eq:Fpdef}) in 
terms of Wigner functions of the reduced density matrices $\rho_{\rm c}=\tre{\rho}$,
\begin{equation}
\Fr=\int{W_{\rho_{\rm c}(0)} W_{\rho^{\rm M}_{\rm c}(t)} {\rm d}\Omega_{\rm c}},
\label{eq:Frwig}
\end{equation}
and echo purity
\begin{equation}
\Fp=\int{\left[ W_{\rho^{\rm M}_{\rm c}(t)} \right]^2  {\rm d}\Omega_{\rm c}}.
\label{eq:Fpwig}
\end{equation}
All these expressions are exact. The classical quantity analogous to Wigner function is just 
the classical density in phase space, therefore these expressions are handy for comparison with 
classical quantities. But note that the Wigner function is not necessarily positive
whereas the classical density is (the positivity of Wigner function is a necessary 
but not sufficient condition for the classicality of quantum state). In the next 
subsection we are going to illustrate the decay of fidelity and echo purity in terms of 
Wigner functions. Specifically, we shall compare chaotic and regular dynamics to see how is 
different decay of fidelity reflected in the corresponding Wigner functions.

\subsubsection{Illustration with Wigner functions}

\label{eq:illustwig}
\begin{figure}[htb!]
\centerline{\includegraphics[width=0.75\textwidth]{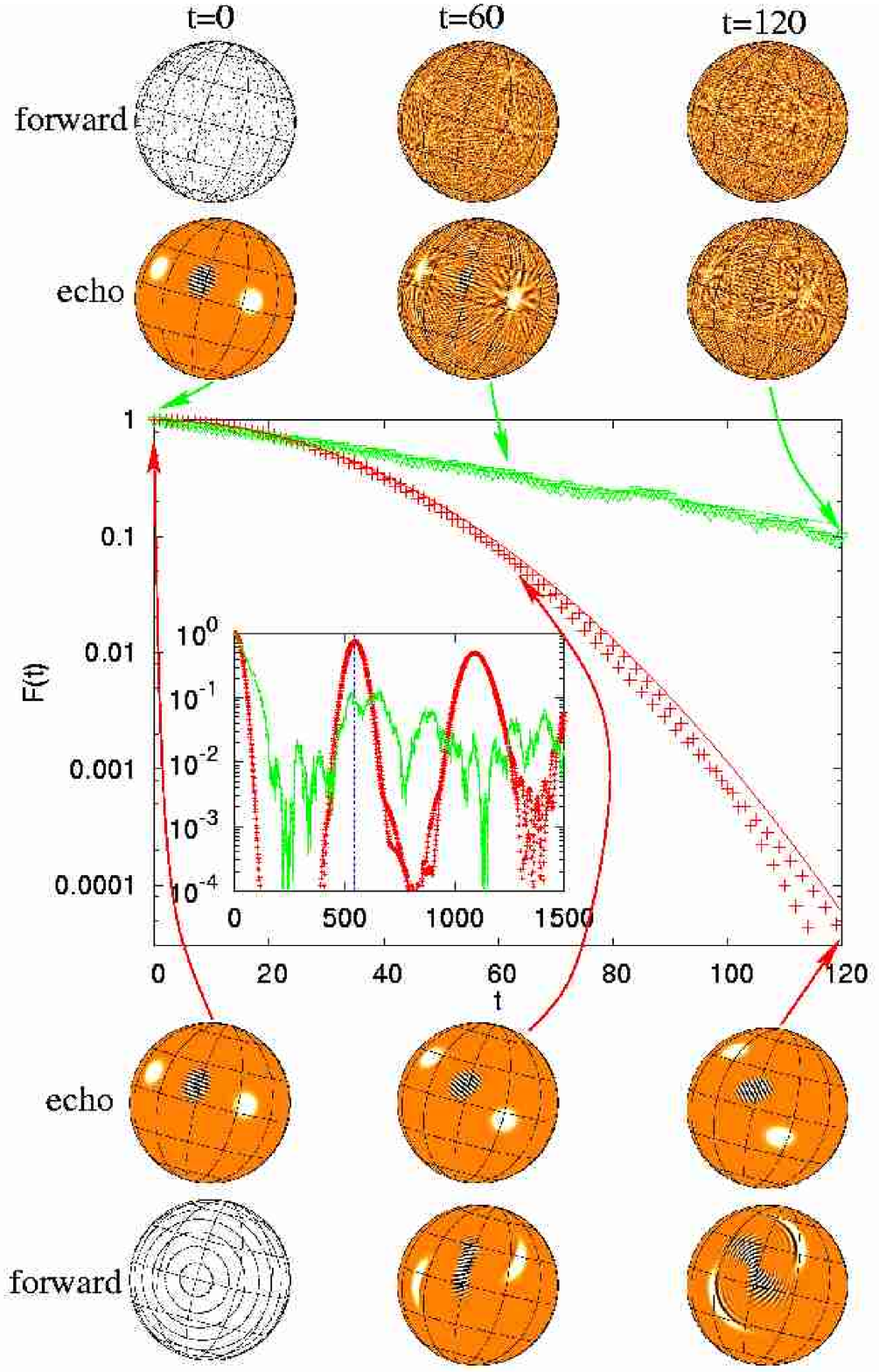}
%\centerline{\includegraphics[width=0.75\textwidth]{parad_wig.eps}
\includegraphics[angle=90,width=5mm]{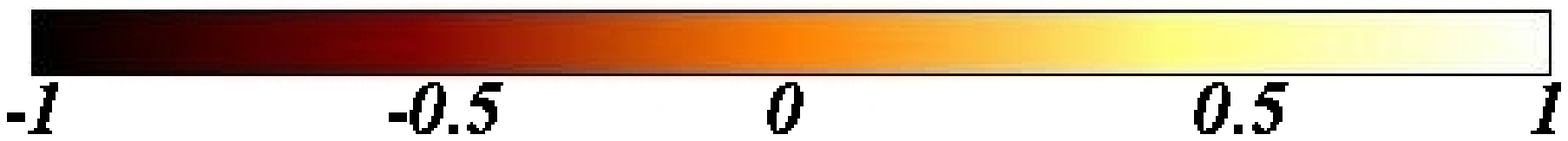}}
\caption{Fidelity decay for chaotic (top curve and pictures) and regular (bottom curve and pictures) 
kicked top (\ref{eq:KTdef}). Initial conditions and the perturbation are the same in both cases 
(see text for details). Wigner functions after forward and echo evolution are shown.}
\label{fig:paradwig}
\end{figure}

Again we will take our standard kicked top (\ref{eq:KTdef}) with parameters $\gamma=\pi/2$ and 
$\alpha=30$ for the chaotic case and $\alpha=0.1$ for the regular one. The spin size is $S=100$ and we 
choose a coherent initial state  at $(\vartheta^*,\varphi^*)=\pi(1/\sqrt{3},1/\sqrt{2})$. The perturbation 
strength is $\eps=1.5\cdot 10^{-2}$. In Fig.~\ref{fig:paradwig} we show the fidelity decay for both cases 
(regular case is the same as in Fig.~\ref{fig:ch4regcoh}) and an illustration of states in 
terms of Wigner functions. For the chaotic (triangles in the fidelity plot and pictures above) 
and the regular case (pluses in the fidelity plot and pictures below) we show two series of 
Wigner functions at times $t=0,60,120$: the Wigner function after the unperturbed forward
evolution (row labeled ``forward'') and the Wigner function after the echo 
(row labeled ``echo''). In the first frame we also show the
structure of the classical phase space being either regular with KAM tori or chaotic. 
In the inset the data for the fidelity decay is shown on a longer time scale and the vertical
line shows the theoretical position of the beating time $t_{\rm b}$ (\ref{eq:tb}). 
In terms of Wigner functions the fidelity can be visualized as 
the overlap between the echo Wigner function and the initial Wigner
function. The initial Wigner function shows two maxima (white regions of high value) 
because we have projected the initial coherent state to the invariant OE subspace, 
resulting in a certain symmetry of the resulting Wigner function. For chaotic dynamics the
forward Wigner function develops negative values around the Ehrenfest
time after which the quantum-classical correspondence is lost. For
regular dynamics this correspondence persist much longer, namely until
the Ehrenfest time of regular dynamics $\sim \hbar^{-1/2}$ after which the
initial wave packet of size $\sim \hbar^{1/2}$ spreads over the phase
space. For a detailed study of Wigner functions in chaotic systems
see Refs.~\cite{Horvat:03,Lombardi:93} and references therein.
The echo Wigner function for regular dynamics moves ballistically from the initial position, 
causing the Gaussian decay of fidelity. We also see that for regular dynamics the echo Wigner 
function does not necessarily have negative values even if they occur in the forward 
Wigner function. In our case quantum fidelity agrees with the classical one for 
regular dynamics. In a chaotic case on the other hand, the echo image stays at the initial 
position and "diffusively" decays in amplitude, causing the fidelity to decay slower than 
in the regular case. Classical fidelity follows quantum fidelity in the chaotic regime only up 
to Ehrenfest time. 

\begin{figure}[htb!]
\centerline{\includegraphics[width=0.75\textwidth]{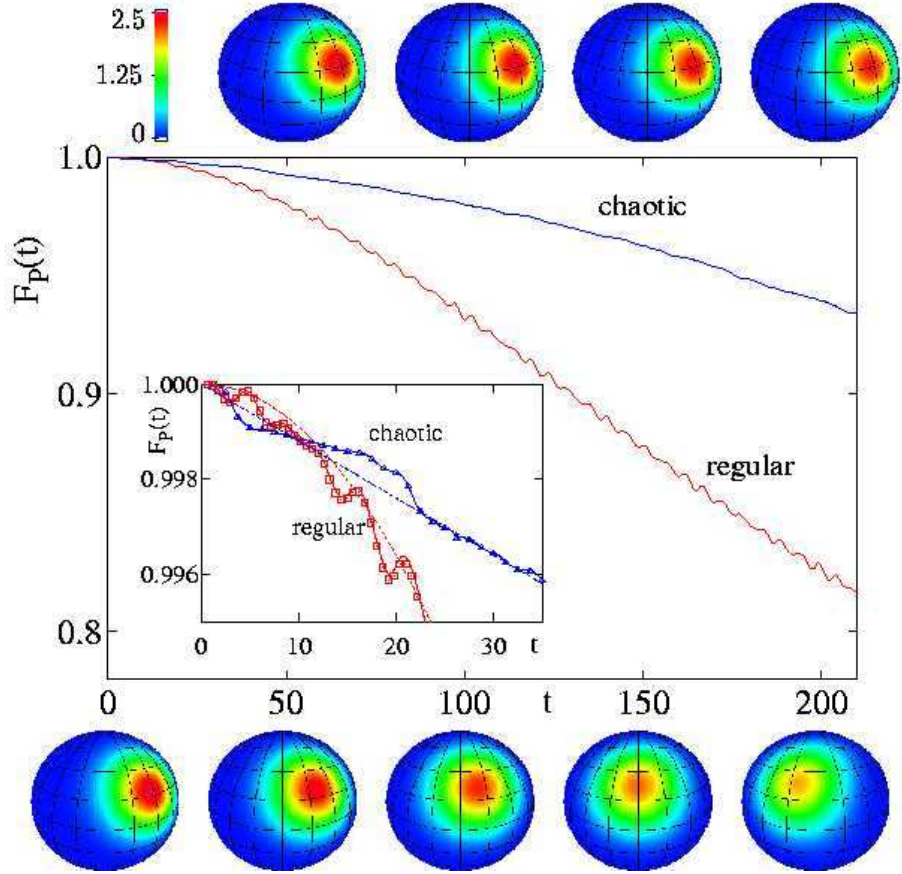}}
%\centerline{\includegraphics[width=0.75\textwidth]{bigwig.eps}}
\caption{Echo purity decay in the Jaynes-Cummings model for regular and 
chaotic dynamics [from~\cite{Prosen:03evol}]. In the inset the same data is shown for smaller times. 
Full curves are numerics while dashed ones are theoretical predictions. 
For small $\eps=0.005$ and large $\hbar=1/4$ used here, echo purity decay can be faster 
in regular than in chaotic system. Top set of pictures shows the square of the Wigner function of 
the reduced density matrix $\rho_{\rm c}^{\rm M}(t)$ for chaotic dynamics and the bottom one for 
regular dynamics. Its integral gives echo purity.}
\label{fig:bigwig}
\end{figure}

The previous figure demonstrated the possibility of faster fidelity decay in regular than in 
chaotic systems, provided inequality (\ref{eq:deltacon}) is fulfilled. Let us now show that 
the same can happen also for echo purity. From inequality (\ref{eq:ecI}) we see, that 
either $\hbar$ has to be large ({\em i.e.} strong quantum regime) or the perturbation has 
to be weak. As a numerical model we use the Jaynes-Cummings model~\cite{Jaynes:63,Tavis:68}, 
describing a system of harmonic oscillator coupled to a spin. As opposed to kicked top systems,
the Jaynes-Cummings model has a time independent Hamiltonian, and time is a continuous variable. 
Regular (integrable) dynamics is 
obtained if only the co-rotating coupling term is present, while predominantly chaotic dynamics takes 
place in the presence of both co- and counter-rotating terms, see Ref.~\cite{Prosen:03corr} 
for details about numerical parameters. In Fig.~\ref{fig:bigwig} we show the decay of echo purity for
relatively small perturbation $\eps=0.005$ in spin energy, {\it i.e.} perturbation is the 
so-called detuning, and large $\hbar=1/4$, so that the condition 
$\eps < \eps_{\rm c}^{\rm I}$ (\ref{eq:ecI}) is satisfied. In addition, we show a set 
of pictures for times $t=50,100,150$ and $200$ representing square of the Wigner function 
of the reduced density matrix $\rho^{\rm M}_{\rm c}(t)$ after an echo. 
The integral of this quantity directly gives 
echo purity (\ref{eq:Fpwig}). The central subsystem is chosen to be the spin degree of freedom, 
so the phase space is the surface of a sphere. The top set of pictures is for chaotic and 
the bottom one for regular dynamics. From the Wigner function plots one can see that 
the blob of maximal value of 
the Wigner function changes its position with time for regular dynamics while it stays at the same place 
for chaotic dynamics. More importantly, the height of this peak decays faster for regular dynamics than 
for chaotic one, reflected in a faster decay of echo purity. 
We should not forget that this regime of faster decay for regular dynamics is reached only for 
sufficiently large $\hbar$ (or small $\eps$), see Eq.~(\ref{eq:ecI}), and that in general, 
echo purity in regular systems will decay very slowly due to its $\hbar$ independence (\ref{eq:Imain}).

\subsubsection{Wigner function approach}
\label{Q2W}

We have seen that the Wigner function often gives an alternate
and sometimes very physical view of echo evolution, particularly
if we start with a coherent state as an initial state. Note
that for a random state the Wigner function is essentially a
mess and does not present, to our knowledge, any useful insight.
In this context we wish to discuss two papers that reach similar results.
In Ref.~\cite{Cucchietti:03I} the authors develop an analogy to the formalism   
used in the treatment of decoherence, despite of the fact, that
they do not consider coupling to some environment. For this
purpose the authors consider an ensemble of perturbations. Yet
they proceed quite differently than in the case of the RMT
treatment offered in Section~\ref{R}, in that they do not consider
unitary evolution and average over the resulting fidelity amplitudes
or fidelities. The Wigner function is the density
operator in the phase space representation of quantum mechanics.
In this picture we obtain a clear distinction between the evolution
of the diminishing coherent part of the Wigner function and the one of
the emerging rapidly oscillating part.
The fidelity is then given as the trace of the product of the
average of the perturbed density matrices $\la\rho(t)\ra$ with the 
$\rho_0(t)$ evolving according to the unperturbed evolution
\begin{equation}
  \la F_z(t)\ra =  \tr(\la\rho(t)\ra\; \rho_0(t)\, ) =
  \int\d \ve{p}\d \ve{q}\;  \la W(\ve{q},\ve{p},t)\ra \; W(\ve{q},\ve{p},t),
\end{equation}
which in phase space representation can be written as an integral
over the Wigner function $W(\ve{q},\ve{p},t)$ and the average Wigner 
function $\la W(\ve{q},\ve{p},t)\ra$, averaged over an ensemble of
perturbations.
        
The use of an average density matrix approaches us to concepts
usually used in decoherence and induces the authors of this
paper to introduce an ensemble of perturbations solely in terms
of the time dependence where they assume white noise. This allows
them to propose a master equation which can be solved. 
It is important to note, that the fidelity decay for each element
of the ensemble is dominated by the decay of correlations in
time through the white noise.
        
As an initial condition a coherent state is used, and they conclude
that with perturbation strength above some threshold, they recover
the exponential decay by limiting the integration to the area
where the Wigner function varies slowly. The rate of this decay is 
perturbation independent and is determined by the (maximal) Lyapunov 
exponent of the underlying classical mechanics. An integration
over the remaining area, where we have the well known fast oscillations
developing rapidly, will produce a second term, that yields
the perturbation dependent Fermi golden rule decay. The Gaussian
perturbative regime is never reached in this analysis, 
because strong perturbations are 
assumed from the outset.
        
While the approximations in this paper are uncontrolled and several
assumptions are very special, it provides very important qualitative
insight: First the somewhat surprising fact that the Lyapunov
decay is independent of the perturbation strength becomes intuitively
clear, and second a close relation to the dynamics of decoherence
is established.  
A similar picture was used in Ref.~\cite{Cucchietti:04} to describe
the transition between the two regimes in more detail.

Regarding fast oscillations of Wigner functions for sufficiently extended states, it has been argued in Refs.~\cite{Zur01,Karkuszewski:02} that such rapid oscillations enhance sensitivity of quantum systems to perturbations. When an initial state $\psi$ has been prepared by a ``preparation'' evolution starting from a localized wave packed $\psi_0$, $\psi=\exp{(-\ii H_0 t)}\psi_0$, an enhanced sensitivity has been observed when increasing the preparation time $t$~\cite{Karkuszewski:02}. This has been explained as being due to the small structure in Wigner function of the initial state $\psi$. The observed dependence though is just due to the dependence of fidelity decay on the initial state and can be explained by classical Lyapunov exponents~\cite{Jacquod:02}. Therefore, it can not depend on quantum intereference effects caused by the rapid oscillations in Wigner function, see also discussion in Ref.~\cite{Srednicki:01}.

\section{Application to Quantum Information}

\label{sec:AppQI}
Quantum information theory is a relatively recent endeavor, for a review
see Refs.~\cite{Ekert:96,Steane:98,BEZ:00,Nielsen:01}. Its beginnings go back to the 1980's
and in recent years theoretical concepts have been demonstrated in experiments. 
While quantum cryptography, a method of provably secure communication, is already 
commercially available, quantum computation is still limited to small laboratory experiments.

Errors and error correction are among the main concerns in quantum information. 
In the present chapter we are going to discuss, how the techniques and results 
developed for fidelity decay can be used to design quantum operations that 
are less prone to errors. We shall divide the whole exposition into two parts. In the 
first part we shall show how a straightforward application of fidelity theory can help 
us understand the dependence of errors on various parameters. In the second part we are 
going to describe how one can actually decrease errors.   

\subsection{Fidelity studies}

\label{sec:fidstudies}
We present a partial overview of results obtained by 
numerically  studying fidelity in quantum computation for various kinds of 
errors. Note that due to the
extensiveness of the literature on the subject the list is by no means 
exhaustive. 

One of the first studies of the sensitivity to errors of quantum algorithms 
appeared shortly after Shor's discovery of the factoring algorithm. 
In Ref.~\cite{Miquel:96} the influence of Markovian errors ({\it e.g.} due 
to the coupling with the environment) on the workings of Shor's algorithm has been studied. 
At random instances of time a random qubit was chosen to decay with a
certain probability to its zero state. In subsequent work~\cite{Miquel:97} 
the stability of Shor's algorithm running on an ion trap quantum computer is 
analyzed in the presence of random phase drift errors due to pulse length 
inaccuracies. Fidelity is found to depend exponentially on the perturbation 
strength $\eps^2$ and time $t$ as one expects for uncorrelated errors, see 
Section~\ref{sec:noisypert} or Eq.~(\ref{eq:Fnoise}). 
In a series of papers~\cite{Banacloche:98,Banacloche:99} 
the influence of unwanted intrinsic qubit-qubit couplings has been studied. 
The dependence of fidelity on the computation time has been found to be 
Gaussian. This agrees with a Gaussian fidelity decay for perturbations with 
non-decaying correlation function. Note that such ``regular'' ({\it i.e.} Gaussian) 
decay of fidelity is not unexpected as the individual one or two-qubit gates, 
due to their simplicity, can decrease correlation function only in a very 
limited way, thus causing a slow decay of correlations and consequently a fast 
Gaussian-like decay of fidelity.

A number of numerical studies has been done by the groups of 
Shepelyansky and Casati. In Ref.~\cite{Georgeot:00,Georgeot:00b} the question was addressed, how inter-qubit 
couplings change the eigenstate structure of a quantum computer, see 
also Refs.~\cite{Benenti:02qc,Benenti:02qc1,Benenti:03qc}. Above some threshold coupling chaos 
sets in and the authors advocated that quantum computation fails. This 
result looks like the opposite to our conclusions, where more chaos can mean 
more stability. One should bear in mind though, that there is no 
straightforward connection between eigenvector statistics and the actual 
dynamical fidelity which is a natural quantity measuring the stability. In~\cite{Georgeot:00b} as well as in~\cite{Flambaum:00} survival probability of register states (i.e., fidelity of an unperturbed eigenstate) has been studied. The decay rate has been found to be given by the local density of states. That the eigenstate mixing is not the most relevant quantity is confirmed also by 
data from nuclear experiments~\cite{Flores:05}. In Ref.~\cite{Benenti:01} the stability of 
a quantum algorithm simulating saw-tooth map has been numerically considered for static 
and random errors, pointing out that static imperfection are more dangerous. Similar study has been performed in Ref.~\cite{Levi:03} for 
the quantum kicked rotator. For random errors the fidelity has been found to 
decay exponentially with $\eps^2$ and the number of gates. Such 
exponential decay can be shown to be general for uncorrelated (in time) 
errors~\cite{Prosen:01}, independent of the quantum algorithm, see also Eq.~(\ref{eq:Fnoise}). 
For static errors a faster Gaussian decay has been 
observed, which can again be understood by the linear response formalism. 
In Ref.~\cite{Terraneo:03} the influence of random and static imperfection has 
been studied, confirming that static imperfections can be more 
dangerous~\cite{Prosen:01,Benenti:01}. The influence of errors on the working of quantum computer has 
been studied also in Ref.~\cite{Frahm:04}. Static imperfections can be decreased by doing rotations between the gates, {\em i.e.}, introducing ``randomizing'' gates during the evolution~\cite{Prosen:01}. Similar idea is used in the so-called PAREC (Pauli Random Error Correction) method~\cite{Kern:05,Kern:05a}.

The group of G.~P.~Berman studied in detail the stability of the Ising quantum computer, 
see Refs.~\cite{Berman:01,Berman:02,Berman:02b,Celardo:03} and references therein. 
The Ising quantum computer consists of a series of spin $\frac{1}{2}$ particles placed in 
a magnetic field with a strong gradient. The magnetic gradient allows
selective addressing of individual qubits having different resonant 
frequencies. The nearest neighbor Ising coupling on 
the other hand enables the execution of two-qubit gates. All gates are 
performed by applying rectangular pulses of a circularly polarized 
electro-magnetic field lying in the plane 
perpendicular to the applied magnetic field. Different gates can be 
applied by choosing appropriate frequencies, phases and lengths of pulses. 
Perturbation theory has been used to explain the dependence of fidelity on 
errors in various physical parameters of the Ising quantum computer.

In Ref.~\cite{rossini:04b} the stability of a quantum algorithm simulating 
the quantum 
saw-tooth map is studied with respect to errors in the parameters of the 
map (classical errors) and with respect to uncorrelated random unitary errors 
of gates (quantum errors). As expected from the theory~\cite{Prosen:01}, the 
fidelity decay for uncorrelated errors is found to be exponential and 
independent of the dynamics of the classical map. Similar results have been obtained 
also for the decay of concurrence in the presence of uncorrelated unitary 
errors~\cite{Rossini:04}. In Ref.~\cite{Bettelli:04} 
noisy unitary errors were also considered theoretically. A quadratic bound 
(in the number of applied errors) on the error is obtained, the result 
being a trivial consequence of the linear response expression for fidelity, 
Eq.~(\ref{eq:fidlr}). Unitary noise in Grover's algorithm has been studied 
in Ref.~\cite{Shapira:03}.

Summarizing, the plethora of numerical investigations has shown that the 
dependence of errors on various parameters like {\it e.g.} perturbation strength 
$\eps$, the number of gates etc., can be readily understood within 
the linear response formalism of fidelity decay. Furthermore, as one is predominantly 
interested in large values of fidelity, linear response is all that is needed.

\subsection{Decreasing the errors}

\label{sec:decerror}
Once we have identified the errors and their dependence on various 
parameters, we can try to decrease or even better to entirely eliminate 
such errors. We shall more closely discuss two possibilities: 
(i) We can change 
the algorithm, {\it i.e.} apply a different set of quantum gates, so that the 
resulting sequence of gates is more resistant against a given kind of errors; 
(ii) If one is able to perform certain unitary operations with sufficiently 
small errors, 
one can do the actual quantum calculation in a transformed frame, called 
the {\em logical} frame, in which the influence of errors is suppressed. Such 
procedure is known as {\em dynamical decoupling}~\cite{Viola:99}. 

There are other approaches to reduce errors like 
{\em quantum error correction}, {\em decoherence-free subspaces} and 
procedures using the {\em quantum Zeno effect}. In the limit of ideal and 
infinitely short measurements quantum Zeno procedures have certain 
similarities~\cite{Facchi:04} to the so-called ``bang-bang''~\cite{Viola:98} 
limit of dynamical decoupling. One should be careful though as dynamical 
decoupling consists of unitary manipulations of dynamics while quantum 
Zeno effect involves quantum measurements. Fidelity theory could help to 
understand errors present in these methods in the non-ideal limit, for 
some studies see Refs.~\cite{Viola:05,Facchi:05}. Another possibility to combat 
errors is by {\em quantum error correction}~\cite{Shor:95,Calderbank:96,Steane:96,Steane:96a}. 
We won't discuss it in detail as we have to first be able to 
implement individual gates sufficiently well for the error correction 
to work. Also, quantum error correction can efficiently cure only errors 
that are correlated over just a few qubits. Its success relies on the fact 
that errors are local, so that information can be ``hidden'' in nonlocal 
states, whose change then serves to diagnose the error and correct them. The efficiency 
therefore sharply decreases for correlated multi-qubit errors. For study of 
error correcting codes and decoherence-free subspaces in the presence of errors see Ref.~\cite{Clemens:04,Silvestrov:01}. Using decoherence-free 
subspaces~\cite{Zanardi:97,Duan:98,Lidar:01a,Lidar:01b} one exploits 
certain symmetries of the coupling (like {\it e.g.} permutation symmetry) 
enabling a part of the Hilbert space to be free of decoherence for a 
given coupling.

In the next section we shall show an illustrative, though not very realistic example 
of the use of fidelity theory 
 to improve the stability of the Quantum Fourier transformation 
against a certain kind of perturbations.

\subsubsection{Modifying the algorithm}

\label{sec:modalgo}
We shall first briefly present the language of quantum computation which 
is slightly different than the so far used framework of continuous-time 
Hamiltonian dynamics. We shall focus on a standard quantum computation where the propagator is decomposed into quantum gates. There are other approaches to quantum computation, for instance, utilizing the ground state to perform computation, like in 
adiabatic quantum computation~\cite{Farhi:00,MIT:01} or using holonomies in degenerate subspaces, as in holonomic quantum computation~\cite{Zanardi:99}. For applications of the linear response approach to the stability of holonomic quantum computation see~\cite{Kuvshinov:03,Buividovich:06,Kuzmin:06}. 

A quantum computer is composed of $n$ elementary 
two-level quantum systems --- called {\em qubits}. The union of all
$n$ qubits is called a quantum register $\ket{\rm r}$. The size of the
Hilbert space $\mathcal{N}$ and therefore the number of different states of a
register grows with the number of qubits as $\mathcal{N}=2^n$. A quantum 
algorithm is a unitary transformation $U$ acting on register states. In 
the standard 
framework of quantum computation 
a quantum algorithm $U$ which acts on $\mathcal{N}$ dimensional space is 
decomposed into simpler units called {\em quantum gates}, $U_t$, where $t$ 
is a discrete time counting gates. Typically, quantum gates act on either 
a single qubit or on two-qubits. Ideal quantum computation with no errors 
then consists of a series of ideal gates:
\begin{equation}
U(T)=U_T\cdots U_2 U_1.
\label{eq:ch7U}
\end{equation} 
A quantum algorithm is called {\em efficient} if the number of 
needed elementary gates $T$ grows at most {\em polynomially} in 
$n = \log_2 \mathcal{N}$, and only in this case it can generally be 
expected to outperform the best classical (digital) algorithm. Decomposition of a 
given algorithm $U$ into individual gates $U_t$ is of course not unique. 
There are many decompositions with different $T$´s, resulting in the same algorithm $U=U(T)$. 
At present only few efficient quantum algorithms are known.

The above ideal situation without errors cannot be achieved in practice. 
Generally, at each time step there will be errors caused by perturbations 
({\it i.e.} unwanted evolution). These can result from the coupling of qubits with external 
degrees of freedom or from the non-perfect evolution of qubits 
within the real algorithm in contrast to the ideal one. We shall model both perturbations 
on a given gate $U_t$ by a unitary 
perturbation~\footnote{We treat only unitary errors here. Quantum 
computation is expected to require probability conservation. To treat 
non-unitary effects we would have to include ``environment'' into our 
picture.} of the form 
\begin{equation}
U^\eps_t=\exp{(-\ii \eps V(t) )}U_t.
\label{eq:pert}
\end{equation}
We set $\hbar=1$ and use the superscript $\eps$ to denote a
perturbed gate. We allow for different perturbations $V(t)$ at different
gates. The perturbed algorithm is a product of perturbed gates,
$U^\eps(T)=U^\eps_T\cdots U^\eps_1$. Please note that $U_j$ 
(subscript) is a single gate, that is a transformation acting between time 
$t=j$ and $t=j+1$, while we use $U(t)$ for a propagator from the beginning 
to time $t$, {\it i.e.} $U(t)=U_t\cdots U_1$. Fidelity will again serve as a
measure of stability so we have
\begin{equation}
F(T)=|\ave{M_\eps(T)}|^2,\qquad M_\eps(T)=U^{\dagger}(T) U^\eps(T).
\label{eq:deffid}
\end{equation}
Note that here the discrete total time $T$ is not a free parameter but is 
fixed by the number of gates, although we could in principle also look 
at the fidelity as it changes during the course of exectution of the algorithm. 

We have seen in Section~\ref{sec:timescales} that we can generally have higher 
fidelity in chaotic systems than in regular ones. For chaotic evolution fidelity 
decays linearly in time, while it decays quadratically for regular 
dynamics. Therefore, one might expect that an algorithm will be the more 
stable (have higher fidelity) against a given perturbation the 
more ``chaotic'' it is. A nontrivial important question then is, for which 
decomposition of the algorithm $U$ shall we have the highest fidelity? 
In the present 
section we give a concrete example of such optimization. The 
presentation closely follows Ref.~\cite{Prosen:01}.

As quantum computer will work adequately only if fidelity is extremely high, 
the linear response expansion of $F(T)$ is all we need. Writing the echo operator 
in the interaction picture we have,
\begin{equation}
M_\eps={\rm e}^{-\ii \eps \tilde{V}(T)}\cdots{\rm e}^{-\ii \eps 
\tilde{V}(2)}{\rm e}^{-\ii \eps \tilde{V}(1)},
\label{eq:prodfid}
\end{equation}
where $\tilde{V}(t)=U^\dagger(t)V(t) U(t)$ is the perturbation of the $t$-th
gate $V(t)$ propagated with the unperturbed gates $U(t)=U_t\cdots
U_1$. Be aware that $\tilde{V}(t)$ is time dependent for two reasons;
due to the use of the interaction picture (propagation with $U(t)$) and due to the 
explicit time dependence of the perturbation itself, {\it i.e.} the fact that different
perturbations occur at different gates. To assess the typical performance of 
a quantum algorithm independent of a particular initial state of the register,
we average over Hilbert space of initial states as 
$\ave{\bullet}=\frac{1}{\mathcal{N}}\tr{(\bullet)}$. Expanding 
fidelity we obtain to lowest order
\begin{equation}
F(T)=1-\eps^2 \sum_{t,t'=1}^T{C(t,t')}, \qquad C(t,t')=\frac{1}{\mathcal{N}} \tr{\lbrack \tilde{V}(t) \tilde{V}(t')\rbrack}.
\label{eq:Fcor}
\end{equation}
Two situation can now be distinguished depending on the time dependence 
of the perturbation $V(t)$ at different gates. 

The extreme case of time-dependent $V(t)$ is {\em uncorrelated noise}. 
This can result from  
coupling to a many body chaotic system, for which the matrix elements of 
$V(t)$ may be assumed to be 
{\em Gaussian random} variables, uncorrelated in time,
\begin{equation}
\ave{V_{jk}(t) V_{lm}(t')}_{\rm noise} = \frac{1}{\mathcal{N}}\delta_{jm}\delta_{kl}\delta_{tt'}.
\end{equation} 
Hence one finds $\ave{C(t,t')}_{\rm noise} = \delta_{t t'}$, where we
have averaged over noise. In fact the average of the product in
$M_\eps$ (\ref{eq:prodfid}) equals the product of the average and 
yields the noise-averaged fidelity to all orders
\begin{equation} 
\ave{F(T)}_{\rm noise} = \exp(-\eps^2 T),
\label{eq:Fnoise}
\end{equation} 
which is {\em independent} of the quantum algorithm $U(T)$. This result 
is completely general provided that the correlation time of the 
perturbation $V(t)$ is {\em smaller} than the duration of a single gate.

\begin{figure}[h!]
\centerline{\includegraphics[width=\figw\textwidth]{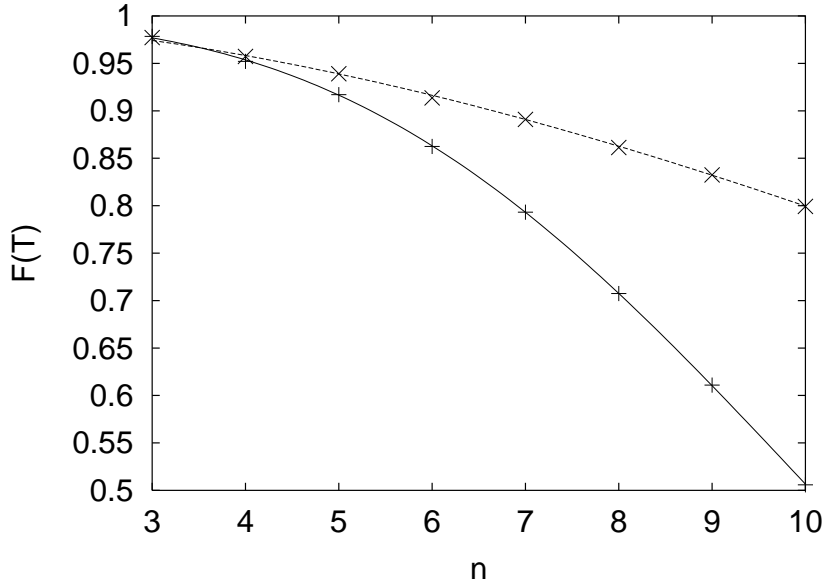}}
\caption{Dependence of the fidelity $F(T)$ on the number of 
qubits $n$ for the QFT (pluses) and the IQFT algorithms (crosses), for
fixed $\eps=0.04$ [from~\cite{Prosen:01}]. The full curve is 
$\exp{(-\eps^2 \{ 0.47 n^3 -0.76 n^2+2.90 n \})}$ and the dashed one 
$\exp{(-\eps^2 \{ 1.22 n^2+1.78 n\})}$. An "improved" IQFT algorithm, 
being more random, has a higher fidelity, which furthermore decreases slower with 
$n$ than for the original QFT algorithm. All is for a static random matrix perturbation.}
\label{fig:ft}
\end{figure}
On the other hand, for a {\em static} perturbation, $V(t)\equiv V$, one may 
expect {\em slower} correlation decay, depending on the ``regularity'' of the 
evolution operator $U(T)$, and hence faster decay of fidelity. 
Note that in a physical situation, where the perturbation is expected to 
be a combination $V(t) = V_{\rm static} + V_{\rm noise}(t)$, the 
fidelity drop due to a static component
is expected to {\em dominate} long-time quantum computation $T\to\infty$ 
(i.e. large numbers of qubits $n$) as compared to the
noise component, as soon as the quantum algorithm exhibits 
long time correlations of the operator $V_{\rm static}$. The fact that static imperfections can be more dangerous than time-dependent ones ({\em e.g.} noise) has been also observed in numerical experiments done in Ref.~\cite{Benenti:01}. Therefore, in the 
following we shall concentrate on static perturbations as they are more dangerous. 
We concentrate on the Quantum Fourier Transformation algorithm 
(QFT)~\cite{Shor:94,Coppersmith:94} and shall consider its stability 
against static random
   perturbations. The perturbation $V(t)\equiv V$ will be a random
   hermitian matrix from a Gaussian unitary ensemble (GUE)~\cite{Mehta:91}. 
The GUE matrices have been extensively used to model quantum statistical 
properties of classically chaotic Hamiltonians without time-reversal 
invariance~\cite{Haake:91,Guhr:98,Stockmann:99}.
Note that a GUE matrix acts on the whole Hilbert space and therefore 
represents a perturbation that affects correlations 
between all $n$ qubits. Quantum error correction methods for instance do 
not work for this type of errors. Second moments of a GUE matrix $V$ are normalized as
\begin{equation}
\ave{V_{jk}V_{lm}}_{\rm GUE}=\delta_{jm}\delta_{kl}/\mathcal{N},
\end{equation}
where the averaging is done over a GUE ensemble. 
Using this in the general expression for the 
correlation function (\ref{eq:Fcor}) we arrive at the correlation function 
for a static GUE perturbation,
\begin{equation}
\ave{C(t,t')}_{\rm GUE} = \left|\frac{1}{\mathcal{N}} \tr{U(t,t')}\right|^2,
\label{eq:Canal}
\end{equation}
where $U(t,t')$ is the unperturbed propagator from gate $t'+1$ to $t$, 
$U(t,t')=U_t\cdots U_{t'+1}$, with the convention $U(t,t)\equiv \mathbbm{1}$. 
The correlation sum must then be evaluated for the QFT algorithm. For the 
usual decomposition of the QFT algorithm into gates we have $T\sim n^2/2$ 
gates, namely $n$ one-qubit Hadamard gates and $n^2/2$ diagonal 
two qubit ``B'' 
gates, ${\rm B}_{jk}={\rm diag}\{ 1,1,1,\exp{(\ii \theta_{jk})} \}$, 
with $\theta_{jk}=\pi/2^{k-j}$. The correlation sum $\nu=\sum_{t,t'=1}^T C(t,t')$ 
scales as $\nu_{\rm QFT} \propto n^3$. Cubic growth can be understood from 
the special block structure of the QFT. There are $n$ blocks, each containing 
one Hadamard gate and $j$ ``B''-gates, $j=0,\ldots,n-1$. These 
``B''-gates, which are almost identities, are responsible for the correlation 
function to decay very 
slowly, resulting in $\nu \propto n^3$, see Ref.~\cite{Prosen:01} for details. 
The idea how to modify the QFT algorithm is to get rid of {\em bad} 
``B''-gates. This can in fact be done~\cite{Prosen:01}, obtaining 
the so-called improved QFT (IQFT) algorithm. The IQFT has twice as many 
gates as the QFT, but retains the same scaling $T_{\rm IQFT}\sim n^2$. Although the perturbation is 
applied twice as often during the evolution of IQFT the 
correlation sum is smaller, because the correlation function is greatly 
decreased. More importantly, the correlation sum  grows only quadratically 
with $n$, $\nu_{\rm IQFT} \propto n^2$. The dependence of fidelity $F(T)$ 
on the number of qubits $n$ can be seen in Fig.~\ref{fig:ft}. Both the QFT and 
the IQFT are represented, and we see the considerable advantage of the 
modification for large $n$. 

A few comments are in place at this point. The high fidelity of 
the IQFT algorithm 
depends on the perturbation being a GUE matrix. The optimization is 
therefore somewhat perturbation specific. We should point out that the 
optimization of the QFT algorithm becomes more difficult if the perturbation 
results from a two-body (two-qubit) GUE ensemble~\cite{Guhr:98,Benet:03}. 
This is connected with the fact that quantum gates are two-body operators 
and can perform only a very limited set of rotations on full Hilbert space 
and consequently have a limited capability of reducing the
correlation function in a single step. For such errors fidelity will typically 
decay with the square of the number of errors $T$, {\it i.e.} the number of gates, 
like $\sim \exp(-\eps^2 T^2)$, that is the same as for regular systems. This 
means that the very fact that the algorithm is efficient, having a polynomial 
number of gates, makes it very hard to reduce the correlation function and therefore 
causes fast fidelity decay. 

In this section we 
presented a result for optimizing the QFT if the perturbation acts after each 
individual gate. In experimental implementations each gate is usually composed of 
many simple pulses. Provided the perturbation is GUE, the IQFT will be more 
stable even if the perturbation acts after each pulse ({\it i.e.} many times during 
a single gate). This is for instance 
the case in Ref.~\cite{Celardo:03} where the same gain as here has been numerically 
verified for a QFT algorithm running on the Ising quantum computer.

\subsubsection{Dynamical decoupling}

\label{sec:dyndec}
Our approach to the analysis of fidelity consisted in exploring how $F(t)$ 
decays for different perturbations and/or initial states for a given 
unperturbed dynamics $H_0$. The question can be turned around 
though. One can ask, what we can infer about the fidelity $F(t)$ if we have a given 
perturbation $V$, but are free to choose the unperturbed evolution $H_0$. 
This question is of immediate interest if we want to suppress 
the unwanted evolution caused by $V(t)$. By appropriately 
choosing $H_0(t)$ we want to suppress unwanted effects of $V(t)$ as much as possible. 
Of course, just setting $H_0(t)=-V(t)$ will do the trick. But the rule 
of the game is, that we are not able to generate an arbitrary $H_0$, but 
just some subset. One can, for instance, imagine that $V(t)$ is an unwanted 
coupling with the environment. In such case $V(t)$ includes also 
environmental degrees of freedom which we are not able to control. 
So one can allow $H_0$ to act only on the 
the central system and not on the environment.

Such a dynamical suppression of unwanted evolution is known as         
{\em a dynamical decoupling}~\cite{Viola:99,Viola:03,Viola:05,Viola:98,Santos:05}. 
Let us assume that we are able to perform operations $U_0(t)$ from a 
group $\mathcal{G}$ without errors and infinitely fast~\footnote{The condition 
of infinitely fast, so-called ``bang-bang''~\cite{Viola:98} execution, can be 
relaxed~\cite{Viola:03}.}. In the case of a discrete group $\mathcal{G}$ 
time $t$ is a discrete index counting the elements of a group. 
$U_0$ can be thought of as generated by some 
Hamiltonian $H_0(t)$. Usually cyclic conditions are assumed, 
$U_0(t+T_{\rm c})=U_0(t)$, where $T_{\rm c}$ is the period. Due to the 
presence of the unwanted term $V(t)$, the evolution is actually given by 
$U_\eps(t)$ applying both $H_0(t)$ and $V(t)$. 
Because $U_0(t)$ can 
be executed exactly, we can define a new computational frame called 
the {\em logical frame} (interaction frame in fidelity language), in which 
there is no evolution due to $U_0$. Mathematically, the propagator in 
the logical frame is
\begin{equation}
U_{\rm log}(t)=U_0^\dagger(t)U_\eps(t).
\label{eq:logical}
\end{equation}
We can see, that $U_{\rm log}(t)$ is nothing but the echo operator 
$M_\eps(t)$ (\ref{eq:eodef}), so we have
\begin{equation}
U_{\rm log}(t) = \TT\exp\left(-\frac{\ii}{\hbar}\eps\int_0^t
\dd t' \tilde{V}(t')\right),
\end{equation}
where $\TT$ denotes a time-ordering and $\tilde{V}(t) = U_0^\dagger V(t) U_0(t)$. 
The goal of dynamical decoupling is to make evolution in the 
logical frame $U_{\rm log}$ as close to identity $\mathbbm{1}$ as 
possible or more generally, to make it a tensor product of evolution of 
the central system and the environment, 
$U_{\rm log}(t) \rightarrow U_{\rm sys} \otimes U_{\rm bath}$, {\it i.e.} to 
decouple our system from the environment. While 
in the laboratory frame there are 
contribution due to $H_0$ and $V(t)$, in the logical frame 
there is only a contribution due to logical $\tilde{V}(t)$. By appropriate 
choice of $U_0(t)$ such that $\tilde{V}(t)$ ``averages'' out, suppression 
of evolution due to $V(t)$ can be achieved in the logical frame. Two 
approaches are possible, depending on how we choose the correcting 
dynamics $U_0(t)$:

\begin{itemize}
\item The first one is called a {\em deterministic dynamical decoupling}~\cite{Viola:99}, 
because $U_0(t)$ is chosen to deterministically traverse all the 
elements of the group $\mathcal{G}$. Applying the BCH formula 
(Section~\ref{sec:BCH}) on $U_{\rm log}$ we get
\begin{equation}
U_{\rm log}(t)= \exp\left\{-\frac{\ii}{\hbar}\left(
\eps \oV t + \frac{1}{2}\eps^2\Gamma(t) + 
\ldots\right)\right\},
\end{equation}
with the known expressions for $\oV$ (\ref{eq:Vbardef}) and 
$\Gamma(t)$ (\ref{eq:defGamma}). If we are able to choose $H_0(t)$ such
that $\oV\equiv 0$ dynamical decoupling is achieved and is said to 
be of the 1st order. Dynamical decoupling therefore exactly corresponds to 
the freeze of quantum fidelity, either in integrable systems 
(Section~\ref{sec:ch4Vneq0}) or in chaotic (Section~\ref{sec:chfreeze}). The 
fact that going into a different computational frame can dramatically 
change the influence of errors has been also numerically verified in the Ising 
quantum computer~\cite{Celardo:03}.

\item Even if one might not be able to make $\oV=0$, one might be able to 
reduce it. A standard measure of success of such a suppression is fidelity, 
{\it i.e.} the expectation value of $U_{\rm log}(t)$. Learning from the results 
of fidelity decay, we can expect that the suppression will be the larger, 
the more ``chaotic'' the evolution $U_0(t)$ is~\cite{Prosen:02,Prosen:01,Prosen:02corr}. 
In the maximally random case the $U_0(t)$ at different times are 
totally uncorrelated. Dynamical decoupling for such uncorrelated $U_0(t)$ 
has been recently proposed~\cite{Viola:05} and named {\em random dynamical decoupling}. 
Lowest order of the error probability has been 
estimated~\cite{Viola:05} and is equal to the linear response approximation to fidelity 
for chaotic dynamics (\ref{eq:linearmixing}), see also Section~\ref{sec:noisypert} on fidelity 
decay for uncorrelated perturbations.
\end{itemize}

Summarizing, the idea of dynamical decoupling is to suppress unwanted 
evolution $V(t)$ by applying unitary corrections $U_0(t)$ and doing 
quantum operations in the logical frame defined by $U_0(t)$. This can 
be achieved either by making $\oV=0$ or by making $U_0(t)$ uncorrelated 
in time, so the error correlations are decreased. The idea of applying 
some additional pulses to correct for unwanted evolution is extensively used in NMR. 
In the quantum computer context it has been for instances used by the group of Berman 
to suppress near-resonant errors in the Ising quantum computer, see Ref.~\cite{Berman:03} and 
references therein.

\section{A report on experiments}
\label{E}

The first experiments of echo dynamics go back to
 Hahn~\cite{Hahn50} in nuclear magnetic resonance (NMR) and in the 
introduction we have given a historical rundown of NMR techniques. 
In this section we shall limit our discussion to a few experiments or 
potential experiments, that relate directly to the theory presented in this 
paper. In this context recent experiments with microwaves and elastic waves 
are remarkable, in that they yield very good agreement with random matrix 
theory, but we shall actually start with some recent NMR experiments by the 
Cordoba group, which at this point cannot be explained and provide some 
intriguing riddles. Finally we shall discuss extensively potential experiments 
in atomic optics expanding on an idea by
Gardiner Cirac and Zoller~\cite{GCZ97,Gardiner:98erratum,GPSS04}.

\subsection{Echo experiments with nuclear magnetic resonance}
\label{EE}

The first real {\em interacting} many-body echo
experiment was done in solids by Rhim et al.~\cite{Rhi70}. Time reversal,
{\it i.e.} changing the sign of the interaction, is achieved for a dipolar
interaction whose angular dependence can change sign for a certain
``magic'' angle, that causes the method to be called magic
echo. Still, the magic echo showed strong irreversibility. Much
later a sequence of pulses has been devised enabling a {\em local}
detection of polarization~\cite{Zha92,ZhaCor98} ({\it i.e.} magnetic moment). They used a
molecular crystal, ferrocene ${\rm Fe}({\rm C}_5{\rm H}_5)_2$, in
which the naturally abundant isotope $^{13}{\rm C}$ is used as an
``injection'' point and a probe, while a ring of protons $^1{\rm H}$
constitutes a many-body spin system interacting by dipole forces. The
experiment proceeds in several steps: first the $^{13}{\rm C}$ is
magnetized, then this magnetization is transfered to the neighboring
$^1{\rm H}$. We thus have a single polarized spin, while others are in
``equilibrium''. The system of spins then evolves freely, {\it i.e.} spin
diffusion takes place, until at time $t$ the dipolar interaction is
reversed and by this also spin diffusion. After time $2t$ the echo is
formed and we transfer the magnetization back to our probe $^{13}{\rm
C}$ enabling the detection of the {\em polarization echo} (\ref{eq:defPE}). Note that
in the polarization echo experiments the total polarization is
conserved as the dipole interaction has only ``flip-flop'' terms of the form
$S^j_+S^{j+1}_-$, which conserve the total spin. To detect the spin
diffusion one therefore needs a local probe. With the increase of the
reversal time $t$ the polarization echo -- the fidelity -- decreases in
approximately exponential way. The nature of this decay in spin-diffusion reversal experiment has been furthermore elaborated
by Pastawski et al.~\cite{Pasta95}. The group of Pastawski then performed a
series of NMR experiments where they studied in more detail the dependence of the
polarization echo on various parameters~\cite{LevUsa98,Usaj:98,Pastawski:00}. They were able to control the size of the residual part of the Hamiltonian,
which was not reversed in the experiment and is assumed to be responsible for
the polarization echo decay. 
In presence of strong perturbations, such as magnetic or quadrupolar impurities, 
the Echo decays according to a perturbation dependent exponential law as prescribed 
by the Fermi golden rule. However, in pure systems, where residual interactions 
are made small, the decay enters a Gaussian regime where the decay rate is 
perturbation independent. Notice that the experiments are constrained to move from 
a naturally big perturbation and, while one can reduce it, the full cancellation 
of perturbation can not be achieved. A similar situation was observed in liquid 
crystals where a molecule involves a few tens of spins~\cite{Levstein:04}. This is 
consistent with the original interpretation~\cite{Pasta95,Usaj:98} that the 
many body spectrum is so dense, that almost any small perturbation can place the 
Echo dynamics in a regime where it is perturbation independent. These surprising 
results triggered a number of numerical studies and theoretical investigations.
\par
In Ref.~\cite{Weinstein:02qc} quantum Baker map has been experimentally implemented on a 3-qubit system within NMR context. They also studied the influence of perturbations on the dynamics and could by measuring the resulting density matrices calculate the fidelity. For some other NMR measurements of fidelity see also~\cite{Ryan:05}.

\subsection{Measuring fidelity of classical waves}
\label{ES}

\begin{figure}
\centerline{ \includegraphics[width=0.48\textwidth]{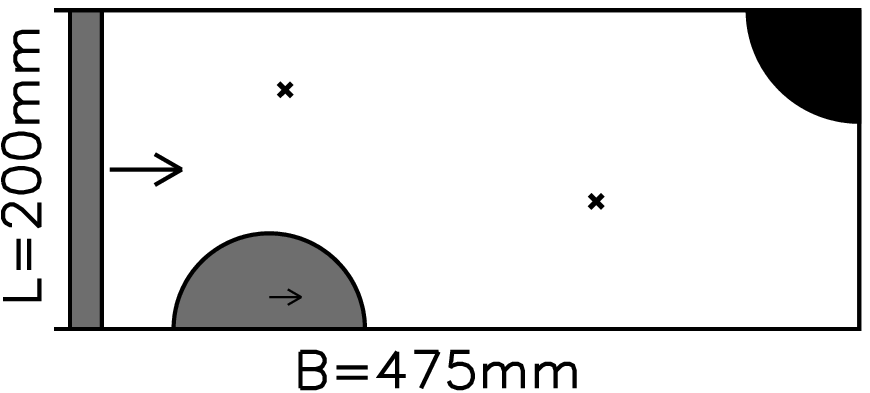}
 \includegraphics[width=0.48\textwidth]{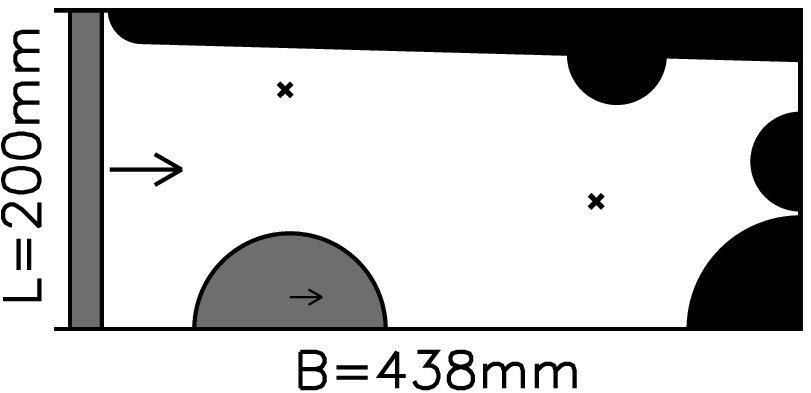}  }
\caption{Geometry of the billiards (figure taken from~\cite{SGSS05}). In the
 billiard in the right figure bouncing ball orbits 
have been avoided by inserting additional
elements.}
\label{ES:f:muwavebills}\end{figure}

\begin{figure}
\centerline{
  \includegraphics[width=0.48\textwidth]{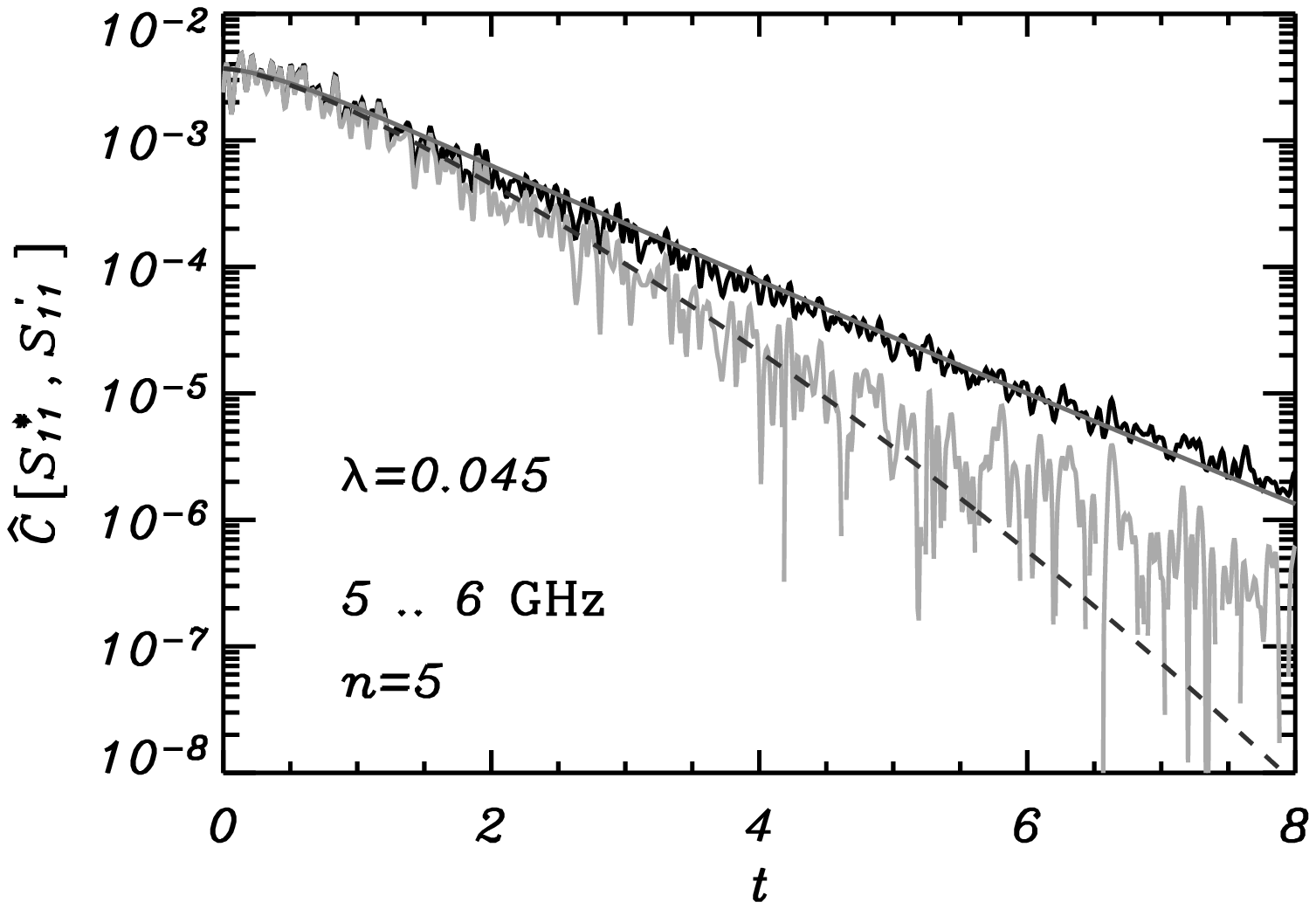}
  \includegraphics[width=0.48\textwidth]{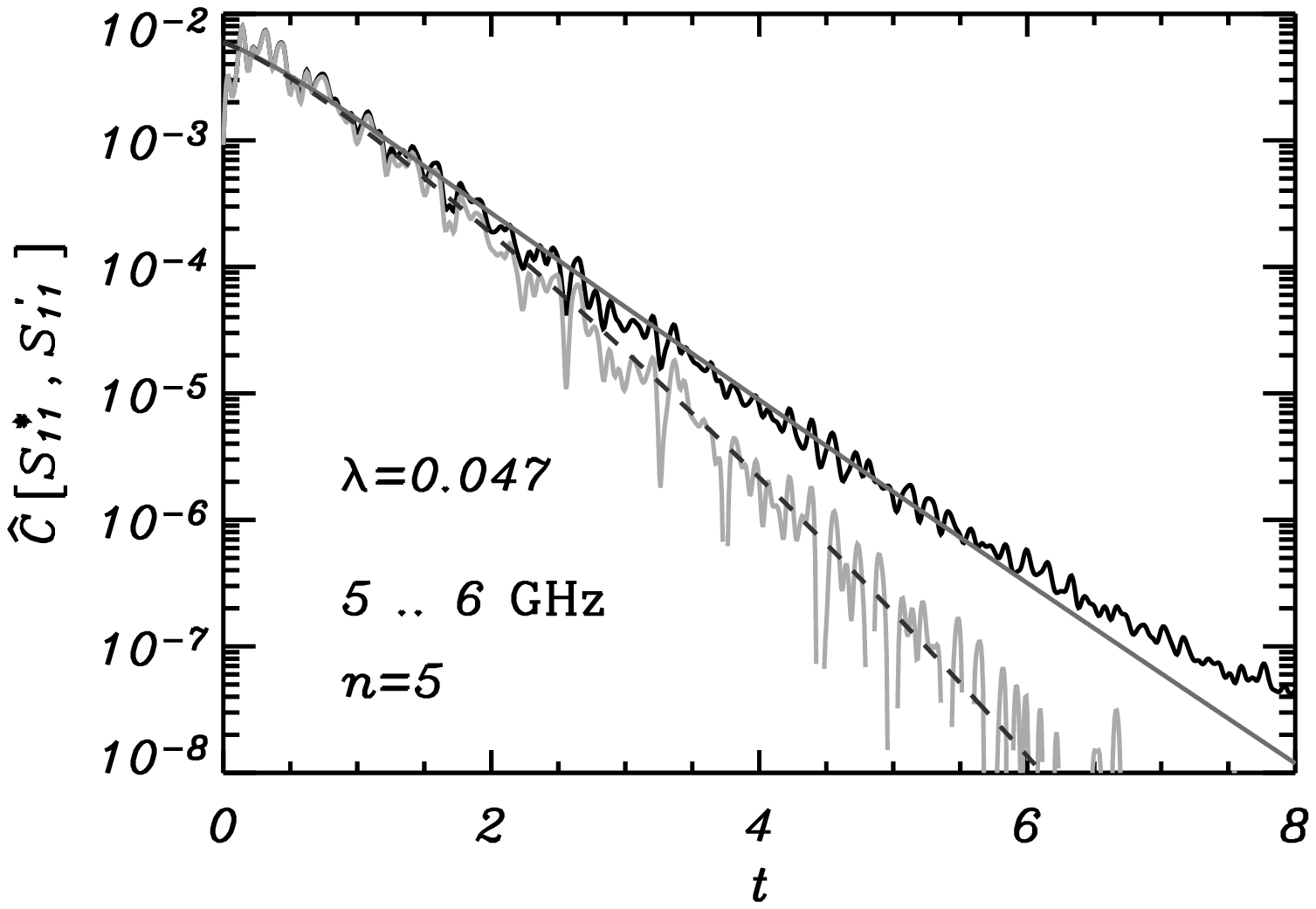} }
\caption{(Figure taken from~\cite{SGSS05})
Logarithmic plot of the correlation function
$\hat C[S^*_{11},S'_{11}]$  for the billiard with bouncing balls
(left), and for the one without bouncing-balls (right). The
experimental results for the auto correlation are shown in black,
while the correlation of perturbed and unperturbed system are
shown in grey. The smooth solid curve corresponds to the
theoretical autocorrelation function, and the dashed curve to the
product of autocorrelation function and fidelity amplitude.}
\label{ES:f:mubillcfun-top}\end{figure}

\begin{figure}
\centerline{
  \includegraphics[width=0.48\textwidth]{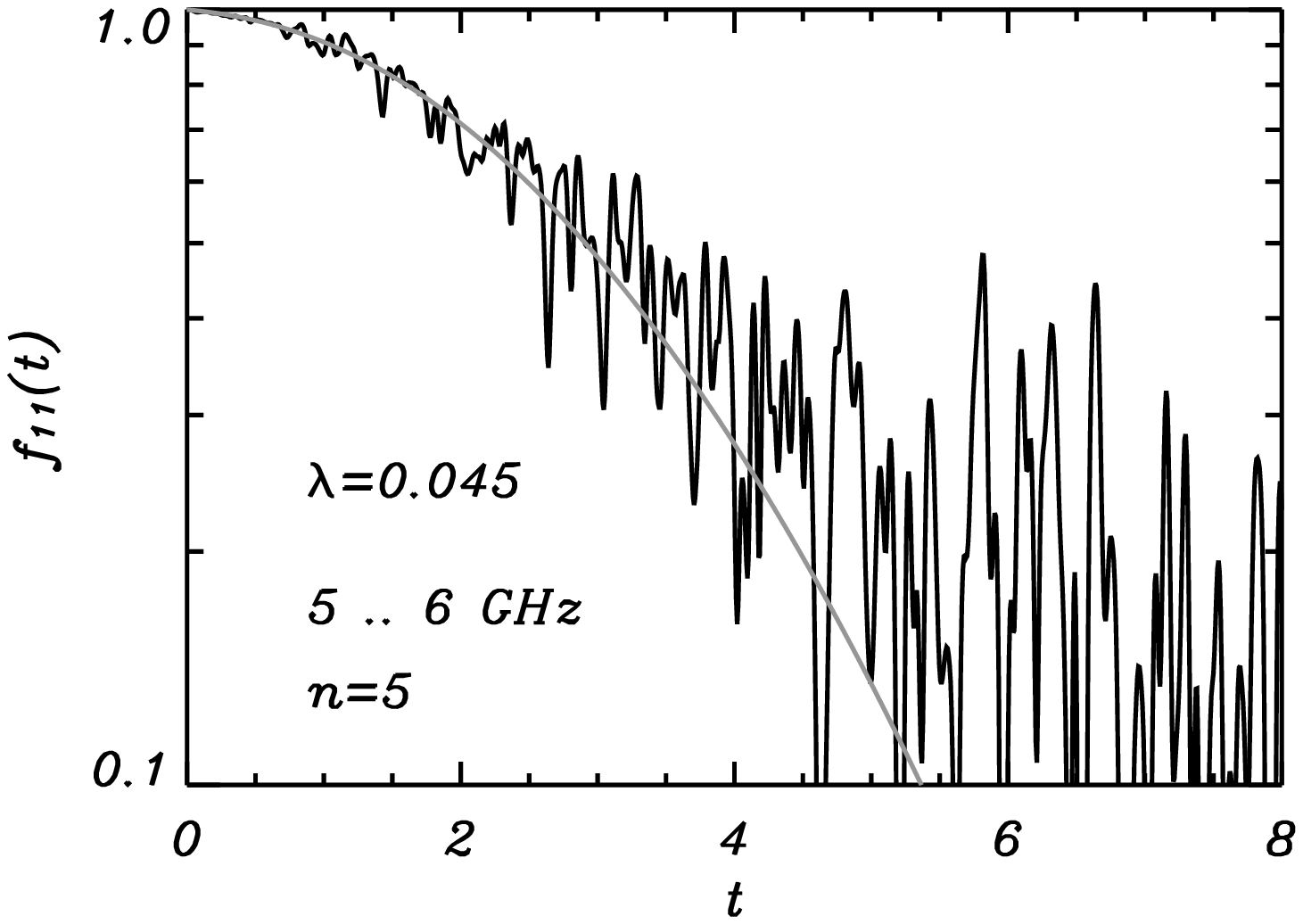}
  \includegraphics[width=0.48\textwidth]{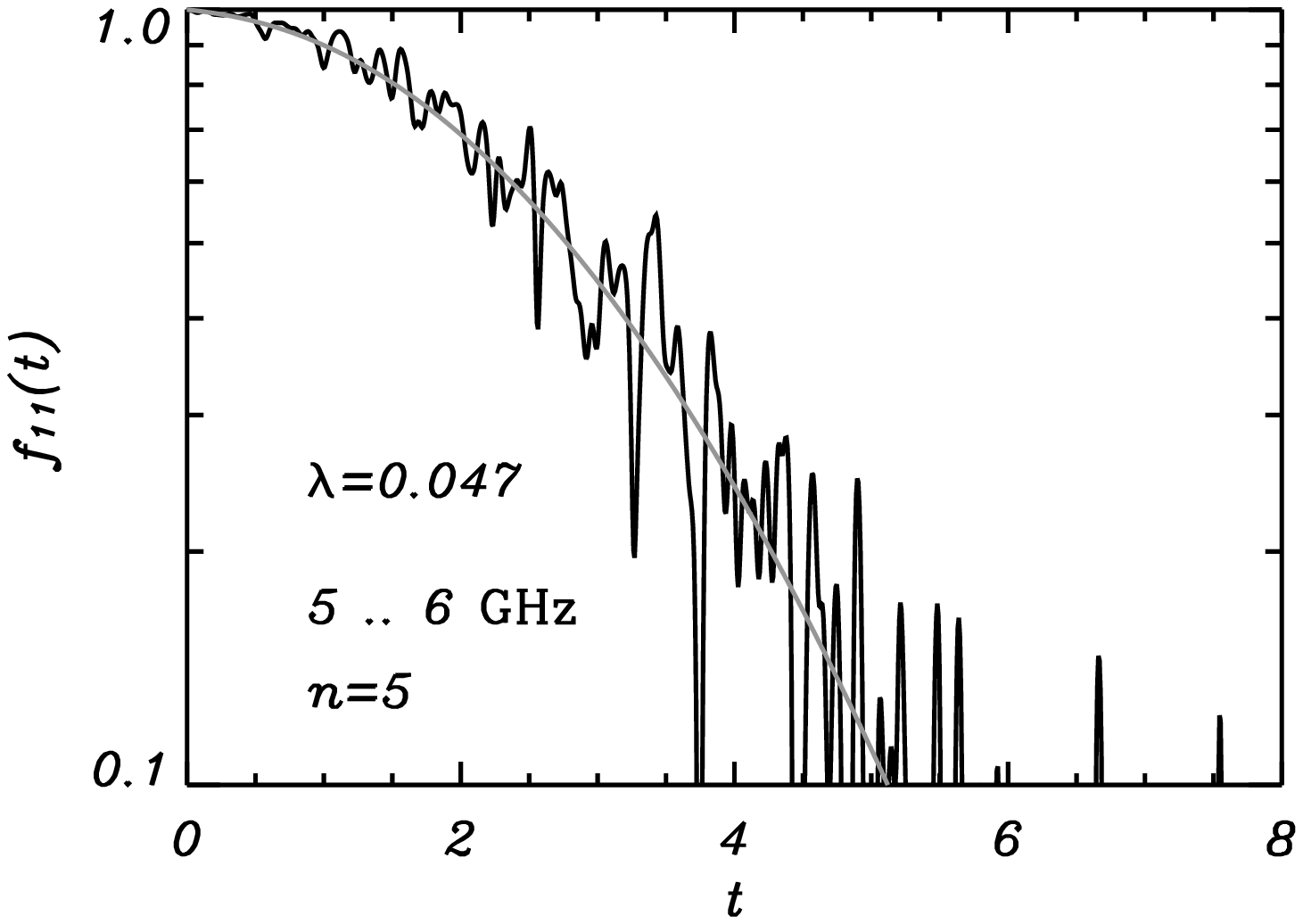} }
\caption{Logarithmic plot of the corresponding fidelity
amplitudes (figure taken from~\cite{SGSS05}).
The smooth curve shows the linear-response result. For the billiard
without bouncing balls, the perturbation parameter $\lambda$ was obtained from
the variance of the level velocities; in the other case it was fitted to the
experimental curve.}
\label{ES:f:mubillcfun-bottom}\end{figure}

\begin{figure}
\centerline{
  \parbox{0.48\textwidth}{ }
  \parbox{0.48\textwidth}{
     \includegraphics[width=0.48\textwidth]{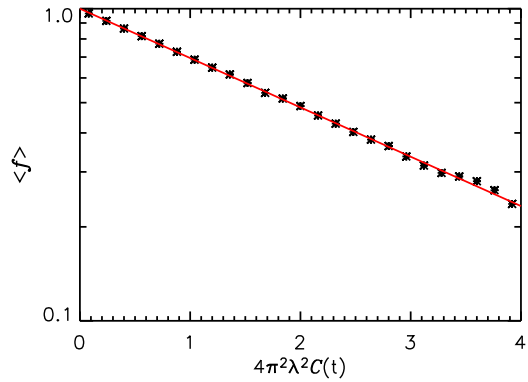} } }
\caption{(Figure taken from~\cite{SGSS05})
Average of the experimental fidelity amplitude on a
rescaled axis $x=4\pi^2\lambda^2C(t)$. The solid line corresponds
to $g(x)=\exp{(-\alpha x)}$ with $\alpha=\lambda_{\rm
exp}^2/\lambda^2=0.36$.}
\label{ES:f:scalingall}\end{figure}

The dynamics of classical electromagnetic waves in a thin resonator is equivalent to a single-particle two dimensional Schr\" odinger equation, {\em i.e.} single particle quantum mechanics. This is exploited in the microwave resonator experiments, where properties of two dimensional quantum billiards can be studied. The Marburg group considered a cavity
consisting of a rectangle with inserts that assure a chaotic ray
behavior, Fig.~\ref{ES:f:muwavebills}. Both situations with and without
parabolic manifolds
leading to so-called bouncing ball states were considered. One wall was
movable in small steps, this change causing the perturbation occurring in an
echo experiment. It is important to note that the shift of the wall changes
the mean level density and thus the Heisenberg time. This trivial
perturbation, which would cause very rapid correlation decay is
eliminated in this case by
measuring all times in proper dimensionless units, {\em i.e.} in terms of the Heisenberg time as proposed in
Section~\ref{R}. Two antennae allow to measure both reflection and transmission channels. The experiment was carried out in the frequency domain. Afterwards the Fourier transform is used to obtain correlation functions in the time domain. Recall (Section~\ref{DOS}) that these functions are used to
define scattering fidelity which, for weak coupling and chaotic dynamics, agrees with standard
fidelity. In Fig.~\ref{ES:f:mubillcfun-top} the correlation functions are
shown for two different systems. Observe that the cross-correlation
agrees with the random matrix prediction
in time up to six times the Heisenberg time and in magnitude of the cross-correlation
function over five orders of magnitude. In the case with bouncing ball states
the agreement with RMT is understandably much poorer. In
Fig.~\ref{ES:f:mubillcfun-bottom},
the scattering fidelity itself is reproduced; as we are in the range of
isolated resonances, {\it i.e.} in a weak coupling situation this should be
standard fidelity, and indeed the agreement with fidelity as obtained from RMT
is excellent. The perturbation strength in this case was determined
independently from the level dynamics and, understandably, for low energies it does not agree with its semi-classical limit. It is more surprising
that the shape of fidelity as obtained for RMT in Eq.~(\ref{Q2LR:expLR}) holds
for some stretch beyond the weak coupling limit. This can be seen in
Fig.~\ref{ES:f:scalingall}, where
a wide range of data for chaotic systems has been averaged over, and the
validity of scaling with $\lambda^2\; C(t)$ is assumed. This in turn implies
scaling with $\lambda^2$ as well as with the time dependence and the dependence on Heisenberg time as it appears in the correlation integral $C(t)$. Indeed if we scale the time
dependence with the function obtained in the linear response approximation of
RMT and average over all results that are available for different systems and
frequency ranges we find that exponentiation of linear response is an excellent
approximation in the range of the present experiment.

\begin{figure}
\centerline{\includegraphics{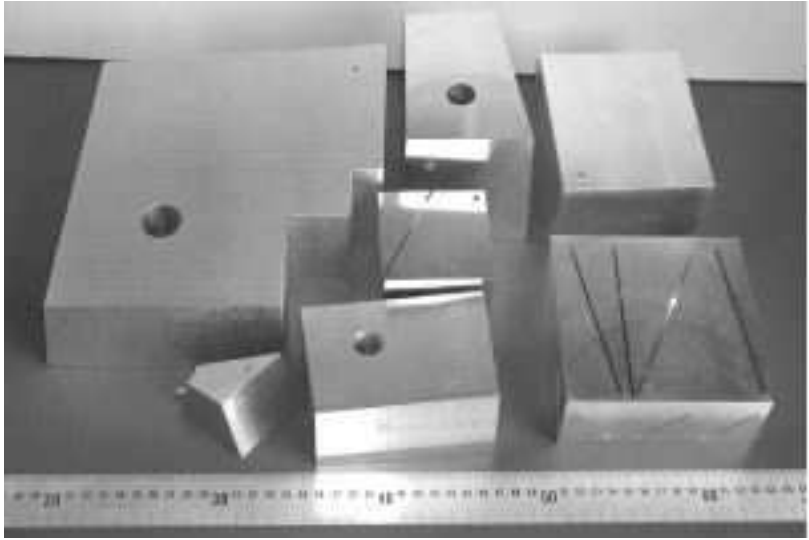}}
\caption{Different aluminum block specimens (figure taken
from~\cite{LobWea03}).}
% Clockwise \ldots}
\label{ES:f:weablocks}\end{figure}

\begin{figure}
\centerline{\includegraphics{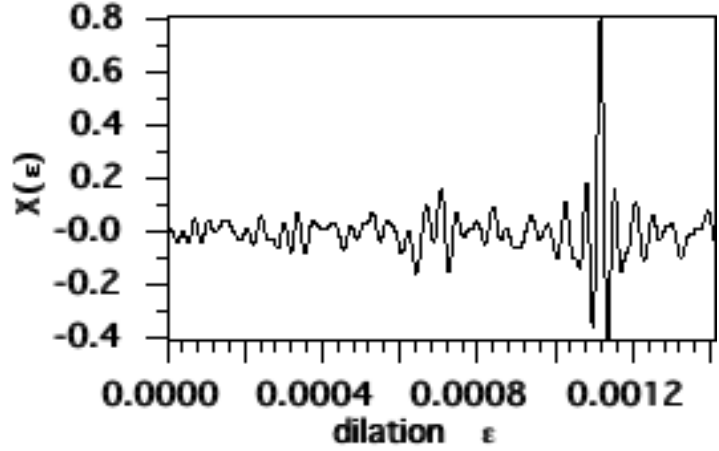}}
\caption{The typical form a a correlation function computed according to
Eq.~(\ref{ES:Xcfun}) (figure taken from~\cite{LobWea03}).}
\label{ES:f:weacfun}\end{figure}

Lobkis and Weaver~\cite{LobWea03} have performed an interesting experiment on
cross-correlations of the long time signal, usually called ``coda'', of elastic 
signals at different temperatures for a number of aluminum blocks, 
Fig.~\ref{ES:f:weablocks}. Both the excitation
and the readout were transmitted with the same piezo-element, that connected
the specimen with the pulse generator and analyzer isolating the latter from
the former by a reed relay. The sample was kept in vacuum and with
strict control of the uniformity of temperature. The purpose of the measurement
was to investigate the normalized cross-correlation of these signals for the same block at different temperatures $T_1$ and $T_2$,
\begin{equation}
X(\varepsilon)= \frac{\int\d \tau\; S_{T_1}(\tau)\; S_{T_2}(\tau(1+\eps))}
  {\sqrt{\int\d \tau\; S_{T_1}^2(\tau)\; \int\d \tau\; S_{T_2}^2(\tau(1+\eps))}}.
\label{ES:Xcfun}\end{equation}
The correlation integral is calculated around some time t, which the authors 
call the ``age'' of the sample.
Temperature steps of $4^\circ$ were chosen. The biggest effect termed dilation is
simply due to changes in volume and in the speeds of propagation of the waves.
This should show in the correlation functions as a maximum of size 1 at a time
difference determined by factors well known for elastic media. In fact
we find a maximum at the predicted time, but it is smaller than 1 (see
Fig.~\ref{ES:f:weacfun}). These maxima as a function of the age $t$ of the
sample, are called distortions $D(t)$.
\begin{figure}
\centerline{\input{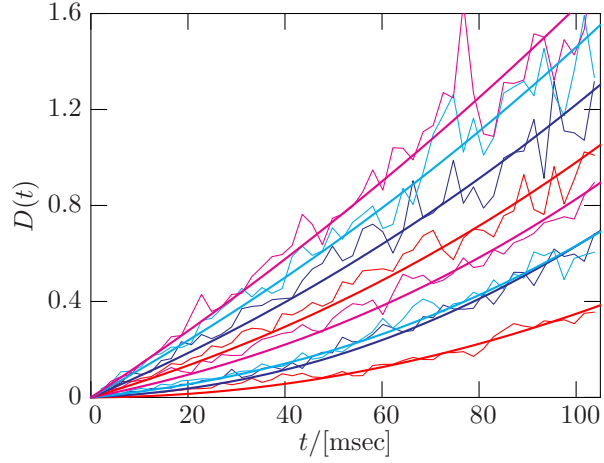}}
\caption{(Color online)
The distortion as a function of time, for the cuboid (figure taken
from~\cite{GSW05}). The thick solid lines
correspond to measurements in the frequency ranges from $100$kHz to  $800$kHz,
in steps of $100$kHz (from bottom to top). The thin solid lines show the best
fits with $-\ln[f(t)]$ where $f(t)$ is given in Eq.~(\ref{Q2LR:expLR}) with $\beta_V=1$. The fit values for $\lambda_0$ can
be found in~\cite{GSW05}.}
\label{ES:f:distrect}\end{figure}

The correlation function~(\ref{ES:Xcfun}) suggests that the distortion can be related to the logarithm of scattering fidelity and indeed it is easy to show~\cite{GSW05} that $D(t) =\;- \ln f_s(t)$ holds. The timeshift in the correlation maximum exactly compensates for the trivial dilation effect, and in this sense is equivalent to measuring time in units of Heisenberg time, as was done in the microwave experiment.

As the piezo-element provides weak coupling and the
environment of the samples is evacuated, we are in a weak coupling limit,
which implies that distortion is also equal to fidelity. In~\cite{GSW05} this
equivalence is shown in detail. Furthermore for a block for which we expect ``chaotic'' split ray behavior good agreement with RMT
results is demonstrated. In this case perturbation strength as obtained from
a fit of RMT  is in fair agreement with an estimate resulting from elastomechanics assuming chaotic split ray dynamics.

Actually RMT also fits very well the results for a regular aluminum brick (upper right sample in Fig.~\ref{ES:f:weablocks}); see Fig.~\ref{ES:f:distrect}. While this 
seems at first sight surprising it is in line with findings of GOE spectral
statistics~\cite{Schaa03} confirming that the corresponding wave system
displays what one may loosely call wave chaos.

\subsection{Fidelity decay and decay of coherences}

Under special circumstances the time evolution of two coupled systems
can show an interesting relation between decay of coherences in one subsystem and fidelity decay in the other one. Basically two types of coupling are
described in \cite{GPSS04} for which this occurs. On one hand we can have
what we might call a dephasing situation, with a Hamiltonian of the form
\begin{equation}
H= H_c +H_e +O_c\;V_e\; , \quad [O_c,H_c]=0 \; .
\end{equation}
Here as usual operators with indices $c$ and $e$ operate only on either of two subsystems termed somewhat arbitrarily central system and environment, and $O_c$
is some operator commuting with the Hamiltonian with eigenvalues $o_j$
when applied to some eigenfunction $\phi_\nu$ with 
$H_c \; \phi_\nu\, = \, \epsilon_\nu\, \phi_\nu$. Then it is easy to prove, 
that for an initial state $(1/\sqrt{2})\, (\phi_j +\phi_k)\otimes \chi_0$ the 
elements of the density matrix of the central system, {\it i.e.} the coherences relate to fidelity decay in the environment as
\begin{equation}
\rho_{j,k}^c(t)= \e^{-\rmi(\epsilon_j-\epsilon_k)}\;
   \langle \chi_0 | \e^{-\rmi t (o_j V_e + H_e)}  \e^{\rmi t (o_k V_e + H_e)} 
   |\chi_0\rangle = \e^{-\rmi(\epsilon_j-\epsilon_k)}\; f(t).
\end{equation}
Up to the phase factor the righthand side is clearly the fidelity amplitude
resulting for the function $\chi_0$ which can be chosen arbitrarily in the 
environment with an unperturbed Hamiltonian $ o_j V_e + H_e $ and a perturbed 
Hamiltonian  $ o_k V_e + H_e $.
% We only have to choose the density matrix $\rho^c$ in an orthogonal basis, 
% where $O$ is diagonal and start with a pure state.

If we can measure the coherences, we can measure fidelity, which does not 
depend on the phases. Moreover, if we know the phases, we actually can measure 
the fidelity amplitude.

A particularly simple situation occurs if the internal evolution of the central
system for some reason is irrelevant. This can occur, if the two states defining the density matrix element we consider, are degenerate, or if the evolution 
of the internal system is so slow, that coherence decays on a much shorter time scale. It has been argued \cite{Braun:01,Strunz:03,Strunz:03a} that this can occur if 
$|o_j -o_k|$ is just large enough. On the other hand Gardiner, Cirac and 
Zoller~\cite{GCZ97,Gardiner:98erratum}  proposed long ago an experiment to test 
the sensitivity to changes of dynamics of chaotic systems. Actually the idea 
they proposed can be generalized and contains much of what was exposed above. 
Experiments along the lines given above can be carried out in atomic optics. 
The non-evolution of the central system as assumed in the original papers by 
the above authors typically holds, though we saw, that it is not essential. 
We shall describe below a proposition for such an experiment for a kicked 
rotor in more detail following Ref.~\cite{Haug:05}.

Yet there is another interesting option presented in \cite{GPSS04}. As usual quadratic Hamiltonians present
additional opportunities. A harmonic oscillator is coupled linearly to a bath of oscillators as
\begin{equation}\label{Hoscis}
H = H^{\rm c} + H^{\rm e} + H^{\rm int}
= \hbar\Omega\; a^{\dagger}a +
   \sum_{\lambda}\hbar\omega_{\lambda}b_{\lambda}^{\dagger}b_{\lambda}+
   \sum\limits_{\lambda}\hbar g_{\lambda}\left( ab_{\lambda}^{\dagger}
    + a^{\dagger}b_{\lambda} \right).
\end{equation}
Here $a,a^\dagger$ are the annihilation and creation operators for the central 
oscillator and $b_\lambda ,b_\lambda^\dagger$ the corresponding operators for 
the bath oscillators. $\Omega$ and $\omega_\lambda$ are the corresponding 
frequencies and $g_\lambda$ the coupling constants.
Now a coherent state basis will produce an adequate factorized basis, which 
will remain factorized for very long times, and allow a similar development.
The distance between complex ``positions'' $z$ of the coherent states in the 
central system will determine the strength of the perturbation. We find 
that the decay of the coherences in the density matrix determines the fidelity 
decay of an initially not excited environment of oscillators. As we are dealing 
with an eigenstate of the unperturbed oscillator we expect Fermi golden rule 
behavior, {\it i.e.} exponential decay with the square of the strength of the 
perturbation, and indeed recover the famous relation 
$|\varrho_{12}^{\rm c}(t)|^2 = \exp(-\gamma t|z_1(0)-z_2(0)|^2)\; 
|\varrho_{12}^{\rm c}(0)|^2$, \cite{Giulini96,Zurek:91,ZurHab93}. 
Historically, it is 
quite interesting to notice, that the Paris experiment~\cite{Bru96} 
that shows decoherence explicitly can be reinterpreted as a measurement of 
fidelity decay of the environment at least for large angles. This follows 
directly from the fact, that the relation between the decoherences and 
fidelity decay is equally true for well separated Gaussian wave packets.

\subsubsection{The Gardiner-Cirac-Zoller approach and the kicked rotor}
\label{EG}

In~\cite{GCZ97,Gardiner:98erratum} the authors proposed an experiment to study 
quantum dynamics
of the center of mass motion of a trapped atom. Indeed they use the internal
degree of freedom of the trapped atom, both to disturb the dynamics and to
measure the effect of this perturbation. While they do not use the word
explicitly,
% as it was then not fashionable, 
they propose a measurement of the fidelity amplitude
using the properties of particular couplings of two-level quantum systems. 
As mentioned above their method is a special case of the general theory
outlined in \cite{GPSS04} and reported above. It is though
important to note that the basic idea is given in this reference. 
%TGout It has since been reported in distinct contexts \cite{?} with or 
%TGout without referring to the original paper. 
A more detailed proposition along these lines to measure
fidelity decay for the kicked rotor {\it e.g.} is given in~\cite{Haug:05}, 
which we shall follow in this subsection.

The kicked rotor can be modeled experimentally by an atom in a standing light
wave~\cite{Moo95,Raizen:99,Arcy01}. For such an experiment typically an
atom of mass $M$ is chosen, which has two electronic levels  with spacing
$\omega_0$. The states shall be denoted by $\ket 1$ and $\ket e$ for the lower
and the excited one. They are driven by two counter--propagating laser fields.
% \commentmarko{Why not $\ket e$ and $\ket g$ or $\ket 0$ and $\ket 1$?}

The dipole $\vec\mu$ of this transition is coupled to the
electromagnetic field $\vec E(x,t)={\mathcal
E}\vec\epsilon\cos(k_{\rm L} x+\varphi_{\rm L})e^{-i\omega t} +
{\rm c.c.}$ of the lasers with wave number $k_{\rm L}$, frequency
$\omega$, complex amplitude ${\mathcal E}$, polarization
$\vec\epsilon$ and phase $\varphi_{\rm L}$. In the rotating wave approximation
the Hamiltonian describing this interaction reads as
\begin{equation}
\hat H_{\rm int} = \frac{\hat p^2}{2M}+\hbar\omega_0\ket e\bra e
 +  \left[\frac{\hbar\Omega}{2} \cos\left(k_{\rm L}
x+\varphi_{\rm L}\right)e^{-i\omega t} \ket e\bra 1 + {\rm
H.c.}\right] \; ,
\end{equation}
with the Rabi--frequency $\Omega=\vec\mu\vec\epsilon{\mathcal
E}/\hbar$. A large detuning $\Delta=\omega_0-\omega$ allows to
adiabatically eliminate the excited state $\ket e$ leading to the
Hamiltonian
\begin{equation}
\hat H_1=\frac{\hat p^2}{2M}+\hbar\kappa_1[\sin(2k_{\rm L}x) + 1]\,\delta_T(t)
\label{eq:atham}
\end{equation}
for the state $\ket 1$ in a frame rotating with the laser
frequency. The interaction strength is given by
$\kappa_1=\Omega^2/(8\Delta)$ and we choose the phase $\varphi_{\rm
L}$ appropriately. Moreover, the periodic kicks theoretically
described by the train of $\delta$--functions in
Eq.~(\ref{eq:atham}) can be approximately realized by rapidly
switching on and off the laser fields with period $T$. Therefore,
the Hamiltonian Eq.~(\ref{eq:atham}) corresponds to the Hamiltonian
of the kicked rotor.

\begin{figure}[ht]
\centerline{\includegraphics[angle=270,width=4cm]{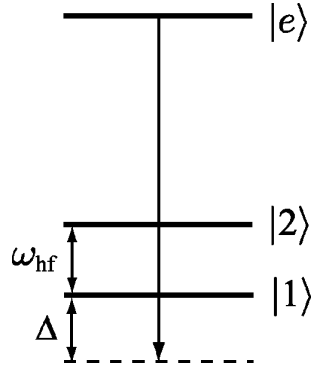}}
\caption{Level scheme for the measurement of the Loschmidt--echo (taken 
from~\cite{Haug:05}).
A classical laser drives the transitions $\ket e \leftrightarrow
\ket 1$ and $\ket e \leftrightarrow \ket 2$ of a three--level atom
with excited electronic state $\ket e$ and two hyperfine ground
states $\ket 1$ and $\ket 2$ which have a level spacing of
$\omega_{\rm hf}$. The laser is detuned by $\Delta$ from the
transition $\ket e \leftrightarrow \ket 1$ and by
$\Delta+\omega_{\rm hf}$ from the transition $\ket e
\leftrightarrow \ket 2$.} \label{fig:scheme}
\end{figure}

In order to measure fidelity a similar setup can be
applied using atom interferometry
\cite{Bie02,Arcy01}. For this reason we take again an
atom with excited electronic state $\ket e$ but two hyperfine
ground states $\ket 1$ and $\ket 2$ separated by the
hyperfine--splitting $\omega_{\rm hf}$, as illustrated in
Fig.~\ref{fig:scheme}.

As above for large detuning $\Delta$
we can eliminate the excited level and find for the lower hyperfine doublet
the Hamiltonian $\hat H_{\rm g}=\hat H_1 \ket 1\bra
1 + \hat H_2 \ket 2\bra 2$ with
\begin{equation}
\hat H_2=\frac{\hat p^2}{2M}+\left\{\hbar\kappa_2[\sin(2k_{\rm
L}x) + 1]+\hbar\omega_{\rm hf}\right\}\,\delta_T(t) \,.
\label{eq:atham2}
\end{equation}
Here we analogously defined a second interaction strength
\begin{equation}
\kappa_2=\frac{\Omega^2}{8(\Delta+\omega_{\rm hf})}=\kappa_1-d \; ,
\end{equation}
where $d=\kappa_1\omega_{\rm hf}/(\Delta+\omega_{\rm hf})$.
We mention that the physical quantity $d$ is then the perturbation strength,
which has to be rescaled if results are to be compared {\it e.g.} with a random
matrix model.

The state for the motion of the atom, prepared in a internal superposition
of states $\ket 1$ and $\ket 2$, propagates in two different
potentials described by the Hamiltonians $\hat H_1$ and $\hat
H_2$, Eqs.~(\ref{eq:atham}) and (\ref{eq:atham2}), respectively.
Indeed, this is exactly the situation we need, in order to realize a 
measurement of the fidelity amplitude.
% with perturbation $d$.
\begin{equation}
\tilde f_N=\bra{\psi_0}\,\hat U_2^{-N}\, \hat U_1^N\,\ket{\psi_0} \; ,
\end{equation}
where $\hat U_i$ describes the
time evolution with the Hamiltonian $\hat H_i$. To achieve this the atomic state must be initially in a superposition of the internal
states $\ket 1$ and $\ket 2$,
\begin{equation}
\ket\Psi=\frac{1}{\sqrt 2}\big (\, \ket 1 + \ket 2\, \big )\; \ket{\psi_0}
\end{equation}
of the composite system of internal and external degrees of
freedom. Here $\ket{\psi_0}$ represents the initial state of the
center--of--mass motion. The time evolution
leads to
\begin{equation} 
\ket{\Psi(t=NT)}=\frac{1}{\sqrt 2}\left[\ket 1\ket{\psi_N} + 
   \ket 2\ket{\phi_N}\right] \; .
\end{equation} 
Fidelity can now be extracted by determining the
probability $W_N(\theta)\equiv \int \! dx\,
|\bra{j(\theta)}\bra{x}\,\Psi_N\rangle|^2$ to find the atom in the
internal state $\ket{j(\theta)}\equiv \frac{1}{\sqrt 2}\left[\ket
1 + e^{-i\theta}\ket 2\right]$. Here we used the position
states $\ket x$ in order to trace over the external degrees of
freedom. We find
\begin{equation} 
W_N(\theta)=1+{\rm Re}\left[e^{-i\theta}\bra{\phi_N}\,\psi_N\,\rangle\right]
\end{equation} 
from which we can finally calculate the real and imaginary part of
the fidelity amplitude
\begin{equation}
{\rm Re} \tilde f_N =  W_N(0)-1 \qquad
{\rm Im} \tilde f_N =  W_N(\pi/2)-1 \; .
\end{equation}
Due to the constant potential terms in the Hamiltonians $\hat H_1$
and $\hat H_2$, Eqs.~(\ref{eq:atham}) and (\ref{eq:atham2}), the
measured fidelity amplitude $\tilde f_N$ picks up a constant phase
factor $\exp[-i(\omega_{\rm hf}-d)N]$. Thus the desired fidelity
amplitude as defined in Eq.~(\ref{eq:fiddef}) is related to the
measured fidelity amplitude via
\begin{equation} 
f_N=e^{i(\omega_{\rm hf}-d)N} \tilde f_N \; .
\end{equation} 
%The setup described is similar to previous experiments using sodium
%atoms~\cite{bib:raizen}. 
For a specific implementation of the above scheme, the authors in
~\cite{Raizen:99} propose to use the D2 level of Sodium.
More technical details are given in~\cite{Haug:05}.
It is important to note, that we have specifically selected the example of the
kicked rotor, because it has recently received 
attention~\cite{Raizen:99,Wimberger:06} and because of the paradigmatic 
character 
it acquired in chaos theory due to many important contributions of Chirikov 
and others. Yet the original proposal referred to the movement in some
trap~\cite{GCZ97,Gardiner:98erratum} and due to the flexibility that lasers permit, many
different Hamiltonians can be studied in this way. Also the perturbation
strength can easily be varied.

Similar
experiments have been performed \cite{Arcy03,AndDav03,Kuh03}, using certain
variants of the idea of Gardiner, Cirac and Zoller, leading to related
quantities.

\section{Summary}
\label{S}

Echo-dynamics and studies about fidelity decay have received considerable 
additional attention as we were writing this paper, and we apologize, if we 
missed some recent developments, though we believe that the basic ideas are 
covered and we are more likely to miss some applications. Echo-dynamics is 
basically defined by the forward evolution with some unitary dynamics, followed 
by a backward evolution with a perturbed one. We have focused on Hamiltonian 
time evolution and based our treatment on the fact that the two-point time 
correlation function in the interaction picture essentially describes the 
process, at least for the functions or observables that are usually considered. 
The most common object of research is fidelity, {\it i.e.} the autocorrelation 
function of some initial state under echo dynamics. This quantity is 
particularly important, because it is equivalent to the cross correlation 
function of some initial state evolved with an ideal and a perturbed 
Hamiltonian, which often serves as a benchmark for the reliability of quantum 
information processes.  The behaviour of other quantities, such as expectation 
values of operators, purity or S-matrix elements, were also considered. The 
basis of the entire analysis resides in the observation, that a system is the 
more stable under perturbations, the shorter the temporal correlations of the 
perturbation operator in the interaction picture. This leads directly to the 
relative harmlessness of noisy perturbations, but also to the slightly 
counterintuitive result that typically a time independent or time periodic 
perturbation will affect stability more strongly in integrable systems then in 
chaotic ones. It also leads to a way of stabilizing quantum information 
processes by introducing chaotic or random elements, or outright engineering 
rapid decay of the correlations of unavoidable errors by adequate gates. 

Most of the standard results are obtained from linear response approximation or 
in other words from the fist non-vanishing term in a Born expansion. The 
remarkable fact arises, that in most instances the simple exponentiation of 
this term gives a result of wide validity, which can be proven in some 
instances, and in others is confirmed by comparison with numerics. Our 
treatment considers integrable and chaotic situations, as well as random matrix 
models for the latter. The concept of integrability and chaos carries over to 
situations with no classical analogue by considering the behaviour of 
correlation functions but is supported by spectral statistics. 

When surveying experiments, we find that measurements of the fidelity amplitude 
are possible in quantum optics due to a proposal by Gardiner Cirac and Zoller, which is somewhat 
dated, but has never been strictly implemented, though related more complicated 
quantities have been measured. On the other hand, fidelity of S-matrix elements 
has been tested directly or via Fourier transform of measurements in the energy 
domain in elastic and microwave experiments, respectively. For chaotic systems scattering fidelity is a good approximation for the
standard fidelity, particularly if ensemble averages are taken as was the 
case in both experiments. Then a good agreement with the results of random 
matrix theory was observed. 
The experiments performed show clearly, that effects as observed in the energy 
domain have no clear signature, thus justifying the general interest in the 
time domain studies. The experiments are at this point limited to weak and 
intermediate perturbation strength. The intriguing Lyapunov regime, 
which shows decay independent of perturbation strength for some time range and 
sufficiently strong perturbation has not yet been reached, and neither has the 
revival at Heisenberg time, predicted by exact solutions of the random matrix 
models. Both experiments will be quite challenging, as they imply measurements 
of very low fidelities.

While we have concentrated on the correlation function approach, we mention 
other approaches, that were used to obtain results in different regimes. 
Vanicek's semi classical approach seems to be the most flexible, as he is also 
able to obtain essentially all the regimes. Yet we have treated systems such as 
kicked spin chains that have no known classical analogue and nevertheless can 
be treated with the correlation function method.

Fidelity decay of mixed systems has been treated on the margin only, as only 
limited studies exist, and the matter  seems in first approximation to result 
in a separate treatment of integrable and chaotic parts.

For future developments, the evolution of expectation values of relevant 
operators seems to be a promising field, but also scattering fidelity has at 
this point only been used where it is equivalent to fidelity in a chaotic 
system. Perturbations of scattering channels are an interesting alternative. 
Also the use of quantum freeze of fidelity in information processes presents 
interesting perspectives. 

\ack

We acknowledge the hospitality of Dr. Brigitte Ibing and of the Centro
Internacional de Ciencias, Cuernavaca, which hosted us during a significant 
part of the effort to write this article. We are grateful to L. Kaplan, H.M. 
Pastawski, H.-J. Stoeckmann and J. Vanicek who sent us many useful comments 
after reading the manuscript. We also wish to thank G. Benenti, R. Blatt, 
G. Casati, B. Georgeot, R. Jalabert, H. Haefner, H. Kohler, T. Kottos, C. Lewenkopf, 
F. Leyvraz, C. Pineda, M. Raizen, W. Strunz and G. Veble.
Financial support from the Slovenian Research Agency under program P1-0044
and project J1-7437, CONACyT projects \# 41000-F and 48937-F as well as 
UNAM-DAPA project IN-101603F are acknowledged. TG acknowledges financial support from the 
EU Human Potential programme No. HPRN-CT-2000-00156 at the initial 
stage of the work. M\v Z acknowledges a fellowship of 
Alexander von Humboldt Foundation and a hospitality of the Department of 
Quantum Physics of the University of Ulm.

\appendix

\section{Time averaged fidelity}

\label{sect:appTA}

In this appendix we discuss and illustrate the time-averaged fidelity for different
initial states (eigenstates of $H_0$, random pure states, or completely mixed states)
and in the crossover regime between weak and strong perturbation 
\cite{Prosen:02corr,Emerson:03}.

The point of crossover $\epsilon_{\rm rm}$ from weak
(\ref{eq:Fnavgweak}) to strong (\ref{eq:Fnavgstrong}) perturbation regime is system 
dependent and can not be discussed in general apart from expecting it to scale with 
$\hbar$ similarly as a mean level spacing $\epsilon_{\rm rm} \sim \hbar^{d}$.
We will discuss the value of $\bar{F}$ for three different initial states:
\begin{enumerate}
\item[(i)] 
First, let us consider the simplest case when the initial state is an {\em eigenstate} of $U_0$ say, 
$\rho=\ket{E_1}\bra{E_1}$ with matrix elements $\rho_{lm} = \delta_{l,1}\delta_{m,1}$. For {\em weak perturbations} 
this gives (\ref{eq:Fnavgweak}) $\bar{F}_{\rm weak}=1$, therefore the fidelity does not decay at all. This result can be 
generalized to the case when $\rho$ is a superposition or even a mixture of a 
number of eigenstates, say $K$ of them, all with approximately 
the same weight, so that one has diagonal density matrix elements of order $\rho_{ll} \sim 1/K$, resulting in 
$\bar{F}_{\rm weak}\sim 1/K$. On the other hand, for {\em strong perturbations} $\eps\gg\eps_{\rm rm}$ we get 
$\bar{F}_{\rm strong}=(4-\beta)/{N}$ for an initial eigenstate. Here $\beta=1$ for systems with anti-unitary symmetry 
where the eigenstates of $H_0$ can be chosen real and $\beta=2$ for the case where anti-unitary symmetries are absent. 
Summarizing, for an initial eigenstate we have time 
averaged values of fidelity
\begin{equation}
\bar{F}_{\rm weak}=1,\qquad \bar{F}_{\rm strong}=(4-\beta)/{N}.
\label{eq:Fnavgeig}
\end{equation}
With this simple result we can easily explain the numerical result of Peres~\cite{Peres:95} where 
almost no decay of fidelity was found for a coherent initial state
sitting in the center of an elliptic island, thus being a superposition of a very small number of eigenstates 
(it is almost an eigenstate). The behavior in a generic case may be drastically different as described in the 
present work.

\item[(ii)]
Second, consider the case of a {\em random pure} initial state $\ket{\Psi}=\sum_m c_m\ket{E_m}$, giving 
$\rho_{ml}=c_m c_l^*$. The coefficients $c_m$ are independent random complex Gaussian variables with variance 
$1/{N}$, resulting in averages $\< |\rho_{lm}|^2 \>=1/{N}^2$ for $m\neq l$ and $\< \rho_{ll}^2 \>=2/{N}^2$ 
(average is over Gaussian distribution of $c_m$). Using this in the expressions for average 
fidelity (\ref{eq:Fnavgweak}) and (\ref{eq:Fnavgstrong}) we get
\begin{equation}
\bar{F}_{\rm weak}=2/{N},\qquad \bar{F}_{\rm strong}=1/{N}.
\label{eq:Fnavgrandom}
\end{equation}
% For random initial state there is therefore only a factor of $2$ difference 
% between finite size fluctuating plateau for 
% weak and for strong perturbation. 
The result for weak perturbations agrees with case (i) where we had 
$\bar{F}_{\rm weak} \sim 1/K$ if there were $K$ participating eigenvectors.

\item[(iii)]
Taking a non-pure initial density matrix $\rho= N^{-1}\, \mathbbm{1}$, we have
\begin{equation}
\bar{F}_{\rm weak}=1/{N},\qquad \bar{F}_{\rm strong}=(4-\beta)/{N}^2.
\label{eq:Fnavgavg}
\end{equation}
As expected, the fluctuating plateau is the smallest for maximally mixed states 
and strong perturbation.
\end{enumerate}

\begin{figure}[h]
\centerline{\includegraphics[width=\figw\textwidth]{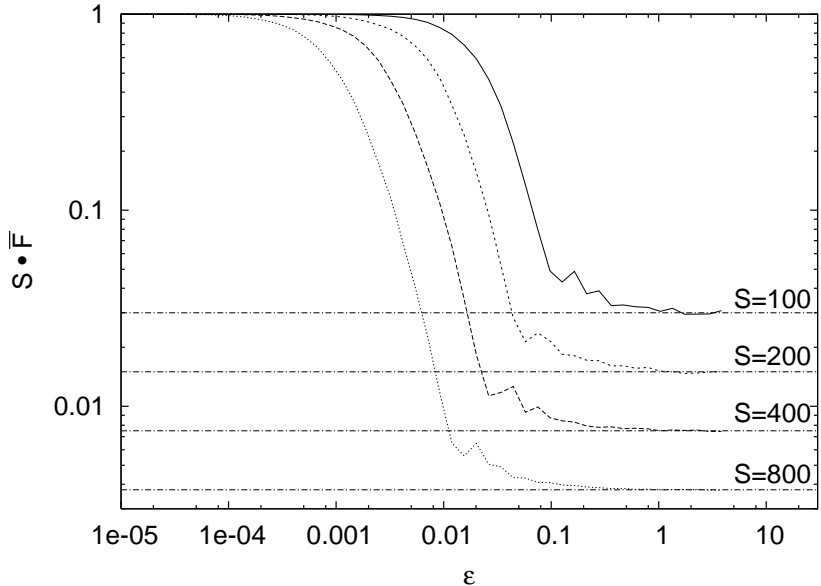}}
\caption{Dependence of time averaged fidelity (multiplied by the
Hilbert space size ${N}=S$) $\bar{F}$ on $\eps$ is shown for a
chaotic kicked top system and Hilbert space average
$\rho=\mathbbm{1}/{N}$, i.e. our case (iii). The transition from
weak to strong perturbation regime is seen
(\ref{eq:Fnavgavg}). Horizontal full lines are the theoretical predictions $\bar{F}_{\rm strong}$ (\ref{eq:Fnavgavg}), 
while the theoretical result for the weak regime corresponds to $1$.} 
\label{fig:avgFchaotic}
\end{figure}

Observe that the average fidelity $\bar{F}$ (\ref{eq:Fnavg}) is of
fourth order in matrix elements of $O$, the same as the inverse
participation ratio (IPR) of the perturbed eigenstates. Actually, in
the case of initial eigenstate, our case (i), the average fidelity
(\ref{eq:Fnavg}) can be rewritten as $\bar{F}=\sum_m{|O_{1m}|^4}$,
which is exactly the IPR. We recall that $O_{mn}=\braket{E_m}{E^\eps_l}$ 
is the overlap matrix between perturbed and unperturbed eigenstates. 
The inverse of the IPR, {\em i.e.} the participation ratio, is a number 
between $1$ and ${N}$ which can be thought of as giving the
approximate number of unperturbed eigenstates represented in the
expansion of a given perturbed eigenstate. For an average over the mixed states, case (iii), we
have instead $\bar{F}=\sum_{l,m}{|O_{lm}|^4}/{N}^2$, i.e. the
average IPR divided by ${N}$. The time averaged fidelity is thus directly related to the
{\em localization} properties of eigenstates of $U_\eps$ in terms of eigenstates of $U_0$. 
% However, except for the pathological case of the initial state being a small 
% combination of eigenstates of $U_0$ with weak perturbation, the
% fidelity fluctuation is always between the limiting values 
% $2/{N}$, and $3/{N}^2$. 
Therefore, fidelity will decay only until it reaches the value of finite size fluctuations and will 
fluctuate around $\bar{F}$ thereafter. The time $t_\infty$ when this happens, 
$F(t_\infty)=\bar{F}$, depends on the decay of fidelity and is discussed in Section~\ref{sec:timescales}.

\begin{figure}[ht]
\centerline{\includegraphics[width=\figw\textwidth]{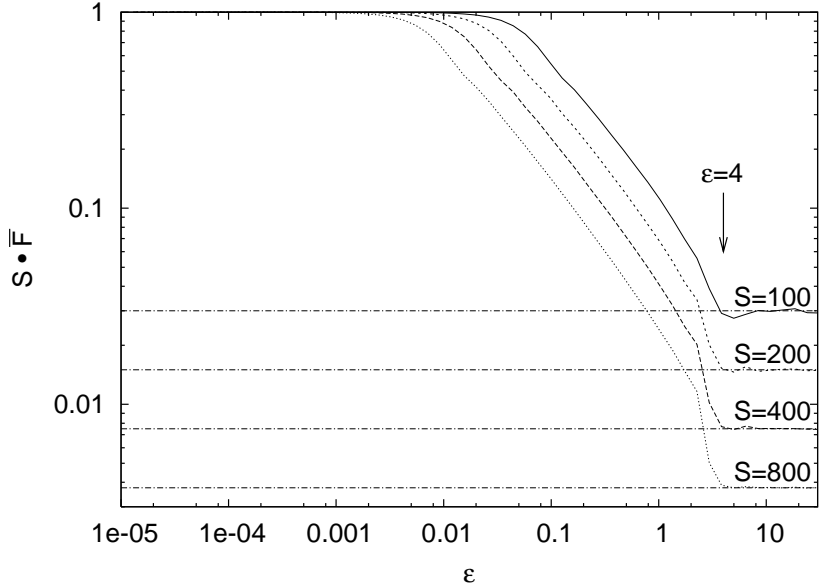}}
\caption{The same as Fig.~\ref{fig:avgFchaotic} but for a regular kicked top.}
\label{fig:avgFregular}
\end{figure}

To illustrate the above theory we have calculated the average fidelity (\ref{eq:Fnavg}) for a kicked top
with a propagator (\ref{eq:KTdef}). As an initial state we used $\rho=\mathbbm{1}/{N}$, i.e. the case (iii), 
where the dimension of the Hilbert space is determined by the spin magnitude: ${N}=S$ (OE subspace). We calculated the 
dependence of $S\, \bar{F}$ on $\eps$ for two cases: a chaotic one for kicked top parameters $\alpha=30$, $\gamma=\pi/2$ 
shown in Fig.~\ref{fig:avgFchaotic} and a regular one for $\alpha=0.1$, $\gamma=\pi/2$ shown in 
Fig.~\ref{fig:avgFregular}. 
In both cases one can see a transition from the weak perturbation regime $\bar{F}_{\rm weak}=1/{N}$ to the strong regime 
$\bar{F}_{\rm strong}=3/{N}^2$ for large $\eps$. In the chaotic case the critical $\eps_{\rm rm}$ can be seen to scale 
as $\eps_{\rm rm} \sim \hbar=1/S$. In the regular situation, the strong perturbation regime is reached only for a  
strong perturbation $\eps \sim 4$, where the propagator $U_\eps$ itself becomes chaotic. (The transition from 
the regular to chaotic regime in the kicked top happens at around $\alpha=3$, see e.g.~\cite{Peres:95}.) 
Still, if one defines 
$\eps_{\rm rm}$ as the points where the deviation from the weak regime starts (point of departure from $1$ in 
Fig.~\ref{fig:avgFregular}) one has scaling $\eps_{\rm rm}\sim 1/S$ also in the regular regime.  

\section{Models for numerics}
\label{sec:models}

Here we shall give a brief overview of models we used to perform numerical experiments to demonstrate various theoretical results. Except for several occasions, we use kicked systems, for which the Hamiltonian is time periodic, consisting of free time evolution, interrupted periodically by instantaneous ``kicks''. For such systems, time is a discrete variable, given by the number of kicks. These systems can frequently be simulated much more efficiently, than time-independent systems, if chaoticity is required. Most extensively, we use the kicked top (Section~\ref{sec:ktop}) and the kicked Ising chain (Section~\ref{sec:KI}). In some occasions, we use the sawtooth map~\cite{CarMei81} and the famous 
kicked rotor~\cite{chirikov}. In addition to kicked models we also use the Jaynes-Cummings system, well known in quantum optics~\cite{Qoptics}, and a system of coupled anharmonic oscillators. These are the only specific numerical models with a time independent Hamiltonian we are discussing in this review. Jaynes-Cummings model is used only once (Fig.~\ref{fig:bigwig}), and for further details, we refer the interested reader to Ref.~\cite{Prosen:03evol}, from which Fig.~\ref{fig:bigwig} has actually been taken. Anharmonic oscillators are used in Section~\ref{sec:compInt} with more details available in the original Ref.~\cite{Znidaric:05b}.

\subsection{The Kicked Top}
\label{sec:ktop}

The kicked top has been introduced by Haake, Ku\' s and Scharf~\cite{Haake:87} and has served as a numerical model in numerous studies ever since~\cite{Haake:91,Shack:94,Fox:94,Alicki:96,Miller:99,Breslin:99}. The kicked top might also be experimentally realizable~\cite{Haake:00}. We shall use different versions of the kicked top leading to different propagators depending on the phenomenon we want to illustrate. All of them are composed as a product of unitary propagators, each depending on standard spin operators, $S_{\rm x,y,z}$. The latter fulfill the following commutator relations: $[S_k,S_l]=\ii \varepsilon_{klm}\,S_m$ ($\varepsilon_{klm}$ is the Levi-Civita symbol). The half-integer (integer) spin $S$ determines the size of the Hilbert space, since ${\mathcal N}=2S+1$, and therefore the value of the effective Planck constant $\hbar=1/S$. The semiclassical limit corresponds to the limit of large spins $S\to\infty$. 

Mostly we use the standard kicked top whose one step propagator (Floquet operator) is given by
\begin{equation}
U_\eps=\exp{(-\ii \gamma S_{\rm y})} \exp{\left(-\ii (\alpha+\eps) \frac{S_{\rm z}^2}{2S} \right) }.
\label{eq:KTdef}
\end{equation}
Here $\eps$ determines the perturbation strength, while the parameters $\alpha$ and $\gamma$ are two parameters, which allow to change the dynamical properties of the system. The propagator for $t$ steps is simply a power $(U_\eps)^t$. The Hamiltonian which generates $U_\eps$ is given by
\begin{equation}
H_\eps=(\alpha+\eps)\frac{1}{2}\left(\frac{S_{\rm z}}{S}\right)^2+\gamma \frac{S_{\rm y}}{S} \sum_{k=-\infty}^{\infty}{\delta(t-k)},
\end{equation} 
where the perturbation is
\begin{equation}
V=\frac{1}{2}\left( \frac{S_{\rm z}}{S} \right)^2,
\label{eq:KTV}
\end{equation}
which has the classical limit $V \to v=z^2/2$ (we use lower case letters for corresponding classical observables). In that limit $S \to \infty$, and the area preserving map corresponding to $U_\eps$ can be written as a map on the unit sphere. Its explicit form can be obtained from the Heisenberg equations for the spin operators. The angle $\gamma$ in Eq.~(\ref{eq:KTdef}) is usually set to $\pi/2$, whereas we shall either use $\gamma=\pi/2$ or $\gamma=\pi/6$. For these two cases, the decay of the classical correlation function displays two different behaviors, monotonic decay for $\gamma=\pi/6$ and oscillatory decay for $\gamma=\pi/2$, see Fig.~\ref{fig:classcor}. In addition, the symmetries are different. For $\gamma=\pi/2$ the propagator $U_0$ commutes with the operator $\exp{(-\ii \pi S_{\rm y})}$, which describes a $\pi$ rotation around the ${\rm y}$-axis. The Hilbert space can be decomposed into three invariant subspaces. Following Peres's book~\cite{Peres:95} we denote them with ${\mathcal H}_{\rm EE}$, ${\mathcal H}_{\rm OO}$ and ${\mathcal H}_{\rm OE}$, with the basis states 
\begin{equation}
\begin{array}{lll}
{\rm EE :} & \ket{0},\left\{ \ket{2m}+\ket{-2m}\right\}/\sqrt{2} & {\mathcal N}_{\rm EE}=S/2+1 \\
{\rm OO :}& \left\{ \ket{2m-1}-\ket{-(2m-1)}\right\}/\sqrt{2} & {\mathcal N}_{\rm OO}=S/2  \\
{\rm OE :}& \left\{ \ket{2m}-\ket{-2m}\right\}/\sqrt{2},\left\{\ket{2m-1}+\ket{-(2m-1)} \right\}/\sqrt{2}\; & {\mathcal N}_{\rm OE}=S,
\end{array}
\label{eq:KTsubspaces}
\end{equation}
where we assume $S$ to be even. Here, $m$ runs from $1$ to $S/2$ and $\ket{m}$ are standard eigenstates of $S_{\rm z}$. For $\gamma \neq \pi/2$ the subspaces ${\mathcal H}_{\rm EE}$ and ${\mathcal H}_{\rm OO}$ coalesce and we have just two invariant subspaces. Unless stated otherwise, we always work in the subspace ${\mathcal H}_{\rm OE}$. Regardless of the angle $\gamma$, the kicked top propagator (\ref{eq:KTdef}) has an anti-unitary symmetry. The parameter $\alpha$ determines the degree of chaoticity of the corresponding classical map. The classical map is mostly regular for small values of $\alpha$ (by regular we mean close to integrable), at $\alpha \sim 3$ (see {\it e.g.} Ref.~\cite{Peres:95}) most of the tori disappear while for larger $\alpha > 3$ the system becomes numerically fully chaotic. We use $\gamma=\pi/2$ and $\alpha=0.1$ to simulate regular dynamics. The small value of $\alpha$ allows to use the integrable dynamics for $\alpha=0$, to compute $\bar{v}$ in Eq.~(\ref{eq:Vbardef}) in an approximate way: $\bar{v}\approx (1-z^2)/4$. To simulate chaotic dynamics, we use $\gamma= \pi/2$ or $\pi/6$ and $\alpha=30$. For $\gamma=\pi/6$ the integral of the classical correlation function is $\sigma_{\rm cl}=0.0515$ while it is much smaller, $\sigma_{\rm cl}=0.00385$, for $\gamma=\pi/2$ due to the oscillating nature of the correlation function.

In Section~\ref{sec:ch4avgF} we describe fidelity decay averaged over the position of initial coherent states for regular dynamics and a perturbation with non-zero time average. For that purpose we use a different perturbation (as compared to the propagator in Eq.~(\ref{eq:KTdef})):
\begin{equation}
U_\eps=\exp{(-\ii (\gamma+\varepsilon) S_{\rm y})} \exp{\left(-\ii \alpha \frac{S_{\rm z}^2}{2S} \right) },
\label{eq:KT1def}
\end{equation}
with $\alpha=0.1$ and $\gamma=\pi/2$ as before.

A third version of the kicked top is used to demonstrate the fidelity freeze. For the chaotic case discussed in Section~\ref{sec:chfreeze} we use the following propagator
\begin{equation}
U_\eps=\exp{\left(-\ii \alpha \frac{S_{\rm z}^2}{2S} \right)} \exp{\left(-\ii \frac{\pi}{2} S_{\rm y} \right)} \exp{\left(-\ii \eps \frac{S_{\rm x}^2-S_{\rm z}^2}{2S} \right)},
\label{eq:KT2def}
\end{equation}
with $\alpha=30$. For that perturbation, there is an associated operator $W$ such that $V$ is given as a time-derivative [see Eq.~(\ref{eq:Wdef})] of
\begin{equation}
W=\frac{1}{2} \left(\frac{S_{\rm z}}{S}\right)^2 .
\label{eq:W1}
\end{equation}
Note that in contrast to the time-independent case, considered in Section~\ref{D}, for kicked systems one has~\cite{Prosen:05} $V=W_1-W_0$, where $W_t=U_0^{-t}WU_0^t$. For the semiclassical calculation of the plateau height the classical limit $w=z^2/2$ is needed.
The long time decay beyond $t_2$, the duration of the freeze, is governed by the operator $R$ (\ref{eq:GR}). In the present case, it is given by
\begin{equation}
R=-\frac{S_{\rm x}S_{\rm y}S_{\rm z}+S_{\rm z}S_{\rm y}S_{\rm x}}{2S^3}.
\label{eq:R1}
\end{equation}
The classical correlation integral corresponding to this operator is $\sigma_{\rm R}=5.1\cdot 10^{-3}$.

In Section~\ref{sec:ch4Veq0}, fidelity freeze in regular systems is discussed. There, we use the propagator
\begin{equation}
U_\eps=\exp{\left(-\ii \alpha \frac{S_{\rm z}^2}{2S} \right)} \exp{(-\ii \eps S_{\rm x} )},
\label{eq:KT3def}
\end{equation}
with $\alpha=1.1$. For the calculation of the plateau, one needs to represent the unperturbed system in action angle variables. In a spherical coordinate system,these variables are given by
\begin{equation}
x = \sqrt{1-j^2}\cos\theta, \qquad y = \sqrt{1-j^2}\sin\theta, \qquad z = j.\qquad
\label{eq:can}
\end{equation}
The classical limit of the perturbation, expressed in terms of the action $j$ and the angle $\theta$ is $v=\sqrt{1-j^2}\cos{\theta}$, while the classical limit $w$ of the associated operator $W$ reads as
\begin{equation}
w(j,\theta)=\frac{1}{2}\sqrt{1-j^2}\,
\frac{\sin{(\theta-\omega/2)}}{\sin{(\omega/2)}}, 
\label{eq:wclas}
\end{equation}
where $\omega=\alpha j$ is the frequency of the unperturbed system.

In Section~\ref{sec:ktopRMTfreeze}, discussing quantum freeze in RMT, we need a propagator having an anti-unitary or a unitary symmetry, {\em i.e.}, belonging to COE or CUE class. So far all kicked top models described had an orthogonal symmetry. Specifically, to demonstrate freeze for a perturbation  that breaks time-reversal symmetry we used
\begin{equation}
U_\eps^{\rm sym}=P^{1/2} \exp{(-\ii \pi S_{\rm y}/2.4)} P^{1/2} \exp{(-\ii \eps S_{\rm x})}, \quad P=\exp{\left(-\ii \alpha \frac{S_{\rm z}^2}{2S}-\ii S_{\rm z}\right)}, 
\label{eq:ktopGOEsym}
\end{equation}
with standard $\alpha=30$. For $\eps=0$ the propagator belongs to COE symmetry class while only unitary symmetry remains for nonzero perturbation $\eps$. Symmetrized version of the propagator is chosen (half of twist $P$ before and half after the rotation) in order to have zero time average of the perturbation $S_{\rm x}$. In the eigenbasis of unperturbed $U_0^{\rm sym}$ the matrix of $S_{\rm x}$ is complex antisymmetric. The classical correlation integral of the perturbation is $\sigma_{\rm cl}=0.16$. To see the radical difference between the freeze and non-freeze fidelity decay we also show numerics for an unsymmetrized propagator,
\begin{equation}
U_\eps^{\rm nosym}=P \exp{(-\ii \pi S_{\rm y}/2.4)} \exp{(-\ii \eps S_{\rm x})},
\label{eq:ktopGOEnosym}
\end{equation}
for which the matrix for $S_{\rm x}$ does not have zeroes on the diagonal in the eigenbasis of $U_0^{\rm nosym}$. Evolution with $U_\eps^{\rm nosym}$ will therefore not result in a quantum freeze, whereas for a symmetrized version $U_\eps^{\rm sym}$ we are going to have freeze.

To show the agreement between RMT for the unitary case and numerics we used the following one step propagator 
\begin{equation}
U_\eps=P \exp{\left(-\ii \frac{\pi}{2.4} S_{\rm y}\right)} \exp{\left(-\ii 10 \frac{S_{\rm x}^2}{2S}-\ii(1+\eps)S_{\rm x}\right)},
\label{eq:ktopguefreeze}
\end{equation}
with the same $P$ as for the orthogonal case (\ref{eq:ktopGOEsym}). In this case the integral of the correlation function is slightly larger, $\sigma_{\rm cl}=0.17$. Note that in order to see freeze for this case we have to set the diagonal elements of the perturbation to zero by hand.

\subsubsection{Two coupled kicked tops}

On several occasions we will use a double kicked top. Its propagator is composed of the coupling term and two single kicked top propagators.

To illustrate the dependence of fidelity freeze in chaotic systems (described in Section~\ref{sec:chfreeze}) on the number of degrees of freedom we use two coupled kicked tops with the one step propagator
\begin{equation}
U_\eps=\exp{(-\ii \kappa S_{\rm z}^{\rm c}S_{\rm z}^{\rm e})} \exp{(-\ii \pi S_{\rm y}^{\rm c}/2)}\exp{(-\ii \pi S_{\rm y}^{\rm e}/2)} \exp{(-\ii \eps V S)},
\label{eq:2KTdef}
\end{equation}
with a strong coupling $\kappa=20$ to ensure chaotic dynamics. The perturbation $V$ is chosen similar to that of a single kicked top (\ref{eq:KT2def}) as
\begin{equation}
V=\frac{(S_{\rm x}^{\rm c})^2-(S_{\rm z}^{\rm c})^2+(S_{\rm x}^{\rm e})^2-(S_{\rm z}^{\rm e})^2}{2S^2}.
\end{equation}
This perturbation can be generated from
\begin{equation}
W=\frac{1}{2}\left(\frac{S_{\rm z}^{\rm c}}{S}\right)^2+\frac{1}{2}\left(\frac{S_{\rm z}^{\rm e}}{S}\right)^2.
\label{eq:2W}
\end{equation}
With the superscripts ${\rm c}$ and ${\rm e}$, we denote the Hilbert spaces (``central'' system or ``environment''), on which the respective operator acts. For numerical simulations we desymmetrized the system. For that purpose, we project the initial state to an invariant subspace of dimension ${\mathcal N}=S(S+1)$ spanned by $\{ {\mathcal H}_{\rm OE} \otimes {\mathcal H}_{\rm r} \}_{\rm sym}$, where ${\mathcal H}_{\rm r}={\mathcal H} \setminus {\mathcal H}_{\rm OE}$ and $\{\cdot\}_{\rm sym}$ is a subspace symmetric with respect to the exchange of the two tops. The integral of the classical correlation function corresponding to the operator $R$ is $\sigma_{\rm R}=9.2\cdot 10^{-3}$.

 To study the reduced fidelity and echo purity for regular systems (Section~\ref{sec:compInt}), we use
\begin{equation}
U_\eps=\exp{(-\ii \gamma_{\rm c} S_{\rm y}^{\rm c})}\exp{(-\ii \gamma_{\rm e} S_{\rm y}^{\rm e})} \exp{(-\ii \eps (S_{\rm z}^{\rm c})^2 (S_{\rm z}^{\rm e})^2/S^3)},
\label{eq:3KTdef}
\end{equation}
where it is important to choose incommensurate parameters, $\gamma_{\rm c}=\pi/2.1$ and $\gamma_{\rm e}=\pi/\sqrt{7}$ to avoid resonant behavior. Note that here the unperturbed dynamics is uncoupled. Therefore echo purity and purity coincide. For this system, the classical actions are given by the coordinates $y_{\rm c,e}$, so the classical time averaged perturbation is
\begin{equation}
\bar{v}=\frac{1}{4}(1-j_{\rm c}^2)(1-j_{\rm e}^2).
\label{eq:3vbar}
\end{equation}

In Sections~\ref{sec:timescales} and~\ref{sec:chcomposite} we use
\begin{equation}
U_\eps=U_{\rm c} U_{\rm e} 
\exp{(-\ii (\kappa+\eps) S_{\rm z}^{\rm c} S_{\rm z}^{\rm e}/S)},
\label{eq:4KTdef}
\end{equation}
with standard single kicked top propagators
\begin{equation}
U_{\rm c,e}=\exp{(-\ii \gamma_{\rm c,e} S_{\rm y}^{\rm c,e})} \exp{\left(-\ii \alpha_{\rm c,e} \frac{1}{2}\left(\frac{S_{\rm z}^{\rm c,e}}{S}\right)^2 \right)}.
\end{equation}
In Section~\ref{sec:timescales} where we compare timescales of fidelity decay in regular and chaotic system we use $\alpha_{\rm c,e}=0$, $\gamma_{\rm c,e}=\pi/2$ and the coupling $\kappa=5$ for chaotic and $\kappa=1$ for regular dynamics. For chaotic dynamics the integral of the correlation function is $\sigma_{\rm cl}=0.058$. On the other hand in Section~\ref{sec:chcomposite} demonstrating purity decay for chaotic dynamics we use $\alpha_{\rm c,e}=30$ and $\gamma_{\rm c,e}=\pi/2.1$ and uncoupled unperturbed system, $\kappa=0$, giving the integral of the correlation function $\sigma_{\rm cl}=0.056$.

In all single kicked top simulations, the initial state $\ket{\psi(0)}$ used for the computation of fidelity decay, is either a random state with the expansion coefficients $c_m=\braket{m}{\psi(0)}$ being independent Gaussian complex numbers or a coherent state. For double kicked top the initial coherent states are products of coherent states for each top. A coherent state centered at the position 
$\mathbf{r}^*=(\sin{\vartheta^*}\cos{\varphi^*},\sin{\vartheta^*}\sin{\varphi^*},\cos{\vartheta^*})$ is given by
\begin{align}
\ket{\vartheta^*,\varphi^*} &=\sum_{m=-S}^S{ {2S \choose S+m}^{1/2} \cos^{S+m}{(\vartheta^*/2)}\sin^{S-m}{(\vartheta^*/2)} {\rm e}^{-\ii m \varphi^*}\ket{m} } \notag \\
&= \frac{{\rm e}^{-\ii \varphi^* S}}{(1+|\tau|^2)^S} \exp{(\tau S_-)}\ket{S},\qquad \tau={\rm e}^{\ii \varphi^*}\tan{(\vartheta^*/2)}, 
\label{eq:SU2coh}
\end{align}
with $S_\pm=S_{\rm x}\pm \ii S_{\rm y}$. In the semiclassical limit of large spin $S$ the expansion coefficients of the above state tend to $c_m= \la m|\vartheta^*,\varphi^*\ra 
\asymp \exp{(-S(m/S-z^*)^2/2(1-z^{*2}))}{\rm e}^{-\ii m \varphi^*}$, therefore the squeezing parameter is
\begin{equation}
\Lambda=\frac{1}{\sin^2\vartheta^*} \; .
\label{eq:squeeze}
\end{equation}
Coherent states have a well defined classical limit and this enables us to compare quantum fidelity for coherent initial states with the corresponding classical fidelity. The initial classical phase space density corresponding to a coherent state is~\cite{Fox:94}
\begin{equation}
\rho_{\rm clas}(\vartheta,\varphi)=\sqrt{\frac{2S}{\pi}} \exp{\{-S \lbrack (\vartheta-\vartheta^*)^2+(\varphi-\varphi^*)^2 \sin^2{\vartheta} \rbrack \}}.
\label{eq:rhoclasSU2}
\end{equation}
The above density is square-normalized as $\int{\!\rho_{\rm clas}^2 {\rm d}\Omega}=1$.

\subsection{The kicked Ising-chain}

\label{sec:KI}

As an example of a generic interacting quantum many body model, which may 
be of some interest also for illustrations relevant for quantum information,
we consider the kicked Ising (KI) model~\cite{Prosen:02,Prosen:02spin} with 
the Hamiltonian
\begin{equation}
H_{\rm KI}(t) = \sum_{j=0}^{L-1} \left\{
J_z\, \sigma^j_z\sigma_z^{j+1} + \delta_p(t)\; (h_x \sigma_x^j 
+ h_z \sigma_z^j)\right\}
\label{eq:hamKI}
\end{equation}
where $\delta_p(t)=\sum_m\delta(t-m p)$ is a periodic train of delta functions 
with period $p$. The operators $\sigma^j_{x,y,z}$ are the standard Pauli spin 
matrices satisfying the canonical commutation relations 
$[\sigma^j_\alpha,\sigma^k_\beta]= 2\ii\, \delta_{jk}\,
\epsilon_{\alpha\beta\gamma} \, \sigma^j_\gamma$. The model represents a
chain of $L$ interacting spins 1/2 with periodic boundary conditions
$L\equiv 0$, subject to a tilted magnetic field. Integrating over one period 
$p$ we arrive at the Floquet operator:
\begin{equation}
U=\exp(-\ii J_z\sum_j \sigma_z^j \sigma_z^{j+1})
\exp(-\ii\sum_j(h_x\sigma_x^j + h_z\sigma_z^j))
\end{equation}
where we chose units such that $p=\hbar=1$. The KI model depends on three 
independent parameters $(J_z,h_x,h_z)$. It is integrable for longitudinal 
($h_x=0$) and transverse ($h_z=0$) magnetic fields. The KI model
is likely to be non-integrable everywhere else.
Numerical investigations of the general case of a tilted magnetic field have
revealed finite regions of parameter space where the system has an 
ergodic and mixing behavior, or non-ergodic quasi-integrable behavior, 
when approaching the thermodynamic limit.
For the numerical illustrations of this paper 
we consider only three typical
cases - points on a line in 3d parameter space with
fixed parameters $J=1,h_x=1.4$ and a
varying parameter $h_z$ exhibiting three different types of 
dynamics~\cite{Prosen:02}: 
$h_z=0$ ({\em integrable}),
$h_z=0.4$ ({\em intermediate}, {\em i.e.} non-integrable and non-ergodic), 
and $h_z=1.4$ ({\em ergodic} and {\em mixing} or simply ``chaotic'').

The non-trivial integrability of a transverse kicking 
field, which somehow inherits the solvable dynamics of its well-known 
autonomous version~\cite{Nie67,CapPer77}, is quite remarkable since it was 
shown~\cite{ProsenPTPS} that the Heisenberg dynamics can be calculated 
explicitly for observables which are bilinear in the Fermi operators 
$c_j = (\sigma^y_j-i\sigma^z_j)\prod_{j'}^{j'<j}\sigma^x_{j'}$ with time 
correlations decaying to the non-ergodic stationary values as 
$ \sim t^{-3/2}$ \cite{ProsenPTPS}.

\section{Scattering fidelity in different approximations}

\subsection{Limit of weak coupling, diagonal S-matrix elements}

For non-diagonal S-matrix elements, the weak coupling limit has been treated in
Section~\ref{DOS}, with the result, Eq.~(\ref{DOS:sfres}). That was the 
important case, since it really makes sure that the scattering fidelity 
$f^s_{ab}(t)$ tends to the standard fidelity of a closed chaotic system.

For $a=b$, however, we find:
\begin{align}
\la\hat S_{aa}(t)^*\, \hat S'_{aa}(t)\ra &\sim 4\pi^2\; w_a\, w_b\; \theta(t)\;
  \e^{-\gamma_W\, t}\left\{ \sum_{j,l} \la |v_{ja}|^4\ra\; O_{jl}^2\;
  \e^{2\pi\rmi (E_j-E'_l) t} + \sum_{j\ne k,l} \la |v_{ja}|^2\ra\;
  \la |v_{ka}|^2\ra\; O_{kl}^2\; \e^{2\pi\rmi (E_j-E'_l) t} \right\}
\notag\\
&= 4\pi^2\; \frac{w_a\, w_b}{N}\; \theta(t)\; \e^{-\gamma_W\, t}\;
  \frac{1}{N}\sum_{j} \left\{ \sum_l \e^{2\pi\rmi (E_j-E'_l) t}
  \left( 3 O_{jl}^2 + \sum_{k\ne j} O_{kl}^2\right)\right\} \notag\\
&= 4\pi^2\; \frac{w_a\, w_b}{N}\; \theta(t)\; \e^{-\gamma_W\, t}\;
  \left\{ 3\; f(t) + \frac{1}{N}\sum_{jl} \e^{2\pi\rmi (E_j-E'_l) t}\;
  \sum_{k\ne j} O_{kl}^2 \right\} \; ,
\end{align}
whereas
\begin{equation}
\la |\hat S_{aa}(t)|^2\ra \sim 4\pi^2\; w_a\, w_b\; \theta(t)\;
  \e^{-\gamma_W\, t}\left\{ 3+ \frac{1}{N}\sum_j\sum_{l\ne j}
  \e^{2\pi\rmi (E_j-E_l) t} \right\} \; .
\end{equation}
While the first term separates again as desired into an auto-correlation part 
and the fidelity amplitude, the second term, in general, does not. Only in the
perturbative regime, where me may assume  $O_{kl}^2 = \delta_{kl}$ and
treat the perturbed spectrum to first order in $\lambda$, we again find
the factorization into fidelity amplitude and autocorrelation function.

%There are good reasons to expect the deviations from the factorization to be
%small, however, a numerical study seems to be necessary. For instance, it is
%not at all clear whether one might find the partial recovery in the scattering
%fidelity.

\subsection{Rescaled Breit-Wigner approximation and the perturbative regime}

For this purpose we refer to the rescaled Breit-Wigner approximation~\cite{GS02}, which amounts to a treatment of the
anti-Hermitian part in $H_{\rm eff}$ in first order perturbation theory. The
main point is that due to the rescaling this gives useful results even at
relatively strong coupling to the continuum. Here, we also treat the
perturbation $\lambda\, W$ in first order perturbation theory.
In the eigenbasis of $H_0$ we have
\begin{equation}
\la \hat S_{ab}(t)^*\, \hat S'_{ab}(t)\ra = 4\pi^2\theta(t)
  \sum_{jk} V_{ja}^*\; \e^{2\pi\rmi (E_j +\rmi\Gamma_j/2) t}\; V_{jb}\;
  V_{kb}^*\left( \sum_l O_{kl}^2\; \e^{-2\pi\rmi (E'_l-\rmi\Gamma_l/2) t}
  \right) V_{ka} \; ,
\end{equation}
where $O$ is again the orthogonal matrix that transforms the eigenbasis of
$H_0'$ into the eigenbasis of $H_0$. The numbers $\Gamma_j$ and $\Gamma'_j$
denote the diagonal element of the anti-Hermitian part of $H_{\rm eff}$ and
$H'_{\rm eff}$ in the respective bases. As we treat $\lambda\, W$ in first order perturbation theory we have $O_{kl}^2 = \delta_{kl}$, $\Gamma_j' = \Gamma_j$, and
$E_l'= E_l + \lambda W_{ll}$. Therefore
\begin{align}
\la \hat S_{ab}(t)^*\, \hat S'_{ab}(t)\ra &= 4\pi^2\theta(t)
  \sum_{jk} V_{ja}^*\, V_{jb}\, V_{kb}^*\, V_{ka}\;
  \e^{-\pi (\Gamma_j+\Gamma_k) t}\; \e^{-2\pi\rmi (E'_k - E_j) t}
\notag\\
 &= \sum_k \e^{-2\pi\rmi\lambda\, W_{kk}\, t} \sum_j
  V_{ja}^*\, V_{jb}\, V_{kb}^*\, V_{ka}\;
  \e^{-\pi (\Gamma_j+\Gamma_k) t}\; \e^{-2\pi\rmi (E_k - E_j) t}
  \; .
\end{align}
Under the assumption that $W$ is a random perturbation, uncorrelated with
the dynamics of the unperturbed system, we may average over
$\exp[2\pi\rmi\lambda\, W_{kk}\, t]$ independently, which yields the fidelity
amplitude $f(t)$. What remains is just
the autocorrelation function of the S-matrix element $S_{ab}$ and we find
\begin{equation}
\la \hat S_{ab}(t)^*\, \hat S'_{ab}(t)\ra =
f(t)\;  \la |\hat S_{aa}(t)|^2\ra \; .
\end{equation}
Note that is has not been necessary to actually perform the rescaling (of the
Breit-Wigner approximation). It just serves as an argument for the enlarged
region of validity of the approximation presented.

\subsection{Rescaled Breit-Wigner approximation and linear response}

Here, we use the rescaled Breit-Wigner approximation to obtain the decaying
dynamics for the unperturbed system. We then compute the scattering fidelity
in linear response approximation.
\begin{align}
&\qquad \hat S_{ab}(t)^*\, \hat S'_{ab}(t) = 4\pi^2\; w_a\, w_b\; \theta(t)\;
 \la v_b|\, U_0^\dagger\, |v_a\ra \; \la v_a|\, U_1\, |v_b\ra \notag\\
&\qquad U_1= \e^{-2\pi\rmi\, H'_{\rm eff}\, t} \qquad
U_0= \e^{-2\pi\rmi\, H_{\rm eff}\, t} = \sum_j |j\ra\;
  \e^{-2\pi\rmi (E_j- \rmi\Gamma_j/2) t}\; \la j| \; ,
\end{align}
Only $U_1$ depends on the perturbation $W$ (assumed random), so we may write
$U_1$ in linear response approximation and average the result over $W$:
\begin{equation}
U_1= U_0 \left\{ \one -2\pi\rmi\lambda\int_0^t\d\tau\; \tilde W(\tau)
 - 4\pi^2\lambda^2\int_0^t\d\tau\int_0^\tau\d\tau'\; \tilde W(\tau)\;
  \tilde W(\tau') \right\} \; ,
\label{U1}\end{equation}
where $\tilde W(t)= U_0(t)^{-1}\, W\, U_0(t)$. Note that $U_0$ is not unitary,
so that $U_0(t)^{-1} \ne U_0(t)^\dagger$. Averaging over $W$ yields:
\begin{equation}
\overline{U_1} = U_0 \left[ \one -4\pi^2\lambda^2\; \mathcal{C}(t) \right]
\qquad
\mathcal{C}(t) = \int_0^t\d\tau\int_0^\tau\d\tau'\;
  {\rm diag}\left[ 1+ \textstyle\sum_k \e^{-2\pi\rmi (E_j-E_k)\, \tau'}
     \e^{\pi (\Gamma_j - \Gamma_k)\, \tau'} \right] \; ,
\label{cint}\end{equation}
where we assumed that $W$ is taken from the Gaussian orthogonal ensemble,
such that
$\la W_{ij} W_{kl}\ra = \delta_{ik}\delta_{jl} + \delta_{il}\delta_{kj}$.
Apart from the additional factor $\e^{\pi (\Gamma_j - \Gamma_k)\, \tau'}$
we have here precisely the linear response expression of the fidelity
amplitude as obtained in~\cite{GPS04}. This factor, which depends on the
widths of the resonances, tends to smooth out possible level correlations.
However, the two dominant features, the quadratic decay for small $\lambda$
(perturbative regime) and the linear decay for larger $\lambda$ (Fermi golden
rule regime) remain present. In~\cite{GPS04}, we found that the level
correlations have at most a 15\% effect on the fidelity decay curve.

Note that both, $U_0$ and $\overline{U_1}$ are diagonal in the eigenbasis of
$H_0$. Therefore, we obtain for the cross correlation function above:
\begin{equation}
\hat S_{ab}(t)^*\, \hat S'_{ab}(t) = 4\pi^2\; w_a\, w_b\; \theta(t)\;
  \sum_{jk} v_{jb}^*\; \e^{2\pi\rmi (E_j +\rmi\Gamma_j/2)\, t}\;
     v_{ja}\; v_{ka}^*\; \e^{-2\pi\rmi (E_k -\rmi\Gamma_k/2)\, t}\;
  \left[ 1 -4\pi^2\lambda^2\; \mathcal{C}_k(t) \right]\;  v_{kb} \; .
\end{equation}
In the case $a\ne b$ this simplifies to:
\begin{equation}
\hat S_{ab}(t)^*\, \hat S'_{ab}(t) =
  4\pi^2\; w_a\, w_b\; \theta(t)\; \sum_j |v_{jb}|^2\, |v_{ja}|^2\;
  \e^{-2\pi\Gamma_j\, t}\; \left[ 1 -4\pi^2\lambda^2\; \mathcal{C}_j(t)
     \right]
\label{transa}\end{equation}
If we neglect the dependence of $\mathcal{C}(t)$ on the resonance widths, we
may average independently over the ``autocorrelation part'' in each term of he sum.
In this way, we obtain
\begin{equation}
\la \hat S_{ab}(t)^*\, \hat S'_{ab}(t)\ra = \la |\hat S_{ab}(t)|^2\ra \;
  \left[\textstyle 1 -4\pi^2\lambda^2\; N^{-1}\sum_j\mathcal{C}_j(t) \right]
= \la |\hat S_{ab}(t)|^2\ra \; f(t)\; ,
\end{equation}
i.e. a product of the autocorrelation function and the fidelity amplitude.
Note that again it was not necessary to actually perform the rescaling (of the
Breit-Wigner approximation).

For the case $a=b$ we obtain an additional term in Eq.~(\ref{transa}):
\begin{align}
\hat S_{aa}(t)^*\, \hat S'_{aa}(t) &= 4\pi^2\; w_a^2\; \theta(t)\;
  \sum_j |v_{ja}|^4\; \e^{-2\pi\Gamma_j\, t}\; \left[ 1 -4\pi^2\lambda^2\;
     \mathcal{C}_j(t) \right] \notag\\
&\qquad + 4\pi^2\; w_a\, w_b\; \theta(t)\; \sum_{j\ne k} |v_{ja}|^2\,
     |v_{ka}|^2\; \e^{-2\pi\rmi (E_j-E_k)\, t}\;
     \e^{-\pi (\Gamma_j+\Gamma_k)\, t}\; \left[ 1 -4\pi^2\lambda^2\;
     \mathcal{C}_k(t) \right] \; .
\end{align}
As long as level correlations are not important in the fidelity
amplitude $f(t)$, we may average over the autocorrelation part and the
fidelity amplitude part independently, and obtain the desired result:
\begin{equation}
\hat S_{aa}(t)^*\, \hat S'_{aa}(t) = \la |\hat S_{aa}(t)|^2\ra \; f(t)\; .
\end{equation}
It will be interesting to investigate situations where the separate averages
are still possible, but effects of the width fluctuations on the fidelity
amplitude, Eq.~(\ref{cint}) become noticeable.

\section{\label{app:EP} Echo purity in RMT: technicalities}

In Eq.~(\ref{Q2PE:echoop}), $\one$, $\tilde I$, and $\tilde J$ denote
operators (matrices) on the Hilbert space $\mathcal{H}$. Note that in general
$\tilde I$ is hermitian while $\tilde J$ is not.
Keeping only terms of order up to $(2\pi\eps)^2$, we find
\begin{align}
&\varrho^M(t)\otimes\varrho^M(t) = \big[\, (
  \one -2\pi\rmi\eps\; \tilde I - (2\pi\eps)^2\; \tilde J )\; \varrho_0\;
  ( \one +2\pi\rmi\eps\; \tilde I -
  (2\pi\eps)^2\; \tilde J^\dagger )\, \big ]\; \otimes\; \big [ \ldots \big ]
  + \mathcal{O}(\eps^3)\notag\\
&\qquad = \varrho_0\otimes\varrho_0 - 2\pi\rmi\eps \big[ \tilde I\, \varrho_0
  \otimes\varrho_0 - \varrho_0\, \tilde I \otimes \varrho_0 +
  \varrho_0\otimes \tilde I\, \varrho_0 - \varrho_0\otimes\varrho_0\, \tilde I
  \big ] \notag\\
&\qquad\quad + (2\pi\eps)^2\big[
  \tilde J\, \varrho_0 \otimes \varrho_0 +
  \varrho_0\, \tilde J^\dagger \otimes \varrho_0 +
  \varrho_0 \otimes \tilde J\, \varrho_0 +
  \varrho_0 \otimes \varrho_0\, \tilde J^\dagger \big] \notag\\
&\qquad\quad + (2\pi\eps)^2 \big[
  \tilde I\, \varrho_0\, \tilde I \otimes \varrho_0 -
  \tilde I\, \varrho_0 \otimes \tilde I\, \varrho_0 +
  \tilde I\, \varrho_0 \otimes \varrho_0\, \tilde I +
  \varrho_0\, \tilde I \otimes \tilde I\, \varrho_0 -
  \varrho_0\, \tilde I \otimes \varrho_0\, \tilde I +
  \varrho_0 \otimes \tilde I\, \varrho_0\, \tilde I \big] \notag\\
&\qquad\quad + \mathcal{O}(\eps^3) \; ,
\notag\end{align}
where $\varrho_0$ is completely general. If the initial state is separable:
$\varrho_0= \varrho_{\rm c}\otimes \varrho_{\rm e}$ then it can be shown
that the linear terms vanish. If, in addition, the initial state is pure
(in the full Hilbert space) than the above expression can be reduced to
Eq.~(\ref{eq:LRecho}) in Section~\ref{sec:LRE}. In what follows the linear 
terms are ignored, because they become zero when averaged over the
perturbation $V$.

The tensor form $p[\, .\, ]$ obeys a number of symmetry relations, which may be
used to reduce the number of terms in the above expression. By writing this
form, given in Eq.~(\ref{Q2PE:psym}), explicitely in the product basis of
the full Hilbert space $\mathcal{H}$, we find
\begin{equation}
p[A\otimes B] = p[A^\dagger\otimes B^\dagger]^* = p[B\otimes A]
  = p[B^\dagger \otimes A^\dagger]^* \; .
\label{appEP:sym}\end{equation}
Therefore we have
\begin{align}
&F_P(t) \approx p[\varrho_0\otimes\varrho_0] -2\, (2\pi\eps)^2
  \big(\, A_J - A_I \, \big) \notag\\
&\qquad A_J= 2\, {\rm Re}\; p[\tilde J\, \varrho_0 \otimes \varrho_0]
\label{appEP:LRbas}\\
&\qquad A_I= p[\tilde I\, \varrho_0\, \tilde I \otimes \varrho_0]
  - {\rm Re}\; p[\tilde I\, \varrho_0 \otimes \tilde I\, \varrho_0] +
  p[\tilde I\, \varrho_0 \otimes \varrho_0\, \tilde I] \; .
\notag\end{align}
If the initial state is pure and separable, this expression reduces to the one
in Eq.~(\ref{eq:LRecho}) in Section~\ref{sec:LRE} term by term.

We start by averaging over $V$. In the case of $A_J$, this is particularly
simple. Let us use Greek letters $\mu,\nu,\ldots$ as indeces which run over
the basis states in the full Hilbert space $\mathcal{H}$ (these basis states
need not be product states). In view of Eq.~(\ref{Q2PE:IuJ}) and applying
the rule $\la V_{\alpha\beta}\, V_{\gamma\delta}\ra = \delta_{\alpha\gamma}
\delta_{\beta\delta} + \delta_{\alpha\delta}\delta_{\beta\gamma}$ we obtain
\begin{equation}
J_{\mu\nu}= \delta_{\mu\nu}\; C_\mu(t)\qquad
C_\mu(t)=  \int_0^t\d\tau\int_0^\tau\d\tau' \left( 1 + \sum_\xi
  \e^{-2\pi\rmi\, (E_\mu - E_\xi)\, (\tau-\tau')} \right) \; .
\label{appEP:J}\end{equation}
We finally average $\tilde J$ over $O\in \mathcal{O}(N)$ the orthogonal group
and obtain:
\begin{equation}
\la \tilde J\ra = \sum_{\mu\nu} |\mu\ra \sum_\xi O_{\xi\mu}\; C_\xi(t)\;
  O_{\xi\nu}\; \la\nu| = \sum_\mu |\mu\ra \left(\textstyle
  \frac{1}{N}\sum_\xi C_\xi(t) \right) \la\mu| = \left(\textstyle
  \frac{1}{N}\sum_\xi C_\xi(t) \right) \; \one\; ,
\end{equation}
such that
\begin{equation}
A_J= 2\, \mathcal{C}(t)\; p[\varrho_0\otimes \varrho_0] \qquad
\mathcal{C}(t)= {\rm Re}\; \frac{1}{N}\sum_\xi C_\xi(t) =
\frac{1}{N}\sum_\xi C_\xi(t) \; ,
\end{equation}
as this sum is always real.
This is precisely the same correlation integral, which appeared in the linear
response study of the fidelity amplitude, Section~\ref{Q2LR}. As in
Eq.~(\ref{R:dav}), we may replace the sum of phases with the
spectral form factor. This leads to the expression in squared brackets
of Eq.~(\ref{R:rmt-lr}) for $\mathcal{C}(t)$. As
$I(0)= p[\varrho_0\otimes\varrho_0]$ factors out from $A_J$, we find:
\begin{equation}
F_P(t)= I(0) [1-4\, (2\pi\eps)^2\; \mathcal{C}(t)\, ]
  + 2\, (2\pi\eps)^2\; A_I \; .
\label{appEP:bas2}\end{equation}
If we could neglect the term $A_I$, the echo-purity decay of an initially pure state would be
given by the decay of the fourth power of the fidelity amplitude.

For the second part of the echo purity, we have to average the matrix
$\tilde I\, \varrho_0\, \tilde I$, as well as the higher rank tensors:
$\tilde I\, \varrho_0 \otimes \tilde I\, \varrho_0$ and
$\tilde I\, \varrho_0 \otimes \varrho_0\, \tilde I$. The average
over $V$ yield
\begin{align}
I_{\alpha\beta} \, I_{\gamma\delta} &=
  C_{\alpha\beta}^+\; \delta_{\alpha\gamma}\, \delta_{\beta\delta} +
  C_{\alpha\beta}^-\; \delta_{\alpha\delta}\, \delta_{\beta\gamma}\, \big) \; ,
\qquad C_{\alpha\beta}^\pm = \int_0^t\d\tau\int_0^t\d\tau'\;
  \e^{2\pi\rmi\, (E_\alpha-E_\beta)\, (\tau\pm\tau')}\; ,
\label{Q2PE:avI}\\
\tilde I_{\alpha\beta} \, \tilde I_{\gamma\delta} &=
  \sum_{\mu\nu\xi\rho} O_{\mu\alpha}\, I_{\mu\nu}\, O_{\nu\beta} \,\ldots\,
  O_{\xi\gamma}\, I_{\xi\rho}\, O_{\rho\delta} \; .
\end{align}
Generally, these terms appear within sums which run over all indeces
$\alpha, \beta, \gamma, \delta$, and for that case we find
\begin{align}
&\sum_{\alpha\beta\gamma\delta} \tilde I_{\alpha\beta}\,
\tilde I_{\gamma\delta}\; X_{\alpha\beta\gamma\delta} =
  \sum_{\alpha\beta\gamma\delta} \sum_{\mu\nu}
  O_{\mu\alpha}\, O_{\nu\beta} \big( C_{\mu\nu}^+\; O_{\mu\gamma}\,
  O_{\nu\delta} + C_{\mu\nu}^-\; O_{\nu\gamma}\, O_{\mu\delta}\, \big)\;
  X_{\alpha\beta\gamma\delta}
\notag\\
&\qquad = A \sum_{\alpha\ne\gamma} X_{\alpha\alpha\gamma\gamma}\; +\;
  \sum_{\alpha\ne\beta} \big ( \, B\; X_{\alpha\beta\alpha\beta}
  + C\; X_{\alpha\beta\beta\alpha} \, \big ) +
  D \sum_\alpha X_{\alpha\alpha\alpha\alpha} \notag\\
&\qquad = \sum_{\alpha\beta} \big ( \, A\; X_{\alpha\alpha\beta\beta} + B\;
  X_{\alpha\beta\alpha\beta} + C\; X_{\alpha\beta\beta\alpha} \, \big )
  + (D-A-B-C) \sum_\alpha X_{\alpha\alpha\alpha\alpha}\; ,
\end{align}
where $X$ is an arbitrary tensor, and
\begin{alignat}{2}
A &= \sum_{\mu\nu} \big(\, C_{\mu\nu}^+ +C_{\mu\nu}^-\, \big) \;
  O_{\mu\alpha}\, O_{\mu\gamma}\, O_{\nu\alpha}\, O_{\nu\gamma} \qquad
& B &= \sum_{\mu\nu} \big(\, C_{\mu\nu}^+\; O_{\mu\alpha}^2\, O_{\nu\beta}^2
  + C_{\mu\nu}^-\; O_{\mu\alpha}\, O_{\mu\beta}\, O_{\nu\alpha}\,
  O_{\nu\beta} \, \big) \notag\\
D &= \sum_{\mu\nu} \big(\, C_{\mu\nu}^+ +C_{\mu\nu}^-\, \big) \;
  O_{\mu\alpha}^2\, O_{\nu\alpha}^2 \qquad
& C &= \sum_{\mu\nu} \big(\, C_{\mu\nu}^+\; O_{\mu\alpha}\,
  O_{\mu\beta}\, O_{\nu\alpha}\, O_{\nu\beta} + C_{\mu\nu}^-\;
  O_{\mu\alpha}^2\, O_{\nu\beta}^2\, \big) \; .
\label{Q2PE:ABCD}\end{alignat}
Distinguishing between the cases $\mu=\nu$ and $\mu\ne\nu$ allows to
evaluate the group integrals. In this way, we obtain
\begin{alignat}{2}
A &= \frac{1}{N+2}\left[ 2\, t^2 - \frac{1}{N-1}\big(\,
  \mathcal{C}_+ + \mathcal{C}_- \,\big) \right] \; ,\qquad &
B &= \frac{1}{N+2}\left[ 2\, t^2 + \frac{1}{N-1}\big(\,
  (N+1)\; \mathcal{C}_+ - \mathcal{C}_- \,\big) \right] \; , \notag\\
D &= \frac{1}{N+2}\left[ 6\, t^2 + \mathcal{C}_+ + \mathcal{C}_- \right] \; ,
  \qquad &
C &= \frac{1}{N+2}\left[ 2\, t^2 +\frac{1}{N-1}\big(\, (N+1)\; \mathcal{C}_-
  - \mathcal{C}_+\,\big) \right] \; .
\end{alignat}
The new coefficients $\mathcal{C}_\pm$ are related to double integrals over
the spectral form factor as follows:
\begin{equation}
\mathcal{C}_\pm = \frac{1}{N}\sum_{\mu\ne\nu} C^\pm_{\mu\nu}
  \sim - B_\pm(t) + \begin{cases} 0 &: +\,\text{case}\\ t &: -\,\text{case}
  \end{cases} \; , \qquad
B_\pm(t) = \int_0^t\d\tau\int_0^t\d\tau'\; \la b_2(E,\, \tau\pm\tau')\ra
  \; .
\end{equation}
Due to the symmetric sums, the coefficients $\mathcal{C}_\pm$ are strictly
real quantities -- and so are $A, B, C$ and $D$. As the two point form factor
may depend on the energy range, we include here a global average over the full
length of the spectrum.
% {\bf THS I do not understand this}

Due to $D= A+B+C$ we have 
\begin{equation}
\sum_{\alpha\beta\gamma\delta} \tilde I_{\alpha\beta}\,
\tilde I_{\gamma\delta}\; X_{\alpha\beta\gamma\delta} =
\sum_{\alpha\beta} \big ( \, A\; X_{\alpha\alpha\beta\beta} + B\;
  X_{\alpha\beta\alpha\beta} + C\; X_{\alpha\beta\beta\alpha} \, \big ) \; .
\label{Q2PE:tenav}\end{equation}

Let us start with $\tilde I\, \varrho_0\, \tilde I$. With the help of
Eq.~(\ref{Q2PE:tenav}), we find
\begin{align}
\tilde I\, \varrho_0\, \tilde I &=
  \sum_{\alpha\beta\gamma\delta} |\alpha\ra\; \tilde I_{\alpha\beta}\,
  \varrho^0_{\beta\gamma}\, \tilde I_{\gamma\delta}\; \la\delta|
= \sum_{\alpha\beta} \big ( \, A\; |\alpha\ra\; \varrho^0_{\alpha\beta}\;
  \la\beta| + B\; |\alpha\ra\; \varrho^0_{\beta\alpha}\; \la\beta| +
  C\; \la\alpha|\; \varrho^0_{\beta\beta}\; \la\alpha| \, \big ) \notag\\
&= A\; \varrho_0 + B\; \varrho_0^T + C\; \one \; .
\end{align}
Next we consider
\begin{align}
&\tilde I\, \varrho_0 \otimes \tilde I\,\varrho_0 =
  \sum_{\alpha\beta\gamma\delta}\sum_{\eps\xi} |\alpha\ra\;
  \tilde I_{\alpha\beta}\, \varrho^0_{\beta\eps}\; \la\eps| \;\otimes\;
  |\gamma\ra\; \tilde I_{\gamma\delta}\, \varrho^0_{\delta\xi}\; \la\xi|
\notag\\
&\qquad= \sum_{\eps\xi}\sum_{\alpha\beta} \big ( \, A\; |\alpha\ra\;
  \varrho^0_{\alpha\eps}\; \la\eps| \;\otimes\; |\beta\ra\;
  \varrho^0_{\beta\xi}\; \la\xi| + B\; |\alpha\ra\; \varrho^0_{\beta\eps}\;
  \la\eps| \;\otimes\; |\alpha\ra\; \varrho^0_{\beta\xi}\; \la\xi| \notag\\
&\qquad\qquad + C\;
  |\alpha\ra\; \varrho^0_{\beta\eps}\; \la\eps| \;\otimes\; |\beta\ra\;
  \varrho^0_{\alpha\xi}\; \la\xi| \, \big ) \notag\\
&\qquad = A\; \varrho_0\otimes\varrho_0 + \sum_{\alpha\beta} \big ( \,
  B\; |\alpha\ra\, \la\beta|\; \varrho_0 \;\otimes\; |\alpha\ra\,
  \beta|\; \varrho_0 + C\; |\alpha\ra\, \la\beta|\; \varrho_0 \;\otimes\;
  |\beta\ra\, \la\alpha|\; \varrho_0 \, \big ) \; .
\end{align}
Finally we find 
\begin{align}
&\tilde I\, \varrho_0 \otimes \varrho_0\, \tilde I =
  \sum_{\alpha\beta\gamma\delta}\sum_{\eps\xi} |\alpha\ra\;
  \tilde I_{\alpha\beta}\; \varrho^0_{\beta\eps}\; \la\eps| \;\otimes\;
  |\xi\ra\; \varrho^0_{\xi\gamma}\, \tilde I_{\gamma\delta}\; \la\delta|
\notag\\
&\qquad = \sum_{\eps\xi}\sum_{\alpha\beta} \big ( \, A\; |\alpha\ra\;
  \varrho^0_{\alpha\eps}\; \la\eps| \;\otimes\; |\xi\ra\;
  \varrho^0_{\xi\beta}\; \la\beta| + B\; |\alpha\ra\; \varrho^0_{\beta\eps}\;
  \la\eps| \;\otimes\; |\xi\ra\; \varrho^0_{\xi\alpha}\; \la\beta|
\notag\\ 
&\qquad\qquad + C\;
  |\alpha\ra\; \varrho^0_{\beta\eps}\; \la\eps| \;\otimes\; |\xi\ra\;
  \varrho^0_{\xi\beta}\; \la\alpha| \, \big ) \notag\\
&\qquad = A\; \varrho_0\otimes\varrho_0 + \sum_{\alpha\beta} \big ( \, B\;
  |\alpha\ra\, \la\beta| \; \varrho_0 \;\otimes\; \varrho_0\; |\alpha\ra\,
  \la\beta| + C\; |\alpha\ra\, \la\beta|\; \varrho_0 \;\otimes\; \varrho_0\;
  |\beta\ra\, \la\alpha| \, \big ) \; .
\end{align}
Collecting all terms, we obtain
\begin{align}
A_I &= p[X]\notag\\
X &= A\; \varrho_0\otimes\varrho_0  + B\; \varrho_0^T\otimes\varrho_0 + C\;
  \one\otimes\varrho_0\notag\\
&\qquad - {\rm Re}\left[ |\alpha\ra\,\la\beta|\; \varrho_0 \otimes\;
  \big (\, B\; |\alpha\ra\,\la\beta|\; \varrho_0 + C\;
     |\beta\ra\, \la\alpha|\; \varrho_0 \, \big )\right] \notag\\
&\qquad
 + |\alpha\ra\,\la\beta|\; \varrho_0 \otimes\; \big (\, B\; \varrho_0\;
  |\alpha\ra\,\la\beta| + C\; \varrho_0\; |\beta\ra\, \la\alpha|\, \big ) \; ,
\end{align}
where summation over $\alpha$ and $\beta$ is assumed. There are seven different
tensors, appearing in the expression for $X$. We compute the purity functional
for each of them:
\begin{align}
p[\varrho_0\otimes\varrho_0] &= I(0) \\
p[\varrho_0^T\otimes\varrho_0] &= I'(0) \\
p[\one\otimes\varrho_0] &= n_{\rm e} = {\rm dim}\, \mathcal{H}_{\rm e} \\
p\big [\, \textstyle \sum_{\alpha\beta} |\alpha\ra\,\la\beta|\; \varrho_0
  \otimes |\alpha\ra\,\la\beta|\; \varrho_0\, \big ]
&= \sum_{\alpha\beta} (\varrho^0_{\alpha\beta})^2
 = {\rm tr}\big (\, \varrho_0^T\, \varrho_0\, \big )
\label{appEP:PFres:d}\\
p\big [\, \textstyle \sum_{\alpha\beta} |\alpha\ra\,\la\beta|\; \varrho_0
  \otimes |\beta\ra\,\la\alpha|\; \varrho_0\, \big ]
&= {\rm tr}_{\rm e}\big [\, ( {\rm tr}_{\rm c}\, \varrho_0)^2\, \big ]
 = p_{\rm dual}[\varrho_0\otimes\varrho_0] = I_{\rm dual}(0) \\
p\big [\, \textstyle \sum_{\alpha\beta} |\alpha\ra\,\la\beta|\; \varrho_0
  \otimes \varrho_0\; |\alpha\ra\,\la\beta|\, \big ]
&= \sum_{ij,kl}\; \la ij|\, \varrho_0\, |kl\ra\; \la kj|\, \varrho_0\, |il\ra
 = I_2(0) \label{appEP:PFres:f}\\
p\big [\, \textstyle \sum_{\alpha\beta} |\alpha\ra\,\la\beta|\; \varrho_0
  \otimes \varrho_0\; |\beta\ra\,\la\alpha|\, \big ]
&= n_{\rm c}\; {\rm tr}\, \varrho_0^2 \qquad m_{\rm c}
  = {\rm dim}\, \mathcal{H}_{\rm c} \; ,
\end{align}
where we have used nothing but the fact that $\varrho_0$ is a density
matrix, {\it i.e.} $\varrho_0 = \varrho_0^\dagger$ and
${\rm tr}\, \varrho_0 = 1$. Latin indices are used to indicate basis states of
the factor spaces $\mathcal{H}_{\rm c}$ and $\mathcal{H}_{\rm e}$. With the
help of the symmetry relations of the
purity functional, Eq.~(\ref{appEP:sym}), we can show that all quantities
except for~(\ref{appEP:PFres:d}) are real. We find that
\begin{equation}
A_I= A\; I(0) + B\; \big [\, I'(0) - {\rm Re}\, {\rm tr}(\varrho_0^T\,
  \varrho_0) + I_2(0)\, \big ] + C\; \big [\, n_{\rm c} + n_{\rm e}\;
  {\rm tr}\, \varrho_0^2 + I_{\rm dual}(0)\, \big ] \; .
\end{equation}
This expression must be invariant under arbitrary unitary transformations in
the factor spaces $\mathcal{H}_{\rm c}$ and $\mathcal{H}_{\rm e}$. Note that
$p[\varrho_0\otimes\varrho_0] = p_{\rm dual}[\varrho_0\otimes\varrho_0]$ only
if $\varrho_0$ is a pure state. Therefore $A_I$ needs not be invariant under
exchange of the factor spaces, as long as $\varrho_0$ is not a pure state.

If $\varrho_0$ is a pure state, {\it i.e.} if $\varrho_0= |\Psi\ra\, \la\Psi|$,
some of above quantities simplify, and we obtain
\begin{equation}
A_I = (A-C)\; I(0) + C\; (n_{\rm c} + n_{\rm e})
  + B\; \big [\, I'(0) + I_2(0) -
  {\rm Re}\; {\rm tr}(\, \varrho_0^T\, \varrho_0\, )\, \big ] \; .
\end{equation}
Further simplifications rely not so much on the separability of the initial
state, but rather on the possibility to choose a product basis, in which
the initial state $\varrho_0$ has real elements. If this is possible,
$I'(0)= I_2(0) = I(0)$ and ${\rm tr}(\, \varrho_0^T\, \varrho_0\, ) = 1$, such
that
\begin{equation}
A_I = (A+2B-C)\; I(0) + C\; (n_{\rm c} + n_{\rm e}) - B\; .
\end{equation}
If, finally, the initial state is also separable, we have $I(0)=1$. Then using
$A+B=D-C$, we find:
\begin{equation}
A_I= D + (n_{\rm c} + n_{\rm e}-2)\; C \; .
\end{equation}
In the limit of large $N$, $D$ and some parts of $C$ may be neglected. We then
find
\begin{equation}
A_I= \frac{n_{\rm c} + n_{\rm e}-2}{N+2}\, \big(\, 2t^2 + t - B_-(t)\, \big) \; ,
\qquad B_-(t)= 2\int_0^t\d\tau\int_0^\tau\d\tau'\; \la b_2(E,\, \tau)\ra \; .
\end{equation}
Due to the latter relation, we obtain for the purity echo decay:
\begin{equation}
F_P(t)= 1 - 4\lambda_0^2\, \left(1
 - \frac{n_{\rm c} + n_{\rm e}-2}{N+2}\right)\; \left(
  t^2 + t/2 - \int_0^t\d\tau\int_0^\tau\d\tau'\; \la b_2(E,\, \tau)\ra
  \right) \; .
\end{equation}
This decay is of the same form as the decay of the fourth power of the 
absolute value of the fidelity amplitude but faster by
a factor of $ [1- (n_{\rm c} + n_{\rm e}-2)/(N+2)]$.

\subsection{\label{appEPD} The decoupled case}

For that case, we work in the eigenbasis of $H_0$. It might be strictly
separable, but another interesting situation would be a weak coupling,
treatable by first order perturbation theory and thus breaking the separability
of the Hamiltonian without affecting its separable eigenbasis. For echo purity
in the linear response approximation, Eq.~(\ref{appEP:LRbas}) remains
valid, but when calculating $A_J$ and $A_I$ we have to replace $\tilde J$ and
$\tilde I$ with their bare counterparts $J$ and $I$. We find that
Eq.~(\ref{appEP:bas2}) remains valid (if we may perform an independent
spectral average), and that only $A_I$ needs to be modified. Due to the 
(almost) separable Hamiltonian $H_0$ it is not realistic to assume that
$H_0$ corresponds to chaotic dynamics. The best we can then do with RMT, is
to assume that $H_0$ has an uncorrelated Poisson spectrum. In our approach 
this amounts to setting the
spectral two-point form factor to zero. For $A_J$, this yields:
\begin{equation}
A_J= 2\; I(0)\; (t^2 + t/2) \; .
\end{equation}

In order to compute $A_I$ we use Eq.~(\ref{Q2PE:avI}), which gives
the average of the product of two matrix elements of $I$ and find
\begin{equation}
\sum_{\alpha\beta\gamma\delta} I_{\alpha\beta}\; I_{\gamma\delta}\;
  X_{\alpha\beta\gamma\delta} = \sum_{\alpha\beta} \big ( \,
  C_{\alpha\beta}^+\; X_{\alpha\beta\alpha\beta} + C_{\alpha\beta}^-\;
  X_{\alpha\beta\beta\alpha} \, \big ) \; .
\end{equation}
This gives
\begin{align}
I\, \varrho_0\, I &= \sum_{\alpha\beta} \big ( \, C_{\alpha\beta}^+\;
  |\alpha\ra\; \varrho^0_{\beta\alpha}\; \la\beta| + C_{\alpha\beta}^-\;
  |\alpha\ra\; \varrho^0_{\beta\beta}\; \la\alpha| \, \big ) \notag\\
I\, \varrho_0\; \otimes\; I\, \varrho_0 &= \sum_{\alpha\beta} \big ( \,
  C_{\alpha\beta}^+\; |\alpha\ra\, \la\beta|\; \varrho_0\; \otimes\;
  |\alpha\ra\, \la\beta|\; \varrho_0 + C_{\alpha\beta}^-\;
  |\alpha\ra\, \la\beta|\; \varrho_0\; \otimes\; |\beta\ra\, \la\alpha|\;
  \varrho_0 \, \big ) \notag\\
I\, \varrho_0\; \otimes\; \varrho_0\, I &= \sum_{\alpha\beta} \big ( \,
  C_{\alpha\beta}^+\; |\alpha\ra\, \la\beta|\; \varrho_0\; \otimes\;
  \varrho_0\; |\alpha\ra\, \la\beta| + C_{\alpha\beta}^-\;
  |\alpha\ra\, \la\beta|\; \varrho_0\; \otimes\; \varrho_0\;
  |\beta\ra\, \la\alpha| \, \big ) \; .
\end{align}
Neglecting level correlations in the spectrum of $H_0$, we may set
$C_{\alpha\beta}^+= \delta_{\alpha\beta}\, t^2$ and
$C_{\alpha\beta}^-= \delta_{\alpha\beta}\, t^2+(1-\delta_{\alpha\beta})\, t/N$. Note however, in the case of
separable $H_0$, the matrices $C_{\alpha\beta}^+$ and $C_{\alpha\beta}^-$ are
also separable. That might need special treatment. Disregarding this
possibility, we find
\begin{align}
I\, \varrho_0\, I &= \big (\, 2t^2- \frac{t}{N}\, \big )\; \sum_\alpha
  |\alpha\ra\, \la\alpha| \; \varrho_0\; |\alpha\ra\, \la\alpha|
 + \frac{t}{N}\; \one \notag\\
I\, \varrho_0\; \otimes\; I\, \varrho_0 &= \big (\, 2t^2-\frac{t}{N}\, \big )\;
  \sum_\alpha |\alpha\ra\, \la\alpha| \; \varrho_0 \otimes
  |\alpha\ra\, \la\alpha| \; \varrho_0 + \frac{t}{N}\; \sum_{\alpha\beta}
  |\alpha\ra\, \la\beta| \; \varrho_0 \otimes
  |\beta\ra\, \la\alpha| \; \varrho_0 \notag\\
I\, \varrho_0\; \otimes\; \varrho_0\, I &= \big (\, 2t^2-\frac{t}{N}\, \big )\;
  \sum_\alpha |\alpha\ra\, \la\alpha| \; \varrho_0 \otimes
  \varrho_0\; |\alpha\ra\, \la\alpha| + \frac{t}{N} \sum_{\alpha\beta}
  |\alpha\ra\, \la\beta| \; \varrho_0 \otimes \varrho_0\; |\beta\ra\,
  \la\alpha| \; .
\end{align}
Collecting all terms, and applying the purity functional, we get
\begin{equation}
A_I= \big (\, 2t^2-\frac{t}{N}\, \big ) \left[\textstyle \sum_i(\,
  {\rm tr}_{\rm e}\, \varrho_0\, )_{ii}^2
  - \sum_\alpha (\varrho_0)_{\alpha\alpha}^2 + \sum_{ijk}
  |\la ij|\varrho_0|kj\ra|^2 \right]
  + \frac{t}{N} \left[ n_{\rm e} - I_{\rm dual}(0) + n_{\rm c}\;
    {\rm tr}\,\varrho_0^2 \right] \; .
\end{equation}
If $\varrho_0$ is a pure initial state this simplifies to
\begin{equation}
A_I= \big (\, 2t^2-\frac{t}{N}\, \big ) \left[ {\rm Ipr}\big (\,
  {\rm tr}_{\rm e}\, \varrho_0\, \big ) - {\rm Ipr}\, \varrho_0 +
  {\rm Ipr}\big (\, {\rm tr}_{\rm c}\, \varrho_0\, \big )
  \right] + \frac{t}{N} \left[ n_{\rm c}+n_{\rm e} - I(0) \right] \; ,
\end{equation}
where ${\rm Ipr}\, \varrho$ is the sum of the diagonal elements squared of
$\varrho$. As such it is a certain generalization of the inverse
participation ratio from pure to mixed states.
For the purity, we obtain
\begin{align}
\la I(t)\ra &= 1 - 2\, (2\pi\eps)^2\; \big [\, A_J - A_I\, \big ] \notag\\
&= 1 - 4\, (2\pi\eps)^2\; \left\{ t^2\; \big [\, I(0) - {\rm Ipr}_2\, \varrho_0
  \, \big ] + \frac{t}{2}\;
  \big [\, 1 - \frac{n_{\rm c}+n_{\rm e} - I(0) - {\rm Ipr}_2\, \varrho_0}{N}
  \big ] \right\} + \mathcal{O}(\eps^4)\; ,
\end{align}
where ${\rm Ipr}_2\, \varrho = {\rm Ipr}(\, {\rm tr}_{\rm e}\, \varrho\, )
+ {\rm Ipr}(\, {\rm tr}_{\rm e}\, \varrho\, ) - {\rm Ipr}\, \varrho$.
If the initial state is a basis state of the product basis, this gives
\begin{equation}
I(t)= 1- 4\; (2\pi\eps)^2\; \frac{t}{2}\left(1
 - \frac{n_{\rm c}+n_{\rm e}-2}{N}\right) + \mathcal{O}(\eps^4)\; .
\end{equation}
If the initial state is a product state of two random states in
$\mathcal{H}_c$ and $\mathcal{H}_e$, we find:
\begin{equation}
I(t)= 1- 4\; (2\pi\eps)^2\left\{ t^2\; \big [\,
  1 - {\rm Ipr}_2\, \varrho_0\, \big ] + \frac{t}{2}\; \big [\,
  1 - \frac{n_{\rm c}+n_{\rm e}-1- {\rm Ipr}_2\,\varrho_0}{N}\, \big ]
  \right\} + \mathcal{O}(\eps^4)\; ,
\label{appEPD:LRres}\end{equation}
where ${\rm Ipr}_2\, \varrho_0 = 3/(n_{\rm c}+2) + 3/(n_{\rm e}+2)
- 9/((n_{\rm c}+2)(n_{\rm e}+2))$.

\section{Higher order terms in the Born series}
\label{app:HBS}

We will calculate averages of products of real symmetric matrices with zero diagonal and imaginary antisymmetric matrices with Gaussian independently distributed off-diagonal elements. Both
ensembles have been used in the RMT formulation of the fidelity freeze in
Section~\ref{Q2QF}.

Nonzero matrix elements of the perturbation $V$ are independent random numbers
with Gaussian distribution, zero mean, and second moments equal to
\begin{equation}
\langle V_{ij} V_{kl} \rangle=
\frac{1}{N}
\begin{cases}
\delta_{il}\delta_{jk}+\delta_{ik}\delta_{jl}-2\delta_{ij}\delta_{ik}\delta_{il} &; \text{symmetric,} \\
\delta_{il}\delta_{jk}-\delta_{ik}\delta_{jl} &; \text{antisymmetric,}
\end{cases}
\label{eq:2point}
\end{equation}
for the two cases considered. Equivalently, for $i \neq k$ the last equation
can be stated as
\begin{equation}
\langle V_{ij} V_{ji} \rangle = \frac{1}{N},\qquad
\langle V_{ij} V_{ij} \rangle =
\frac{1}{N}\begin{cases}
1 &; {\rm symmetric,} \\
-1 &; {\rm antisymmetric.}
\end{cases}
\label{eq:d2}
\end{equation}
The bracket $\langle \bullet \rangle$ denotes an average over a Gaussian distribution of matrix elements.

Using the Born expansion of the echo operator (\ref{eq:born}) the fidelity
amplitude is expressed as a sum of integrals of $m$-time correlation
functions, {\it i.e.}, terms of the form
\begin{equation}
\frac{1}{N} \tr{\tilde{V}(t_1)\cdots \tilde{V}(t_m)},
\label{eq:trcor}
\end{equation}
where we use a trace for the initial state average.
What we would like to do is to
get an estimate for the difference between both ensembles for the above
average (\ref{eq:trcor}). Using the eigenbasis of the unperturbed evolution $U_0(t)$
and averaging over the ensemble of $V$'s, the above correlation function is
\begin{equation}
\frac{1}{N} \langle V_{jk} V_{kl} \cdots V_{rj} \rangle \exp{\left\{ \ii
   (E_j-E_k)t_1/\hbar +\cdots \right\} },
\label{eq:trcor1}
\end{equation}
where a summation over indices $j,k,\ldots,r$ is implied and we did not write
in full the exponential factor involving eigenvalues of $U_0$, as it is unimportant in the present context.

As far as the second order correlation function is concerned, we have already seen that it is the
same for real symmetric and for complex antisymmetric ensembles. We shall
show that the same holds also for the 4th order term.
Forgetting about the exponential phase term,
we would like to calculate
\begin{equation}
\langle V_{ik} V_{kl} V_{lp} V_{pi} \rangle.
\label{eq:4th}
\end{equation}
Note that no summation is implied here. We use Wick contraction, giving three contribution,
\begin{equation}
\wick{11}{ <1V_{ik} >1V_{kl} <2V_{lp} >2V_{pi} },\qquad \wick{12}{ <1V_{ik} <2V_{kl} >1V_{lp} >2V_{pi} },\qquad \wick{12}{ <1V_{ik} <2V_{kl} >2V_{lp} >1V_{pi} },
\label{eq:4thwick}
\end{equation}
denoted terms (i), (ii) and (iii), respectively. By renaming indices as $i \to
k$, $k \to l$, $l \to p$ and $p \to i$, we see that terms (iii) and (i) are
equal. We are mainly interested whether there is a difference between
the symmetric and the antisymmetric case. From Eq.~(\ref{eq:d2}) we see, that the
difference could come only from the terms of form $\langle V_{ij} V_{ij}
\rangle$, differing in sign between the two ensembles. Let us look if there
are any such terms present in (i) or (ii) (\ref{eq:4thwick}). In term (i) the
presence of such contraction would immediately mean that we have also diagonal
element, e.g. $V_{ii}$, which though are zero by construction for our
perturbations. Similarly, in term (ii) we see that the nonzero term is
$\langle V_{ij} V_{ij}\rangle \langle V_{ji} V_{ji} \rangle$. For the
antisymmetric case the minus sign occurs twice. Thus the average is the same for
both ensembles. Altogether there are no terms involving one minus sign,
{\it i.e.} of the form $\langle V_{ij} V_{ij} \rangle \langle V_{kl} V_{lk}
\rangle$, and therefore the 4th order average is the same for both
ensembles. For the 4th order term (\ref{eq:4th}) one can actually relatively quickly explicitly calculate the average by using (\ref{eq:2point}). The result is
\begin{align}
(\delta_{il}-\delta_{il}\delta_{ip}-\delta_{il}\delta_{ik}+\delta_{ik}\delta_{il}\delta_{ip})/N^2 \qquad&; \text{term (i),} \nonumber \\
(\delta_{il}\delta_{kp}-\delta_{ik}\delta_{il}\delta_{ip})/N^2 \qquad&; \text{term (ii),}
\end{align}
regardless of the ensemble.

For the higher order terms we suspect, that the two ensembles
differ in terms of order $1/N$, because such is the case for
${\rm tr}\, V^6$.

\bibliographystyle{plain}
\bibliography{lorev}

\end{document}